\documentclass[twoside,12pt]{article}
\usepackage{epsfig}
\usepackage{latexsym}
\usepackage{slashed}
\def\Journal#1#2#3#4{{#1} {#2} (#4) #3 }
\def\NC{{\em Nuovo Cimento}}
\def\NCA{{\em Nuovo Cimento} A}

\def\PRO{{\em Prog. Theor. Phys.}}
\def\NPB{{\em Nucl. Phys.} B}

\def\PLB{{\em Phys. Lett.} B}

\def\PL{{\em Phys. Lett.}}
\def\PRL{\em Phys. Rev. Lett.}

\def\PREP{\em Phys. Rep.}
\def\PR{{\em Phys. Rev.}}

\def\PRD{{\em Phys. Rev.} D}

\def\PREP{\em Phys. Rep.}

\def\INTA{{\em Int. J. Mod. Phys.} A}
\def\FP{{\em Found. Phys.}}
\def\ZETF{{\em Zh. Eksp. Teor. Fiz.}}
\def\SPJETP{{\em Sov. Phys. JETP}}
\def\MPLA{{\em Mod. Phys. Lett. A}}
\def\PPNP{{\em Prog. Part. Nucl. Phys.}}
\def\NAT{{\em Nature}}
\def\MZ{{\em Math. Zeit.}}
\def\APJ{{\em Astrophys. J.}}
\def\GRG{{\em Gen. Rel. Gravit.}}
\def\MONRAS{{\em Mon. Not. R. Astron. Soc.}} 
\def\JPCS{{\em J. Phys. Conf. Ser.}}
\def\IJMPD{{\em Int. Jour. Mod. Phys. D}}
\def\ANP{{\em Annalen. Physik}}
\def\SPAW{\em Sitzsungber. Preuss. Akad. Wiss.}
\def\CMP{\em Comm. Math. Phys.}
\begin{document}
\title{ \vspace{1cm} Mass Generation, the Cosmological Constant Problem, Conformal Symmetry, and  the Higgs Boson}
\author{Philip D. Mannheim 
\\
Department of Physics, University of Connecticut, Storrs, CT 06269, USA\\
email: philip.mannheim@uconn.edu\\}
\date{January 29, 2017} 
\maketitle

\begin{abstract}
In 2013 the Nobel Prize in Physics was awarded to Francois Englert and Peter Higgs for their work in 1964 along with the late Robert Brout on the mass generation mechanism (the Higgs mechanism) in local gauge theories. This mechanism requires the existence of a massive scalar particle, the Higgs boson, and in 2012 the Higgs boson was finally discovered at the Large Hadron Collider after being sought for almost half a century. In this article  we review the work that led to the discovery of the Higgs boson and discuss its implications. We approach the topic from the perspective of a dynamically generated  Higgs boson that is a fermion-antifermion bound state rather than an  elementary field that appears in an input Lagrangian. In particular, we emphasize the connection with the  Barden-Cooper-Schrieffer theory of superconductivity. We identify the double-well Higgs potential not as a fundamental potential but as  a mean-field effective Lagrangian with a dynamical Higgs boson being generated through a residual interaction that accompanies the mean-field Lagrangian. We discuss what we believe to be the key challenge raised by the discovery of the Higgs boson, namely determining whether it is elementary or composite, and through study of a conformal invariant field theory model as realized with critical scaling and anomalous dimensions, suggest that the width of the Higgs boson might serve as a suitable diagnostic for discriminating between an elementary Higgs boson and a composite one. We discuss the implications of Higgs boson mass generation for the cosmological constant problem, as the cosmological constant receives contributions from the very mechanism that generates the Higgs boson mass in the first place.  We show that the contribution to the cosmological constant due to a composite Higgs boson is more tractable and under control than the contribution due to an elementary Higgs boson, and is potentially completely under control if there is an underlying conformal symmetry not just in a critical scaling matter sector (which there would have to be if all mass scales are to be dynamical), but equally in the gravity sector to which the matter sector couples.
\end{abstract}
 
\newpage
\tableofcontents

\newpage

\section{Introduction}
\subsection{Preamble}

The 2013 Nobel Prize in Physics was awarded to Francois Englert and Peter Higgs for their work in 1964 on the Higgs mechanism, work that led in 2012 to the discovery  at the CERN Large Hadron Collider (LHC) of the Higgs Boson after its being sought for almost 50 years. It is great tragedy that Robert Brout, the joint author with Francois Englert of one of the papers that led to the 2013 Nobel Prize, died in 2011, just one year before the discovery of the Higgs boson and two years before the awarding of the Nobel prize for it. (For me personally this is keenly felt since my first post-doc was with Robert and Francois in Brussels 1970 - 1972.) 

The paper coauthored by Englert and Brout appeared on August 31, 1964 in Physical Review Letters \cite{Englert1964} after being submitted on June 26, 1964 and was two and one half pages long. Higgs wrote two papers on the topic. His first paper appeared on September 15, 1964 in Physics Letters \cite{Higgs1964a}  after being submitted on July 27, 1964 and was one and one half pages long, and the second paper appeared on October 19, 1964 in Physical Review Letters \cite{Higgs1964b} after being submitted on August 31, 1964 and was also one and one half pages long. Thus a grand total of just five and one half pages.
 
The significance of the Higgs mechanism introduced in these papers and the Higgs Boson identified by Higgs is that they are tied in with the theory of the origin of mass, and of the way that mass can arise through collective effects (known as broken symmetry) that only occur in systems with a large number of degrees of freedom. Such collective effects are properties that a system of many objects collectively possess that each one individually does not -- the whole being greater than the sum of its parts. A typical example is temperature. A single molecule of $H_2O$ does not have a temperature, and one cannot tell if it was taken from ice, water or steam. These different phases are collective properties of large numbers of $H_2O$ molecules acting in unison. Moreover, as one changes the temperature all the $H_2O$ molecules can act collectively to change the phase (freezing water into ice for instance), with it being the existence of such phase changes that is central to broken symmetry.

I counted at least 21 times that Nobel Prizes in Physics have in one way or another been given for aspects of the problem: 
Dirac (1933); Anderson (1936); Lamb (1955); Landau (1962); Tomonaga, Schwinger, Feynman (1965); Gell-Mann (1969); Bardeen, Cooper, Schrieffer (1972); Richter, Ting (1976); Glashow, Salam, Weinberg (1979); Wilson (1982); Rubbia, van der Meer (1984); Friedman, Kendall, Taylor (1990); Lee, Osheroff, Richardson (1996); 't Hooft,Veltman (1999); Abrikosov, Ginzburg, Leggett  (2003); Gross, Politzer, Wilczek (2004); Mather, Smoot (2006); Nambu, Kobayashi, Maskawa (2008); Perlmutter, Schmidt, Reiss (2011); Englert, Higgs (2013); Kajita, McDonald (2015). And this leaves out Anderson who made major contributions to collective aspects of mass generation and Yang who (with Mills) developed non-Abelian Yang-Mills gauge theories but got Nobel prizes (Anderson 1977, Yang 1957) for something else. While the Nobel prizes to Mather and Smoot and to Perlmutter, Schmidt, and Reiss were for cosmological discoveries (cosmic anisotropy and cosmic acceleration), because they tie in with the cosmological constant problem, a problem which itself ties in with mass generation and the Higgs mechanism, I have included them in the list. Since in the absence of any mass scales one has an underlying conformal symmetry, the interplay of mass generation with the cosmological constant problem, with this underlying conformal symmetry, and with its local conformal gravity extension will play a central role in this article, as I review the work and ideas that led up to the discovery of the Higgs boson and then discuss its implications.

\subsection{Ideas About Mass and Implications for the Cosmological Constant Problem}

As introduced by Newton mass was mechanical. The first ideas on dynamical mass were due to Poincare (Poincare stresses needed to stabilize an electron all of whose mass  came from its own electromagnetic self-energy according to $mc^2=e^2/r$). However this was all classical.

With quantum field theory, the mass of a particle is able to change through self interactions (radiative corrections to the self-energy -- Lamb shift) to give $m=m_0+\delta m$, or through a change in vacuum (Bardeen, Cooper, Schrieffer -- BCS theory)  according to $E=p^2/2m-\Delta$ where $\Delta$ is the self-consistent gap parameter. 
Then through Nambu (1960) and Goldstone (1961)  the possibility arose that not just some but in fact all of the mass could come from self interaction, and especially so for gauge bosons, viz. Anderson (1958, 1963), Englert and Brout  (1964), Higgs (1964), Guralnik, Hagen, and Kibble (1964). This culminated in the Weinberg (1967), Salam (1968), and Glashow (1961, 1970)  renormalizable $SU(2)\times U(1)$ local gauge theory of electroweak interactions, and the confirming discoveries first of weak neutral currents (1973), then charmed particles (1974), then the intermediate vector bosons of the weak interactions (1983), and finally the Higgs boson (2012). All of this is possible because of Dirac's Hilbert space formulation of quantum mechanics in which one sets $\psi(x)=\langle x|\psi\rangle$, with the physics being in the properties of the states $|\psi\rangle$. We thus live in Hilbert space and not in coordinate space, and not only that, there is altogether more in Hilbert space than one could imagine, such as half-integer spin and collective macroscopic quantum systems such as superconductors and superfluids. In this Hilbert space we find an infinite Dirac sea of negative energy particles. Because of their large number these degrees of freedom can collectively act to provide the dynamics needed to produce mass generation and the Higgs boson. Since the dynamical generation of mass leads to contributions to the cosmological constant (i.e. to the energy density of the vacuum), mass generation and the cosmological constant problem are intimately connected. Moreover, since gravity couples to energy density and not just to energy density difference, gravity knows where the zero of energy is, to thus be sensitive to the mass generation mechanism. Gravity and the mass generation mechanism are thus intimately connected to each other.

\section{The Higgs Boson Discovery}

The discovery of the Higgs boson was announced by CERN on July 4, 2012, accompanied by simultaneous announcements by the experimental groups ATLAS and CMS at the Large Hadron Collider at CERN, as then followed by parallel publications that were simultaneously submitted to Physics Letters B on July 31, 2012 and published on September 17, 2012, one publication by the ATLAS Collaboration \cite{ATLAS2012}, and the other by the CMS Collaboration \cite{CMS2012}. The ATLAS paper had 2924 authors, and the CMS paper had 2883. The Higgs boson signature used by both collaborations was to look for the lepton pairs that would be found in the decay products of any Higgs boson that might be produced in high energy  proton proton collisions at the Large Hadron Collider.

From amongst a set of $10^{15}$ proton-proton collisions produced at the Large Hadron Collider, of the order of 240,000 collisions produce a Higgs boson. Of them just 350 decay into pairs of gamma rays, and of those gamma rays just 8 decay into a pair of leptons. 
The search for the Higgs boson is thus a search for some very rare events. Thus to see Higgs bosons one needs an energy high enough to produce them and the sensitivity to see such rare decays when they are produced. In searches over the years it was not known in what energy regime to look for Higgs particles, with the Large Hadron Collider proving to be the collider whose energy was high enough that one could finally explore in detail the 125 GeV energy domain where the Higgs boson was ultimately found to exist.

\section{Background Leading to the Higgs Mechanism and Higgs Boson Papers in 1964}

In order to characterize macroscopic ordered phases in a general way Landau introduced the concept of a macroscopic order parameter $\phi$. For a ferromagnet for instance $\phi$ would represent the spontaneous magnetization $M$ and would be a matrix element of a field operator $\hat{\phi}$ in an ordered quantum state that described the ordered magnetic phase. Building on this approach Ginzburg and Landau \cite{Ginzburg1950} wrote down a Lagrangian for such a $\phi$ for a superconductor, with kinetic energy $\vec{\nabla}\phi\cdot \vec{\nabla}\phi/2$ and potential $V(\phi)=\lambda\phi^4/4+m^2(1-T_C/T) \phi^2/2$, where $T_C$ is the critical temperature, and $m$ is a real constant.  For temperatures above the critical temperature the potential would have the shape of a single well, viz. like the letter U, with the coefficient of the $\phi^2$ term being positive, and with the potential minimum being at $\phi=0$. For temperatures below the critical temperature  the potential would have the shape of a double well, viz. like the letter W, with the coefficient of the $\phi^2$ term being negative, and with the potential minimum being at $\phi=m(T_C/T-1)^{1/2}/\lambda^{1/2}$. Above  the critical temperature the order parameter would be zero at the minimum  of the potential  (normal phase with state vector $|N\rangle$ in which $\langle N|\hat{\phi}|N\rangle=0$). Below the critical temperature the order parameter would be nonzero (superconducting state $|S\rangle$ in which $\phi=\langle S|\hat{\phi}|S\rangle=m(T_C/T-1)^{1/2}/\lambda^{1/2}$ is nonzero).

In 1957 Bardeen, Cooper, and Schrieffer  \cite{Bardeen1957} developed a microscopic theory of superconductivity (BCS)  based on Cooper pairing of electrons  in the presence of a filled Fermi sea of electrons, and explicitly constructed the state $|S\rangle$. In this state the matrix element $\langle S|\psi(x)\psi(x)|S\rangle$ was equal to a spacetime-independent  function $\Delta$, the gap parameter, which led to a mass shift to electrons propagating in a superconductor of the form $E=p^2/2m-\Delta$. The gap parameter $\Delta$ would be temperature dependent ($\sim (T_C-T)^{1/2}$), and would only be nonzero below the critical temperature. In 1959 Gorkov \cite{Gorkov1959} was able to derive the Ginzburg-Landau Lagrangian starting from the BCS theory  and identify the order parameter as the spacetime-dependent $\phi(x)=\langle C|\psi(x)\psi(x)|C\rangle$ where $|C\rangle$ is a coherent state in the Hilbert space based on $|S\rangle$. In the superconducting case then $\phi$ is not itself a quantum-field-theoretic operator (viz. a q-number  operator that would have a canonical conjugate with which it would not commute) but is instead a c-number matrix element  of a q-number field operator $\psi\psi$ in a macroscopic coherent quantum state.

In 1958 Anderson \cite{Anderson1958} used the BCS theory to explain the Meissner effect, an effect in which electromagnetism becomes short range inside a superconductor, with photons propagating in it becoming massive. The effect was one of spontaneous breakdown of local gauge invariance, and was explored in detail by Anderson \cite{Anderson1958} and Nambu \cite{Nambu1960a}.

In parallel with these studies Nambu \cite{Nambu1960b},  Goldstone \cite{Goldstone1961}, and Nambu and Jona-Lasinio \cite{Nambu1961} explored the spontaneous  breakdown of some continuous global symmetries and showed that collective massless excitations (Goldstone bosons) were generated, and that the analog gap parameter would provide for dynamically induced fermion masses. In 1962  Goldstone, Salam, and Weinberg \cite{Goldstone1962} showed that there would always be massless Goldstone bosons in any Lorentz invariant theory in which a continuous global symmetry was spontaneously broken. While one could avoid this outcome if the symmetry was also broken in the Lagrangian, as was, through the weak interaction,  thought to be the case for the pion, a non-massless but near Goldstone particle (i.e. one with broken symmetry suppressed couplings to matter at low energies), in general the possible presence of massless Goldstone bosons was a quite problematic outcome because it would imply the existence of non-observed long range forces.  

In 1962 Schwinger \cite{Schwinger1962a,Schwinger1962b} raised the question of whether gauge invariance actually required  that photons be massless, and noted for the  photon propagator $D(q^2)=1/[q^2-q^2\Pi(q^2)]$  that if the vacuum polarization  $\Pi(q^2)$ had a massless pole of the form $\Pi(q^2)=m^2/q^2$, $D(q^2)$  would behave as the massive particle $D(q^2)=1/[q^2-m^2]$. A massless Goldstone boson could thus produce a massive vector boson. 

With Anderson having shown that a photon would become massive in a superconductor, there was a spirited discussion in the literature (Anderson \cite{Anderson1963}, Klein and Lee \cite{Klein1964}, Gilbert \cite{Gilbert1964}) as to whether an effect such as this might hold in a relativistic  theory as well or whether it might just have been an artifact of the fact that the BCS theory was non-relativistic. With the work of Englert and Brout \cite{Englert1964} and Higgs \cite{Higgs1964a,Higgs1964b}, and then Guralnik, Hagen, and Kibble \cite{Guralnik1964}, the issue  was finally resolved, with it being established that in the relativistic case the Goldstone theorem did not in fact hold if there was a spontaneous breakdown of a continuous local theory, with the would-be Goldstone boson no longer being an observable massless particle but instead combining with the initially massless vector boson to produce a massive vector boson. Technically, this mechanism should be known as the Anderson, Englert, Brout, Higgs, Guralnik, Hagen, Kibble mechanism, and while it has undergone many name variations over time, it is now commonly called the Higgs mechanism. What set Higgs' work apart from the others was that in his 1964 Physical Review Letters paper Higgs noted that as well as a massive gauge boson there should also be an observable massive scalar boson, this being the Higgs boson.

At the time of its development  in 1964 there was not much interest in the Higgs mechanism, with all of the Englert and Brout, Higgs, and Guralnik, Hagen and Kibble papers getting hardly any citations during the 1960s at all. The primary reason for this was that at the time there was little interest in non-Abelian Yang-Mills gauge theories in general, broken or unbroken, and not only was there no experimental indication at all that one should consider them, it was not clear if Yang-Mills theories were even quantum-mechanically viable. All this changed in the early 1970s when 't Hooft and Veltman showed that these theories were renormalizable,  and large amounts of data started to point in the direction of  the relevance of non-Abelian gauge theories to physics, leading to the $SU(3)\times SU(2)\times U(1)$ picture of strong, electromagnetic and weak interactions, which culminated in the discoveries of the $W^{+}$, $W^{-}$ and $Z_0$ intermediate vector bosons (1983) with masses that were generated by the Higgs mechanism, and then finally the Higgs boson itself (2012). What gave the Higgs boson the prominence that it ultimately came to have was the realization that in the electroweak $SU(2)\times U(1)$  theory the Higgs boson not only gives masses to the gauge bosons while maintaining renormalizability, but through its Yukawa couplings to the quarks and leptons of the theory it gives masses to the fermions as well. The Higgs boson is thus  responsible not just for the masses of the gauge bosons then but for the masses of all the other fundamental particles as well, causing it to be dubbed the  ``god particle".\footnote{For the reader who wishes to delve further into the subject, a nice discussion of the  history and development of the Higgs mechanism may be found in \cite{Close2013}.}

\section{Broken Symmetry}
\label{S0}
\subsection{Global Discrete Symmetry -- Real Scalar Field -- Goldstone (1961)}

\begin{figure}[htpb]
\begin{center}
\begin{minipage}[t]{8 cm}
\epsfig{file=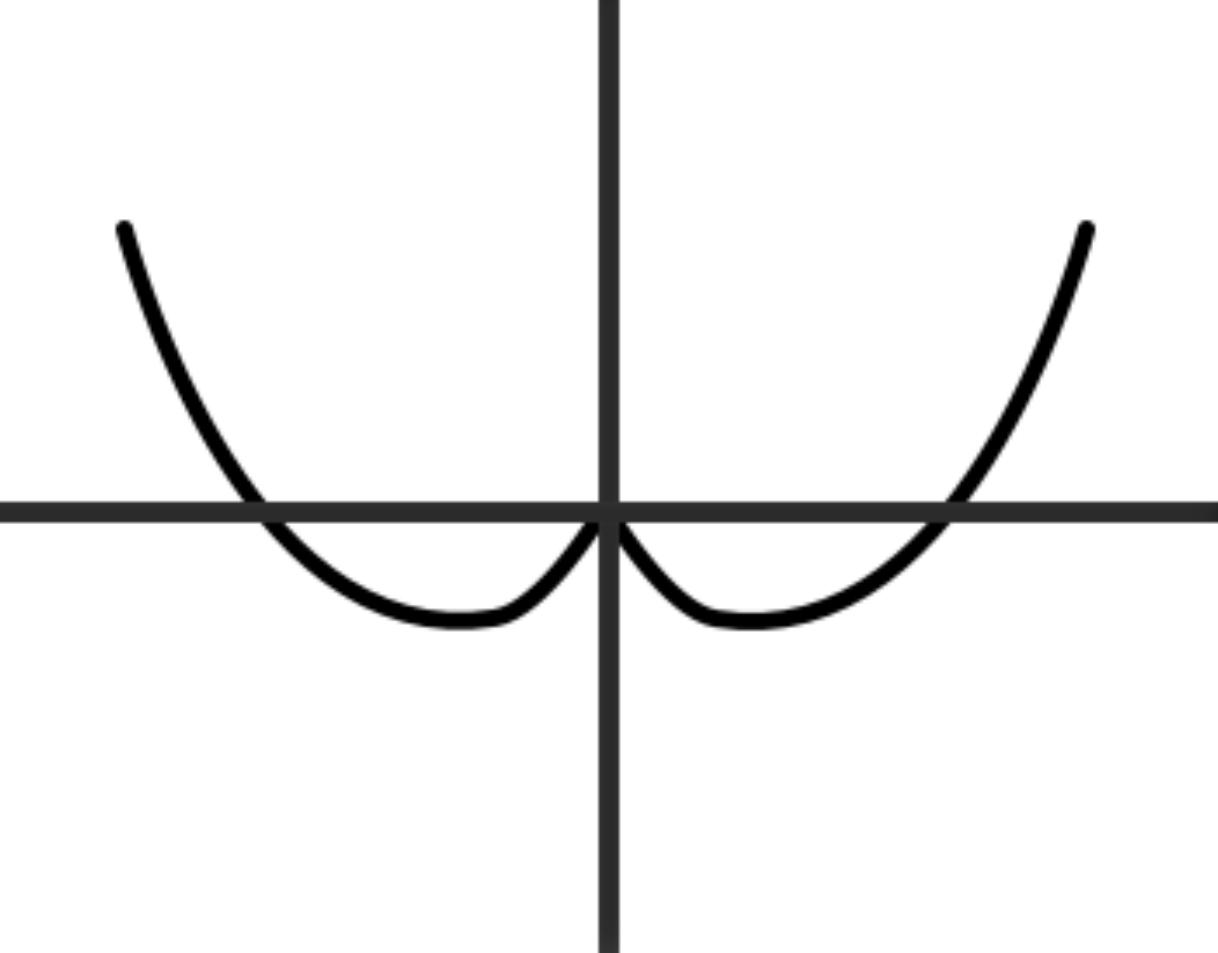,scale=0.5}
\end{minipage}
\begin{minipage}[t]{16.5 cm}
\caption{Discrete Double Well Potential  -- $V(\phi)$ plotted as a function of  $\phi$.}
\label{mhd}
\end{minipage}
\end{center}
\end{figure}

Even though broken symmetry is an intrinsically quantum-mechanical phenomenon, one can illustrate many of  the aspects needed to understand the Higgs boson by considering properties not of a q-number quantum field such as a  quantum scalar field $\hat{\phi}(x)$ but of its c-number vacuum matrix element $\phi(x)=\langle \Omega|\hat{\phi}(x)|\Omega\rangle$. Below we shall explore the quantum mechanics  of broken symmetry, but for pedagogical purposes it is convenient to first proceed as though one could treat the problem classically by studying properties of the c-number $\phi(x)$ without needing to delve into its origin or significance.

Thus consider a real classical scalar field (just one degree of freedom) with a potential energy in the form of the  double-well potential shown in Fig. (\ref{mhd}), viz. a potential shaped like a letter W with two wells:
\begin{eqnarray}
V(\phi)=\frac{1}{4}\lambda^2\phi^4-\frac{1}{2}\mu^2\phi^2.
\label{M1}
\end{eqnarray}
This potential has a discrete symmetry under $\phi \rightarrow -\phi$, with its first two derivatives being given by
\begin{eqnarray}
\frac{dV(\phi)}{d\phi}=\lambda^2\phi^3-\mu^2\phi,~~~\frac{d^2V(\phi)}{d\phi^2}=3\lambda^2\phi^2-\mu^2.
\label{M2}
\end{eqnarray}
The potential has a local maximum at $\phi=0$ where $V(\phi=0)$ is zero and $d^2V(\phi=0)/d\phi^2=-\mu^2$ is negative, and two-fold degenerate global minima at $\phi=+\mu/\lambda$  and $\phi=-\mu/\lambda$ where $V(\phi=\pm \mu/\lambda)$ is equal to $-\mu^4/4\lambda^2$ and $d^2V(\phi=\pm \mu/\lambda)/d\phi^2=2\mu^2$ is positive. Since $\phi=0$ is a local maximum, if we consider small oscillations  around $\phi=0$ of the form $\phi =0+\chi$ we generate a negative quadratic term $-(1/2)\mu^2\chi^2$ and a thus negative squared mass, viz. $m^2=-\mu^2$, a so-called tachyon. The tachyon signals an instability of the configuration with $\phi=0$ (i.e. we roll away from the top of the hill).

However if we fluctuate around either global minimum (i.e. we oscillate in the vertical around either of the two valleys) by setting $\phi=\pm \mu/\lambda +\chi$ we get 
\begin{eqnarray}
V(\phi)=-\frac{\mu^4}{4\lambda^2}+\mu^2\chi^2\pm \mu\lambda\chi^3+\frac{1}{4}\lambda^2\chi^4.
\label{M3}
\end{eqnarray}
The field $\chi$ now has the positive squared mass $m^2=d^2V(\phi=\pm \mu/\lambda)/d\phi^2=2\mu^2$ ($=-2\times  m^2({\rm tachyon})$). Thus (Goldstone \cite{Goldstone1961}) the would-be tachyonic particle becomes  a massive  particle, and with one scalar field we obtain one particle. This particle is the Higgs boson in embryo.

The $-\mu^4/4\lambda^2$ potential term contributes to the cosmological constant, and would be unacceptably large ($10^{60}$ times too large) if the boson has the 125 GeV mass that the Higgs boson has now been found to have.

All this arises because the minimum is two-fold degenerate, and picking either one breaks the symmetry spontaneously, since while one could just as equally be in either one minimum or the other, one could not be in both. This is just like people at a dinner. Each one can take the cup to their left or their right, but once one person has done so, the rest have no choice. However, a person at the opposite end of the table may not know what choice was made at the other end of the table and may make the opposite choice of cup, and thus persons in the middle could finish up with no cup. To ensure that this does not happen we need long range correlations -- hence massless Goldstone bosons.

\subsection{Global Continuous Symmetry -- Complex Scalar Field -- Goldstone (1961)}

\begin{figure}[htpb]
\vskip-0,5in
\begin{center}
\begin{minipage}[t]{15 cm}
\epsfig{file=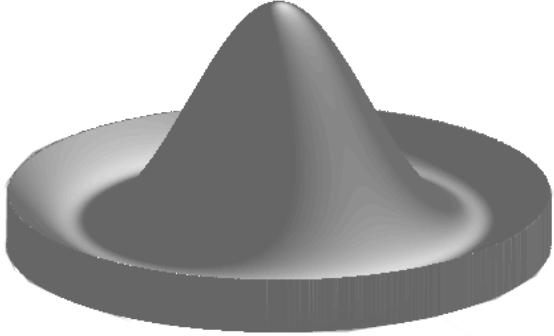,scale=0.5}
\end{minipage}
\vskip-0.6in
\begin{minipage}[t]{16.5 cm}
\caption{Continuous Double Well Mexican Hat Potential  -- $V(\phi)$ plotted as a function of  $\phi$.} 
\label{mhc}
\end{minipage}
\end{center}
\end{figure}

Consider a complex scalar field (two degrees of freedom) $\phi=\phi_1+i\phi_2=re^{i\theta}$, $\phi^*\phi=\phi_1^2+\phi_2^2=r^2$, with a potential energy in the shape of a rotated letter W or a broad-brimmed, high-crowned Mexican Hat as exhibited in Fig. (\ref{mhc}), viz.   

\begin{eqnarray}
V(\phi)=\frac{1}{4}\lambda^2(\phi^*\phi)^2-\frac{1}{2}\mu^2\phi^*\phi
=\frac{1}{4}\lambda^2(\phi_1^2+\phi_2^2)^2-\frac{1}{2}\mu^2(\phi_1^2+\phi_2^2).
\label{M4}
\end{eqnarray}
The potential has a continuous global symmetry of the form  $\phi \rightarrow e^{i\alpha} \phi$ with constant $\alpha$, with derivatives
\begin{eqnarray}
\frac{dV(\phi)}{d\phi_1}=\lambda^2\phi_1^3+\lambda^2\phi_1\phi_2^2 -\mu^2\phi_1,\qquad
\frac{dV(\phi)}{d\phi_2}=\lambda^2\phi_2^3+\lambda^2\phi_1^2\phi_2 -\mu^2\phi_2.
\label{M5}
\end{eqnarray}
This potential has a local maximum at $\phi_1=0,\phi_2=0$ where $V(\phi=0)$ is zero, and infinitely degenerate global minima at $\phi_1^2+\phi_2^2=\mu^2/\lambda^2$ (the entire 360 degree valley or trough between the brim and the crown of the Mexican hat). Again we would have  a tachyon if we expand around the local maximum, only this time we would get two. However, if we fluctuate around any one of the global minima by setting $\phi_1=\mu/\lambda +\chi_1$, $\phi_2=\chi_2$, we get 
\begin{eqnarray}
V(\phi)=-\frac{\mu^4}{4\lambda^2}+\mu^2\chi_1^2+\mu\lambda\chi_1^3+\frac{1}{4}\lambda^2\chi_1^4+\mu\lambda\chi_1\chi_2^2+\frac{\lambda^2}{2}\chi_1^2\chi_2^2.
\label{M6}
\end{eqnarray}
The (embryonic) Higgs boson field is now $\chi_1$ with $m^2=+2\mu^2$. However, the field $\chi_2$ no has no mass at all \cite{Goldstone1961} (it corresponds to horizontal oscillations along the valley floor), and is called a Goldstone boson or, because of Nambu's related work, a Nambu-Goldstone boson. Thus from a complex scalar field we obtain two particles. Since the Goldstone boson is massless, it travels at the speed of light. It is thus intrinsically relativistic, and being massless can provide for long range correlations. Moreover, if such particles exist then they could generate fermion masses  entirely dynamically (Nambu \cite{Nambu1960b}, and  Nambu and Jona-Lasinio \cite{Nambu1961}), with the pion actually serving this purpose.

The $-\mu^4/4\lambda^2$ potential term remains and the cosmological constant problem is just as severe as before.

Now if we have massless particles, we would get long range forces (just like photons), and yet nuclear and weak forces are short range. So what can we do about such Goldstone bosons. Two possibilities - they could get some mass because the symmetry is not exact (pion mass), or we could get rid of them altogether (the Higgs mechanism).

\subsection{Local Continuous Symmetry -- Complex Scalar Field and Gauge Field -- Higgs (1964)}

Consider the model studied by Higgs \cite{Higgs1964b} consisting of a complex scalar field (two degrees of freedom) coupled to a massless vector gauge boson (another two degrees of freedom) for a total of four degrees of freedom in all, viz. $\phi=re^{i\theta}$, $\phi^*\phi=r^2$, $F_{\mu\nu}=\partial_{\mu}A_{\nu}-\partial_{\nu}A_{\mu}$. The system has  kinetic energy $K$ and potential energy $V(\phi)$:
\begin{eqnarray}
K&=&\frac{1}{2}(-i\partial_{\mu}+eA_{\mu})(re^{-i\theta})(i\partial^{\mu}+eA^{\mu})(re^{i\theta})-\frac{1}{4}F_{\mu\nu}F^{\mu\nu}
\nonumber\\
&=&\frac{1}{2}\partial_{\mu}r\partial^{\mu}r+\frac{1}{2}r^2(eA_{\mu}-\partial_{\mu}\theta)(eA^{\mu}-\partial^{\mu}\theta)
-\frac{1}{4}F_{\mu\nu}F^{\mu\nu},
\nonumber \\
V(\phi)&=&\frac{1}{4}\lambda^2(\phi^*\phi)^2-\frac{1}{2}\mu^2\phi^*\phi
=\frac{1}{4}\lambda^2r^4-\frac{1}{2}\mu^2r^2,
\label{M7}
\end{eqnarray}
and because of the gauge boson the system is now invariant under continuous local gauge transformations of the form $\phi \rightarrow e^{i\alpha(x)} \phi$, $eA_{\mu}\rightarrow eA_{\mu}+\partial_{\mu}\alpha(x)$ with spacetime dependent  $\alpha(x)$. With derivatives
\begin{eqnarray}
\frac{dV(\phi)}{dr}=\lambda^2r^3-\mu^2r,\qquad \frac{d^2V(\phi)}{dr^2}=3\lambda^2r^2-\mu^2,
\label{M8}
\end{eqnarray}
the potential has a local maximum at $r=0$ where $V(r=0)$ is zero and degenerate global minima at $r=\mu/\lambda$ (infinitely degenerate since independent of $\theta$). Again we would have two tachyons if we expand around the local maximum. So fluctuate around the global minimum by setting $r=\mu/\lambda +\chi_1$, $\theta_2=\chi_2$. On defining $B_{\mu}=A_{\mu}-(1/e)\partial_{\mu}\chi_2$ we obtain
\begin{eqnarray}
K&=&\frac{1}{2}\partial_{\mu}\chi_1\partial^{\mu}\chi_1+\frac{e^2\mu^2}{2\lambda^2}B_{\mu}B^{\mu}
-\frac{1}{4}(\partial_{\mu}B_{\nu}-\partial_{\nu}B_{\mu})(\partial^{\mu}B^{\nu}-\partial^{\nu}B^{\mu})
+\frac{e^2}{2}\left(\frac{2\mu}{\lambda}\chi_1+\chi_1^2\right)B_{\mu}B^{\mu},
\nonumber\\
V(\phi)&=&-\frac{\mu^4}{4\lambda^2}+\mu^2\chi_1^2+\mu\lambda\chi_1^3+\frac{1}{4}\lambda^2\chi_1^4.
\label{M9}
\end{eqnarray}
There is again a Higgs boson field $\chi_1$ with $m^2=+2\mu^2$. However, the field $\chi_2$ has disappeared completely. Instead the vector boson now has a nonzero mass given by $m=e\mu/\lambda$. Since a massive gauge boson has three degrees of freedom (two transverse and one longitudinal) while a massless gauge boson such as the photon only has two transverse degrees of freedom, the would-be massless Goldstone boson is absorbed into the now massive gauge boson to provide its needed longitudinal degree of freedom. Hence a massless Goldstone boson and a massless gauge boson are replaced by one massive gauge boson, with two long-range interactions being replaced by one short range interaction. This is known as the Higgs mechanism though it was initially found by Anderson \cite{Anderson1958} in his study of the Meissner effect in superconductivity. The remaining fourth of the original four degrees of freedom becomes the massive Higgs boson, and its presence is an indicator that the Higgs mechanism has taken place. However, with the presence of $-\mu^4/4\lambda^2$ term in $V(\phi)$, the cosmological constant problem remains as severe as before.

\section{The Physics Behind Broken Symmetry}
\subsection{The Effective Action}

To discuss broken symmetry in quantum field theory it is convenient to introduce local sources. In the Gell-Mann-Low adiabatic switching procedure one introduces a quantum-mechanical Lagrangian density  $\hat{{{\cal L}}}_0$ of interest, switches on a real local c-number source $J(x)$ for some Hermitian quantum field $\hat{\phi}(x)$ at time $t=-\infty$, and switches $J(x)$  off at $t=+\infty$. While the source is active the Lagrangian density of the theory is given by $\hat{{{\cal L}}}_J=\hat{{{\cal L}}}_0+J(x)\hat{\phi}(x)$. Before the source is switched on the system is in the ground state $|\Omega_0^-\rangle$ of the Hamiltonian $\hat{H}_0$ associated with $\hat{{{\cal L}}}_0$, and after the source is switched off the system is in the state $|\Omega_0^+\rangle$. While $|\Omega_0^-\rangle$ and $|\Omega_0^+\rangle$ are both eigenstates of  $\hat{H}_0$, they differ by a phase, a phase that is fixed by $J(x)$ according to 
\begin{eqnarray}
\langle \Omega_0^+|\Omega_0^-\rangle |_J=\langle \Omega_J|T\exp\left[i\int d^4x(\hat{{{\cal L}}}_0+J(x)\hat{\phi}(x))\right]|\Omega_J\rangle=e^{iW(J)},
\label{M10}
\end{eqnarray}
with this expression serving to define the functional $W(J)$. As introduced, $W(J)$ serves as the generator of the connected $J=0$ theory Green's functions $G^{n}_0(x_1,...,x_n)=\langle\Omega_0|T[\hat{\phi}(x_1)...\hat{\phi}(x_n)]|\Omega_0\rangle$ according to
\begin{eqnarray}
W(J)=\sum_n\frac{1}{n!}\int d^4x_1...d^4x_nG^{n}_0(x_1,...,x_n)J(x_1)...J(x_n).
\label{M11}
\end{eqnarray}
On Fourier transforming the Green's functions, we can expand $W(J)$  about the point where all momenta vanish, to obtain
\begin{eqnarray}
W(J)&=&\sum_n\frac{1}{n!}\int d^4x_1...d^4x_n \int d^4p_1...d^4p_nJ(x_1)...J(x_n)e^{ip_1\cdot x_1}...e^{ip_n\cdot x_n} (2\pi)^4\delta^4(\sum p_i)
\nonumber\\
&\times&\bigg{[}G^{n}_0(p_i=0)+\sum p_ip_j\frac{\partial}{\partial p_i}\frac{\partial}{\partial p_j}G^{n}_0(p_k)|_{p_k=0}+...\bigg{]}
\nonumber\\
&=&\int d^4x\left[-\epsilon(J)+\frac{1}{2}Z(J)\partial_{\mu}J\partial^{\mu}J+....\right],
\label{M12}
\end{eqnarray}
with the first two terms in the last expression for $W(J)$ being in the standard $-V+K$ form required for actions. With $\hat{H}_J$ being the Hamiltonian associated with $\hat{{{\cal L}}}_J$, the physical significance of $\epsilon(J)$ is that when $J$  is spacetime independent, $\epsilon(J)$ is the energy-density difference
\begin{eqnarray}
\epsilon(J)=\frac{1}{V}\left(\langle \Omega_J|\hat{H}_J|\Omega_J\rangle-\langle \Omega_0|\hat{H}_0|\Omega_0\rangle\right)
\label{M13}
\end{eqnarray}
in a volume $V$.\footnote{$\epsilon(J)$ would have to be an energy density difference rather than an absolute energy density since it is not sensitive to the $J$-independent energy density of $|\Omega_0\rangle$, this being an absolute energy density that, as we explore below, only gravity is sensitive to.} On taking both $\hat{H}_0$ and $J(x)\hat{\phi}(x)$ to be Hermitian,  with constant $J$ the energy density difference $\epsilon(J)$ would be real, something that will prove to be of significance below when we study models of symmetry breaking. Given $W(J)$, via functional variation we can construct the so-called classical (c-number) field $\phi_C(x)$ 
\begin{eqnarray}
\phi_C(x)=\frac{\delta W}{\delta J(x)}=\frac{\langle \Omega^+|\hat{\phi}(x)|\Omega^-\rangle}{\langle \Omega^+|\Omega^-\rangle}\bigg{|}_J
\label{M14}
\end{eqnarray}
and the effective action functional
\begin{eqnarray}
\Gamma(\phi_C)=W(J)-\int d^4x J(x)\phi_C(x)=\sum_n\frac{1}{n!}\int d^4x_1...d^4x_n\Gamma^{n}_0(x_1,...,x_n)\phi_C(x_1)...\phi_C(x_n),
\label{M15}
\end{eqnarray}
with the $\Gamma^{n}_0(x_1,...,x_n)$ being the one-particle-irreducible, $\phi_C=0$, Green's functions of $\hat{\phi}(x)$.  Functional variation of $\Gamma(\phi_C)$ then yields
\begin{eqnarray}
\frac{\delta \Gamma(\phi_C)}{\delta \phi_C}=\frac{\delta W}{\delta J }\frac{\delta J}{\delta \phi_C}-J-\frac{\delta J}{\delta \phi_C} \phi_C=-J,
\label{M16}
\end{eqnarray}
to relate $\delta \Gamma(\phi_C)/\delta \phi_C$ back to the source $J$.

On expanding in momentum space around the point where all external momenta vanish, we can write $\Gamma(\phi_C)$ as
\begin{eqnarray}
\Gamma(\phi_C)=\int d^4x\left[-V(\phi_C)+\frac{1}{2}Z(\phi_C)\partial_{\mu}\phi_C\partial^{\mu}\phi_C+....\right].
\label{M17}
\end{eqnarray}
The quantity 
\begin{eqnarray}
V(\phi_C)=\sum_n\frac{1}{n!}\Gamma^{n}_0(q_i=0)\phi_C^n
\label{M18}
\end{eqnarray}
is known as the effective potential as introduced in \cite{Goldstone1962,Jona-Lasinio1964} (a potential that is spacetime independent if $\phi_C$ is), while the $Z(\phi_C)$ term  serves as the  kinetic energy of $\phi_C$.\footnote{In going from (\ref{M15}) to (\ref{M17}) to (\ref{M18}) a relative minus sign is engendered by the Jacobian involved in changing to the center of mass coordinates, as needed to implement the total momentum conservation delta function. (For two coordinates and constant $\phi_{C}$ for instance, on setting $X=(x_1+x_2)/\surd{2}$, $x=(x_1-x_2)/\surd{2}$, $P=(p_1+p_2)/\surd{2}$, $p=(p_1-p_2)/\surd{2}$, the Jacobian is equal to minus one, and we obtain $\int dx_1dx_2\exp(ip_1x_1+ip_2x_2)\Gamma^2_0(x_1-x_2)\phi_C^2=-\int dX dx\exp(iPX+ipx)\Gamma^2_0(x)\phi_C^2=-2\pi\delta(P) \int dx\exp(ipx)\Gamma^2_0(x)\phi_C^2$.)} The $\Gamma^{n}_0(q_i=0)$ Green's functions can contain two kinds of contributions, tree approximation graphs that involve vertex interactions but no loops, and radiative correction graphs that do contain loops.\footnote{An early analysis of cases where the only contributions are due to loops alone may be found in \cite{Coleman1973}.} For constant $\phi_C$ and $J$ the effective potential is related to the source via $dV/d\phi_C=J$, so that $J$ does indeed break any symmetry that $V(\phi_C)$ might possess. The significance of $V(\phi_C)$ is that when $J$ is zero and $\phi_C$ is spacetime independent, we can write $V(\phi_C)$ as 
\begin{eqnarray}
V(\phi_C)=\frac{1}{V}\left(\langle S|\hat{H}_0|S\rangle-\langle N|\hat{H}_0|N\rangle\right)
\label{M19}
\end{eqnarray}
in a volume $V$, where $|S\rangle$ and $|N\rangle$ are spontaneously broken and normal vacua in which $\langle S|\hat{\phi}|S\rangle$ is nonzero and $\langle N|\hat{\phi}|N\rangle$ is zero. In the analyses of classical potentials such as $V(\phi)=\lambda^2\phi^4/4-\mu^2\phi^2/2$ and classical kinetic energies such as $K=(1/2)\partial_{\mu}\phi\partial^{\mu}\phi$ presented above, the classical field $\phi$ represented $\phi_C$, the potential $V(\phi)$ represented $V(\phi_C)$, the kinetic energy represented $(1/2)Z(\phi_C)\partial_{\mu}\phi_C\partial^{\mu}\phi_C$, and in  the $\Gamma^{n}_0(q_i=0)$ Green's functions only tree approximation graphs were included (with $Z(\phi_C)$ then being equal to one). In this way the search for non-trivial minima of $V(\phi)$ is actually a search for states $|S\rangle$ in which $V(\phi_C)=\left(\langle S|\hat{H}_0|S\rangle-\langle N|\hat{H}_0|N\rangle\right)/V$ would be negative. Thus while the analyses presented above in Sec. (4) looked to be classical they actually had a quantum-mechanical underpinning with the classical field being a c-number vacuum matrix element of a q-number quantum field. It is in this way that the classical analyses presented above are to be understood.

\subsection{The Nature of Broken Symmetry}

To understand the nature of a broken symmetry vacuum it is instructive to reconsider $\epsilon(J)$. It is associated with a system $\hat{H}_0$ to which an external field has been added. This external field breaks the symmetry by hand at the level of the Lagrangian since an effective potential such as  $\epsilon(J)=V(\phi)-J\phi=\lambda^2\phi^4/4-\mu^2\phi^2/2-J\phi$ would be lopsided with one of its minima lower than the other, and would not have any $\phi \rightarrow-\phi$ symmetry. Such a situation is analogous to that found in a ferromagnet. In the presence of an external magnetic field (cf. $J$) all the spins line up in the direction of the magnetic field. If one is above the critical temperature, then when one removes the magnetic field the spins flop back into a configuration in which the net magnetization is zero. However, if one is below the critical point, the spins stay aligned and remember the direction of the magnetic field after it has been removed (hysteresis). Moreover, if the magnetic field is taken to point in some other direction, below the critical point the spins will remember that direction instead, and will remain aligned in that particular direction after the magnetic field is removed. For a spherically symmetric ferromagnetic system at a temperature below the critical point one can thus align the magnetization at any angle $\theta$ over a full $0$ to $2\pi$ range, and have it remain aligned after the magnetic field is removed.

Given all the different orientations of the magnetization that are possible below the critical point, we need to determine in what way we can  distinguish  them. So consider a single spin pointing in the $z$-direction with spin up, and a second spin pointing at an angle $\theta$ corresponding to a rotation through an angle $\theta$ around the $y$ axis. For these two states the overlap is given by 
\begin{eqnarray}
\langle 0|\theta \rangle=(1,0)e^{-i\theta\sigma_y/2}\left(\matrix{1\cr 0\cr}\right)=(1,0)\left(\matrix{\cos(\theta/2)& \sin(\theta/2)\cr -\sin(\theta/2)&\cos(\theta/2)\cr}\right)\left(\matrix{1\cr 0\cr}\right)=\cos(\theta/2),
\label{M20}
\end{eqnarray}
with the overlap being nonzero and with the two states thus necessarily being in the same Hilbert space. Suppose we now take an $N$-dimensional ensemble of these same sets of spin states and evaluate the overlap of the state with all spins pointing in the z-direction and the state with all spins pointing at an angle $\theta$. This gives the overlap 
\begin{eqnarray}
\langle 0, N|\theta , N\rangle=\cos^N(\theta/2).
\label{M21}
\end{eqnarray}
In the limit in which $N$ goes to infinity this overlap goes to zero. With this also being true of excitations built out of these states, the two states are now in different Hilbert spaces. Thus broken symmetry corresponds to the existence of different, inequivalent  vacua, and even though the various vacua  all have the same energy (i.e. degenerate vacua), the vacua are all in different Hilbert spaces. The quantum Hilbert spaces associated with the various minima of the effective potential (i.e. the differing states $|\Omega\rangle$ in which the vacuum expectation values $\langle \Omega|\hat{\phi}(x)|\Omega\rangle$ are evaluated) become distinct in the limit of an infinite number of degrees of freedom, even though they would not be distinct should $N$ be finite. Broken symmetry is thus not only intrinsically quantum-mechanical, it is intrinsically a many-body effect associated with an infinite number of degrees of freedom.

Whether or not a Hamiltonian $\hat{H}_0$ possesses such a set of degenerate vacua is a property of $\hat{H}_0$ itself. It is not a property of the external field $J$. The role of the external field is solely to pick one of the vacua, so that the system will then remain in that particular vacuum after the external field is removed. Whether or not the system is actually able to remember the direction of the external field after it has been removed is a property of the system itself and not of the external field.

To underscore the need for an infinite number of degrees of freedom, consider a system with a finite number of degrees of freedom such as the one-dimensional, one-body, quantum-mechanical system with potential $V(x)=\lambda^2 x^4/4-\mu^2x^2/2$ and Hamiltonian $H=-(1/2m)\partial^2/\partial x^2+V(x)$. Like the field-theoretic $V(\phi)=\lambda^2\phi^4/4-\mu^2\phi^2/2$, the potential $V(x)$ has a double-well structure, with minima  at $x=\pm \mu/\lambda$. However the eigenstates of the Hamiltonian cannot be localized around either of these two minima. Rather, since the Hamiltonian is symmetric under $x \rightarrow -x$, its eigenstates can only be even functions or odd functions of $x$, and must thus take support in both of the two wells. Wave functions localized to either of the two wells are in the same Hilbert space, as are then linear superpositions of them, with it being the linear combinations that are the eigenstates. Thus with a finite number of degrees of freedom, wave functions localized around the two minima are in the same Hilbert space. It is only with an infinite number of degrees of freedom that one could get inequivalent Hilbert spaces.

\subsection{Broken Symmetry and Multiplevaluedness}

Now if the role of $J$ is only to pick a vacuum and not to make the chosen state actually be a vacuum,  we need to inquire what is there about the $J$ dependence of the theory that might tell us whether or not we do finish up in a degenerate vacuum when we let $J$ go to zero. The answer to this question  is contained in $\epsilon(J)$, with $\epsilon(J)$ needing to be a multiple-valued function of $J$, with $\langle \Omega_J|\hat{\phi}|\Omega_J\rangle$ vanishing on one branch of $\epsilon(J)$ in the limit in which $J$ goes to zero, while not vanishing on some other one.  As a complex function of $J$ the function $\epsilon(J)$ has to have one or more branch points in the complex $J$ plane, and thus has to have some inequivalent determinations as $J$ goes to zero. These different determinations correspond to different phases, with it being the existence of such inequivalent determinations that is the hallmark of phase transitions.

To appreciate the point consider the two-dimensional Ising model of a ferromagnet in the presence of an external magnetic field $B$ at temperature $T$. In the mean-field approximation the free energy per particle is given by (see e.g. \cite{Brout1965}) 
\begin{eqnarray}
\frac{F(B,T)}{N}=\frac{1}{2}kT_CM^2-kT\ln\left[\cosh\left(\frac{T_CM}{T}+\frac{B}{kT}\right)\right],
\label{M22}
\end{eqnarray}
where $T_C$ is the critical temperature and $M$ is the magnetization. At the minimum where $dF/dM=0$  
the magnetization obeys
\begin{eqnarray}
M=\tanh\left(\frac{T_CM}{T}+\frac{B}{kT}\right).
\label{M23}
\end{eqnarray}
Given the structure of (\ref{M23}), it follows that when $T$ is greater than $T_C$ the magnetization can only be nonzero if $B$ is nonzero. However, if $T$ is less than $T_C$ one can have a nonzero $M$ even if $B$ is zero, and not only that, for every non-trivial $M$ there is another solution with $-M$. Since $\cosh(T_CM/T)$ is an even function of $M$, solutions of either sign for $M$ have the same  free energy. Symmetry breaking is thus associated with a degenerate vacuum energy. If we take $B$ to be complex, set $B=B_R+iB_I$, and set $\alpha=T_CM/T+B_R/kT$, $\beta=B_I/kT$, then when $B$ is nonzero we can set $\cosh(\alpha+i\beta)=\cosh\alpha\cos\beta+i\sinh\alpha\sin\beta$. With the logarithm term in the free energy having branch points in the complex $B$ plane whenever $\cosh(\alpha+i\beta)=0$, we see that branch points occur when $\alpha=0$, $\beta=\pi/2, 3\pi/2, 5\pi/2,...$. Thus as required, the free energy is a multiple-valued function in the complex $B$ plane, with branch points on the imaginary $B$ axis. While $M=\tanh(T_CM/T)$ only has two real solutions for any given $T<T_C$, it has an infinite number of pure imaginary solutions, and these are reflected in the locations of the branch points of $F(B,T)$.

A second example of multiplevaluedness  may be found in the double-well potential  $V(\phi)=\lambda^2\phi^4/4-\mu^2\phi^2/2$ given in Sec. (4) in the presence of a constant source $J$. Solutions to the theory are constrained to obey 
\begin{eqnarray}
\frac{dV(\phi)}{d \phi}=\lambda^2\phi^3-\mu^2\phi=J,
\label{M24}
\end{eqnarray}
and are of the form
\begin{eqnarray}
\phi=i^{1/3}[p(J)+iq(J)]^{1/3}+[i^{1/3}[p(J)+iq(J)]^{1/3}]^*,
\label{M25}
\end{eqnarray}
where 
\begin{eqnarray}
p(J)=\left(\frac{\mu^6}{27\lambda^6}-\frac{J^2}{4\lambda^2}\right)^{1/2},\qquad q(J)=-\frac{J}{2\lambda^2}.
\label{M26}
\end{eqnarray}
If we set $i^{1/3}=\exp(-i\pi/2)$, $i^{1/3}=\exp(i\pi/6)$, or $i^{1/3}=\exp(5i\pi/6)$, then when $J=0$, the solutions are given by $\phi_1=0$, $\phi_2=\mu/\lambda$, $\phi_3=-\mu/\lambda$, just as found in Sec (4.1). 

However, suppose instead we fix $i^{1/3}=\exp(-i\pi/2)$, and treat $\phi$ as a multiple-valued function of $J$. Then, because of the cube root in the $[p(J)+ iq(J)]^{1/3}$ term, as we set $J$ to zero we obtain three determinations of $p^{1/3}$, viz. $p_1=\mu/\lambda\surd{3}$, $p_2=\exp(2\pi i/3)\mu/\lambda\surd{3}$, $p_3=\exp(4\pi i/3)\mu/\lambda\surd{3}$. With $J=0$, these determinations then precisely give the previous $\phi_1=0$, $\phi_2=\mu/\lambda$, $\phi_3=-\mu/\lambda$ solutions. With this multiplevaluedness then propagating to $V(\phi)$ and $\epsilon(J)=V(\phi)-J\phi$ when they are evaluated in these three solutions, i.e. when we set $dV(\phi)/d\phi=J$, $d\epsilon(J)/dJ=-\phi$ and obtain 
\begin{eqnarray}
\epsilon(J)&=&-\frac{3\lambda^2}{4}\phi^4+\frac{\mu^2}{2}\phi^2=-\frac{\mu^2}{4}\phi^2-\frac{3}{4}\phi J
\nonumber\\
&=&-\frac{3J}{4}\left[i^{1/3}[p(J)+iq(J)]^{1/3}+{\rm c.~c.}\right]-\frac{\mu^2}{4}\left[i^{2/3}[p(J)+iq(J)]^{2/3}+{\rm c.~c.}\right]-\frac{\mu^4}{6\lambda^2},
\label{M27}
\end{eqnarray}
we see that  in any solution $\epsilon(J)$ is indeed a multiple-valued function of $J$, and see that from any one solution we can derive the others by analytic continuation, with the limit $J \rightarrow 0$ having multiple determinations. 

\subsection{Cooper Pairing in Superconductivity}

The binding of electrons into bound state pairs (Cooper pairing \cite{Cooper1956}) due to attractive forces induced by their interactions with the positive charged ions in a  crystal is responsible for the phenomenon of superconductivity. As such it is a beautiful example of a many-body effect, one than even admits of an exact treatment. In a quantum-mechanical bound state 
Schr\"odinger equation for a standard two-body system, the potential energy $V(r)$ of an attractive potential is minimized by having the particles be close, while the kinetic energy $p^2/2m$ is minimized by having the particles be far apart (minimization of the momentum). In a system with three spatial dimensions competition between the kinetic energy and the potential energy can lead to bound states only if the potential strength is above some (potential-dependent) minimum value.\footnote{For a particle of mass $m$ in a 3-dimensional well of depth $V_0$ and width $a$ for instance,  binding only occurs if $V_0a^2 \geq\pi^2\hbar^2/8m$.}  Now  in a superconductor the attractive force between two electrons is very weak, and in and of itself is not big enough to produce binding. However, there are not just two electrons in a superconductor but a large number $N$  of them. Because of the Pauli principle the electrons of mass $m$ are distributed in differing momentum and energy states up to the Fermi momentum $k_{\rm F}$ and Fermi energy $E_{\rm F}=k_{\rm F}^2/2m$.  Thus electrons that attempt to bind must be in  high momentum states since the low momentum states are occupied. Consequently, now the kinetic energy does not have to prefer widely separated electrons, and even a very weak attractive potential can then bind them. Moreover, no interaction is required between the two electrons in a Cooper pair and all the $N-2$ other electrons in the superconductor, with the only role required of the $N-2$ other electrons being to block off momentum states  (Pauli blocking). In this way Cooper pairing is a many-body effect and not a two-body one.

To discuss the pairing phenomenon in more detail we follow \cite{Brout1965}. Because of Pauli blocking up to the Fermi surface momentum $k_{\rm F}$, we take the pairing wave function to be of the form $\psi({\bf  r})=\sum_{q>k_{\rm F}} a_{{\bf  q}}\exp(i{\bf  q}\cdot{\bf  r})$, where ${\bf  r}$ is the relative radius vector of the pair. With a potential $V$, which for simplicity we take to be constant, the momentum space Schr\"odinger equation takes the form 
\begin{eqnarray}
(E_{k}-E)a_{{\bf  k}}+\sum_{q>k_{\rm F}}\langle{\bf  k}|V|{\bf  q}\rangle a_{{\bf  q}}=0,
\label{M28}
\end{eqnarray}
where $E_{k}=k^2/2m$. We now set 
\begin{eqnarray}
\langle{\bf  k}|V|{\bf  q}\rangle&=&\lambda~~{\rm when}~~ E_{\rm F}\leq E_k, E_q\leq E_{\rm F}+D;
\nonumber\\
\langle{\bf  k}|V|{\bf  q}\rangle&=&0~~{\rm when}~~ E_k, E_q > E_{\rm F}+D,
\label{M29}
\end{eqnarray}
where the constant $\lambda$ is the strength of the potential and $D$ is the bandwidth (typically of order the Debye frequency). Solutions to the  Schr\"odinger equation thus obey
\begin{eqnarray}
a_{{\bf  k}}=-\frac{\lambda}{E_k-E}\sum_{E_{\rm F}}^{E_{\rm F}+D} a_{{\bf  q}}, 
\label{M30}
\end{eqnarray}
with a summation over ${\bf  k}$ yielding
\begin{eqnarray}
f(E)=\sum_{E_{\rm F}}^{E_{\rm F}+D} \frac{1}{E_k-E}=-\frac{1}{\lambda},
\label{M31}
\end{eqnarray}
with (\ref{M31}) serving to define $f(E)$.
For $E<E_{\rm F}$ the function $f(E)$ is positive definite. Thus with $\lambda$ negative (i.e. attractive potential) there is a bound state with energy below the Fermi surface no matter how small in magnitude $\lambda$ might be. 

For such a bound state with energy $E$ the denominator in $f(E)$ has no singularities, and so we can pass to the continuum limit, with the integration then yielding
\begin{eqnarray}
f(E)={\rm ln}\left(\frac{E_{\rm F}+D-E}{E_{\rm F}-E}\right).
\label{M32}
\end{eqnarray}
The binding energy is thus given by
\begin{eqnarray}
\Delta=E_{\rm F}-E=\frac{D}{\exp(-1/\lambda)-1},
\label{M33}
\end{eqnarray}
with electrons now having energies of the shifted form $E_k=k^2/2m-\Delta$ as they propagate in the superconducting medium. For small $\lambda$ the binding energy is given by the so-called gap equation
\begin{eqnarray}
\Delta=D\exp(1/\lambda).
\label{M34}
\end{eqnarray}
Now in this discussion and in its full BCS generalization \cite{Bardeen1957} we note that there are no elementary scalar fields in the theory, just electrons and ions. The symmetry breaking is due to the difermion pairing condensate operator $\psi\psi$ acquiring a non-zero vacuum expectation value in a state $|S \rangle$ according to  $\langle S|\psi\psi|S\rangle\neq 0$. The BCS theory thus provides  a well-established, working model in which all the breaking is done by condensates.  Thus in the following we shall explore whether the Higgs boson might be generated by condensate dynamics too, with no elementary scalar Higgs field being present in the Lagrangian that is to describe elementary particle physics.

Even though one might expect, and can of course find, bound states that are associated with strong coupling rather than weak coupling, as  constructed, we obtain Cooper pairing no matter how weak the coupling $\lambda$ might be, with the driver being the filled Fermi sea not the strength of the coupling. In the relativistic models of dynamical symmetry breaking that we discuss in the following we shall find models in which the coupling needs to be strong, but shall  also find models in which the coupling can be weak.

The form for $\Delta$ has an essential singularity when $\lambda=0$, and thus the superconducting phase where $\Delta$ is nonzero cannot be reached perturbatively starting from the normal conductor. The normal and superconducting phases thus have vacua $|N\rangle$ and $|S\rangle$ that are in different Hilbert spaces.   They can be related by a Bogoliubov transform to the particle-hole basis, and while this was done by BCS themselves to give a wave function that described all pairs at once, for our purposes here it is more instructive to describe the relativistic generalization, with the filled negative energy sea of a Dirac fermion replacing the filled positive energy Fermi sea of the superconductor.

\subsection{Degenerate Fermion Vacua}

To construct the relativistic analog of the superconducting vacuum and illustrate the distinction between the normal and the spontaneously broken vacua, we follow \cite{Nambu1961} and, using the notation of \cite{Mannheim1975}, consider free massless and massive fermions that obey 
\begin{eqnarray}
i\gamma^{\mu}\partial_{\mu}\psi^{(0)}(x)=0,\qquad (i\gamma^{\mu}\partial_{\mu}-m)\psi^{(m)}(x)=0.
\label{M35}
\end{eqnarray}
With $i$ denoting $0$ or $m$, we can expand both the cases in a standard Fourier decomposition of the form 
\begin{eqnarray}
\psi^{(i)}({\bf  x},t=0)=\frac{1}{V^{1/2}}\sum_{{\bf  p},s}\left(u^{(i)}({\bf  p},s)b^{(i)}({\bf  p},s)e^{i{\bf  p}\cdot {\bf  x}}+v^{(i)}({\bf  p},s)d^{(i)\dagger}({\bf  p},s)e^{-i{\bf  p}\cdot {\bf  x}}\right),
\label{M36}
\end{eqnarray}
in a volume $V$ as summed over up and down spins $s$ and an infinite set of momentum states ${\bf  p}$.  Here each spinor is restricted to its own mass shell ($E^{(0)}_p=p$, $E^{(m)}_p=(p^2+m^2)^{1/2}$, $p=|{\bf p}|$) and normalized according to $u^{\dagger}u=1$, $v^{\dagger}v=1$. With each set of creation and annihilation operators obeying canonical anticommutation relations 
\begin{eqnarray}
\{b^{(i)}({\bf  p},s),b^{(i) \dagger}({\bf  p}^{\prime},s^{\prime})\}=\delta^3({\bf  p}-{\bf  p}^{\prime})\delta_{s,s^{\prime}},\qquad
\{d^{(i)}({\bf  p},s),d^{(i) \dagger}({\bf  p}^{\prime},s^{\prime})\}=\delta^3({\bf  p}-{\bf  p}^{\prime})\delta_{s,s^{\prime}},
\label{M37}
\end{eqnarray}
they must be related by a canonical Bogoliubov transformation. On introducing 
\begin{eqnarray}
\lambda^{\pm}_{p}=\left[\frac{1}{2}\left(1\pm \frac{p}{(p^2+m^2)^{1/2}}\right)\right]^{1/2},
\label{M38}
\end{eqnarray}
and on normalizing the spinors so that $\psi^{(0)}(x)=\psi^{(m)}(x)$ at $t=0$, we obtain 
\begin{eqnarray}
b^{(m)}({\bf  p},s)=\lambda_p^{+}b^{(0)}({\bf  p},s)+\lambda_p^{-}d^{(0)\dagger}(-{\bf  p},s),\quad
d^{(m)}({\bf  p},s)=\lambda_p^{+}d^{(0)}({\bf  p},s)-\lambda_p^{-}b^{(0)\dagger}(-{\bf  p},s),
\label{M39}
\end{eqnarray}
viz. a transformation to the particle-hole basis. If we now define normalized vacua that obey
\begin{eqnarray}
a^{(0)}({\bf  p},s)|\Omega_0\rangle=0,\qquad a^{(m)}({\bf  p},s)|\Omega_m\rangle=0, \qquad \langle \Omega_0|\Omega_0\rangle=1,\qquad \langle \Omega_m|\Omega_m\rangle=1
\label{M40}
\end{eqnarray}
we find that
\begin{eqnarray}
|\Omega_m\rangle= \prod_{{\bf  p},s}\left[\lambda_p^{+}-\lambda_p^{-}b^{(0)\dagger}({\bf  p},s)d^{(0)\dagger}(-{\bf  p},s)\right]|\Omega_0\rangle,  
\label{M41}
\end{eqnarray}
with the massive vacuum being given as an infinite superposition of pairs created out of the massless vacuum. 

Given their relation, the overlap of the two vacua evaluates to
\begin{eqnarray}
\langle \Omega_0|\Omega_m\rangle= \exp\left(\sum_{{\bf  p},s}{\rm ln}\lambda_p^{+}\right).
\label{M42}
\end{eqnarray}
With each $\lambda_p^{+}$ being less than one, the overlap vanishes in the limit of an infinite number of modes. Thus while the two vacua would be in the same Hilbert space if the number of modes were to be finite, in the limit of an infinite number of modes the two vacua can no longer overlap and their respective Hilbert spaces become distinct, with there being no measurement that could then connect the two spaces. This disconnecting of the two Hilbert spaces is central to broken symmetry, with it being a specific many-body effect that is expressly generated by the presence of an infinite number of degrees of freedom. 

Since the Bogoliubov transformation preserves the fermion field anticommutation relations it must be unitary. Thus we must be able to write $|\Omega_m\rangle=U|\Omega_0\rangle$ with $U^{\dagger}U=I$. If we introduce a complete basis of states $|n^{(m)}\rangle=(a^{(0)\dagger})^n|\Omega_m\rangle$ in the massive vacuum Hilbert space, we obtain $\langle \Omega_0|\Omega_0\rangle=\sum_{n}\langle \Omega_0|n^{(m)}\rangle\langle n^{(m)}|\Omega_0\rangle=1$. But we had just established that there were no overlaps between the massless and massive Hilbert space. Thus each $\langle \Omega_0|n^{(m)}\rangle$ matrix element must vanish. But nonetheless the sum $\sum_{n}\langle \Omega_0|n^{(m)}\rangle\langle n^{(m)}|\Omega_0\rangle$ is nonzero. The way that it gets to be nonzero is by a very delicate interplay between an infinite number of vanishing matrix elements and a summation over an infinite complete set of states such that $0\times \infty=1$. It is in this way that the massless and massive Hilbert spaces are disconnected.

The above analysis allows us to compare $|\Omega_m\rangle$ with $|\Omega_0\rangle$, and below we will show in the self-consistent Hartree-Fock approximation to a four-fermion Nambu-Jona-Lasinio model that the state  $|\Omega_m\rangle$ has lower energy than the state  $|\Omega_0\rangle$, to thus be preferred. To show that $|\Omega_m\rangle$ is one of an infinite number of degenerate vacua we make a global chiral transformation in the massless theory through an angle $\alpha$ of the form 
\begin{eqnarray}
\psi^{(0)}\rightarrow e^{i\alpha\gamma^5}\psi^{(0)},\qquad b^{(0)}({\bf  p},\pm) \rightarrow e^{\pm i\alpha}b^{(0)}({\bf  p},\pm),\qquad d^{(0)\dagger}({\bf  p},\pm) \rightarrow e^{\pm i\alpha}b^{(0)\dagger}({\bf  p},\pm),
\label{M43}
\end{eqnarray}
a transformation that leaves the massless Dirac equation $i\gamma_{\mu}\partial^{\mu}\psi^{(0)}(x)=0$ invariant. We can thus construct a new canonical transformation
\begin{eqnarray}
b_{\alpha}^{(m)}({\bf  p},\pm)&=&\lambda_p^{+}e^{\mp i\alpha}b^{(0)}({\bf  p},\pm)+\lambda_p^{-}e^{\pm i\alpha}d^{(0)\dagger}(-{\bf  p},\pm),\quad
\nonumber\\
d_{\alpha}^{(m)}({\bf  p},\pm)&=&\lambda_p^{+}e^{\mp i\alpha}d^{(0)}({\bf  p},\pm)-\lambda_p^{-}e^{\pm i\alpha}b^{(0)\dagger}(-{\bf  p},\pm),
\label{M44}
\end{eqnarray}
and a new vacuum
\begin{eqnarray}
|\Omega^{\alpha}_m\rangle= \prod_{{\bf  p},\pm}\left[\lambda_p^{+}-\lambda_p^{-}e^{\pm 2i\alpha}b^{(0)\dagger}({\bf  p},\pm)d^{(0)\dagger}(-{\bf  p},\pm)\right]|\Omega_0\rangle.  
\label{M45}
\end{eqnarray}
The overlaps of $ |\Omega^{\alpha}_m\rangle$  with $|\Omega_0\rangle$ and $|\Omega_m\rangle$ evaluate to
\begin{eqnarray}
\langle \Omega_0|\Omega^{\alpha}_m\rangle= \exp\left(\sum_{{\bf  p},\pm}{\rm ln}\lambda_p^{+}\right),\qquad
\langle \Omega^{\alpha}_m|\Omega_m\rangle=\exp\left[\sum_{{\bf  p},\pm}{\rm ln}\left[1+(e^{\pm 2i\alpha}-1)(\lambda_p^{-})^2\right]\right].
\label{M46}
\end{eqnarray}
Both of these overlaps vanish in the limit of an infinite number of modes. The Hilbert spaces built on $|\Omega_0\rangle$, $|\Omega_m\rangle$, and $|\Omega^{\alpha}_m\rangle$ are all distinct, with there being an infinity of such $|\Omega^{\alpha}_m\rangle$ states for all values of the continuous variable $\alpha$. 

The Hamiltonians and vacuum energies associated with these various vacua are given by 
\begin{eqnarray}
H_0&=&\sum_{{\bf  p},s}\left[p\left(b^{(0)\dagger}({\bf  p},s)b^{(0)}({\bf  p},s)-d^{(0)}({\bf  p},s)d^{(0)\dagger}({\bf  p},s)\right)\right],
\nonumber\\
H_m&=&\sum_{{\bf  p},s}\left[(p^2+m^2)^{1/2}\left(b^{(m)\dagger}({\bf  p},s)b^{(m)}({\bf  p},s)-d^{(m)}({\bf  p},s)d^{(m)\dagger}({\bf  p},s)\right)\right],
\nonumber\\
H^{\alpha}_m&=&\sum_{{\bf  p},s}\left[(p^2+m^2)^{1/2}\left(b_{\alpha}^{(m)\dagger}({\bf  p},s)b_{\alpha}^{(m)}({\bf  p},s)-d_{\alpha}^{(m)}({\bf  p},s)d_{\alpha}^{(m)\dagger}({\bf  p},s)\right)\right],
\label{M47}
\end{eqnarray}
\begin{eqnarray}\langle \Omega_0|H_0|\Omega_0\rangle&=&-2\sum_{{\bf  p}}p,
\nonumber\\
\langle \Omega_m|H_m|\Omega_m\rangle&=&-2\sum_{{\bf  p}}(p^2+m^2)^{1/2},
\nonumber\\
\langle \Omega^{\alpha}_m|H^{\alpha}_m|\Omega^{\alpha}_m\rangle
&=&-2\sum_{{\bf  p}}(p^2+m^2)^{1/2},
\label{M48}
\end{eqnarray}
with the negative signs of the various vacuum energies being due to the filled fermionic negative energy Dirac sea.
We thus confirm that both $|\Omega_m\rangle$ and $|\Omega^{\alpha}_m\rangle$ lie lower than $|\Omega_0\rangle$, while being degenerate with each other for all $\alpha$. The massive vacuum is thus infinitely degenerate. In the following we analyze the four-fermion theory in order to establish the dynamical relevance of what for the moment is just a study of a free fermion system.

\subsection{Symmetry Breaking by Fermion Bilinear Composites}

In studying symmetry breaking in a theory with  action $\int d^4x \hat{{{\cal L}}}_0(x)$, in order to construct an effective potential  we first introduced a local source term $\int d^4x J(x)\hat{\phi}(x)$ that depended on a single quantum field $\hat{\phi}(x)$, and studied the action $\int d^4x \hat{{{\cal L}}}_J(x)=\int d^4x \hat{{{\cal L}}}_0(x)+\int d^4x J(x)\hat{\phi}(x)$ that is obtained in the presence of the source. In studying symmetry breaking by fermion bilinear composites we would need to introduce a source term that depends on two fields. To do this there are two options, the two fields could be at different spacetime points, or the two fields could be at the same point. The first of these options was explored in detail in \cite{Cornwall1974} and the second in \cite{Mannheim1974,Mannheim1975,Mannheim1978}. 

For the approach in which the two fermionic fields are at different spacetime points, rather than generalize the q-number effective action treatment that is based on matrix elements of the quantum field operators as given in (\ref{M10}), it is more convenient to work with a purely c-number path integral approach. With a c-number action $I(\psi,\bar{\psi})=\int d^4x {{\cal L}}_0(x)$ and a bilinear c-number source $K(x,y)$ we define the vacuum to vacuum functional as the path integral
\begin{eqnarray}
e^{iW(K)}=\langle \Omega_0^+|\Omega_0^-\rangle|_K=\int D[\psi]D[\bar{\psi}]
\exp\left[iI(\psi,\bar{\psi})+i\int d^4xd^4y \psi(x)\bar{\psi}(y)K(x,y)\right].
\label{M49}
\end{eqnarray}
Functional variation with respect to $K(x,y)$ allows us to generate the fermion propagator Green's function $G(x,y)=\langle \Omega|T[\psi(y)\bar{\psi}(x)]|\Omega \rangle$ according to 
\begin{eqnarray}
\frac{\delta W(K)}{\delta K(x,y)}=G(x,y),
\label{M50}
\end{eqnarray}
so that we can then construct 
\begin{eqnarray}
\Gamma(G)=W(K)-\int d^4xd^4y G(x,y)K(x,y).
\label{M51}
\end{eqnarray}
In the same way that the functional $\Gamma(\phi_C)$ is the generator of one-particle irreducible diagrams, the functional $\Gamma(G)$ is the generator of two-particle irreducible diagrams. 

In analog to our discussion of $\Gamma(\phi_C)$, we can construct the functional variation of $\Gamma(G)$ with respect to $G(x,y)$, to obtain
\begin{eqnarray}
\frac{\delta \Gamma(G)}{\delta G(x,y)}=-K(x,y),
\label{M52}
\end{eqnarray}
and can then set $K(x,y)=0$ to obtain the stationarity condition $\delta \Gamma(G)/\delta G(x,y)=0$ associated with switching off $K(x,y)$.  However, there is a difference between the one-point and two-point cases. While the one-point function $\phi_C(x)\sim\langle\Omega|\hat{\phi}(x)|\Omega\rangle$ is time independent in a translation invariant vacuum $|\Omega\rangle$,  the two-point function $G(x,y)=\langle \Omega|T[\psi(y)\bar{\psi}(x)]|\Omega \rangle$ is not in general time independent even if $|\Omega\rangle$ is translation invariant. Consequently, unlike in the one-point $\Gamma(\phi_C)$ case where one can compare energy densities of differing candidate vacua at stationary points $\delta \Gamma(\phi_C)/\delta \phi_C=0$, for $\delta \Gamma(G)/\delta G(x,y)=0$ one is not comparing the energy densities of different candidate vacua (unless one is considering $G(x,y)$ that are taken to be static), and thus unlike $\Gamma(\phi_C)$, in the non-static case $\Gamma(G)$ cannot be thought of as being a potential.

In fact the stationarity condition for $\Gamma(G)$ has a quite different significance. Specifically, as shown in  \cite{Cornwall1974},  the condition $\delta \Gamma(G)/\delta G(x,y)=0$ is the Schwinger-Dyson equation for the fermion propagator. In this regard, the approach of \cite{Cornwall1974} actually allows one to derive the Schwinger-Dyson equation without recourse to a perturbative graphical analysis. Since, as we discuss in detail below in Sec. (7),  dynamical symmetry breaking is associated with self-consistent solutions to the Schwinger-Dyson equation, the search for such solutions is facilitated by looking for stationary solutions  to $\delta \Gamma(G)/\delta G(x,y)=0$, and in fact greatly so since, as shown in \cite{Cornwall1974},  the functional approach based on $\Gamma (G)$ organizes the graphs needed for the Schwinger-Dyson equation far more efficiently than in a perturbative graphical expansion. 

Now despite the fact that $\Gamma(G)$ is not a potential, one can still use it to study stability. Specifically, for stability we need to study fluctuations around the stationary solutions. As discussed for instance in \cite{Fukuda1987,Fukuda1988,Miransky1993}, one can associate the second functional derivative $\delta \Gamma(G)/\delta G(x,y)\delta G(x^{\prime},y^{\prime})$ with the Bethe-Salpeter equation for potential dynamical bound states. Stability is then achieved if the Bethe-Salpeter equation has no bound states that are tachyonic.

One can also consider the approach where one takes the source to be a local one, with both of the fermions in the $\bar{\psi}(x)\psi(x)$ bilinear then being at the same point. Since the two fermions are at the same point, matrix elements of the form $\langle\Omega|\bar{\psi}(x)\psi(x)|\Omega\rangle$ will be time independent if the vacuum $|\Omega\rangle$ is translation invariant. And so this approach does allow us to identify a potential and compare the energy densities of different candidate vacua. This approach was not developed as a general approach but as an approach that was tailored to theories with a four-fermion interaction.\footnote{In \cite{Mannheim1974,Mannheim1975} the four-fermion term was introduced as a vacuum energy density counterterm, and in \cite{Mannheim1978} it was introduced in order to construct a Hartree-Fock mean-field theory.} However we will argue below when we couple to gravity that in fact such four-fermion terms are always needed. While we need to use the effective potential $V(\phi_C)$ to explore symmetry breaking by an elementary scalar field, it turns out that even though $W(J)$ was only introduced as an intermediate step in order to get to $\Gamma(\phi_C)$, for  symmetry breaking by Hermitian fermion $\bar{\psi}(x)\psi(x)$ bilinear composites we need to use its $W(m)$ generalization directly, where
\begin{eqnarray}
e^{iW(m)}=\langle \Omega_0^+|\Omega_0^-\rangle|_m=\langle \Omega_m|T\exp\left[i\int d^4x(\hat{{{\cal L}}}_0-m(x)\bar{\psi}(x)\psi(x))\right]|\Omega_m\rangle,
\label{M53}
\end{eqnarray}
with $\bar{\psi}(x)\psi(x)$ being Hermitian and the c-number $m(x)$ being real.  To be specific, consider the chiral-symmetric (${\rm CS}$) action $I_{\rm CS}$  of the form
\begin{eqnarray}
I_{\rm CS}=\int d^4x \left[\hat{{{\cal L}}}_0-\frac{g}{2}[\bar{\psi}\psi]^2-\frac{g}{2}[\bar{\psi}i\gamma^5\psi]^2\right],
\label{M54}
\end{eqnarray}
in which a chirally-symmetric four-fermion interaction with real $g$ (as required to enforce Hermiticity) has been added on to some general chirally-symmetric  $\hat{{{\cal L}}}_0$ (a typical example of which would be the massless fermion QED theory that we study below). As such $I_{\rm CS}$ is a relativistic generalization of the BCS model. Even though there is no fermion mass term present in $I_{\rm CS}$, in the mean-field, Hartree-Fock  approximation one introduces a trial wave function parameter $m$ that is not in the original action, and then decomposes  the action into two pieces,  a mean-field piece and a residual-interaction piece according to:
\begin{eqnarray}
I_{\rm CS}&=&I_{\rm MF}+I_{\rm RI},
\nonumber\\
I_{\rm MF}&=&\int d^4x \left[\hat{{{\cal L}}}_0-m\bar{\psi}\psi +\frac{m^2}{2g}\right],
\nonumber\\
I_{\rm RI}&=&\int d^4x \left[-\frac{g}{2}\left(\bar{\psi}\psi-\frac{m}{g}\right)^2-\frac{g}{2}\left(\bar{\psi}i\gamma^5\psi\right)^2\right].
\label{M55}
\end{eqnarray}
One then tries to show that in the mean-field sector a nonzero $m$ is energetically favored. To this end we recall that in (\ref{M13}) we had identified $\epsilon(m)$ as $\epsilon(m)=(\langle \Omega_m|\hat{H}_m|\Omega_m\rangle-\langle \Omega_0|\hat{H}_0|\Omega_0\rangle)/V$, where $\hat{H}_0$ and $\hat{H}_m$ are respectively associated with $\hat{{{\cal L}}}_0$ and  $\hat{{{\cal L}}}_m=\hat{{{\cal L}}}_0-m\bar{\psi}\psi$, as determined with a general Lagrangian density $\hat{{{\cal L}}}_0$. If the state $|\Omega_0\rangle$  is one in which $\langle \Omega_0|\bar{\psi}\psi|\Omega_0\rangle$ is zero, we can therefore write $\epsilon(m)$ as  
\begin{eqnarray}
\epsilon(m)=\frac{1}{V}\left(\langle \Omega_m|\hat{H}_m|\Omega_m\rangle-\langle \Omega_0|\hat{H}_m|\Omega_0\rangle\right),
\label{M56}
\end{eqnarray}
to thus enable us to compare two candidate vacua for $\hat{H}_m$, with a view to determining whether a vacuum with nonzero $\langle \Omega_m|\bar{\psi}\psi|\Omega_m\rangle$ is energetically favored. For the simple case where $\hat{{{\cal L}}}_0=i \bar{\psi}\gamma^{\mu}\partial_{\mu}\psi$  and $\hat{{{\cal L}}}_m=i \bar{\psi}\gamma^{\mu}\partial_{\mu}\psi-m\bar{\psi}\psi$, then according to (\ref{M48}) $\epsilon(m)$ is given by $\epsilon(m)=-2\sum_{{\bf  p}}[(p^2+m^2)^{1/2}-p]$, to thus be negative definite, just as required for dynamical symmetry breaking, and just as needed to establish the physical relevance of what had previously appeared to be a free fermion theory (i.e. the free fermion theory is the mean-field approximation to $I_{\rm CS}$ when $\hat{{{\cal L}}}_0=i \bar{\psi}\gamma^{\mu}\partial_{\mu}\psi$).\footnote{The quantity $\epsilon(m)=-2\sum_{{\bf  p}}[(p^2+m^2)^{1/2}-p]$ is negative because the summations in (\ref{M48}) are over fermionic negative energy modes. For bosons the analog quantity would be equal to $+2\sum_{{\bf  p}}[(p^2+m^2)^{1/2}-p]$ and thus be positive, not negative. With the expansion of $\epsilon(m)$ as given in (\ref{M57}) being an expansion in loop diagrams, the difference in sign between fermions and boson vacuum energy densities is due to the difference between Fermi and Bose statistics, with a fermion loop and a boson loop having opposite overall signs.}

No matter what choice we make for $\hat{{{\cal L}}}_0$, on comparing (\ref{M11}) and (\ref{M12}) with (\ref{M15}), (\ref{M17}),  and (\ref{M18}), the relevant $\epsilon(m)$ when $m$ is constant is given by 
\begin{eqnarray}
\epsilon(m)=\sum_n\frac{1}{n!}G^{n}_0(q_i=0)m^n,
\label{M57}
\end{eqnarray}
where the $G^{n}_0(q_i=0)$ are the connected $\bar{\psi}\psi$ Green's functions as evaluated in the theory in which $m=0$, viz. that associated with $\hat{{{\cal L}}}_0$. The utility of (\ref{M57}) is that it generates massive fermion theory Green's functions as infinite sums of massless fermion theory Green's functions, and massless fermion Green's functions are easier  to calculate than massive fermion ones, and especially so if the massless theory has an underlying scale or conformal symmetry, something we consider below. Since functional variation with respect to the  $m\bar{\psi}\psi$ source term in (\ref{M53}) generates the $\bar{\psi}\psi$ Green's functions in the massive $\hat{H}_m$ theory, we can identify $\epsilon^{\prime}(m)=d\epsilon(m)/dm$ as the one-point (tadpole diagram) function $\langle \Omega_m|\bar{\psi}\psi|\Omega_m\rangle$, a quantity whose non-vanishing is the indicator of dynamical symmetry breaking.\footnote{One Green's function in the massive theory is equivalent to  an infinite sum of Green's functions in the massless theory.}

Unlike the elementary scalar field $V(\phi_C)$, $\epsilon(m)$ has no tree approximation contribution and is entirely generated by radiative loops. Thus while we need to use $V(\phi_C)$ to explore symmetry breaking by an elementary scalar field with $V(\phi_C)$ being the energy density difference between different candidate vacua of $H_0$, for symmetry breaking by a fermion composite we use $\epsilon(m)$ instead, with $\epsilon(m)$ being the energy density difference between different candidate vacua of $H_m$. We shall explore this issue in more detail below, while noting now that in distinguishing between an elementary (i.e. ``god given") Higgs boson and a composite one there are even differences in setting up the formalism in the first place.

\section{Is the Higgs Boson Elementary or Composite?}

\subsection{What Exactly is the Higgs Field?}

Given that the existence of the Higgs boson has now been confirmed, we need to ask what exactly the field $\phi$ represents. There are two possibilities. It is either a q-number elementary field $\hat{\phi}$ that appears in the fundamental  $SU(3)\times SU(2)\times U(1)$ Lagrangian of strong, electromagnetic, and weak interactions (to thereby be on an equal footing with the fundamental quarks, leptons and gauge bosons), or it is generated as a dynamical bound state, with the field in a dynamically induced Higgs potential then being the c-number matrix element $\langle S|\bar{\psi}\psi|S\rangle$,  a dynamical bilinear fermion condensate. The Mexican Hat potential is thus either part of the fundamental Lagrangian or it is generated by  dynamics. If the Higgs field is elementary, then while the potential $V(\hat{\phi})=\lambda^2\hat{\phi}^4/4-\mu^2\hat{\phi}^2/2$ would be its full quantum-mechanical potential, the discussion given earlier of the minima of the potential would correspond to a c-number tree approximation analysis with the $\phi$ that appeared there being the c-number $\langle S|\hat{\phi}|S\rangle$. However, in the fermion condensate case there is no tree approximation, with the theory being given by radiative loop diagrams alone. To see how to generate a Mexican Hat potential in this case we consider the Nambu-Jona-Lasinio (NJL) four-fermion model.

\subsection{Nambu-Jona-Lasinio Chiral Model as a Mean-Field Theory}

The NJL model \cite{Nambu1961} is a chirally-symmetric four-fermion model of interacting massless fermions with action $I_{\rm NJL}$ as given below in (\ref{M58}). In the mean-field, Hartree-Fock  approximation one introduces a trial wave function parameter $m$ that is not in the original action, and then decomposes  the $I_{\rm NJL}$ action into two pieces,  a mean-field piece and a residual-interaction piece according to $I_{\rm NJL}=I_{\rm MF}+I_{\rm RI}$, where
\begin{eqnarray}
I_{\rm NJL}&=&\int d^4x \left[i\bar{\psi}\gamma^{\mu}\partial_{\mu}\psi-\frac{g}{2}[\bar{\psi}\psi]^2-\frac{g}{2}[\bar{\psi}i\gamma^5\psi]^2\right],
\nonumber\\
I_{\rm MF}&=&\int d^4x \left[i \bar{\psi}\gamma^{\mu}\partial_{\mu}\psi-m\bar{\psi}\psi +\frac{m^2}{2g}\right],
\nonumber\\
I_{\rm RI}&=&\int d^4x \left[-\frac{g}{2}\left(\bar{\psi}\psi-\frac{m}{g}\right)^2-\frac{g}{2}\left(\bar{\psi}i\gamma^5\psi\right)^2\right].
\label{M58}
\end{eqnarray}
Neither of the two pieces in the decomposition is separately chirally symmetric under $\psi\rightarrow e^{i\alpha \gamma^5}\psi$, only their $I_{\rm NJL}$ sum is. While this remark would seem to be innocuous, below we shall see that it has quite far-reaching consequences. In the mean-field, Hartree-Fock approximation as applied to $I_{\rm RI}$ one sets 
\begin{eqnarray}
&&\langle S|\left[\bar{\psi}\psi-\frac{m}{g}\right]^2|S\rangle=\langle S|\left[\bar{\psi}\psi-\frac{m}{g}\right]|S\rangle^2=0,\quad \langle S|\bar{\psi}\psi|S\rangle=\frac{m}{g}
\nonumber\\
&&\langle S|\left[\bar{\psi}i\gamma^5\psi\right]^2|S\rangle=\langle S|\bar{\psi}i\gamma^5\psi|S\rangle^2=0,\qquad \langle S|\bar{\psi}i\gamma^5\psi|S\rangle=0,
\label{M59}
\end{eqnarray}
to thus give the residual-interaction energy density a zero vacuum expectation value in the state $|S\rangle$. In the mean-field approximation the physical mass $M$ is the value of $m$ that satisfies $\langle S|\bar{\psi}\psi|S\rangle=m/g$. 

\begin{figure}[htpb]
\begin{center}
\begin{minipage}[t]{5 cm}
\epsfig{file=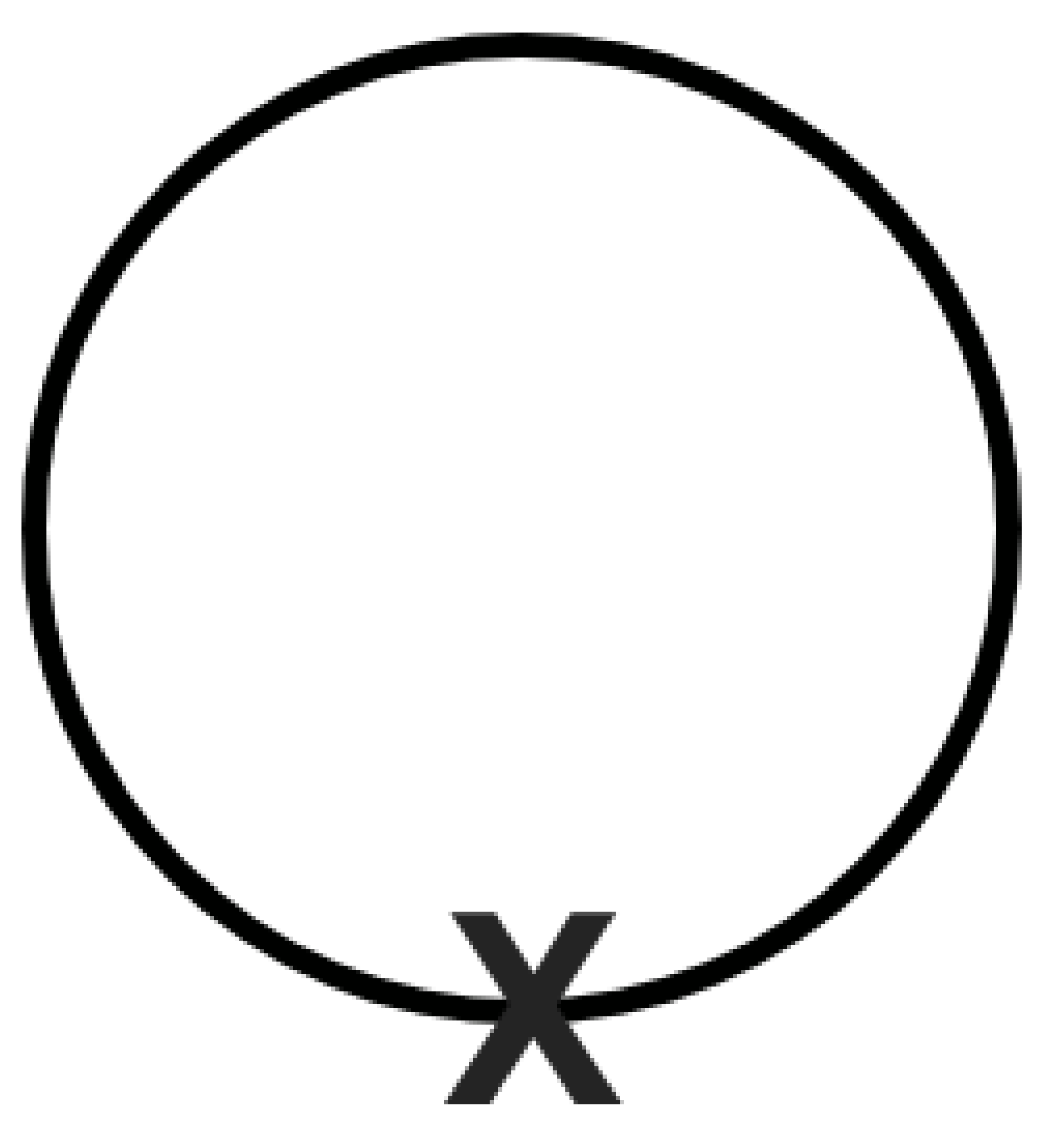,scale=0.2}
\end{minipage}
\begin{minipage}[t]{16.5 cm}
\caption{The NJL tadpole graph for $\langle S|\bar{\psi}\psi|S\rangle$ with a zero-momentum point $m\bar{\psi}\psi$ insertion and an NJL mean-field fermion $1/(\slashed{p}-m+i\epsilon)$ propagator.}
\label{baretadpole}
\end{minipage}
\end{center}
\end{figure}

The one loop contribution of the fermionic negative energy Dirac sea  to the quantity $\langle S|\bar{\psi}\psi|S\rangle$ as given in Fig. (\ref{baretadpole})  yields the gap equation\footnote{One can evaluate $\langle S|\bar{\psi}\psi|S\rangle$ directly or via $\langle S|\bar{\psi}\psi|S\rangle=\epsilon^{\prime}(m)$, where $\epsilon(m)$ is given below in (\ref{M61}).}
\begin{eqnarray}
\langle S|\bar{\psi}\psi|S\rangle=-i\int \frac{d^4p}{(2\pi)^4}{\rm Tr}\left[\frac{1}{\slashed{p}-M+i\epsilon}\right]=
-\frac{M\Lambda^2}{4\pi^2}+\frac{M^3}{4\pi^2}{\rm ln}\left(\frac{\Lambda^2}{M^2}\right)=\frac{M}{g},
\label{M60}
\end{eqnarray}
where $\Lambda$ is an ultraviolet cutoff, as needed since the NJL model is not renormalizable. With $\Lambda^2$ being large, non-trivial solutions to (\ref{M60}) can only exist if $g$ is negative (viz. attractive with our definition of $g$ in (\ref{M58})) and the quantity $-g\Lambda^2/4\pi^2$ is greater than one. One can read this condition as the requirement that $-g>4\pi^2/\Lambda^2$, a condition that would enforce a minimum value for $-g$ since there has to be a finite cutoff in the NJL model, though the required minimum value will however get smaller as $\Lambda$ is made bigger and bigger. Or one can define a new coupling constant $-g^{\prime}=-g\Lambda^2/4\pi^2$ and read (\ref{M60}) as the requirement that $-g^{\prime}$ be greater than one. The NJL gap equation condition given in (\ref{M60}) does not completely parallel the Cooper pair condition given in (\ref{M34}), since the latter condition imposes no minimum value on the coupling constant, requiring only that it be negative. Below we will study a dressed version of the NJL model in which the structure of (\ref{M34}), including its essential singularity at zero coupling, will be recovered.

\begin{figure}[htpb]
\begin{center}
\begin{minipage}[t]{12 cm}
\epsfig{file=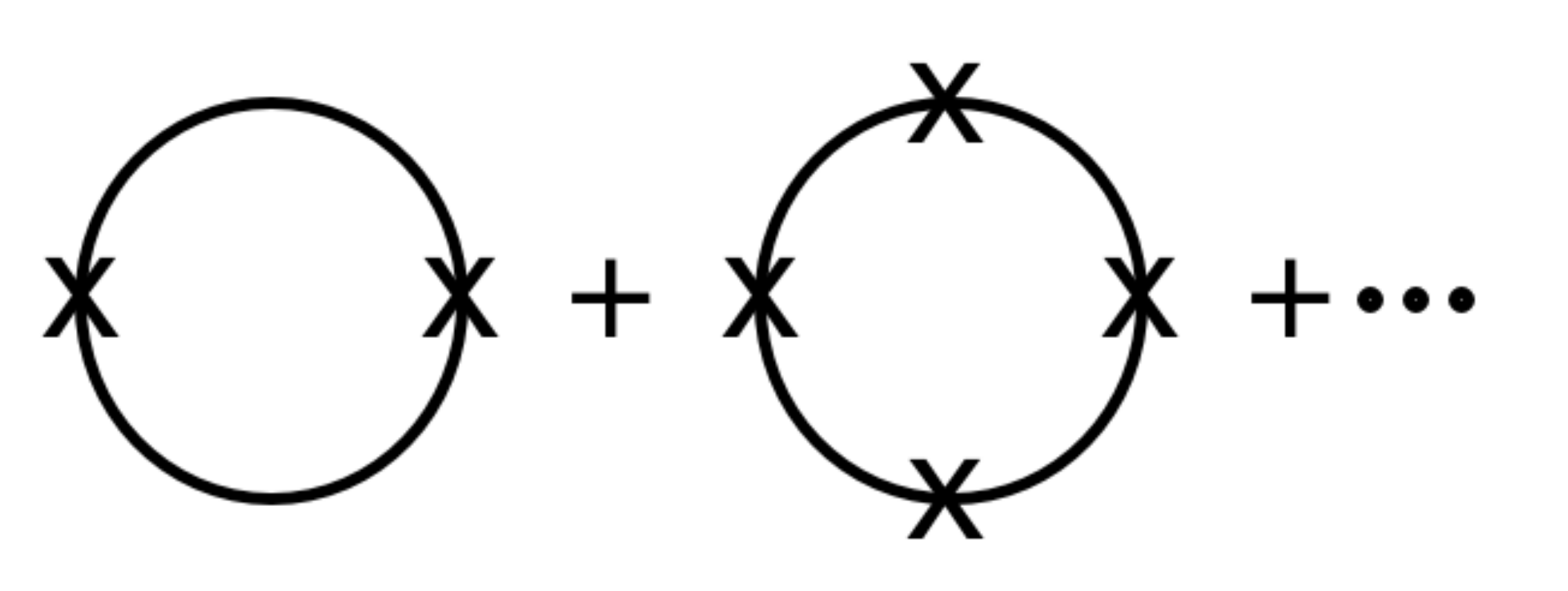,scale=0.5}
\end{minipage}
\begin{minipage}[t]{16.5 cm}
\caption{Vacuum energy density $\epsilon(m)$ via an infinite summation of massless graphs with zero-momentum point $m\bar{\psi}\psi$  insertions.}
\label{lw1}
\end{minipage}
\end{center}
\end{figure}

Given the gap equation (\ref{M60}), for $I_{\rm MF}$ we can calculate the one loop mean-field vacuum energy density difference $\epsilon(m)$ as a function of a constant $m$ via (\ref{M57}) and the graphs of Fig. (\ref{lw1}), a summation in which all massless fermion graphs with an odd number of mass insertions vanish identically. We are able to do the infinite summation in Fig. (\ref{lw1}) analytically, and on incorporating the $m^2/2g$ term in $I_{\rm MF}$ obtain
\begin{eqnarray}
\tilde{\epsilon}(m)&=&\epsilon(m)-\frac{m^2}{2g}=\sum_n\frac{1}{n!}G^{n}_0(q_i=0)m^n-\frac{m^2}{2g}
\nonumber\\
&=&i\int\frac{d^4p}{(2\pi)^4}\sum_{n=1}^{\infty}\frac{(-1)}{2n}
{\rm Tr}\left[(-i)^2\left(\frac{i}{\slashed{ p}+i\epsilon}\right)^2m^2\right]^n-\frac{m^2}{2g}
\nonumber\\
&=&\frac{i}{2}\int \frac{d^4p}{(2\pi)^4}{\rm Tr~ln}\left[1-\frac{m^2}{p^2+i\epsilon}\right]-\frac{m^2}{2g}
\nonumber\\
&=&i\int \frac{d^4p}{(2\pi)^4}\left[{\rm Tr~ln}\left(\slashed{p}-m+i\epsilon\right)-{\rm Tr~ln}\left(\slashed{p}+i\epsilon\right)\right]-\frac{m^2}{2g}
\nonumber\\
&=&-\frac{\Lambda^4}{16\pi^2}{\rm ln}\Lambda^2+\frac{\Lambda^4}{32\pi^2}-\frac{m^2\Lambda^2}{8\pi^2}
+\frac{m^4}{16\pi^2}{\rm ln}\left(\frac{\Lambda^2}{m^2}\right)+\frac{m^4}{32\pi^2}
\nonumber\\
&&-\left[-\frac{\Lambda^4}{16\pi^2}{\rm ln}\Lambda^2+\frac{\Lambda^4}{32\pi^2}\right]+\frac{m^2\Lambda^2}{8\pi^2}-\frac{m^2M^2}{8\pi^2}{\rm ln}\left(\frac{\Lambda^2}{M^2}\right)
\nonumber\\
&=&\frac{m^4}{16\pi^2}{\rm ln}\left(\frac{\Lambda^2}{m^2}\right)-\frac{m^2M^2}{8\pi^2}{\rm ln}\left(\frac{\Lambda^2}{M^2}\right)+\frac{m^4}{32\pi^2}.
\label{M61}
\end{eqnarray}
While the energy density $\langle \Omega_m|\hat{H}_m|\Omega_m\rangle/V=i\int d^4p/(2\pi)^4{\rm Tr~ln}[\gamma^{\mu}p_{\mu}-m]$ of $\hat{H}_m$ has quartic, quadratic and logarithmically divergent pieces, the subtraction of the massless vacuum energy density given as $\langle \Omega_0|\hat{H}_0|\Omega_0\rangle/V=\langle \Omega_0|\hat{H}_m|\Omega_0\rangle/V=i\int d^4p/(2\pi)^4{\rm Tr~ln} [\gamma^{\mu}p_{\mu}]$ removes the quartic divergence, with the subtraction of the self-consistent induced mean-field term $m^2/2g$ then leaving $\tilde{\epsilon}(m)$ only logarithmically divergent. We recognize the resulting logarithmically divergent  $\epsilon(m)$ as having a local maximum at $m=0$, and a global minimum at $m=M$ where $M$ itself is finite.  We thus induce none other than a dynamical double-well Mexican Hat potential, and identify $M$ as the matrix element of a fermion bilinear according to $M/g=\langle S|\bar{\psi}\psi|S\rangle$. In arriving at this result we note the power of dynamical symmetry breaking: it generates a $-m^2/2g$ counterterm automatically, with the quadratic divergence in $\langle \Omega_m|\hat{H}_m|\Omega_m\rangle/V$ being canceled without our needing to introduce a counterterm by hand. This point is particularly significant since for an elementary Higgs field the one loop self-energy contribution is quadratically divergent, to thus naturally be of order some high cutoff scale (the so-called hierarchy problem) rather than the typical weak interaction breaking scale that it is now known to have.

\begin{figure}[htpb]
\begin{center}
\begin{minipage}[t]{17 cm}
\epsfig{file=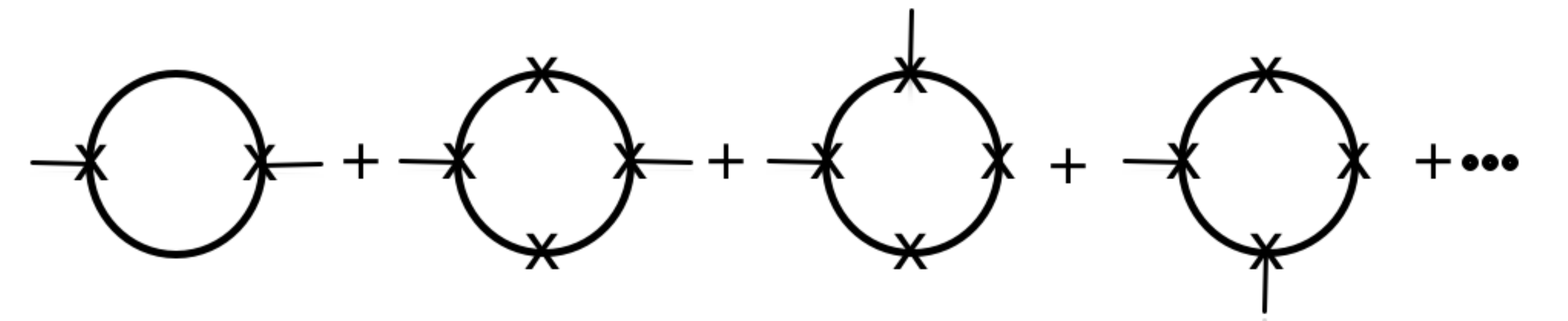,scale=0.5}
\end{minipage}
\begin{minipage}[t]{16.5 cm}
\caption{$\Pi_{\rm S}(q^2,m)$ developed as an infinite summation of massless graphs, each with two point $m\bar{\psi}\psi$ insertions carrying momentum $q_{\mu}$ (shown as external lines), with all other point $m\bar{\psi}\psi$ insertions carrying zero momentum.}
\label{lw5}
\end{minipage}
\end{center}
\end{figure}

If instead of looking at matrix elements in the translationally-invariant vacuum $|S\rangle$ we instead look at matrix elements in coherent states $|C\rangle$  where  $m(x)=\langle C|\bar{\psi}(x)\psi(x)|C\rangle$ is now spacetime dependent,\footnote{Such coherent states can be generated from the self-consistent vacuum $|\Omega_m\rangle$ by a spacetime-dependent Bogoliubov transform  \cite{Bardeen1975}, and lead with elementary scalar field vacuum breaking  \cite{Bardeen1975} or bilinear fermion vacuum breaking \cite{Mannheim1978} to extended, bag-like,  states where a positive energy fermion is localized by its own negative energy sea. (In \cite{Mannheim1978} it was suggested that the bag pressure of such bag-like states could serve as the electrodynamical Poincare stresses mentioned in Sec. (1.2) as now generated dynamically in the vacuum.)  For static, spherically symmetric extended structures $m(x)$ would only depend on the radius and thus be an even function of $x$. With its spatial trace in (\ref{M62}), $I_{\rm EFF}$ can be written as 
$I_{\rm EFF}=i{\rm Tr}{\rm ln}[i\slashed{\partial}_x-m(x)] -i{\rm Tr}{\rm ln}[i\slashed{\partial}_x]$
$=i{\rm Tr}{\rm ln}[i\slashed{\partial}_x-m(x)]/2 -i{\rm Tr}{\rm ln}[i\slashed{\partial}_x]/2$
$+i{\rm Tr}{\rm ln}[-i\slashed{\partial}_x-m(x)]/2-i{\rm Tr}{\rm ln}[-i\slashed{\partial}_x]/2$
$=i{\rm Tr}{\rm ln}[\partial^2_x+m^2(x)]/2 -i{\rm Tr}{\rm ln}[\partial^2_x]/2$. Thus, in analog with (\ref{M12}), in the presence of such extended structures and with $m(x)$ being real, $I_{\rm EFF}$ would be real to all orders in derivatives of $m(x)$. With a constant $m(x)$ also being symmetric, we recover our previous observation that $I_{\rm EFF}$, and thus $\epsilon(m)$, would be real if $m(x)$ is constant.} 
we then obtain \cite{Eguchi1974,Mannheim1976} a mean-field effective action  $I_{\rm EFF}=W(m(x))$ of the form
\begin{eqnarray}
I_{\rm EFF}&=&i{\rm Tr}{\rm ln}\left[\frac{i\slashed{\partial}_x-m(x)}{i\slashed{\partial}_x}\right]
\nonumber\\
&=&\int d^4x\bigg{[}
\frac{1}{8\pi^2}{\rm ln}\left(\frac{\Lambda^2}{M^2}\right)\left(
\frac{1}{2}\partial_{\mu}m(x)\partial^{\mu}m(x)+m^2(x)M^2-\frac{1}{2}m^4(x)\right)+....\bigg{]},
\label{M62}
\end{eqnarray}
where we have explicitly displayed the leading logarithmically divergent part. Here the kinetic energy term is the analog of the $Z(J)$ term given earlier in the expansion of $W(J)$ around the point where all momenta vanish. In terms of the quantity $ \Pi_{\rm S}(q^2,M)$ to be given below in (\ref{M67}), $Z(M)$ is given by $\partial \Pi_{\rm S}(q^2, M)/\partial q^2|_{q^2=0}=(1/8\pi^2){\rm ln}(\Lambda^2/M^2)$.\footnote{As noted in \cite{Mannheim1976}, the full expansion for $I_{\rm EFF}$ involves all higher-order  derivatives of $m(x)$, with each coefficient in the expansion being given by an appropriate derivative of a momentum-space Feynman diagram as calculated with a constant $m$ that is then replaced by $m(x)$ after the integration. $(\Pi_{\rm S}^{\prime\prime}(q^2=0,m))|_{m=m(x)}$ for instance gives the coefficient of $[\Box m(x)]^2$. While such an all-derivative expansion does not violate locality if $m(x)$ is a c-number field since the underlying four-fermion theory that produced it is local, an all-derivative expansion would violate locality  for a q-number scalar field, to thus underscore the distinction between dynamical and elementary Higgs fields.}  If we introduce a coupling $g_{\rm A}\bar{\psi}\gamma_{\mu}\gamma^5A^{\mu}_{5}\psi $ to an axial gauge field $A^{\mu}_{5}(x)$, on setting $\phi(x)=\langle C|\bar{\psi}(1+\gamma^5)\psi|C\rangle$ the effective action becomes 
\begin{eqnarray}
I_{\rm EFF}=\int \frac{d^4x}{8\pi^2}{\rm ln}\left(\frac{\Lambda^2}{M^2}\right)\bigg{[}
\frac{1}{2}|(\partial_{\mu}-2ig_{\rm A}A_{\mu 5})\phi(x)|^2+|\phi(x)|^2M^2
-\frac{1}{2}|\phi(x)|^4-\frac{g_{\rm A}^2}{6}F_{\mu\nu 5}F^{\mu\nu 5}\bigg{]}.
\label{M63}
\end{eqnarray}
We recognize this action as being a double-well Ginzburg-Landau type Higgs Lagrangian, only now generated dynamically. We thus generalize to the relativistic chiral case Gorkov's derivation of the Ginzburg-Landau order parameter action starting from the BCS four-fermion theory. In the $I_{\rm EFF}$ effective action associated with the NJL model there is a double-well Higgs potential, but since the order parameter $m(x)=\langle C|\bar{\psi}(x)\psi(x)|C\rangle$ is a c-number, $m(x)$ does not itself represent a q-number scalar field. And not only that, unlike in the elementary Higgs case, the second derivative of $V(m(x))$ at the minimum where $m=M$ is not the mass of a q-number Higgs boson. Rather, as we now show, the q-number fields are to be found as collective modes generated by the residual interaction, and it is the residual interaction that will fix their masses. Moreover, as we will see in Sec. (8), when we dress the point NJL vertices that are exhibited in Figs. (\ref{baretadpole}), (\ref{lw1}), and (\ref{lw5}), the dynamical Higgs boson will move above the threshold in the fermion-antifermion scattering amplitude, become unstable, and acquire a width. Since this same dressing of the point NJL vertices will lead to a modified effective Ginzburg-Landau $V(m(x))$, and since this $V(m(x))$, and thus its second derivative at its minimum will be real, we see that because of the collective mode Higgs boson width we could not even in principle relate the value of the second derivative of $V(m(x))$ to the collective mode Higgs mass produced by the residual interaction. Since, the (radiatively dressed) Hermitian Higgs potential for an elementary Higgs boson is real, the width of the Higgs boson has the potential to discriminate between an elementary Higgs boson and a dynamical one.

\subsection{The Collective Scalar and Pseudoscalar Tachyon Modes}

To find the collective modes we calculate the scalar and pseudoscalar sector Green's functions $\Pi_{\rm S}(x)=\langle \Omega|T[\bar{\psi}(x)\psi(x)\bar{\psi}(0)\psi(0)]|\Omega\rangle$, $\Pi_{\rm P}(x)=\langle \Omega|T[\bar{\psi}(x)i\gamma^5\psi(x) \bar{\psi}(0)i\gamma^5\psi(0)]|\Omega\rangle$, as is appropriate to a chiral-invariant theory. If first we take the fermion to be massless (i.e. setting $|\Omega\rangle=|\Omega_0\rangle=|N\rangle$ where $\langle N|\bar{\psi}\psi|N\rangle=0$), to one loop order in the four-fermion residual interaction we obtain

\begin{eqnarray}
\Pi_{\rm S}(q^2, m=0)&=&i\int \frac{d^4p}{(2\pi)^4}{\rm Tr}\bigg[
\frac{1}{\slashed{p}+i\epsilon}\frac{1}{\slashed{p}+\slashed{q}+i\epsilon}\bigg]
=-\frac{1}{8\pi^2}\left( 2\Lambda^2+q^2{\rm ln}\left(\frac{\Lambda^2}{-q^2}\right)+q^2\right),
\nonumber\\
\Pi_{\rm P}(q^2, m=0)&=&i\int \frac{d^4p}{(2\pi)^4}{\rm Tr}\bigg[i\gamma^5
\frac{1}{\slashed{p}+i\epsilon}i\gamma^5\frac{1}{\slashed{p}+\slashed{q}+i\epsilon}\bigg]
=-\frac{1}{8\pi^2}\left( 2\Lambda^2+q^2{\rm ln}\left(\frac{\Lambda^2}{-q^2}\right)+q^2\right).
\label{M64}
\end{eqnarray}
On iterating the residual interaction, the scattering matrices in the scalar and pseudoscalar channels are given by 
\begin{eqnarray}
T_{\rm S}(q^2,m=0)&=&\frac{g}{1-g\Pi_{\rm S}(q^2,m=0)}=\frac{1}{g^{-1}-\Pi_{\rm S}(q^2,m=0)},
\nonumber\\
T_{\rm P}(q^2,m=0)&=&\frac{g}{1-g\Pi_{\rm P}(q^2,m=0)}=\frac{1}{g^{-1}-\Pi_{\rm P}(q^2,m=0)}.
\label{M65}
\end{eqnarray}
and with $g^{-1}$ being given by the gap equation (\ref{M60}), near $q^2=-2M^2$ both scattering matrices behave as
\begin{eqnarray}
T_{\rm S}(q^2, M=0)=T_{\rm P}(q^2, M=0)=\frac{Z^{-1}}{(q^2+2M^2)},\qquad
Z=\frac{1}{8\pi^2}{\rm ln}\left(\frac{\Lambda^2}{M^2}\right).
\label{M66}
\end{eqnarray}
We thus obtain degenerate (i.e. chirally-symmetric) scalar and pseudoscalar tachyons at $q^2=-2M^2$ (just like fluctuating around the local maximum in a double-well potential), with $|N\rangle$ thus being unstable.

\subsection{The Collective Goldstone and Higgs Modes}

However, suppose we now take the fermion to have nonzero mass $M$ (i.e. we set $|\Omega\rangle=|\Omega_m\rangle=|S\rangle$ where $\langle S|\bar{\psi}\psi|S\rangle\neq 0$). As per the summation for $\Pi_{\rm S}(q^2,m)$ given in Fig. (\ref{lw5}) and its $\Pi_{\rm P}(q^2,m)$ analog,  we obtain
\begin{eqnarray}
\Pi_{\rm P}(q^2, M)&=&-i\int \frac{d^4p}{(2\pi)^4}{\rm Tr}\bigg[i\gamma^5
\frac{1}{\slashed{p}-m+i\epsilon}i\gamma^5\frac{1}{\slashed{p}+\slashed{q}-m+i\epsilon}\bigg]
\nonumber\\
&=&-\frac{\Lambda^2}{4\pi^2}
+\frac{M^2}{4\pi^2}{\rm ln}\left(\frac{\Lambda^2}{M^2}\right) 
-\frac{q^2}{8\pi^2}{\rm ln}\left(\frac{\Lambda^2}{M^2}\right) 
-\frac{(q^2-4M^2)}{8\pi^2}
\nonumber\\
&-&\frac{(8M^4-8M^2q^2+q^4)}{8\pi^2 q^2}\left(\frac{-q^2}{4M^2-q^2}\right)^{1/2}{\rm ln}\left(\frac{(4M^2-q^2)^{1/2}+(-q^2)^{1/2}}{(4M^2-q^2)^{1/2}-(-q^2)^{1/2}}\right),
\nonumber\\
\Pi_{\rm S}(q^2, M)&=&-i\int \frac{d^4p}{(2\pi)^4}{\rm Tr}\bigg[
\frac{1}{\slashed{p}-m+i\epsilon}\frac{1}{\slashed{p}+\slashed{q}-m+i\epsilon}\bigg]
\nonumber\\
&=&-\frac{\Lambda^2}{4\pi^2}
+\frac{M^2}{4\pi^2}{\rm ln}\left(\frac{\Lambda^2}{M^2}\right)
+\frac{(4M^2-q^2)}{8\pi^2}{\rm ln}\left(\frac{\Lambda^2}{M^2}\right) 
+\frac{(4M^2-q^2)}{8\pi^2}
\nonumber\\
&-&\frac{(4M^2-q^2)}{8\pi^2}\left(\frac{4M^2-q^2}{-q^2}\right)^{1/2}{\rm ln}\left(\frac{(4M^2-q^2)^{1/2}+(-q^2)^{1/2}}{(4M^2-q^2)^{1/2}-(-q^2)^{1/2}}\right),
\label{M67}
\end{eqnarray}
with both of these Green's functions having a branch point at $q^2=4M^2$, which for a fermion of mass $M$, is right at the threshold in the fermion-antifermion scattering amplitude. Given the form for $g^{-1}$,  we find a dynamical pseudoscalar Goldstone boson bound state in  $T_{\rm P}(q^2,M)$ at $q^2=0$ and a  dynamical scalar Higgs boson bound state in  $T_{\rm S}(q^2,M)$ at $q^2=4M^2$ ($=-2\times  M^2({\rm tachyon})$).\footnote{As noted in \cite{Jackiw1973,Cornwall1973}, the Higgs mechanism that holds for an elementary Higgs field generalizes to the dynamical case, with a dynamical Goldstone boson automatically giving a mass to the gauge boson associated with any local current to which the Goldstone boson may couple. It will thus suffice for our purposes here to show that there is a dynamical Goldstone boson to begin with.} Near the respective poles the scattering amplitudes behave as: 
\begin{eqnarray}
T_{\rm S}(q^2, M)=\frac{Z^{-1}_{\rm S}}{(q^2-4M^2)},~~~~T_{\rm P}(q^2, M)=\frac{Z^{-1}_{\rm P}}{q^2},\qquad
Z_{\rm S}=Z_{\rm P}=\frac{1}{8\pi^2}{\rm ln}\left(\frac{\Lambda^2}{M^2}\right).
\label{M68}
\end{eqnarray}
\noindent
The two dynamical bound states are not degenerate in mass (spontaneously broken chiral symmetry), and the dynamical Higgs  scalar mass $2M$ is twice the induced mass of the fermion, to thus lie right at the scattering amplitude threshold. 

Now stationary solutions to $I_{\rm EFF}$ given in (\ref{M62}) obey
\begin{eqnarray}
\Box m(x)-2m(x)M^2+2m^3(x)=0.
\label{M69}
\end{eqnarray}
We can associate this wave equation with a potential
\begin{eqnarray}
V(m)=\frac{1}{2}m^4(x)-m^2(x)M^2,
\label{M70}
\end{eqnarray}
with its first derivative vanishing at $m=M$, and its second derivative being the positive $V^{\prime\prime}(M)=4M^2$ at $m=M$. With fluctuations around $m=M$ taking the form $m=M+\chi$, for such fluctuations the potential takes the form:
\begin{eqnarray}
V(M+\chi)=-\frac{1}{2}M^4+\frac{1}{2}(4M^2)\chi^2+2M\chi^3+\frac{1}{2}\chi^4.
\label{M71}
\end{eqnarray}
to thus describe a scalar fluctuation with positive mass squared equal to $4M^2$. Now while this $4M^2$ mass squared  happens to be equal to the dynamical Higgs boson mass squared as given in (\ref{M68}), such an equality is not generic, since, as we see below,  once one dresses the vertices these squared masses do not remain equal. Thus one cannot use a dressed Ginzburg-Landau action to determine the Higgs mass, one must use the dressed scattering amplitude. 

A very significant aspect of (\ref{M68}) is that the pole in $T_{\rm S}(q^2, M)$ is at $q^2=4M^2$, i.e. at a location that does not depend on the cutoff $\Lambda$ that the NJL model needs. Thus when the Higgs boson is dynamical its mass can naturally be of order the dynamical fermion mass $M$ and not be of order the cutoff scale, as both the fermion and Higgs boson masses are generated by one and the same chiral symmetry breaking mechanism. This is very different from the elementary Higgs case where the Higgs boson self-energy is of order the cutoff scale, and there is no relation between the Higgs boson mass scale and the masses of fermions that arise through their Yukawa couplings to the self-same Higgs boson. 

A second significant feature of (\ref{M68}) is that the residue at the pole in $T_{\rm S}(q^2, M)$, and thus the coupling of a dynamical Higgs boson to a fermion-antifermion pair, is completely determined by the dynamics. This is in sharp contrast to the elementary Higgs boson situation where fundamental Yukawa couplings are totally unconstrained. One of the central shortcomings of the standard elementary Higgs field theory is that the tree approximation minimum to the Higgs potential is determined by the Higgs sector alone, with the Yukawa-coupled fermions playing no role at the tree level. And since the fermions do play no role,  the strengths of the Yukawa coupling terms are totally unconstrained. However, in the dynamical Higgs case it is the interaction of the fermions that determines where the Higgs bound state pole is to lie, while at the same time determining its residue, i.e. its coupling to a fermion-antifermion pair. Thus if one wants to be able to determine the strengths of the Yukawa coupling terms, the fermions would have to play an explicit role in generating the Higgs boson in the first place. 

As a model the NJL model is very instructive since it captures the key features of dynamical symmetry breaking. However its drawback is that the NJL model is not renormalizable. We thus now turn to a discussion of dynamical symmetry breaking in some specific theories that are renormalizable.

\section{The Abelian Gluon Model}

\subsection{The Schwinger-Dyson Equation}

\begin{figure}[htpb]
\vskip-0.2in
\begin{center}
\begin{minipage}[t]{14 cm}
\epsfig{file=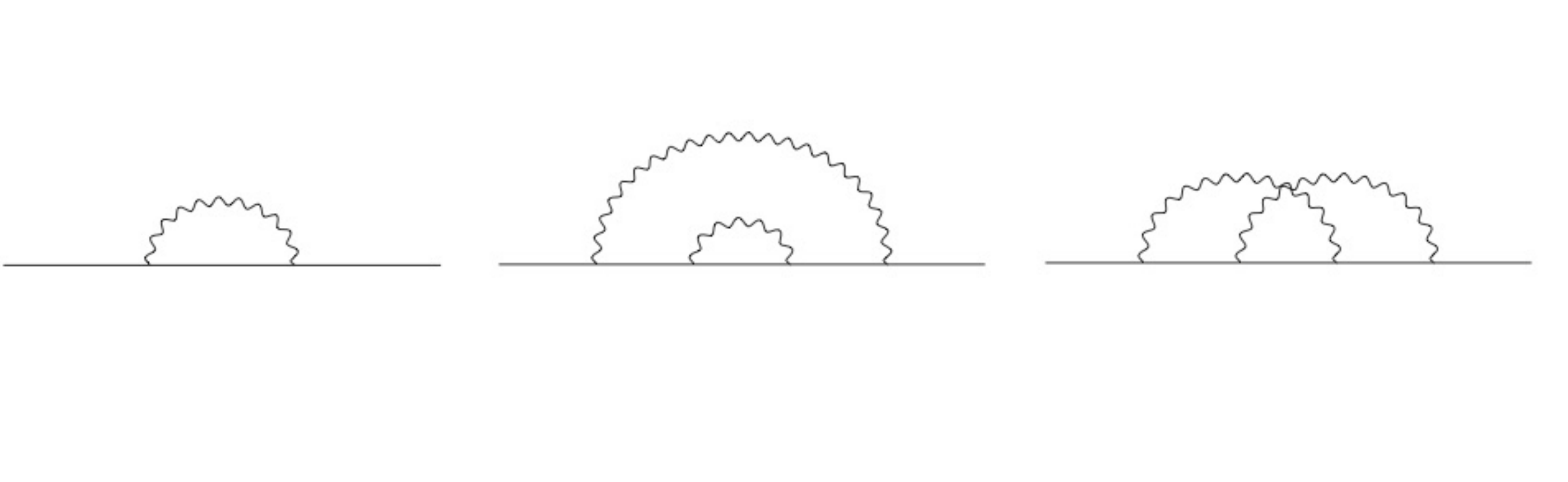,scale=0.5}
\end{minipage}
\vskip-0.6in
\begin{minipage}[t]{16.5 cm}
\caption{The first few graphs in the fermion self-energy Schwinger-Dyson equation}
\label{SD1}
\end{minipage}
\end{center}
\end{figure}

The Abelian gluon model is based on the same action as QED, viz.
\begin{eqnarray}
I_{\rm QED}&=&\int d^4x\left[-\frac{1}{4}F_{\mu\nu}F^{\mu\nu} + \bar{\psi}\gamma^{\mu}(i\partial_{\mu}-e_0A_{\mu})\psi -m_0\bar{\psi}\psi\right],
\label{M72}
\end{eqnarray}
but without the requirement that $e_0$ necessarily be the bare electric charge. This permits consideration of both strong and weak coupling. Though we are interested in dynamical mass generation we have included a bare mass term in (\ref{M72}) as this will enable us to monitor the nature of the mass generation. In the theory the exact inverse fermion propagator $S^{-1}(p)=\slashed{p}-m_0-\Sigma(p)$ obeys the exact, as yet unrenormalized, all-order, Schwinger-Dyson equation
\begin{eqnarray}
\Sigma(p)=ie_0^2\int \frac{d^4q}{(2\pi)^4} D_{\mu\nu}(p-q)\Gamma^{\mu}(p,q)S(q)\gamma^{\nu},
\label{M73}
\end{eqnarray}
where $D_{\mu\nu}(p-q)$ and $\Gamma^{\mu}(p,q)$ are the exact gluon propagator and exact fermion-antifermion-gluon vertex function. Starting with the work of Johnson, Baker, and Willey \cite{Johnson1964,Johnson1967,Baker1969,Baker1971a,Baker1971b,Johnson1973} there has been much interest in finding self-consistent solutions to this Schwinger-Dyson equation. We shall discuss the work of Johnson, Baker, and Willey below but shall first discuss the quenched, planar approximation treatment since it is very instructive. This approximation is referred to as being quenched because the gluon propagator is taken to be the bare propagator (i.e. no charge renormalization), and is planar since the only gluon exchange diagrams that are taken to contribute to the fermion propagator are those in which no two gluon lines cross each other. (Thus one keeps graphs such as the first two graphs in Fig. (\ref{SD1}) and their higher order analogs but not those such as the third graph or its higher order analogs.) The approximation is also called the ladder approximation or the rainbow graph approximation because of the pictorial form of the graphs. Some early studies of the quenched planar approximation may be found in \cite{Maskawa1974,Maskawa1975,Fukuda1976,Fomin1978}, with detailed reviews and full references being provided in  \cite{Miransky1993} and \cite{Yamawaki2010}. 

In general because of its Lorentz structure one can set $S^{-1}(p)=A(p^2)\slashed{p}-B(p^2)$ where $A(p^2)$ and $B(p^2)$ are Lorentz scalars. In the quenched planar approximation it is convenient to initially work in a general  covariant gauge in which $D_{\mu\nu}(k)=(\xi k_{\mu}k_{\nu}/k^2-\eta_{\mu\nu})/k^2$ where $\xi$ is a constant. Then, since  $\Gamma^{\mu}(p,q)$ is equal to $ \gamma^{\mu}$ in this approximation, the Schwinger-Dyson equation takes the form
\begin{eqnarray}
\slashed{p}[1-A(p^2)]+B(p^2)=m_0&+&ie^2\int \frac{d^4q}{(2\pi)^4} \frac{(\xi-4)B(q^2)}{(p-q)^2[A^2(q^2)q^2-B(q^2)]}
\nonumber\\
&+&ie^2\int \frac{d^4q}{(2\pi)^4} \frac{A(q^2)[2\xi(\slashed{p}-\slashed{q})(p\cdot q-q^2)+(2-\xi)\slashed{q}(p-q)^2]}{(p-q)^4[A^2(q^2)q^2-B(q^2)]}.
\label{M74}
\end{eqnarray}
In (\ref{M74}) we have replaced the bare charge $e_0$ by the physical charge $e$ since there is no charge renormalization. 

On transforming to Euclidean space, we introduce angular coordinates $q_1=q\sin\psi\sin\theta\cos\phi$, $q_2=q\sin\psi\sin\theta\sin\phi$, $q_3=q\sin\psi\cos\theta$, $q_4=q\cos\psi$, where $0\leq \psi\leq \pi$, $0\leq \theta \leq \pi$, $0\leq \phi \leq 2\pi$, and where the volume element is $dq d\psi d\theta d\phi q^3\sin^2\psi \sin\theta $. For the dependence on $p_{\mu}$ we conveniently set $p_{\mu}=(0,0,0,p)$. Then, with $p\cdot q=pq\cos\psi$, $(p-q)^2=p^2-2pq\cos\psi+q^2$, we find that on the right-hand side of (\ref{M74}) the only dependence on $\theta$ and $\phi$ is in the $\gamma^1q_1$, $\gamma^2q_2$, and $\gamma^3q_3$ terms. On doing the $d\theta d\phi$ integration all of the integrals that contain these terms are found to vanish identically, doing so in fact for arbitrary $\xi$. For arbitrary $\xi$ the term proportional to $\gamma^4$ does not vanish on doing the remaining angular integration  viz. that on $\psi$, but does conveniently vanish in the Landau gauge where $\xi=1$.\footnote{To show this when $\xi=1$, we note that $2(p_4-q_4)(p\cdot q-q^2)+q_4(p-q)^2=-2pq^2+3q(p^2+q^2)\cos\psi-4pq^2\cos^2\psi$. On introducing $I_0=(1/\pi)\int_0^{\pi}d\psi/(p^2+q^2-2pq\cos\psi)=1/|p^2-q^2|$, 
$I_n=(1/\pi)\int_0^{\pi}d\psi \sin^2\psi\cos^{n-1}\psi/(p^2+q^2-2pq\cos\psi)^2$ with
$I_1=[(p^2+q^2)I_0-1]/4p^2q^2$, 
$I_2=[2(p^2+q^2)^2I_0-4p^2q^2I_0-2p^2-2q^2]/8p^3q^3$, 
$I_3=[3(p^2+q^2)^3I_0-8p^2q^2(p^2+q^2)I_0+2p^2q^2-3(p^2+q^2)^2]/16p^4q^4$, we find that $-2pq^2I_1+3q(p^2+q^2)I_2-4pq^2I_3=0$. With this cancellation occurring identically without needing to specify any actual form for $I_0$, and with $I_0$ actually behaving as $1/|p^2-q^2|$, the cancellation occurs in (\ref{M74}) for all values of the integration variable $q$, i.e. for $q$ greater than, equal to, or lesser than $p$.} With there then being no dependence on $A(p^2)$ in the numerator of the second integral on the right-hand side of (\ref{M74}), from the structure of the left-hand side we conclude that in the Landau gauge $A(p^2)=1$ \cite{Johnson1964}. In the Landau gauge (\ref{M74}) thus takes the form  
\begin{eqnarray}
B(p^2)&=&m_0+\frac{3e^2}{8\pi^3}\int_0^{\infty} q^2dq^2 \int_0^{\pi}d\psi \sin^2\psi \frac{B(q^2)}{(p^2+q^2-2pq\cos\psi)[q^2+B^2(q^2)]}
\nonumber\\
&=&m_0+\frac{3e^2}{8\pi^3}\int_0^{\infty} q^2dq^2 \frac{B(q^2)}{[q^2+B^2(q^2)]}\frac{\pi}{4p^2q^2}\left(p^2+q^2-\frac{(p^2-q^2)^2}{|p^2-q^2|}\right),
\label{M75}
\end{eqnarray}
to thus take the form in which it commonly appears in the quenched ladder approximation literature, viz. 
\begin{eqnarray}
B(p^2)=m_0+\frac{3\alpha}{4\pi}\bigg{[}\int_0^{p^2}dq^2 \frac{q^2B(q^2)}{p^2[q^2+B^2(q^2)]}
+\int_{p^2}^{\infty}dq^2 \frac{B(q^2)}{[q^2+B^2(q^2)]}\bigg{]},
\label{M76}
\end{eqnarray}
where $\alpha=e^2/4\pi$.

On cutting off the $q^2$ integration at $\Lambda^2$, a convergent asymptotic solution  of the form $B(p^2)=m(p^2/m^2)^{(\nu-1)/2}$  is found for (\ref{M76}) with 
\begin{eqnarray}
\nu=\pm\left(1-\frac{3\alpha}{\pi}\right)^{1/2},\qquad m_0=\frac{3\alpha m}{2\pi(1-\nu)}\frac{\Lambda^{\nu-1}}{m^{\nu-1}},
\label{M77}
\end{eqnarray}
with $\nu$ being real provided $\alpha \leq \pi/3$.\footnote{The power solution to (\ref{M75}) of the form $B(p^2)=(p^2)^{(\nu-1)/2}$ with $\nu$ as given in (\ref{M77}) was first presented  in \cite{Johnson1964}.} With $\nu$ having a branch point at $\alpha=\pi/3$, we can thus anticipate that a phase transition might occur at that value of the coupling constant. While the above solution gives a real $\nu$ if $\alpha \leq \pi/3$, if $\alpha>\pi/3$ we would instead obtain $B(p^2)\sim (p^2)^{-1/2}\exp(\pm i(\mu/2){\rm ln}(p^2/m^2))$ where $\mu=(3\alpha/\pi-1)^{1/2}$. This then gives two classes of real solutions, viz.  $(p^2)^{-1/2}\cos[(\mu/2){\rm ln}(p^2/m^2)]$ and $(p^2)^{-1/2}\sin[(\mu/2){\rm ln}(p^2/m^2)]$. Combining them gives the $\alpha> \pi/3$ asymptotic solution
\begin{eqnarray}
B(p^2)=\frac{m\cos[((3\alpha/\pi-1)^{1/2}/2){\rm ln}(p^2/m^2)+\sigma]}{(p^2/m^2)^{1/2}},\qquad m_0=-\frac{3m\alpha\cos[(3\alpha/\pi-1)^{1/2}{\rm ln}(\Lambda/m)+\tau]}{2\pi(\mu^2+1)^{1/2}(\Lambda/m)},
\label{M78}
\end{eqnarray}
where $\sigma$  is a  (possibly $\Lambda^2/m^2$ but not $p^2/m^2$ dependent) phase and $\tau=\sigma+{\rm arctan}\mu$. As required, we see that for both $\alpha \leq \pi/3$ and $\alpha>\pi/3$ the bare mass vanishes in the limit in which the cutoff goes to infinity. However, as we elaborate on below, that does not mean that the bare mass is identically zero, only that it vanishes in the limit of infinite cutoff. For  $\alpha \leq \pi/3$ this is the only option for the bare mass. However for  $\alpha>\pi/3$ there is a second option for the bare mass, since it  will vanish identically  if we set 
\begin{eqnarray}
\left(\frac{3\alpha}{\pi}-1\right)^{1/2}{\rm ln}(\Lambda/m)+\tau=\frac{\pi}{2}.
\label{M79}
\end{eqnarray}
Now  initially this would suggest that as we let $\Lambda$ go to infinity, the only allowed value for $\alpha$ would be $\alpha=\pi/3$. In order to be able to obtain a solution that is to hold for all $\alpha>\pi/3$, we take $\tau$ to be of the form $\tau=\delta{\rm ln}(\Lambda/m)$ where $\delta$ is finite, so that  we obtain 
\begin{eqnarray}
\left(\frac{3\alpha}{\pi}-1\right)^{1/2}+\delta=\frac{\pi}{2{\rm ln}(\Lambda/m)},
\label{M80}
\end{eqnarray}
and thus
\begin{eqnarray}
B(p^2)=\frac{m\cos[((3\alpha/\pi-1)^{1/2}/2){\rm ln}(p^2/\Lambda^2)-{\rm arctan}\mu+\pi/2]}{(p^2/m^2)^{1/2}},\qquad m_0=0.
\label{M81}
\end{eqnarray}
Then, with $\delta$ being an appropriately chosen function of $\alpha$, all values of $\alpha$ greater than $\pi/3$ are allowed in the limit of infinite cutoff.\footnote{A ${\rm ln}(\Lambda/m)$ dependence to the phase $\tau$ seems not to have been considered in the quenched ladder approximation literature, where instead one restricts  \cite{Miransky1985} to $(3\alpha/\pi-1)^{1/2}=\pi/2{\rm ln}(\Lambda/m)$, a quantity that vanishes in the limit of infinite cutoff, to then not permit  $\alpha$ to take any value other than $\pi/3$.} Since for all such values of $\alpha$ the bare mass is identically zero, for $\alpha >\pi/3$ dynamical symmetry breaking will take place for any non-trivial solution to
\begin{eqnarray}
B(p^2)=3e^2\int \frac{d^4q}{(2\pi)^4} \frac{B(q^2)}{(p-q)^2(q^2+B^2(q^2))}
\label{M82}
\end{eqnarray}
that behaves asymptotically as in (\ref{M81}).\footnote{We note that while the asymptotic solution given for $B(p^2)$ in (\ref{M77}) is cutoff independent, the solution given in (\ref{M81}) does depend on the cutoff (as would the solution for $B(p^2)$ given in (\ref{M78}) if we were to set $(3\alpha/\pi-1)^{1/2}=\pi/2{\rm ln}(\Lambda/m)$). Consequently, solutions to the quenched ladder approximation Schwinger-Dyson equation can only be cutoff independent if $\alpha \leq \pi/3$.} Thus for all values of $\alpha$ greater than $\pi/3$ (i.e. strong coupling), the quenched ladder approximation has a chiral symmetry that is broken dynamically, with dynamical Goldstone boson generation taking place \cite{Maskawa1974,Maskawa1975,Fukuda1976,Fomin1978}. With the asymptotic solution possessing  a branch point at $\alpha=\pi/3$, we can anticipate that the behavior of the theory below $\alpha=\pi/3$ will be qualitatively different  from its behavior above $\alpha=\pi/3$. Thus to discuss what happens when $\alpha$ is below $\pi/3$ we turn to a renormalization  group analysis.

\subsection{Renormalization Group Analysis}

In the Abelian gluon model it was shown  \cite{Adler1971,Adler1972} that  the  renormalized inverse fermion propagator $\tilde{S}^{-1}(p,m)$ and the renormalized vertex function $\tilde{\Gamma}_{\rm S}(p,p,0,m)$ associated with the insertion of  the composite operator $\theta=\bar{\psi}\psi$ with zero momentum into the inverse fermion propagator are related by the renormalization group equation
\begin{eqnarray}
\left[m\frac{\partial}{m}+\beta(\alpha)\frac{\partial}{\partial \alpha}\right]\tilde{S}^{-1}(p,m)=-m[1-\gamma_{\theta}(\alpha)]\tilde{\Gamma}_{\rm S}(p,p,0,m)
\label{M83}
\end{eqnarray}
in the limit in which the fermion momentum $p_{\mu}$ is deep Euclidean. In (\ref{M83}) $\beta(\alpha)$ is associated with coupling constant renormalization, and $\gamma_{\theta}(\alpha)$ is the anomalous dimension associated with the operator  
$\bar{\psi}\psi$, defined here so that the total dimension of $\bar{\psi}\psi$ is given by $d_{\theta}(\alpha)=3+\gamma_{\theta}(\alpha)$ where $3$ is the canonical value.\footnote{A possible  anomalous dimension $\gamma_{\rm F}$ term for the fermion has been removed from (\ref{M83}) via a judicious choice of gauge. Also we note that in the critical scaling literature the dimension and anomalous dimension of $d_{\theta}(\alpha)$ are sometimes denoted by $d_m$ and $\gamma_m$, and defined via $d_m=3-\gamma_{m}$.} In any situation in which $\beta(\alpha)=0$, this equation admits of an exact asymptotic solution
\begin{eqnarray}
\tilde{S}^{-1}(p,m)&=& \slashed{ p}-m\left(\frac{-p^2-i\epsilon}{m^2}\right)^{\gamma_{\theta}(\alpha)/2}+i\epsilon,\qquad\tilde{\Gamma}_{\rm S}(p,p,0,m)=\left(\frac{-p^2-i\epsilon}{m^2}\right)^{\gamma_{\theta}(\alpha)/2}, 
\label{M84}
\end{eqnarray}
one that converges asymptotically if $\gamma_{\theta}(\alpha)$ is negative. Asymptotic convergence is thus achieved if dynamically $d_{\theta}(\alpha)$ is less than canonical.

There are two standard ways to achieve  $\beta(\alpha)=0$. The first is to only have $\beta(\alpha)$ vanish for some specific value of $\alpha$, the Gell-Mann-Low eigenvalue condition. It is this particular possibility that was explored by Johnson, Baker, and Willey, and will be discussed below.  The second  is to use the quenched approximation described above since then the coupling constant is not renormalized for any value of $\alpha$, with $\beta(\alpha)$ vanishing for every value of $\alpha$.  Comparing (\ref{M83}) with (\ref{M77}), we see that the asymptotic solution found in the quenched ladder approximation  exactly scales as a power in the $\alpha\leq \pi/3$ region just as the renormalization group requires, to thus yield the identification $\nu-1=\gamma_{\theta}(\alpha)$. At the critical value $\alpha=\pi/3$ we note that $\gamma_{\theta}(\alpha)=-1$, so that $d_{\theta}(\alpha)=2$. We shall have occasion to return to this value for $d_{\theta}(\alpha)$ below while noting now that with it $(\bar{\psi}\psi)^2$ acts as an operator whose dimension is reduced from six to four, i.e. to the value that would make the four-fermion interaction be power-counting renormalizable. 

We can also use the renormalization group equation in the $\alpha>\pi/3$ region, and inserting either $B(p^2)\sim (p^2)^{-1/2}\exp(+i(\mu/2){\rm ln}(p^2/m^2))$ or $B(p^2)\sim (p^2)^{-1/2}\exp(-i(\mu/2){\rm ln}(p^2/m^2))$ into (\ref{M83}) yields
\begin{eqnarray}
{\rm Re}[\gamma_{\theta}(\alpha>\pi/3)]=-1,
\label{M85}
\end{eqnarray}
with the real part of $\gamma_{\theta}(\alpha)$, viz. the part that controls the asymptotic behavior of the propagator,  thus being equal to $-1$ for all $\alpha>\pi/3$.\footnote{As noted above, in \cite{Miransky1985}  (\ref{M79}) was solved with finite $\tau$, to thus lead to $(3\alpha/\pi-1)^{1/2}=\pi/2{\rm ln}\Lambda$. On defining an effective beta function according to $\beta(\alpha)=\Lambda\partial_{\Lambda}\alpha(\Lambda)$, this particular $\beta(\alpha)$ evaluates (see e.g. \cite{Leung1986}) to $\beta(\alpha)=-(2/3)(3\alpha/\pi-1)^{3/2}$, to thus permit use of the $\alpha>\pi/3$ region renormalization group analysis given in studies such as that of \cite{Yamawaki1986}. However, as noted in \cite{Leung1986}, whatever this effective beta function is, it cannot be associated with charge renormalization, since there is no charge renormalization in the quenched ladder approximation even in the $\alpha>\pi/3$ region. Rather, as we indicate here,  one should stay with the vanishing of the standard charge renormalization beta function, and then use the standard renormalization group equation given as (\ref{M83}) in the $\alpha>\pi/3$ region. Curiously though, we note that standard renormalization group procedure and that presented in \cite{Yamawaki1986} both lead to ${\rm Re}[\gamma_{\theta}(\alpha>\pi/3)]=-1$ in the entire $\alpha>\pi/3$ region, though in the discussion presented in \cite{Yamawaki1986} one additionally has ${\rm Im}[\gamma_{\theta}(\alpha>\pi/3)]=0$.} In this respect the phase transition at $\alpha=\pi/3$ is one in which ${\rm Re}[\gamma_{\theta}(\alpha)]$ varies continuously from zero to one as $\alpha$ varies continuously from zero to $\pi/3$, but then stays at one when $\alpha$ is greater than $\pi/3$, with there thus being a discontinuity in $\gamma_{\theta}(\alpha)$ at $\alpha=\pi/3$, even as there is none in $\beta(\alpha)$. One should expect a discontinuity in $\gamma_{\theta}(\alpha)$ rather in $\beta(\alpha)$, since it is $\theta=\bar{\psi}\psi$ that is the mass operator. Thus if any discontinuity is to herald a dynamical mass generation phase transition, it should be in the mass operator sector rather than in the charge operator sector.

\subsection{Johnson-Baker-Willey Electrodynamics}

The objective of the study of Johnson, Baker, and Willey \cite{Johnson1964,Johnson1967,Baker1969,Baker1971a,Baker1971b,Johnson1973} was not to study dynamical symmetry breaking per se, but to determine whether it might be possible for all the renormalization constants of a quantum field theory to be finite. Quantum electrodynamics was a particularly convenient theory to study since its gauge structure meant that two of its renormalization constants (the fermion wave function renormalization constant $Z_2$ and the fermion-antifermion-gauge boson vertex renormalization constant $Z_1$ to which $Z_2$  is equal) were gauge dependent and could be made finite by an appropriate choice of gauge, with the anomalous dimension of the fermion $\gamma_{\rm F}$ associated with $Z_2$ consequently then being zero. Johnson, Baker, and Willey were thus left with the gauge boson wave function renormalization constant $Z_3$ and the fermion bare mass $m_0$ and its shift $\delta m$ to address. 

Now if one were also to consider the coupling of electrodynamics to gravity, one would then have to address another infinity that electrodynamics possesses, namely that of the zero-point vacuum energy density, and we will return to this issue below. However, in the flat spacetime study that  Johnson, Baker, and Willey engaged in, the need to address the vacuum energy density infinity did not arise, and it could be normal ordered away. 

As regards $Z_3$ and $m_0$, Johnson, Baker and Willey showed that $Z_3$ would be finite if the fermion-antifermion-gauge boson coupling constant $\alpha$ was at a solution to the Gell-Mann-Low eigenvalue condition. At this eigenvalue they showed that the fermion propagator would scale asymptotically as in (\ref{M84}), and that the bare mass would scale as 
\begin{eqnarray}
m_0=m\left(\frac{\Lambda^2}{m^2}\right)^{\gamma_{\theta}(\alpha)/2}, 
\label{M86}
\end{eqnarray}
where $\Lambda$ is an ultraviolet cutoff and $m=m_0+\delta m$ is the renormalized fermion mass. Consequently if the power $\gamma_{\theta}(\alpha)$ is negative (which it perturbatively is: $\gamma_{\theta}(\alpha)=-3\alpha/2\pi-3\alpha^2/16\pi^2+{\rm O}(\alpha^3)$), the bare mass would vanish in the limit of infinite cutoff and $\delta m$ would be finite. Thus the mechanism for both the finiteness and vanishing of $m_0$ is to have the dimension of $\bar{\psi}\psi$ be less than canonical.\footnote{If we expand $m_0$ in (\ref{M86}) we obtain $m_0/m=1+\gamma_{\theta}(\alpha) {\rm ln} (\Lambda/m)+(\gamma_{\theta}^2(\alpha)/2){\rm ln}^2(\Lambda/m)+...$. Even though each term in the series diverges, their coefficients are fixed by the vanishing of $\beta(\alpha)$ so that the series exponentiates.} 
As such, the work of Johnson, Baker, and Willey was quite remarkable since it predated the work of Wilson and of Callan and Symanzik on critical scaling, anomalous dimensions, and the renormalization group. 

\subsection{The Baker-Johnson Evasion of the Goldstone Theorem}

With the vanishing of the bare mass and the non-vanishing of the physical mass, it looks as though there should be dynamical mass generation and an associated  Goldstone boson, with JBW electrodynamics then becoming a possible laboratory in which to explore dynamical symmetry breaking. However, this turned out not to be the case due to a hidden renormalization effect in the theory \cite{Baker1971a}, one associated with the renormalization constant $Z^{-1/2}_{\theta}= (\Lambda^2/m^2)^{\gamma_{\theta}(\alpha)/2}$ that renormalizes $\bar{\psi}\psi$ according to $Z^{-1/2}_{\theta}(\bar{\psi}\psi)_0=\bar{\psi}\psi$  \cite{Adler1971}. Specifically,  given (\ref{M86}), it follows that as $m_0$ vanishes the quantity $(\bar{\psi}\psi)_0$ diverges at the same rate so that $m_0(\bar{\psi}\psi)_0=mZ^{-1/2}_{\theta}Z^{1/2}_{\theta}\bar{\psi}\psi=m\bar{\psi}\psi$ is finite. Then if $m$ is non-zero,  the bare $m_0(\bar{\psi}\psi)_0$ is non-zero too. The solution associated with (\ref{M86}) thus corresponds to a theory in which a term $m_0(\bar{\psi}\psi)_0$ is present in the Lagrangian from the outset. The chiral symmetry is thus broken in the Lagrangian and there is no associated Goldstone boson. This then is the evasion of the Goldstone theorem as found by Baker and Johnson \cite{Baker1971a}.

Since the Baker-Johnson evasion of the Goldstone theorem would hold in any theory in which the bare mass has a negative power dependence on the cutoff, it follows that this must also be the case in the quenched ladder approximation to the Abelian gluon model in the  $\alpha \leq \pi/3$ region. Hence the branch point in $\nu$ at $\alpha=\pi/3$ separates two distinct phases. At or below $\alpha=\pi/3$ there is no Goldstone boson, while above $\alpha=\pi/3$ there is. Now suppose that a priori one did not know whether or not the Baker-Johnson evasion of the Goldstone theorem applied below $\alpha=\pi/3$. If one had set the bare mass equal to zero in an equation such as (\ref{M74}) right at the beginning of the calculation, one would have come to the conclusion that with an identically zero $m_0$ and the power-behaved solution given in (\ref{M77}), equation (\ref{M74}) would exist without renormalization and lead to self-consistent mass generation. One would thus have concluded that even in the $\alpha \leq \pi/3$ region there should be a Goldstone boson, with a self-consistently non-zero $B(p^2)$, and a self-consistently non-zero fermion mass,  generating themselves in the Schwinger-Dyson equation.  However, if we first introduce both a bare mass and a cutoff and then solve (\ref{M74}) asymptotically, we would find that for $\alpha\leq \pi/3$ the bare mass as given in (\ref{M77}) would actually be non-zero. Then with $m_0$ as given in (\ref{M77}) vanishing in the limit of infinite cutoff we would know that we are in fact in the Baker-Johnson situation. Thus to check whether or not we might be in the Baker-Johnson situation, even if we are in a renormalizable and seemingly chiral-symmetric theory, we should nonetheless study the Schwinger-Dyson equation with a cutoff.\footnote{We had noted above that one can obtain the Schwinger-Dyson equation via the stationarity condition $\delta \Gamma(G)\delta G(x,y)=0$ associated with the bilinear source $K(x,y)$ approach presented in  \cite{Cornwall1974}.  While this would then permit self-consistent non-trivial solutions to the Schwinger-Dyson equation, this would not necessarily mean that in them the bare mass would be identically zero and that one would have dynamical symmetry breaking, since if one has an asymptotically power-behaved solution to the Schwinger-Dyson equation, one would be in the Baker-Johnson situation where despite appearances the chiral symmetry would actually be broken in the Lagrangian. In and of itself the effective action approach of \cite{Cornwall1974} does not distinguish between a bare mass that is zero and one that only vanishes in the limit of infinite cutoff, as this is determinable not from the form of the unconstrained action itself but from the structure of its stationary solution.}

With the realization that there is no Goldstone boson when $\alpha\leq \pi/3$, the prevailing wisdom that then ensued from the quenched ladder approximation study is that Goldstone boson generation is strictly a strong coupling effect, with the $\alpha \leq \pi/3$ region not being of relevance to dynamical symmetry breaking. Below we shall revisit this issue and show that it is possible to generate Goldstone and Higgs bosons dynamically even if the coupling is weak. To do this we will need to couple a weakly coupled QED to an equally weakly coupled four-fermion interaction. However, before doing this we need to ask how reliable a guide to dynamical symmetry breaking the quenched ladder approximation actually is.

\subsection{The Shortcomings of the Quenched Ladder Approximation}

To determine how relevant the wisdom obtained from the quenched planar approximation to the Abelian gluon model might be for the full theory, we need to include the non-planar graphs and need to dress the gluon propagator. Inclusion of all the Feynman graphs  was the objective of Johnson, Baker, and Willey in their study of quantum electrodynamics. In their study of quantum electrodynamics Johnson, Baker, and Willey initially kept the photon propagator canonical \cite{Johnson1964} (i.e. no internal fermion loops in the photon propagator) with the photon propagator thus being quenched, but they otherwise included both planar and non-planar graphs to all orders, to thus include the full $\Gamma^{\mu}(p,p-k)$ vertex in the Schwinger-Dyson equation rather than just its undressed $ \gamma^{\mu}$ approximation. In \cite{Johnson1964} they then found that the fermion self-energy then scaled asymptotically just as in (\ref{M84}), i.e. as $(-p^2/m^2)^{\gamma_{\theta}(\alpha)/2}$. Moreover, Johnson, Baker and Willey noted that their analysis only depended on the asymptotic momentum behavior of Feynman diagrams and not on the strength of the coupling. Their analysis thus holds for both weak and strong coupling, and as such shows no region in which a power-behaved solution changes into an oscillating one such as the one exhibited in (\ref{M78}), with the power-behaved, and thus non-Goldstone mode, solution holding for all values of the coupling constant. Thus in the quenched but otherwise all-order approximation to the Abelian gluon model, the phase transition at $\alpha=\pi/3$ in the planar approximation is nullified by the non-planar graphs, and one has to conclude that in the full planar plus non-planar quenched gluon approximation there is no Goldstone boson. 

As had been noted above, in the quenched ladder approximation we keep graphs such as the first two graphs in Fig. (\ref{SD1}) and their higher-order planar analogs but do not include the non-planar third graph or its higher-order non-planar analogs. However, the second and the third graphs in Fig. (\ref{SD1}) are of the same order in $\alpha$. With the third graph being a factor of $\alpha $ weaker than the first graph in the figure, and with the three-gluon non-planar graph (not shown) being a factor of $\alpha$ weaker than the second graph in the figure, we see that compared to the planar graphs the non-planar graphs are suppressed by a factor of $\alpha$. Their neglect is thus only valid if the coupling is weak.  The quenched planar approximation is thus strictly a weak coupling approximation, with its application in the strong coupling regime where $\alpha$ is of order $\pi/3$ potentially not being valid, and with the strong coupling wisdom that  is gleaned from the quenched ladder approximation not being reliable.

To see exactly at what point the quenched ladder approximation begins to depart from the quenched, all-order Johnson-Baker-Willey (JBW) study reported in \cite{Johnson1964}, we note that  through second order in $\alpha$ the contribution of all graphs gives \cite{Johnson1964} $\gamma_{\theta}(\alpha)=-3\alpha/2\pi-3\alpha^2/16\pi^2$, whereas, as per (\ref{M77}), the expansion of $\nu-1=(1-3\alpha/\pi)^{1/2}-1$ to second order is given by $-3\alpha/2\pi-9\alpha^2/8\pi^2$. Thus, as can be anticipated from the graphs in Fig. (\ref{SD1}),  already at the level of the two-photon graphs $\gamma_{\theta}(\alpha)$ and $\nu-1$ begin to depart from each other, and yet the compatibility of (\ref{M77}) and (\ref{M84}) would require that $\gamma_{\theta}(\alpha)$ and $\nu-1$ be equal. The extrapolation of the $\nu-1=(1-3\alpha/\pi)^{1/2}-1$ expression beyond lowest order is thus not justifiable, with the branch point obtained in $\nu$ at $\alpha=\pi/3$ in the quenched planar approximation being an indicator not that there is a phase transition, but that the extrapolation is problematic.\footnote{That a lowest order type calculation  could lead to a phase transition that is only apparent is familiar from a study of the four-fermion theory in two dimensions with interaction $-(g/2)(\bar{\psi}\psi)^2- (g/2)(\bar{\psi}i\gamma^5\psi)^2$.  With point $\bar{\psi}\psi$ couplings, one could repeat the four-dimensional NJL calculation and conclude that there is dynamical mass generation and a dynamical pseudoscalar Goldstone boson. However, as was shown by Coleman \cite{Coleman1973b}, there cannot be any Goldstone bosons in two dimensions. Thus the higher order in $g$ radiative corrections must remove the dynamical symmetry breaking.}

In \cite{Johnson1967} Johnson, Baker, and Willey went further and actually justified the restriction to a canonical photon by noting that the photon would indeed be canonical if the coupling constant satisfied the Gell-Mann-Low eigenvalue condition, and thus satisfied $\beta(\alpha)=0$.\footnote{Radiative corrections to the photon propagator lead to the presence of a troublesome negative norm Landau ghost pole in the photon propagator at spacelike momenta. Its residue will vanish at the Gell-Mann-Low eigenvalue condition, a condition that is equivalent \cite{Adler1972} to $\beta(\alpha)=0$.} Thus whether one takes the photon to be quenched or takes it to couple with a charge that obeys $\beta(\alpha)=0$, in either case the power-behaved $(-p^2/m^2)^{\gamma_{\theta}(\alpha)/2}$ asymptotic behavior is the only allowed solution, with there being no sign of any phase transition. Moreover, if one is at the Gell-Mann-Low eigenvalue, the coupling constant would only have one (or possibly a few) discrete values, with there then being no possibility of being able to vary $\alpha$ over the entire coupling constant range and look for any discontinuity in the first place. Other than the fact that a zero of $\beta(\alpha)$ (should it have one) might be at a location other than $\alpha=\pi/3$, the wisdom obtained in the quenched ladder approximation at $\alpha=\pi/3$ does dovetail with that found by Johnson, Baker, and Willey at a Gell-Mann-Low eigenvalue, i.e. there is asymptotic scaling but no Goldstone boson.\footnote{In  \cite{Mannheim1974b} the present author had argued that dynamical symmetry breaking does not occur in an Abelian gluon model.  The present article completes the analysis, by showing that there is no Goldstone boson if the photon is canonical. Since dressing the photon leads to a Landau ghost pole in the photon propagator, QED is potentially inconsistent (and discussion of any Goldstone mode is essentially irrelevant) unless the coupling constant satisfies the Gell-Mann-Low eigenvalue condition, in which case the Landau ghost pole is canceled and the photon is again canonical.} As we will see, the correspondence is even stronger since in both the cases the dimension of $d_{\theta}(\alpha)$ is reduced from three to two. In the quenched ladder approximation this follows directly from (\ref{M77}) at $\alpha=\pi/3$, while in the Johnson, Baker, and Willey case it is, as we show below,  the condition \cite{Mannheim1974,Mannheim1975} that the vacuum spontaneously break.

\begin{figure}[htpb]
\begin{center}
\begin{minipage}[t]{6 cm}
\epsfig{file=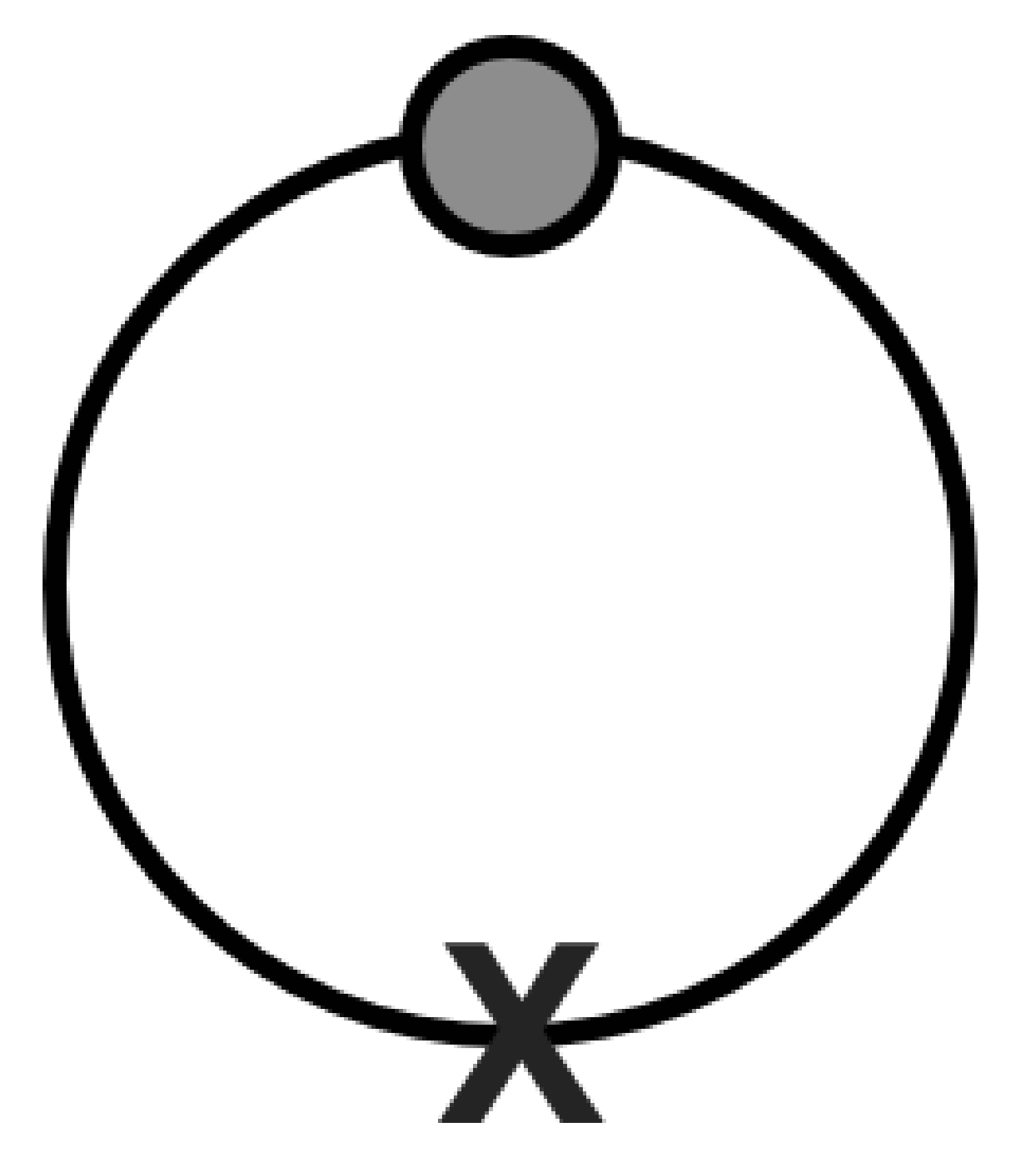,scale=0.25}
\end{minipage}
\begin{minipage}[t]{16.5 cm}
\caption{$\langle\Omega_m|\bar{\psi}\psi|\Omega_m\rangle$ as constructed from the quenched ladder approximation fermion propagator, which is represented as a blob.}
\label{semitadpole}
\end{minipage}
\end{center}
\end{figure}

\begin{figure}[htpb]
\begin{center}
\vskip-0.3in
\begin{minipage}[t]{16 cm}
\epsfig{file=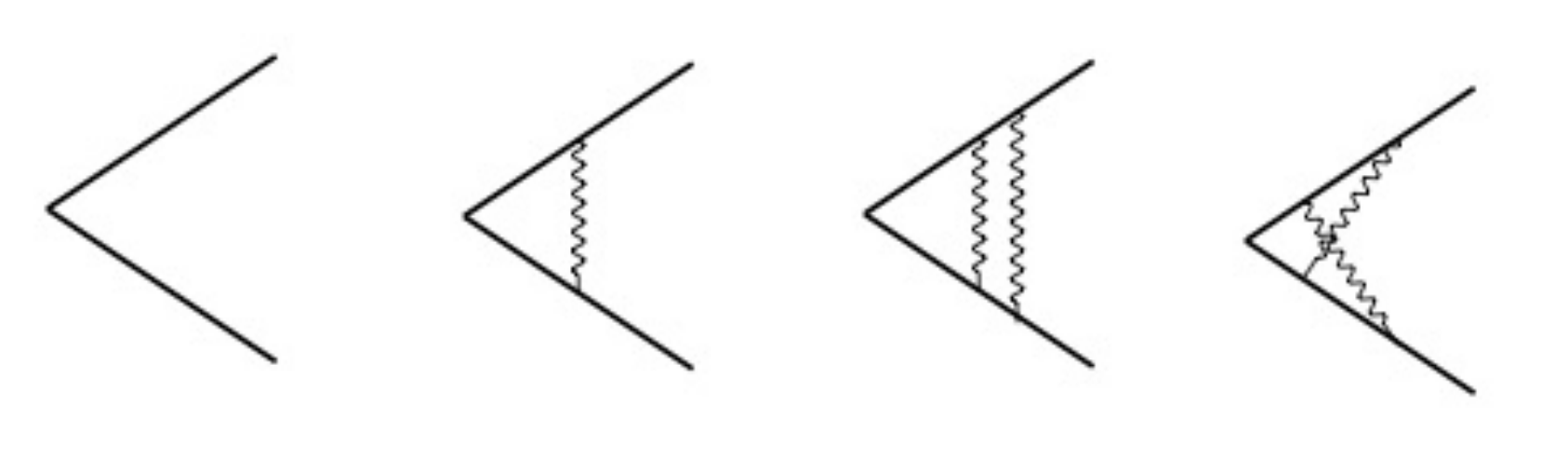,width=6.0in,height=2.0in}
\end{minipage}
\begin{minipage}[t]{16.5 cm}
\caption{The first few graphs that can appear in  $\langle\Omega_m|T[\psi(x)\bar{\psi}(z)\psi(z)\bar{\psi}(y)]|\Omega_m\rangle$.}
\label{SD2}
\end{minipage}
\end{center}
\end{figure}

\begin{figure}[htpb]
\begin{center}
\begin{minipage}[t]{16 cm}
\epsfig{file=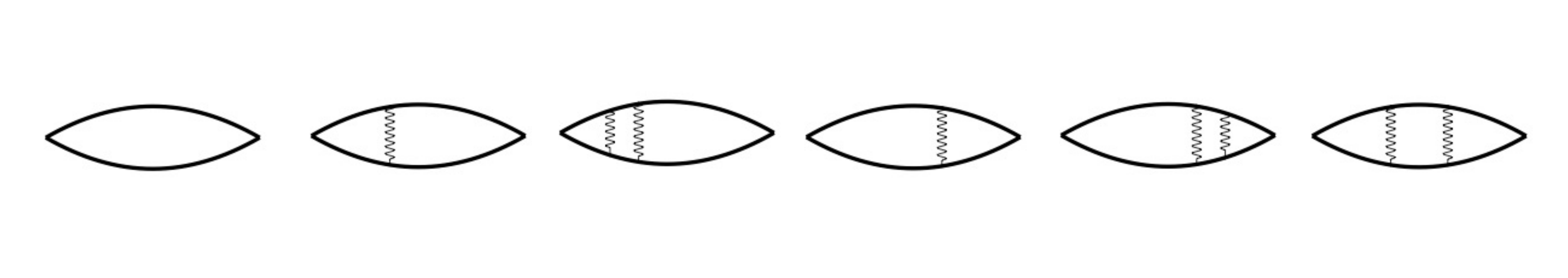,width=6.0in,height=2.0in}
\end{minipage}
\begin{minipage}[t]{16.5 cm}
\caption{The first few graphs that can appear in  $\langle\Omega_m|T[\psi(x)\bar{\psi}(x)\psi(y)\bar{\psi}(y)]|\Omega_m\rangle$.}
\label{SD3}
\end{minipage}
\end{center}
\end{figure}

\begin{figure}[htpb]
\begin{center}
\begin{minipage}[t]{6 cm}
\epsfig{file=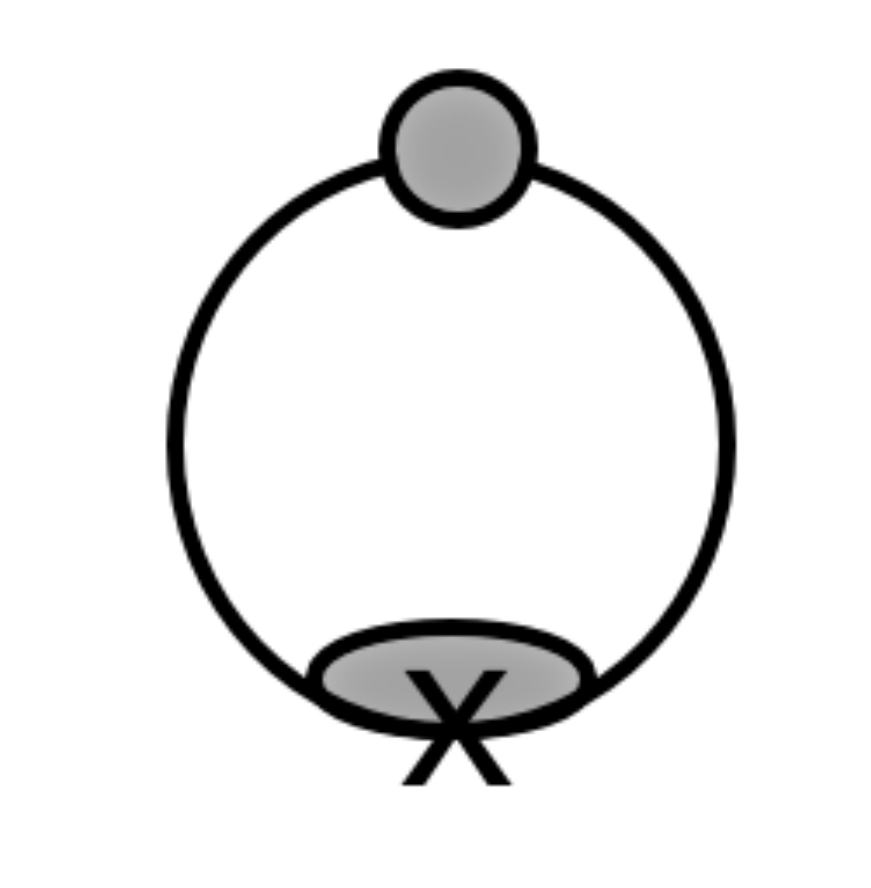,scale=0.5}
\end{minipage}
\begin{minipage}[t]{16.5 cm}
\caption{$\langle\Omega_m|\bar{\psi}\psi|\Omega_m\rangle$ with a dressed propagator and a dressed vertex.}
\label{lw3}
\end{minipage}
\end{center}
\end{figure}

Beyond the issue of the validity of the quenched ladder approximation, there may also be  a possible counting issue. Specifically, if we want to calculate the one-point function expectation value $\langle\Omega_m|\bar{\psi}\psi|\Omega_m\rangle$ in the quenched ladder approximation, we would just sew together the two ends of the fermion propagator (cf. the planar part of Fig. (\ref{SD1})) to produce Fig. (\ref{semitadpole}), this being the tadpole used in \cite{Leung1986} for instance. However, suppose we want to evaluate the three-point function $\langle\Omega_m|T[\psi(x)\bar{\psi}(z)\psi(z)\bar{\psi}(y)]|\Omega_m\rangle$ in this same approximation. This would involve the planar graphs in Fig. (\ref{SD2}) and their higher-order analogs but not the non-planar ones. Suppose we now want to evaluate the two-point function $\langle\Omega_m|T[\bar{\psi}(x)\psi(x)\bar{\psi}(y)\psi(y)]|\Omega_m\rangle$. If we sew the fermion lines in Fig. (\ref {SD2}) to the fermion lines in an identical copy of Fig. (\ref{SD2}) this would yield Fig. (\ref{SD3}), with there being two one-photon exchange graphs and three two-photon exchange graphs, whereas a straightforward ladder graph approximation would only involve one one-photon exchange graph and one  two-photon exchange graph.\footnote{In Figs. (\ref{SD2}) and (\ref{SD3}) we have not drawn a cross to indicate the $\bar{\psi}\psi$ insertion, since the analysis would hold for any type of fermion bilinear insertion ($\bar{\psi}\psi$, $\bar{\psi}i\gamma^5\psi$, $\bar{\psi}\gamma^{\mu}\psi$, $\bar{\psi}\gamma^{\mu}\gamma^5\psi$, $\bar{\psi}[\gamma^{\mu},\gamma^{\nu}]\psi$) and even for no insertion at all, where the figures in Fig. (\ref{SD3}) would then be vacuum to vacuum graphs.}  Similarly, if we sew together the planar graphs of Fig. (\ref{SD1}) with those of Fig. (\ref{SD2}) (as evaluated with a $\bar{\psi}\psi$ insertion) that would yield Fig. (\ref{lw3}) rather than Fig. (\ref{semitadpole}). The enumeration of diagrams thus has to be done carefully, and as we will see in the following it will be the tadpole graph of Fig. (\ref{lw3}) and not that of Fig. (\ref{semitadpole})\ that will prove to be the relevant one.\footnote{One way to avoid counting issues is to work not with the Schwinger-Dyson equation, but with the Bethe-Salpeter equation instead and then approximate the kernel. For the fermion propagator the Bethe-Salpeter equation  takes the form \cite{Johnson1968} $\{\gamma^5,\Sigma(p)\}=\int d^4k K(p,k,0)S(k)\{\gamma^5,\Sigma(k)\}S(k) +2m_0\int d^4k K(p,k,0)S(k)\gamma^5S(k)$, where $K(p,k,0)$ is the scattering kernel. Below we will provide another way to avoid counting issues, one based on the use of conformal invariance.}

As we have seen, because of asymptotic scaling of the form exhibited in (\ref{M84}), there is no phase transition in the all-order, planar plus non-planar Abelian gluon model. Nonetheless,  it is of interest to ask whether one can still find a way to get  dynamical symmetry breaking in the event that there is such scaling. As noted in \cite{Mannheim2015} and \cite{Mannheim2016},  this does occur if a massless fermion electrodynamics  with scaling is coupled to a four-fermion interaction, with the driver being a reduction in the dynamical dimension of $d_{\theta}(\alpha)$ from three to two. To this end we need to reinterpret JBW electrodynamics as the mean-field sector of a massless fermion electrodynamics theory coupled to a four-fermion interaction, since as we saw with the chirally-symmetric NJL model, the mean-field sector contains an explicit non-chirally-symmetric fermion mass term and possesses no Goldstone boson, just as JBW electrodynamics contains an explicit fermion mass term and involves no Goldstone boson. However, as we also saw in the NJL model, a dynamical pseudoscalar Goldstone boson is instead generated by an accompanying residual interaction. We shall find precisely the same residual interaction effect when a massless electrodynamics with scaling is coupled to a four-fermion interaction, and because of the underlying chiral symmetry, we shall obtain a dynamical scalar Higgs boson as well.

\section{Structure of Johnson-Baker-Willey Electrodynamics}

\subsection{JBW Electrodynamics  as a Mean-Field Theory}

JBW electrodynamics with vanishing $\beta(\alpha)$ in the charge sector and scaling with anomalous dimensions in the fermion mass sector (collectively critical scaling) is a theory in which the chiral symmetry is broken in the Lagrangian by the presence of a bare mass $m_0(\bar{\psi}\psi)_0$ term. However, in the above we encountered another situation in which the chiral symmetry is broken in the Lagrangian, namely the mean-field sector of the chirally-symmetric NJL model. And, as we noted then, there are no dynamical Goldstone or Higgs bosons in the mean-field sector. Instead  they are generated by the residual-interaction sector, with neither of the two sectors separately  being chiral symmetric on its own, with only their sum being so. It is thus of interest  to ask whether JBW electrodynamics might also be the mean-field sector of some larger theory that is chirally symmetric, and this turns out to be the case, with Goldstone and Higgs bosons then being generated by the associated residual interaction. 

To establish this result consider a massless fermion electrodynamics coupled to a four-fermion interaction with a chiral-invariant action of the form: 
\begin{eqnarray}
I_{\rm QED-FF}=\int d^4x\left[-\frac{1}{4}F_{\mu\nu}F^{\mu\nu}+\bar{\psi}\gamma^{\mu}(i\partial_{\mu}-eA_{\mu})\psi 
-\frac{g}{2}[\bar{\psi}\psi]^2-\frac{g}{2}[\bar{\psi}i\gamma^5\psi]^2\right].
\label{M87}
\end{eqnarray}
As with the NJL model itself we break the action into two pieces by introducing a mass term that is not present in the action, to obtain $I_{\rm QED-FF}=I_{\rm QED-MF}+I_{\rm QED-RI}$, where
\begin{eqnarray}
I_{\rm QED-MF}&=&\int d^4x\left[-\frac{1}{4}F_{\mu\nu}F^{\mu\nu}+\bar{\psi}\gamma^{\mu}(i\partial_{\mu}-eA_{\mu})\psi 
-m\bar{\psi}\psi +\frac{m^2}{2g}\right],
\nonumber\\
I_{\rm QED-RI}&=&\int d^4x\left[-\frac{g}{2}\left(\bar{\psi}\psi-\frac{m}{g}\right)^2-\frac{g}{2}\left(\bar{\psi}i\gamma^5\psi\right)^2\right].
\label{M88}
\end{eqnarray}
Comparing with (\ref{M58}), our task is to generalize our treatment of the point-coupled NJL model so as to include QED radiative corrections. We thus need to generalize $\epsilon(m)$ and $\Pi_{\rm S}(q^2,m)$ to QED. Now at first this appears to be a quite difficult task, since even with critical scaling, the renormalization group only gives the Green's functions of QED at momenta for which $-p^2 \gg m^2$ rather than at the full range of momenta that flow in the loops that make up $\epsilon(m)$ and $\Pi_{\rm S}(q^2,m)$. But as we noted in Figs. (\ref{lw1}) and (\ref{lw5}), we can construct the massive theory $\epsilon(m)$ and $\Pi_{\rm S}(q^2,m)$ via an infinite summation of massless graphs. However, unlike massive QED, a critical scaling massless QED is scale invariant at all momenta, and thus we do have a way to fix the massless theory graphs at all momenta. In a massless QED with critical scaling we thus set
\begin{eqnarray}
\tilde{S}^{-1}(p,m=0)&=& \slashed{ p}+i\epsilon,\qquad \tilde{\Gamma}_{\rm S}(p,p,0,m=0)=\left(\frac{-p^2-i\epsilon}{\mu^2}\right)^{\gamma_{\theta}(\alpha)/2},
\label{M89}
\end{eqnarray}
at all momenta, with $\mu^2$ being an off-shell subtraction point that is needed in a massless theory. (Below we will set $\mu^2=M^2$, but for tracking purposes it  is instructive to keep them distinct until the end.)

To resolve the counting problem mentioned above, we note that in a critical scaling massless theory we can use conformal invariance to determine two-point and three-point functions exactly at all momenta. Thus the functional form for $\Pi_{\rm S}(x,m=0)$  is fixed entirely by the anomalous dimension $d_{\theta}(\alpha)$ of $\bar{\psi}\psi$ according to
\begin{eqnarray}
&&\langle \Omega_0|T(:\bar{\psi}(x)\psi(x)::\bar{\psi}(y)\psi(y):)|\Omega_0\rangle
=\frac{\mu^{-2\gamma_{\theta}}{\rm Tr}[(\slashed{ x}-\slashed{ y})(\slashed{ y}-\slashed{ x})]}{[(x-y)^2]^{(d_{\theta}+1)/2}[(y-x)^2]^{(d_{\theta}+1)/2}}
=-\frac{4\mu^{-2\gamma_{\theta}}}{[(x-y)^2]^{d_{\theta}}}.
\label{M90}
\end{eqnarray}
With an appropriate normalization, Fourier transforming then gives 
\begin{eqnarray}
\Pi_{\rm S}(q^2,m=0)=-i\int \frac{d^4p}{(2\pi)^4}{\rm Tr}\left[\left(\frac{(-p^2)}{\mu^2}\frac{(-(p+q)^2)}{\mu^2}\right)^{\frac{\gamma_{\theta}(\alpha)}{4}}
\frac {1}{\slashed{ p}}\left(\frac{(-p^2)}{\mu^2}\frac{(-(p+q)^2)}{\mu^2}\right)^{\frac{\gamma_{\theta}(\alpha)}{4}}\frac {1}{\slashed{ p} +\slashed {q}}\right].
\label{M91}
\end{eqnarray}
As well as construct $\Pi_{\rm S}(x,m=0)$ via conformal invariance, we can start with its definition as $\langle \Omega_0|T(:\bar{\psi}(x)\psi(x)::\bar{\psi}(y)\psi(y):)|\Omega_0\rangle$ and make a Dyson-Wick contraction between the fields at $x_{\mu}$ and $y_{\mu}$. At the one-loop level this then yields 
\begin{eqnarray}
&&\Pi_{\rm S}(q^2,m=0)\nonumber\\
&&=-i\int \frac{d^4p}{(2\pi)^4}{\rm Tr}\bigg[\tilde{\Gamma}_{\rm S}(p+q,p,-q,m=0)
\tilde{S}(p,m=0)\tilde{\Gamma}_{\rm S}(p,p+q,q,m=0)\tilde{S}(p+q,m=0)\bigg],
\label{M92}
\end{eqnarray}
where the massless  $\tilde{S}(p,m=0)$ is given in (\ref{M89}), and where, on comparing (\ref{M91}) and (\ref{M92}), we can identify
\begin{eqnarray}
\tilde{\Gamma}_{\rm S}(p,p+q,q,m=0)=\left[\frac{(-p^2)}{\mu^2}\frac{(-(p+q)^2)}{\mu^2}\right]^{\frac{\gamma_{\theta}(\alpha)}{4}}=\tilde{\Gamma}_{\rm S}(p+q,p,-q,m=0).
\label{M93}
\end{eqnarray}
Comparing (\ref{M91}) with the point-coupled (\ref{M64}), we see that we should dress both of the vertices in $\Pi_{\rm S}(q^2,m=0)$ and not just one, just as is to be anticipated given the construction of Fig. (\ref{SD3}) from Fig. (\ref{SD2}). 

\begin{figure}[htpb]
\begin{center}
\begin{minipage}[t]{10 cm}
\epsfig{file=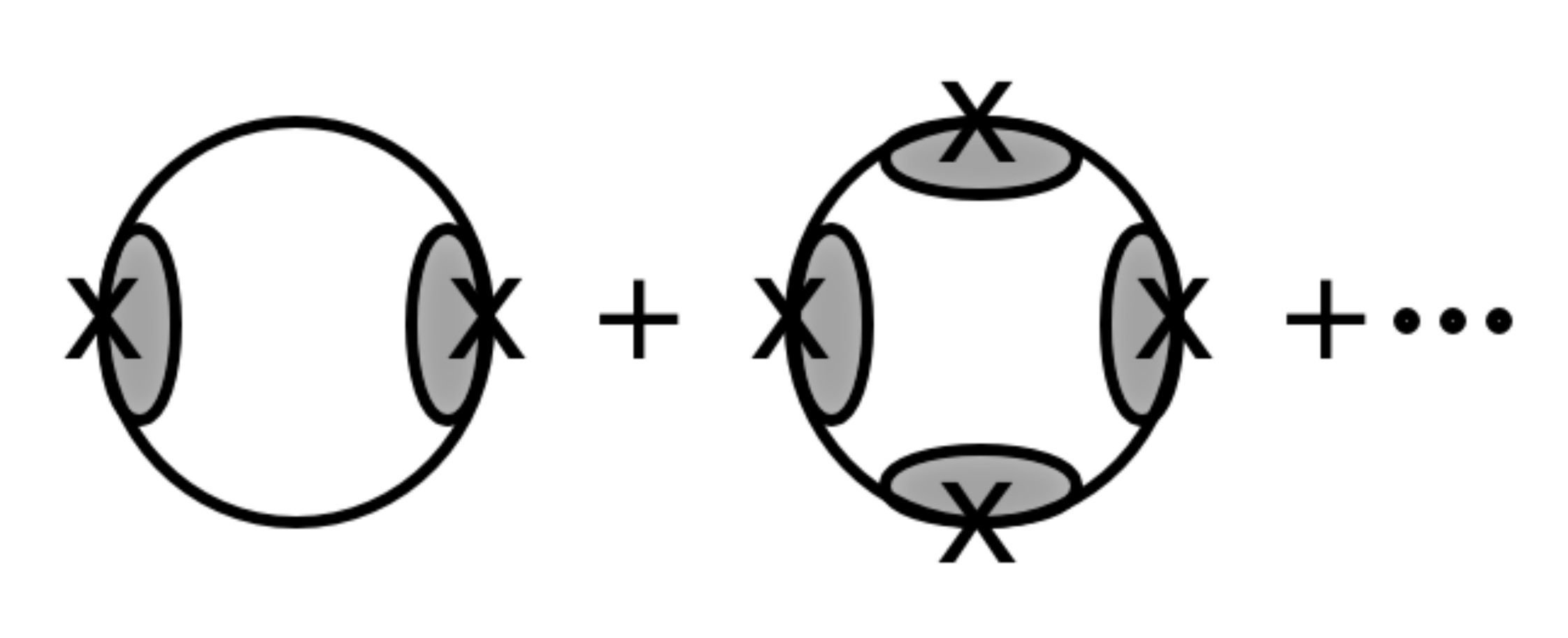,scale=0.5}
\end{minipage}
\begin{minipage}[t]{16.5 cm}
\caption{Vacuum energy density  $\epsilon(m)$ via an infinite summation of massless graphs with zero-momentum dressed $m\bar{\psi}\psi$ insertions.}
\label{lw2}
\end{minipage}
\end{center}
\end{figure}

\begin{figure}[htpb]
\begin{center}
\begin{minipage}[t]{17 cm}
\epsfig{file=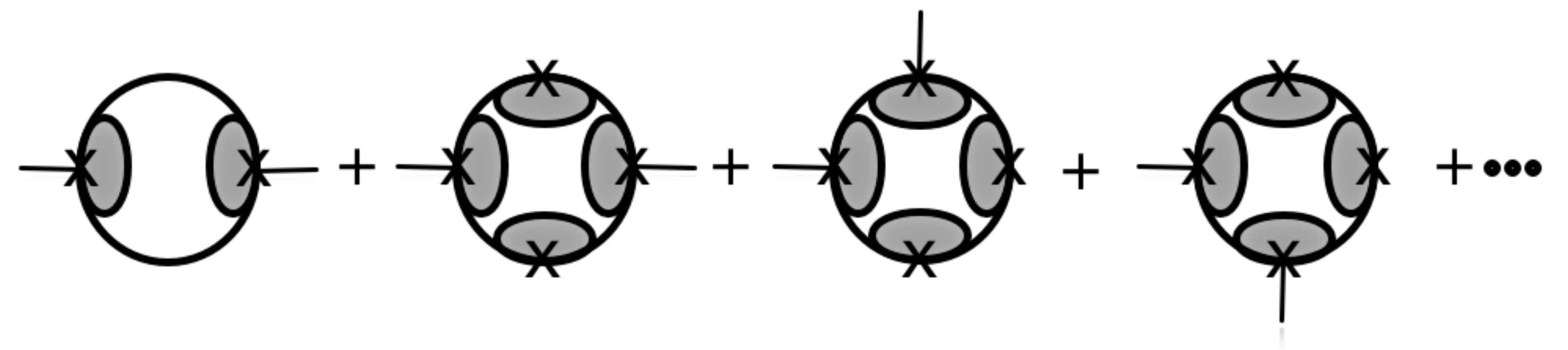,scale=0.5}
\end{minipage}
\begin{minipage}[t]{16.5 cm}
\caption{$\Pi_{\rm S}(q^2,m)$ developed as an infinite summation of massless graphs, each with two dressed $m\bar{\psi}\psi$  insertions carrying momentum $q_{\mu}$ (shown as external lines), with all other dressed $m\bar{\psi}\psi$  insertions carrying zero momentum.}
\label{lw6}
\end{minipage}
\end{center}
\end{figure}

As we see, to generalize NJL, we only need to replace point vertices by the dressed vertices given in (\ref{M89}) and (\ref{M93}), with Figs. (\ref{lw1}), (\ref{baretadpole}), and (\ref{lw5}) being replaced by Figs. (\ref{lw2}), (\ref{lw3}), and (\ref{lw6}), to 
yield \cite{Mannheim1974,Mannheim1975} 
\begin{eqnarray}
\epsilon(m)&=&i\int\frac{d^4p}{(2\pi)^4}\sum_{n=1}^{\infty}\frac{(-1)}{2n}
{\rm Tr}\left[(-i)^2\left(\frac{-p^2-i\epsilon}{\mu^2}\right)^{\gamma_{\theta}(\alpha)}\left(\frac{i}{\slashed{ p}+i\epsilon}\right)^2m^2\right]^n
\nonumber\\
&=&\frac{i}{2}\int \frac{d^4p}{(2\pi)^4}{\rm Tr~ln}\left[1-\frac{m^2}{p^2+i\epsilon}\left(\frac{-p^2-i\epsilon}{\mu^2}\right)^{\gamma_{\theta}(\alpha)}\right]
\nonumber\\
&=&i\int \frac{d^4p}{(2\pi)^4}\left[{\rm Tr~ln}(\tilde{S}^{-1}_{\mu}(p))-{\rm Tr~ln}(\slashed{ p}+i\epsilon)\right],
\label{M94}
\end{eqnarray}
\begin{eqnarray}
\langle \Omega_m|\bar{\psi}\psi|\Omega_m\rangle =\epsilon^{\prime}(m)
=-i\int \frac{d^4p}{(2\pi)^4}{\rm Tr}[\tilde{\Gamma}_{\rm S}(p,p,0,m=0)\tilde{S}_{\mu}(p)],
\label{M95}
\end{eqnarray}
and
\begin{eqnarray}
\Pi_{\rm S}(q^2,m)=-i\int \frac{d^4p}{(2\pi)^4}{\rm Tr}\bigg[\tilde{\Gamma}_{\rm S}(p+q,p,-q,m=0)
\tilde{S}_{\mu}(p)\tilde{\Gamma}_{\rm S}(p,p+q,q,m=0)\tilde{S}_{\mu}(p+q)\bigg],
\label{M96}
\end{eqnarray}
together with the pseudoscalar analog
\begin{eqnarray}
\Pi_{\rm P}(q^2,m)=-i\int \frac{d^4p}{(2\pi)^4}{\rm Tr}\bigg[\tilde{\Gamma}_{\rm S}(p+q,p,-q,m=0)
i\gamma^5\tilde{S}_{\mu}(p)\tilde{\Gamma}_{\rm S}(p,p+q,q,m=0)i\gamma^5\tilde{S}_{\mu}(p+q)\bigg].
\label{M97}
\end{eqnarray}
In the above we have introduced
\begin{eqnarray}
\tilde{S}^{-1}_{\mu}(p)&=& \slashed{ p}-m\left(\frac{-p^2-i\epsilon}{\mu^2}\right)^{\gamma_{\theta}(\alpha)/2}+i\epsilon,
\label{M98}
\end{eqnarray}
a propagator that is distinct from the asymptotic $\tilde{S}^{-1}(p,m)$ propagator that is given in (\ref{M84}). Since $\tilde{\Gamma}_{\rm S}(p,p,0,m)$ obeys the renormalization group equation \cite{Adler1971}
\begin{eqnarray}
&&\left[m\frac{\partial}{m}+\beta(\alpha)\frac{\partial}{\partial \alpha}+\gamma_{\theta}(\alpha)\right]\tilde{\Gamma}_{\rm S}(p,p,0,m)
=m(1-\gamma_{\theta}(\alpha)]\tilde{\Gamma}_{\rm SS}(p,p,0,m),
\label{M99}
\end{eqnarray}
where $\tilde{\Gamma}_{\rm SS}(p,p,0,m)$ involves two soft $\bar{\psi}\psi$ insertion, $\tilde{S}^{-1}(p,m)$ as given in (\ref{M84}) will develop non-leading terms, so it cannot be the exact  propagator of JBW electrodynamics. On the other hand, as constructed $\tilde{S}^{-1}_{\mu}(p)$ is the exact propagator that we need for the mean-field theory associated with the $I_{\rm QED-MF}$ action, and unlike the asymptotic JBW ${S}^{-1}(p,m)=\slashed{ p}-(-p^2/m^2)^{\gamma_{\theta}(\alpha)/2}$ propagator, $\tilde{S}^{-1}_{\mu}(p)$ is to be used at all momenta in  the mean-field (\ref{M94}), (\ref{M95}), (\ref{M96}), and (\ref{M97}). The  pole structure of $\tilde{S}^{-1}_{\mu}(p)$ will then be modified by the residual interaction.\footnote{In \cite{Mannheim1975} it was argued that for the action $I_{\rm QED-MF}$ with critical scaling, the expansions in Figs. (\ref{lw2}) and (\ref{lw6}) that are used for (\ref{M94}), (\ref{M95}), (\ref{M96}), and (\ref{M97}) are potentially exact, since for each massless fermion graph in Fig. (\ref{lw2}) conformal invariance correctly describes the associated infrared divergence structure, while yielding coefficients for each of the graphs such that the infinite sum over all the graphs is infrared finite.}

\subsection{Vacuum Structure of JBW Electrodynamics at $\gamma_{\theta}(\alpha)=-1$.}

While (\ref{M94}), (\ref{M95}), (\ref{M96}), and (\ref{M97}) hold for any value of $\gamma_{\theta}(\alpha)$, it turns out that one particular value of $\gamma_{\theta}(\alpha)$ is preferred, namely $\gamma_{\theta}(\alpha)=-1$. Two independent reasons for this particular value have been identified in \cite{Mannheim1974,Mannheim1975}, the first one being based on the compatibility of the short-distance Wilson expansion with the asymptotic JBW massive propagator $\tilde{S}^{-1}(p,m)\sim \slashed{ p}-(-p^2/m^2)^{\gamma_{\theta}(\alpha)/2}$,  and the second one being based on the behavior of the vacuum energy density.

In a scale-invariant theory such as a critical scaling massless QED the Wilson operator product expansion has a  leading behavior at short distances of the form
\begin{eqnarray}
T(\psi(x)\bar{\psi}(0))&=&\langle \Omega_0|T(\psi(x)\bar{\psi}(0))|\Omega_0\rangle
+(\mu^2x^2)^{\gamma_{\theta}(\alpha)/2}:\psi(0)\bar{\psi}(0):
\label{M100}
\end{eqnarray}
In (\ref{M100}) the normal ordering is done with respect to the unbroken massless vacuum so that $:\psi(0)\bar{\psi}(0):=\psi(0)\bar{\psi}(0)-\langle\Omega_0|\psi(0)\bar{\psi}(0)|\Omega_0\rangle$. Thus while the matrix element of $:\psi(0)\bar{\psi}(0):$ in the $|\Omega_0\rangle$ vacuum would vanish, its matrix element in any other vacuum such as $|\Omega_m\rangle$ need not. On taking $\langle\Omega_m|:\psi(0)\bar{\psi}(0):|\Omega_m\rangle$ to be non-zero, up to numerical coefficients an evaluation of the matrix element of (\ref{M100}) in the spontaneously broken vacuum $|\Omega _m\rangle$ yields an asymptotic propagator for a massive fermion of the form 
\begin{eqnarray}
\tilde{S}(p)=\frac{1}{\slashed{ p}}+(-p^2)^{(-\gamma_{\theta}(\alpha)/2-2)},\qquad
\tilde{S}^{-1}(p)=\slashed{ p}-(-p^2)^{(-\gamma_{\theta}(\alpha)-2)/2}.
\label{M101}
\end{eqnarray}
Now in (\ref{M84}) we constructed the asymptotic massive theory propagator via the renormalization group. Compatibility of (\ref{M101}) with the asymptotic form $\tilde{S}^{-1}(p,m)\sim \slashed{ p}-(-p^2)^{\gamma_{\theta}(\alpha)/2}$ given in (\ref{M84}) then yields
\begin{eqnarray}
\gamma_{\theta}(\alpha)=-\gamma_{\theta}(\alpha)-2,\qquad \gamma_{\theta}(\alpha)=-1.
\label{M102}
\end{eqnarray}
Thus not only do we obtain a unique determination for $\gamma_{\theta}(\alpha)$, we also show that even though there is no Goldstone boson, the vacuum associated with JBW electrodynamics is nonetheless a spontaneously broken one. It thus must be a vacuum associated with a mean-field theory. To confirm this, we need to evaluate the $\epsilon(m)$ that is associated with ${\cal{L}}_{\rm QED-MF}$ at $\gamma_{\theta}(\alpha)=-1$, and note immediately that because the dimension of $\bar{\psi}\psi$ has been  reduced from three to two at this value for  $\gamma_{\theta}(\alpha)$, we are able to add a four-fermion interaction on to QED as in (\ref{M87}) without losing renormalizability.

In regard to $\epsilon(m)$, we note that in  \cite{Mannheim1974,Mannheim1975} $\epsilon(m)$ was evaluated as a function  of $\gamma_{\theta}(\alpha)$, with it taking the form of a single well ($\cup$) in the range $-1< \gamma_{\theta}(\alpha)<0$, with it taking the form of a upside-down single well  ($\cap$) if $ \gamma_{\theta}(\alpha)<-1$, and with it taking the desired form of a double well (shaped like a letter $W$) at $\gamma_{\theta}(\alpha)=-1$.\footnote{While having $\gamma_{\theta}(\alpha)<0$ is sufficient to control the ultraviolet behavior of the theory, as  $\gamma_{\theta}(\alpha)$ is made more negative the theory becomes more divergent in the infrared, with the infrared divergences becoming so severe at $ \gamma_{\theta}(\alpha)=-1$ that the theory is forced into a new vacuum with a minimum away from $m=0$ and a dynamical mass that emerges as a long range order parameter.} At $ \gamma_{\theta}(\alpha)=-1$ $\epsilon (m)$ and $\epsilon^{\prime}(m)$ straightforwardly evaluate to 

\begin{eqnarray}
\epsilon(m)=-\frac{m^2\mu^2}{8\pi^2}\left[ {\rm ln}\left(\frac{\Lambda^2}{m\mu}
\right)+\frac{1}{2}\right],
\label{M103}
\end{eqnarray}
\begin{eqnarray}
\langle \Omega_m|\bar{\psi}\psi|\Omega_m\rangle&=&\epsilon^{\prime}(m)
=-i\int \frac{d^4p}{(2\pi)^4}{\rm Tr}[\tilde{\Gamma}_{\rm S}(p,p,0,m=0)\tilde{S}_{\mu}(p)]
\nonumber\\
&=&4i\int \frac{d^4p}{(2\pi)^4}\frac{m\mu^2}{(p^2+i\epsilon)^2+m^2\mu^2}
=-\frac{m\mu^2}{4\pi^2}{\rm ln}\left(\frac{\Lambda^2}{m\mu}\right). 
\label{M104}
\end{eqnarray}
On recalling the quadratic divergences in (\ref{M61}) and (\ref{M60}) that arise for $\epsilon (m)$ and $\epsilon^{\prime}(m)$ in the point-coupled NJL case, since $\gamma_{\theta}(\alpha)=-1$, this time both $\epsilon(m)$ and $\epsilon^{\prime}(m)$ are only logarithmically divergent.

To now impose the mean-field condition on the residual interaction, we set $\epsilon^{\prime}(M)=M/g$, to thus obtain   
\begin{eqnarray}
-\frac{\mu^2}{4\pi^2}{\rm ln}\left(\frac{\Lambda^2}{M\mu}
\right)=\frac{1}{g},~~~~M=\frac{\Lambda^2}{\mu}\exp\left(\frac{4\pi^2}{\mu^2g}\right). 
\label{M105}
\end{eqnarray}
We recognize (\ref{M105}) as the analog of the BCS  gap equation given in (\ref{M34}), and so unlike in the NJL case (cf. (\ref{M60})), this time we get a solution no matter how small $g$ might be as long as it is negative (viz. attractive), with a BCS-type essential singularity being found at $g=0$. Hence with $\gamma_{\theta}(\alpha)=-1$ dynamical symmetry breaking occurs no matter how weak the coupling $g$ might be.

\begin{figure}[htpb]
\begin{center}
\begin{minipage}[t]{7 cm}
\epsfig{file=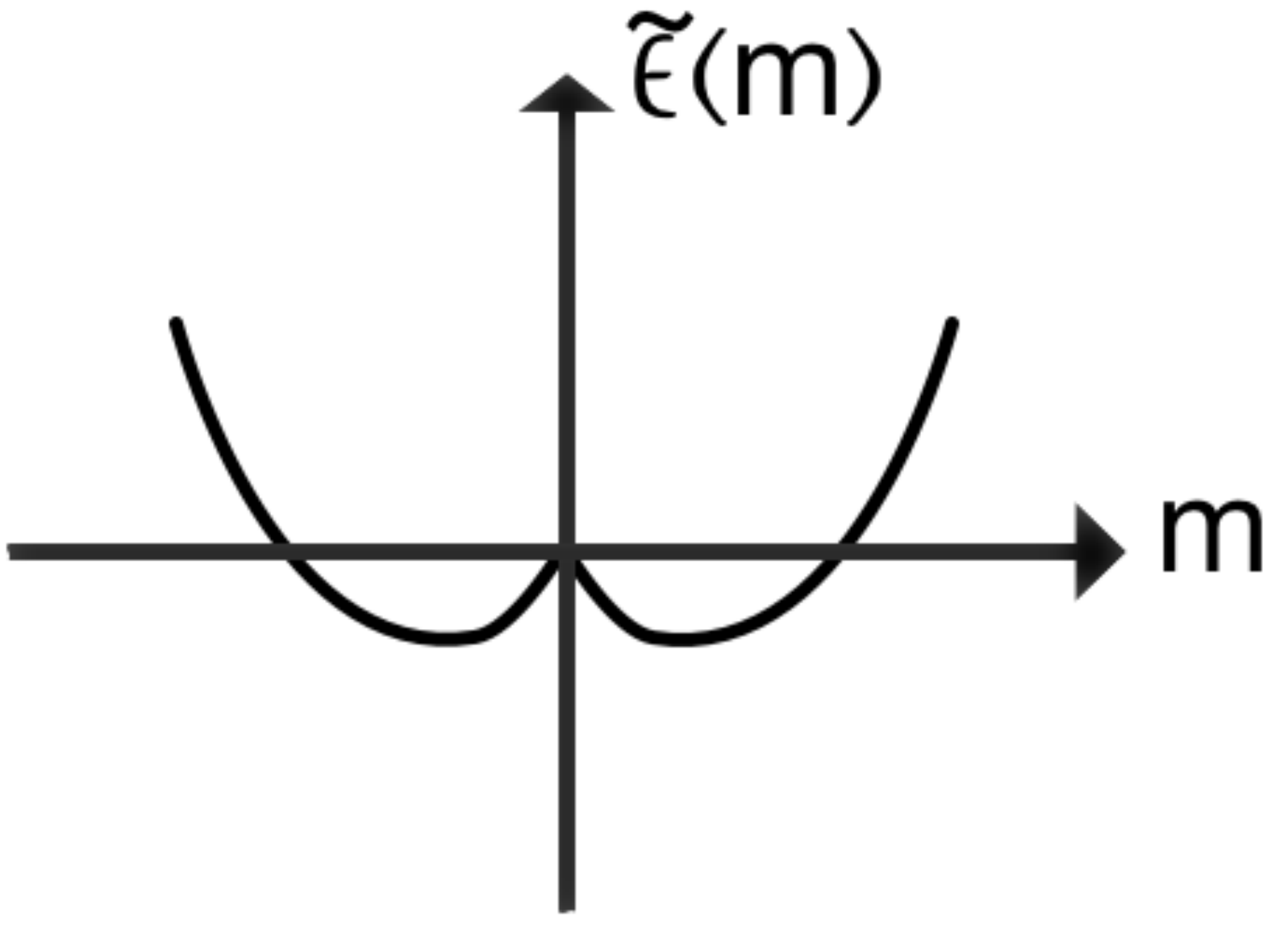,scale=0.5}
\end{minipage}
\begin{minipage}[t]{16.5 cm}
\caption{$\tilde{\epsilon}(m)$ plotted as a function of  $m$ at $\gamma_{\theta}(\alpha)=-1$.}
\label{lw4}
\end{minipage}
\end{center}
\end{figure}

One of the bonuses of the mean-field approach is that it automatically provides us with a vacuum energy density contribution of the form $-m^2/2g$. Incorporating it then yields as the vacuum energy density associated with  ${\cal{L}}_{\rm QED-MF}$
\begin{eqnarray}
\tilde{\epsilon}(m)=\epsilon(m)-\frac{m^2}{2g}=\frac{m^2\mu^2}{16\pi^2}\left[{\rm ln}\left(\frac{m^2}{M^2}\right)-1\right],
\label{M106}
\end{eqnarray}
an expression that is completely finite and has the double-well shape given in Fig. (\ref{lw4}) with a minimum at $m=M$.

In complete parallel to the derivation of (\ref{M62}) in the point-coupled NJL case, on looking for spacetime-dependent matrix elements of $\bar{\psi}\psi$ in coherent states, when $\gamma_{\theta}(\alpha)=-1$ and when accordingly $\Pi^{\prime}_{\rm S}(q^2=0)=-3\mu/128\pi m$, we obtain \cite{Mannheim1978}
\begin{eqnarray}
I_{\rm EFF}&=&\int d^4x\bigg{[}-\tilde{\epsilon}(m(x))
-\frac{1}{2}m(x)[\Pi_{\rm S}(-\partial_{\mu}\partial^{\mu},m(x))-\Pi_{\rm S}(0,m(x))
]m(x)+...\bigg{]}
\nonumber\\
&=&\int d^4 x\bigg{[}-\frac{m^2(x)\mu^2}{16\pi^2}\left[{\rm ln}\left(\frac{m^2(x)}{M^2}\right)-1\right]
+\frac{3\mu}{256\pi m(x)}\partial_{\mu}m(x)\partial^{\mu}m(x)+....\bigg{]}.
\label{M107}
\end{eqnarray}
With $m(x)$ being the order parameter, we thus generate a Ginzburg-Landau effective Lagrangian, one that is completely finite. In this way we can generate an effective Higgs-like Lagrangian, though since $m(x)$ is a c-number, (\ref{M107}) does not describe an elementary Higgs field. Rather, just as with NJL, the scalar Higgs boson and the pseudoscalar Goldstone boson will both emerge as dynamical states generated by the residual interaction.

\subsection{Dynamical Tachyons of  $m=0$ Electrodynamics at $\gamma_{\theta}(\alpha)=-1$}

To establish the presence of collective dynamical modes when the fermion acquires a mass, it is instructive to first show that the theory is unstable if the fermion stays massless. For general  $\gamma_{\theta}(\alpha)$, $\Pi_{\rm S}(q^2,m=0)$ and $\Pi_{\rm P}(q^2,m=0)$ are given by 

\begin{eqnarray}
&&\Pi_{\rm P}(q^2,m=0)=-i\int \frac{d^4p}{(2\pi)^4}{\rm Tr}\bigg[\left[\frac{(-p^2)}{\mu^2}\frac{(-(p+q)^2)}{\mu^2}\right]^{\frac{\gamma_{\theta}(\alpha)}{4}}i\gamma^5
\frac {1}{\slashed {p}}\left[\frac{(-p^2)}{\mu^2}\frac{(-(p+q)^2)}{\mu^2}\right]^{\frac{\gamma_{\theta}(\alpha)}{4}}i\gamma^5\frac {1}{\slashed{ p }+\slashed {q}}\bigg],
\nonumber\\
&&\Pi_{\rm S}(q^2,m=0)=-i\int \frac{d^4p}{(2\pi)^4}{\rm Tr}\bigg[\left[\frac{(-p^2)}{\mu^2}\frac{(-(p+q)^2)}{\mu^2}\right]^{\frac{\gamma_{\theta}(\alpha)}{4}}
\frac {1}{\slashed{ p}}\left[\frac{(-p^2)}{\mu^2}\frac{(-(p+q)^2)}{\mu^2}\right]^{\frac{\gamma_{\theta}(\alpha)}{4}}\frac {1}{\slashed{ p} +\slashed{q}}\bigg].
\label{M108}
\end{eqnarray}
At   $\gamma_{\theta}(\alpha)=-1$ they both evaluate to
\begin{eqnarray}
&&\Pi_{\rm S}(q^2,m=0)=\Pi_{\rm P}(q^2,m=0)
=-\frac{\mu^2}{4\pi^2}\bigg[{\rm ln}\left(\frac{\Lambda^2}{(-q^2)}\right)-3+4~{\rm ln}2\bigg].
\label{M109}
\end{eqnarray}
Due to the iteration of the residual interaction, the scattering amplitudes in the scalar and pseudoscalar channels are given by
\begin{eqnarray}
T_{\rm S}(q^2,m=0)=\frac{1}{g^{-1}-\Pi_{\rm S}(q^2,m=0)},\qquad
T_{\rm P}(q^2,m=0)=\frac{1}{g^{-1}-\Pi_{\rm P}(q^2,m=0)},
\label{M110}
\end{eqnarray}
and even though $\Pi_{\rm S}(q^2,m=0)$,  $\Pi_{\rm P}(q^2,m=0)$, and $g^{-1}$ all diverge logarithmically, they all diverge at precisely the same rate, with both scattering amplitudes then being finite. Both amplitudes are found \cite{Mannheim2015,Mannheim2016} to possess a tachyonic pole at 
\begin{eqnarray}
q^2=-M\mu e^{4{\rm ln}2-3}=-0.797M\mu,
\label{M111}
\end{eqnarray}
while  behaving as 
\begin{eqnarray}
T_{\rm S}(q^2,m=0)=T_{\rm P}(q^2,m=0)=\frac{31.448M\mu}{(q^2+0.797M\mu)}
\label{M112}
\end{eqnarray}
near the poles. The poles in the scalar and pseudoscalar channels are at the same $q^2$, as is to be expected since the chiral symmetry is unbroken if the fermion is massless. Since the poles are found to occur at spacelike $q^2$, the massless theory is unstable. Consequently, the theory is forced into an alternative Hilbert space where the fermion is massive, where both of the two poles  have moved out of the spacelike region, and where the theory is now stable.

\subsection{Dynamical Goldstone Boson of $m\neq 0$ Electrodynamics at $\gamma_{\theta}(\alpha)=-1$}

For the massive fermion case at $\gamma_{\theta}(\alpha)=-1$, we need to evaluate $ \Pi_{\rm S}(q^2,m)$ and $\Pi_{\rm P}(q^2,m)$. On translating $p_{\mu}$ to $p_{\mu}-q_{\mu}/2$, (\ref{M96}) and (\ref{M97}) take the form 
\begin{eqnarray}
\Pi_{\rm S}(q^2,m)=-4i\mu^2\int \frac{d^4p}{(2\pi)^4}\frac{N(q,p)+m^2\mu^2}{D(q,p,m)},
\label{M113}
\end{eqnarray}
\begin{eqnarray}
\Pi_{\rm P}(q^2,m)=-4i\mu^2\int \frac{d^4p}{(2\pi)^4}\frac{N(q,p)-m^2\mu^2}{D(q,p,m)},
\label{M114}
\end{eqnarray}
where
\begin{eqnarray}
N(q,p)&=&(p^2+i\epsilon-q^2/4)(-(p-q/2)^2-i\epsilon)^{1/2}(-(p+q/2)^2-i\epsilon)^{1/2},
\nonumber\\
D(q,p,m)&=&(((p-q/2)^2+i\epsilon)^2+m^2\mu^2)(((p+q/2)^2+i\epsilon)^2+m^2\mu^2).
\label{M115}
\end{eqnarray}
We note that even though  $\Pi_{\rm S}(q^2,m)$,  $\Pi_{\rm P}(q^2,m)$, and $g^{-1}$ all diverge logarithmically, they all diverge at precisely the same rate, with both $T_{\rm S}(q^2,M)=1/(g^{-1}-\Pi_{\rm S}(q^2,M))$ and $T_{\rm P}(q^2,M)=1/(g^{-1}-\Pi_{\rm P}(q^2,M))$ then being completely finite, a highly desirable outcome for a quantum field theory.

With $\Pi_{\rm P}(q^2=0,M)$ taking the form 
\begin{eqnarray}
\Pi_{\rm P}(q^2=0,M)&=&-4i\mu^2\int \frac{d^4p}{(2\pi)^4}\frac{(p^2)(-p^2)-M^2\mu^2}{((p^2+i\epsilon)^2+M^2\mu^2)^2}.
\nonumber\\
&=&4i\mu^2\int \frac{d^4p}{(2\pi)^4}\frac{1}{(p^2+i\epsilon)^2+M^2\mu^2}=-\frac{\mu^2}{4\pi^2}{\rm ln}\left(\frac{\Lambda^2}{M\mu}\right),
\label{M116}
\end{eqnarray}
we recognize $\Pi_{\rm P}(q^2=0,M)$ as being equal to none other than $1/g$ as given in (\ref{M105}). The pseudoscalar scattering amplitude $T_{\rm P}(q^2,M)$ thus has a massless pole at $q^2=0$. And with $\Pi^{\prime}_{\rm P}(q^2=0,M)=-7\mu/128\pi M$, near the pole the amplitude behaves as \cite{Mannheim2015,Mannheim2016}
\begin{eqnarray}
T_{\rm P}(q^2,M)=\frac{128\pi M}{7\mu q^2}=\frac{57.446 }{ q^2},
\label{M117}
\end{eqnarray}
with the last equality following when we set $\mu=M$. As required by dynamical mass generation in a chirally-symmetric theory such as that based on ${\cal {L}}_{\rm QED-FF}$ as given in (\ref{M87}), a massless pseudoscalar Goldstone boson is indeed generated dynamically. Thus by coupling QED to a four-fermion interaction, the Baker-Johnson evasion of the Goldstone theorem is itself evaded, and the mass generation that Johnson, Baker, and Willey had found is accompanied by a Goldstone boson after all. Then, because of the underlying chiral symmetry,  the pseudoscalar Goldstone boson must, as we now show, be accompanied by a dynamical scalar partner, a dynamical Higgs boson.

\subsection{Dynamical Higgs Boson of  $m\neq 0$ Electrodynamics at $\gamma_{\theta}(\alpha)=-1$}

Determining the bound state structure of $\Pi_{\rm S}(q^2,m)$ is not nearly as straightforward as that  of $\Pi_{\rm P}(q^2,m)$. With $\Pi_{\rm S}(q^2,m)$ and $\Pi_{\rm P}(q^2,m)$ being related according to 
\begin{eqnarray}
\Pi_{\rm S}(q^2,m)-\frac{1}{g}=\Pi_{\rm P}(q^2,m)-\frac{1}{g}-4i\mu^2\int \frac{d^4p}{(2\pi)^4}\frac{2m^2\mu^2}{D(q,p,m)},
\label{M118}
\end{eqnarray}
and with the last integral in (\ref{M118}) being readily doable analytically at $q^2=0$, we find that  with $\Pi_{\rm P}(q^2=0,M)=1/g$ the quantity $\Pi_{\rm S}(q^2=0,M)-1/g$ is equal to $\mu^2/4\pi^2$, to thus be non-zero. Consequently $T_{\rm P}(q^2,M)$ cannot have a pole at $q^2=0$, but must instead have one at some non-zero $q^2$.
\begin{figure}[htpb]
\begin{center}
\begin{minipage}[t]{9 cm}
\epsfig{file=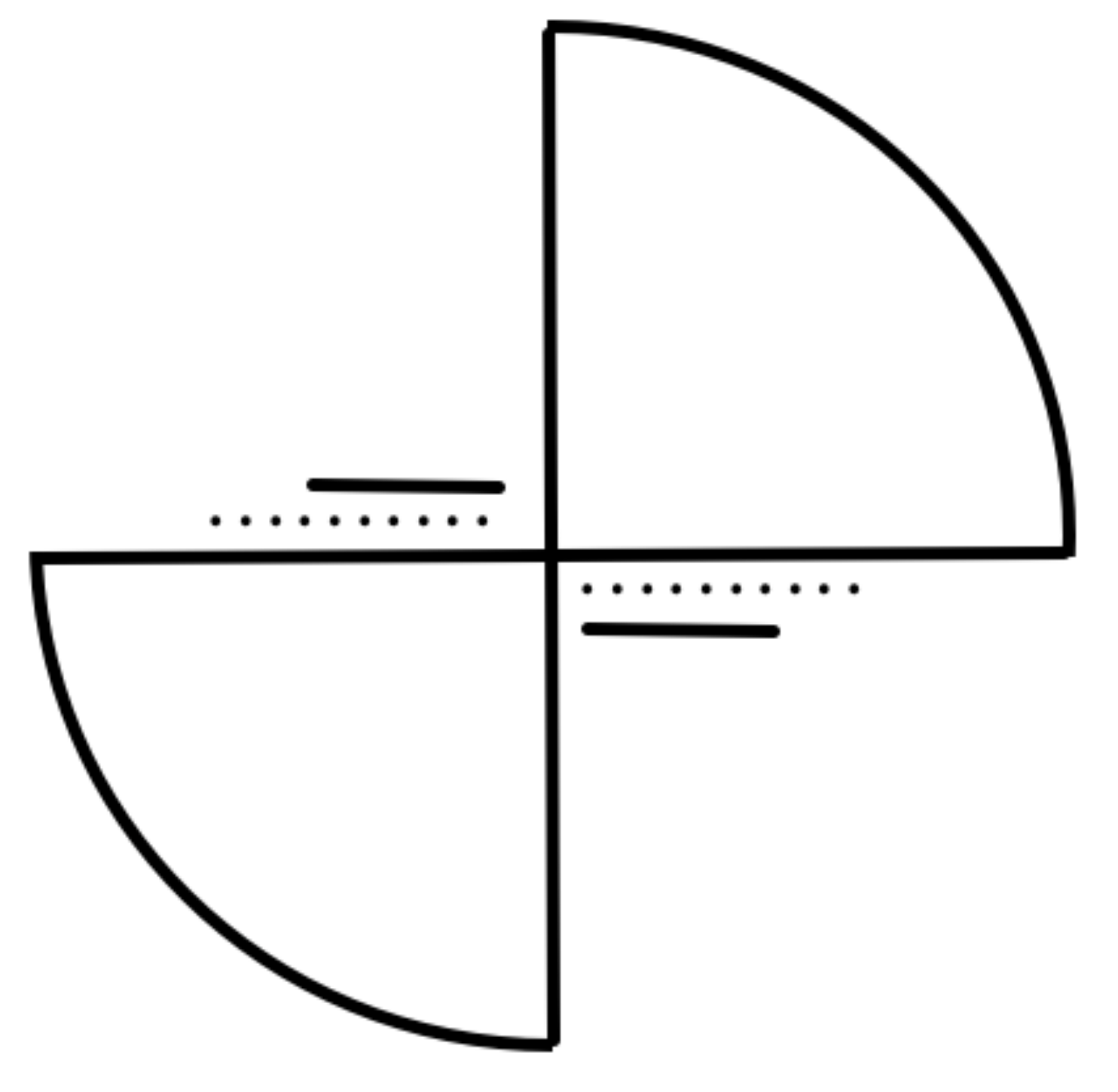,scale=0.5}
\end{minipage}
\begin{minipage}[t]{16.5 cm}
\caption{The Wick contour for $\Pi_{\rm S}(q^2,m=0)$ in the complex $p_0$ plane when $q^2$ is spacelike. The branch cuts are shown as lines and the poles as dots. For $\Pi_{\rm S}(q^2,m\neq 0)$ the poles move into the complex plane but remain in their respective quadrants and do not cross the Wick contour.}
\label{lw7}
\end{minipage}
\end{center}
\end{figure}

\begin{figure}[htpb]
\begin{center}
\begin{minipage}[t]{9 cm}
\epsfig{file=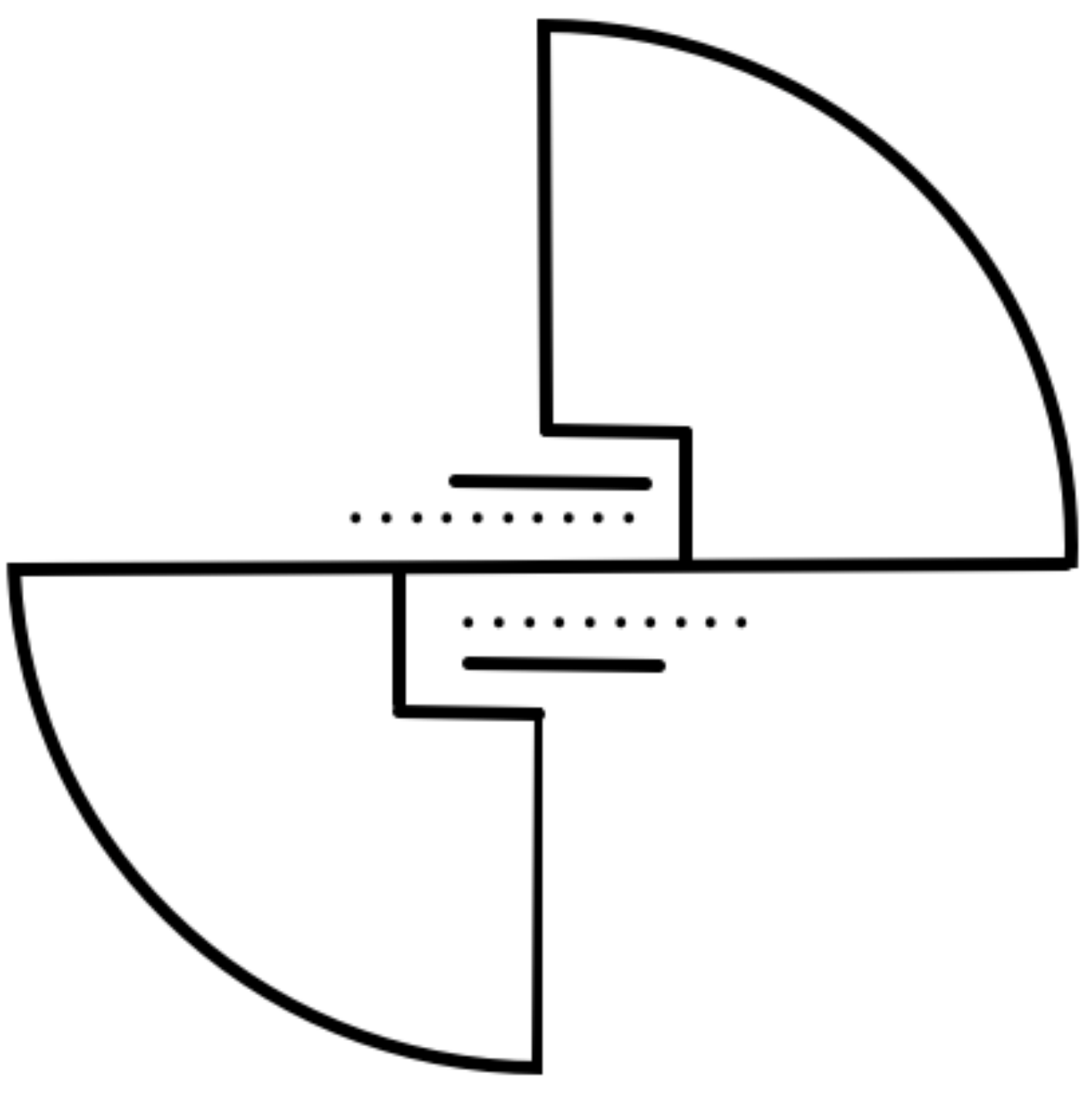,scale=0.5}
\end{minipage}
\begin{minipage}[t]{16.5 cm}
\caption{The migrated Wick contour for $\Pi_{\rm S}(q^2,m=0)$ in the complex $p_0$ plane when $q^2$ is timelike and  $p<q_0/2$. The branch cuts are shown as lines and the poles as dots. For $\Pi_{\rm S}(q^2,m\neq 0)$ the poles move into the complex plane but  do not cross the Wick contour.}
\label{lw8}
\end{minipage}
\end{center}
\end{figure}

The analytic structure of $\Pi_{\rm S}(q^2,m)$ is much more complicated than in the point-coupled case, and can most readily be determined by first looking at the analytic structure of $\Pi_{\rm S}(q^2,m=0)$  \cite{Mannheim2015}. For $\Pi_{\rm S}(q^2,m=0)$  there are analytic features associated with branch points in $N(q,p)$ and with poles due to zeroes in $D(q,p,m=0)$. When $q^2$ is spacelike, the singularity structure is completely familiar, with all the branch points in $N(q,p)$ and the zeroes in $D(q,p,m=0)$ being in the upper left-hand and lower right-hand quadrants in the complex $p_0$ plane just as exhibited in Fig. (\ref{lw7}). For $D(q,p,m\neq 0)$ the zeroes move into the complex $p_0$ plane but remain in their respective quadrants and do not cross the Wick contour. However, the singularity structure changes when $q^2$ is timelike. Specifically, if we set $q_{\mu}=(q_0,0,0,0)$ and $p_1^2+p_2^2+p_3^2=p^2$, we find that the branch points in $N(q,p)$ are located at
\begin{eqnarray}
&&p_0=q_0/2+p-i\epsilon,\qquad p_0=-q_0/2+p-i\epsilon
\nonumber\\
&&p_0=q_0/2-p+i\epsilon,\qquad p_0=-q_0/2-p+i\epsilon. 
\label{M119}
\end{eqnarray}
Thus for $p<q_0/2$ the associated branch points in $N(q,p)$ migrate into the upper right-hand and lower left-hand quadrants in the complex $p_0$ plane. With exactly the same migration happening for the zeroes in $D(q,p,m)$, the  $p<q_0/2$ singularity structure of  $\Pi_{\rm S}(q^2,m)$ with timelike $q^2$ is as shown in Fig. (\ref{lw8}).\footnote{Even though $\Pi_{\rm P}(q^2,m)$ has the same singularity structure as $\Pi_{\rm S}(q^2,m)$, since the migration of singularities does not occur for any lightlike $q^2$ including the tip of light cone where  $q_{\mu}=0$, we are able to identify the $q^2=0$ pole in $T_{\rm P}(q^2,M)$ without needing to take the singularity structure of $\Pi_{\rm P}(q^2,m)$ into account.}

Given the singularity structure in Fig. (\ref{lw8}), we can deform the $p_0$-plane contour for $\Pi_{\rm S}(q^2,m)$ into a Wick piece and a cut discontinuity piece, and with the circle at infinity contributions in Fig. (\ref{lw8}) being negligible, for the $p_0$ integration we symbolically obtain 
\begin{eqnarray}
-i\int_{-\infty}^{\infty}dp_0=\int_{-\infty}^{\infty}dp_4-\left( \int_0^{q_0/2-p}dp_0+\int_{q_0/2-p}^0dp_0+\int_0^{-q_0/2+p}dp_0+\int_{-q_0/2+p}^0dp_0\right),
\label{M120}
\end{eqnarray}
where $p_4=ip_0$. Thus we can set $\Pi_{\rm S}(q^2,m)=I_{\rm Wick}+I_{\rm cut}$, where 
\begin{eqnarray}
I_{\rm cut}=-\frac{4i\mu^2}{\pi^3}\int_0^{q_0/2}dp p^2 \int_0 ^{q_0/2-p}dp_0\frac{N(q_0,p,p_0)}{D(q_0,p,p_0,m)},
\label{M121}
\end{eqnarray}
\begin{eqnarray}
I_{\rm Wick}=\frac{\mu^2}{\pi^3}\int_0^{\infty} dp p^2\int_{\infty}^{\infty}dp_4\frac{N(q_0,p,p_4)+m^2\mu^2}{D(q_0,p,p_4,m)}.
\label{M122}
\end{eqnarray}
To simplify the cut contribution we set $p_0=q_0\lambda/2$, $p=q_0\sigma/2$, and obtain 
\begin{eqnarray}
I_{\rm cut}&=&-\frac{4i\mu^2}{\pi^3}\int_0^1d\sigma\sigma^2\int_0^{1-\sigma}d\lambda \frac{N_{\rm cut}}{D_{\rm cut}},
\nonumber\\
N_{\rm cut}&=&-(\lambda^2-\sigma^2-1)[(\lambda^2-\sigma^2+1)^2-4\sigma^2]^{1/2}q_0^8,
\nonumber\\
D_{\rm cut}&=&256m^4\mu^4+32m^2\mu^2[(\lambda^2-\sigma^2+1)^2+4\sigma^2]q_0^4
+[(\lambda^2-\sigma^2+1)^2-4\sigma^2]^2q_0^8.
\label{M123}
\end{eqnarray}
To simplify the Wick contribution we set $p_4=rz$, $p=r(1-z^2)^{1/2}$, and obtain
\begin{eqnarray}
I_{\rm Wick}&=&\frac{2\mu^2}{\pi^3}\int_0^{\infty}dr r^3\int_0^1dz(1-z^2)^{1/2}
\left[\frac{N(q,r,z)+m^2\mu^2}{D(q,r,z)}\right],
\nonumber\\
N(q,r,z)&=&-(r^2+q_0^2/4)[(r^2-q_0^2/4)^2+r^2z^2q_0^2]^{1/2},
\nonumber\\
D(q,r,z)&=&[(r^2-q_0^2/4)^2+r^2z^2q_0^2-m^2\mu^2]^2
+4m^2\mu^2(r^2-q_0^2/4)^2.
\label{M124}
\end{eqnarray}

Writing the square root factor in $I_{\rm cut}$ as $(\lambda^2-\sigma^2+1-2\sigma)(\lambda^2-\sigma^2+1+2\sigma)=[2+\lambda^2-(\sigma+1)^2][2+\lambda^2-(\sigma+1)^2+4\sigma]$, we see that with $0\leq \sigma\leq 1$ the square root factor is always real in the range of integration. Consequently, for any value of $q_0$ $I_{\rm cut}$ is pure imaginary, and thus to cancel this complex piece so that $g^{-1}-\Pi_{\rm S}(q^2,M)$ vanishes somewhere, we will need some other complex contribution. This additional contribution is provided by $D(q,p,m)$ as it has a branch point of its own at $q^2=2m\mu$ \cite{Mannheim2015}. Specifically, we note that $D(q,r,z)$ will vanish if $r=q_0/2$, $z=m\mu/rq_0$, i.e. if $z=2m\mu/q_0^2$. Since $z$ is less than one there will always be some $r$ and some $z$ for which $D(q,r,z)$ will vanish if $q_0^2\geq 2m\mu$. We thus identify $q_0^2=q^2=2m\mu$ as a threshold, with, as shown explicitly in  \cite{Mannheim2015},  there being a discontinuity in $I_{\rm Wick}$ if $q^2\geq 2m\mu$.\footnote{The existence of a discontinuity at $q^2=2m\mu$  can also be seen by noting that at $p=0$ the $\tilde{S}_{\mu}(p)$ propagator has poles at  $p_0=(1+i)(m\mu)^{1/2}/2^{1/2}$ and $p_0=(1-i)(m\mu)^{1/2}/2^{1/2}$ when $\gamma_{\theta}(\alpha)=-1$, to thus give the $\tilde{S}_{\mu}(p)$-based $\Pi_{\rm S}(q^2,m)$ a particle-antiparticle threshold at $q^2=((1+i)(m\mu)^{1/2}/2^{1/2}+(1-i)(m\mu)^{1/2}/2^{1/2})^2=2m\mu$.} Thus $I_{\rm Wick}$ with its seemingly real integrand  actually develops an imaginary part when $q^2\geq 2m\mu$, and it is this imaginary part that will then cancel the pure imaginary $I_{\rm cut}$. Solutions to $T_{\rm S}(q^2,M)=g^{-1}-\Pi_{\rm S}(q^2,M)=0$ must thus lie above the $q^2=2M\mu$ threshold, to thus correspond to resonances rather than bound states. Since none of the discontinuity structure affects the ultraviolet behavior of $T_{\rm S}(q^2,M)$, as with $T_{\rm P}(q^2,M)$, $T_{\rm S}(q^2,M)$ is completely finite when $\gamma_{\theta}(\alpha)=-1$.

Thus once we dress the point vertices of the original point-coupled NJL model the dynamical scalar bound state pole in $T_{\rm S}(q^2,M)$ must move into the complex $q^2$ plane and become an above-threshold resonance. This is a key point of our study, and it provides a quite sharp contrast with an elementary Higgs boson since the mass of an elementary Higgs boson is given by the magnitude of the second derivative of the Higgs potential at its minimum, a quantity that must be real if the potential itself is.

With $I_{\rm Wick}$ being expressible in terms of elliptic integrals \cite{Mannheim2015}, an actual analytic evaluation of the discontinuity in $I_{\rm Wick}$ can be obtained \cite{Mannheim2015}, and with a little numerical work, the dynamical Higgs boson associated with $T_{\rm S}(q^2,M)$ as evaluated at $\gamma_{\theta}(\alpha)=-1$ is found to lie at \cite{Mannheim2015,Mannheim2016}
\begin{eqnarray}
q_0=(1.48-0.02i)(M\mu)^{1/2},\qquad q^2=(2.19-0.05i)M\mu,
\label{M125}
\end{eqnarray}
with $T_{\rm S}(q^2,M)$ having the Breit-Wigner structure
\begin{eqnarray}
T_{\rm S}(q^2,M)=\frac{46.14+1.03i}{q^2-2.22M\mu+0.05iM\mu}
\label{M126}
\end{eqnarray}
near the resonance. The Higgs boson thus lies just above the $q^2=2M^2$ threshold (on setting $\mu=M$) with an expressly negative imaginary part just as required for decay, with the associated decay width being fairly narrow ($0.05/2.19=0.02$). Such a width could potentially serve to distinguish a dynamical Higgs boson from an elementary one. 

Since the solution to $\beta(\alpha)=0$ could possibly be the physical electric charge rather than the bare charge  \cite{Adler1972},\footnote{As noted in \cite{Adler1972}, if $\beta(\alpha)$ has a zero it must be an infinite order one. To understand this we note that if $\beta(\alpha)$ is a power series in $\alpha$ then it follows from the Callan-Symanzik equations that the charge renormalization constant $Z_3$ will be a power series in ${\rm ln }(\Lambda^2/m^2)$. However, if $\beta(\alpha)$ has a zero at $\alpha=\alpha_0$, then at that zero $Z_3$ will be cutoff independent. Hence the coefficients of the ${\rm ln }(\Lambda^2/m^2),~{\rm ln }^2(\Lambda^2/m^2),...$ terms must all vanish identically, an infinite amount of information. If we now Taylor series $\beta(\alpha)=(\alpha-\alpha_0)\beta^{\prime}(\alpha_0)+(1/2)(\alpha-\alpha_0)^2\beta^{\prime\prime}(\alpha_0)+...$, then since each power of $\alpha-\alpha_0$ generates differing combinations of  powers of ${\rm ln}(\Lambda/^2m^2)$, it follows that all derivatives of $\beta(\alpha)$ must vanish at $\alpha=\alpha_0$.} with $\alpha$ then being small, and with, as noted above, the four-fermion $g$ also being able to be small, we are able to obtain dynamical symmetry breaking even with weak coupling, even though the prevailing wisdom based on the quenched ladder approximation always having been that dynamical symmetry breaking is strictly a strong-coupling effect. Thus as with Cooper pairing, the driver is not the strength of the coupling but the existence of a filled sea of fermions as needed to make the vacuum energy density be negative.

As a final comment on the ideas presented here, we note that their extension to a non-Abelian gauge theory coupled  a four-fermion interaction is (in principle at least) straightforward, with it requiring that one have critical scaling as realized with the dimension of the mass operator being reduced from three to two. Having critical scaling in a non-Abelian gauge theory is not ordinarily considered in the literature because it would mean giving up asymptotic freedom. However, this loss of asymptotic freedom may not be as problematic as it may at first sound. As shown  in \cite{Mannheim1975}, albeit somewhat heuristically, the residual-interaction-generated fluctuations around the self-consistent Hartree-Fock vacuum turn out to be asymptotically free. So we use critical scaling with anomalous dimensions to get into the self-consistent vacuum in the first place, with the fluctuations around it then being asymptotically free.

\subsection{Some Other Approaches}

Beyond the work described here of a critical scaling JBW electrodynamics coupled to a four-fermion interaction, there have been other studies of an Abelian gluon model coupled to a four-fermion interaction, though in them the Abelian gluon sector has been treated in the quenched ladder approximation, see e.g. \cite{Bardeen1986,Leung1986,Yamawaki1986,Kondo1989,Bardeen1990,Bardeen1992,Carena1992,Gusynin1992,Kondo1993,Harada1994,Hashimoto1998,Gusynin1998}. Studies of non-Abelian gauge theories, where one studies the implications of renormalization group $\beta$ functions that are everywhere negative or have regions that are negative and regions that are positive,\footnote{In the non-Abelian $SU(3)$ quantum chromodynamics (QCD) case with $N_f$ fermions and coupling constant $\alpha_s$, the first two terms in the expansion of the coupling constant renormalization $\beta$ function are given by $\beta(\alpha_s)=-\beta_1\alpha_s^2-\beta_2\alpha_s^3$, where $\beta_1=(33-2N_f)/6\pi$, $\beta_2=(306-38N_f)/24\pi^2$. While both $\beta_1$ and $\beta_2$ have the same positive sign for small enough $N_f$, $\beta_2$ can change sign if $N_f>8.05$. There is thus a window in which $\beta_1$ is positive and $\beta_2$ is negative, viz. $8.05\leq N_f\leq 16.5$, with $\beta(\alpha_s)$ having a zero in this window at which it changes sign. Additionally, the anomalous dimension of the fermion mass operator is given by (see e.g. \cite{Ryttov2011} and references therein) $\gamma_{\theta}(\alpha_s)=-2\alpha_s/\pi-\alpha_s^2(303-10N_f)/36\pi^2$, and like in the Abelian case starts off negative in lowest order, and would remain negative in second order if $N_f< 30.3$, with propagators being asymptotically damped.} may be found in e.g. \cite{Holdom1981,Miransky1981,Holdom1985,Appelquist1986,Holdom1987,Appelquist1991,Miransky1993,Appelquist2010a}. General reviews of dynamical symmetry breaking and composite Higgs bosons may be found in e.g. \cite{Hill2003,Miransky2010,Contino2010,Appelquist2010b,Yamawaki2010,Panico2015}. While it is straightfoward to apply the ladder approximation in the non-Abelian case since it effectively duplicates the Abelian calculation (one simply replaces the charge $\alpha$ by a non-Abelian combinatoric factor), unlike in the Abelian case in order to go beyond the ladder approximation one needs to incorporate the three-gluon vertex, with a graph in which a quark emits a gluon that breaks up into two gluons that both then rejoin the fermion line is still quenched, still planar, but not a ladder graph. Some analysis of non-Abelian theories beyond the ladder approximation may be found in \cite{Appelquist1988,Holdom1988,Holdom1989,Mahanta1989}.

\begin{figure}[htpb]
\begin{center}
\begin{minipage}[t]{9 cm}
\epsfig{file=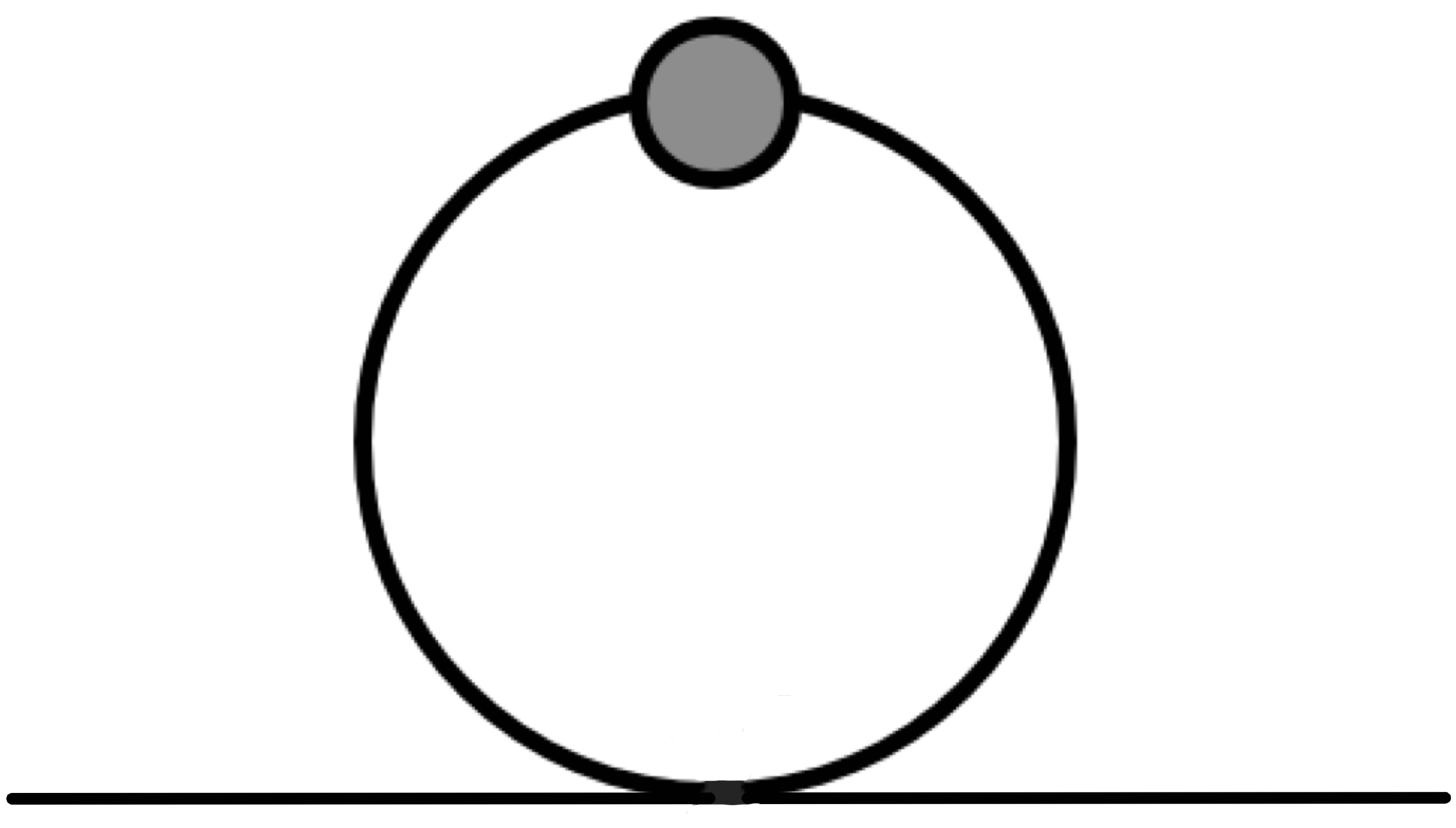,scale=0.3}
\end{minipage}
\begin{minipage}[t]{17.5 cm}
\caption{The four-fermion interaction tadpole contribution to the fermion self energy in the Abelian gluon model quenched ladder approximation. The blob represents the quenched ladder approximation contribution to the fermion propagator.}
\label{semitadpoleselfenergy}
\end{minipage}
\end{center}
\end{figure}

It is instructive to compare and contrast  our work on JBW electrodynamics coupled to a four-fermion interaction with the studies of a quenched ladder approximation Abelian gluon model coupled to a four-fermion interaction. In the JBW case one evaluates the $\langle \Omega_m|\bar{\psi}\psi|\Omega_m\rangle$ expectation value using the fully dressed tadpole given in Fig. (\ref{lw3}), whereas, in the quenched ladder approximation one evaluates $\langle \Omega_m|\bar{\psi}\psi|\Omega_m\rangle$ using the partially dressed tadpole given in 
Fig. (\ref{semitadpole}). Then, to determine the fermion propagator, in the quenched ladder approximation to the Schwinger-Dyson equation one adds on to the planar graphs contained in Fig. (\ref{SD1}) the partially dressed tadpole contribution given in Fig. (\ref{semitadpoleselfenergy}).\footnote{In contrast, in the JBW case, to determine the fermion propagator one uses only the QED contribution to the Schwinger-Dyson equation without the addition of any tadpole graph contribution, with the tadpole contribution to the fermion mass  being generated by the residual interaction as per $m=g\langle \Omega_m|\bar{\psi}\psi|\Omega_m\rangle$. The mean field approach thus organizes the Feynman graphs very differently than the quenched ladder approximation approach.} In the quenched ladder approach one thus replaces the NJL point-coupled $ \langle \Omega_m|\bar{\psi}\psi|\Omega_m\rangle$ given in (\ref{M60}) by the Landau gauge quenched ladder approximation propagator contribution to $ \langle \Omega_m|\bar{\psi}\psi|\Omega_m\rangle$ , to give
\begin{eqnarray}
\langle \Omega_m|\bar{\psi}\psi|\Omega_m\rangle=-i\int \frac{d^4q}{(2\pi)^4}{\rm Tr}\frac{1}{\slashed{q}-B(q^2)}
=-4i\int \frac{d^4q}{(2\pi)^4}\frac{B(q^2)}{q^2-B^2(q^2)},
\label{M127}
\end{eqnarray}
where $B(p^2)$ is to be self-consistently determined from the Schwinger-Dyson equation
\begin{eqnarray}
B(p^2)=g\langle \Omega_m|\bar{\psi}\psi|\Omega_m\rangle+\frac{3\alpha}{4\pi}\bigg{[}\int_0^{p^2}dq^2 \frac{q^2B(q^2)}{p^2(q^2+B^2(q^2))}
+\int_{p^2}^{\infty}dq^2 \frac{B(q^2)}{(q^2+B^2(q^2))}\bigg{]}
\label{M128}
\end{eqnarray}
that is to replace (\ref{M76}). With the bare mass $m_0$ now being taken to be zero identically, non-trivial solutions to (\ref{M128}) correspond to dynamical symmetry breaking. As before we look for an asymptotic solution, and since in the quenched ladder approximation the tadpole has not been quenched enough so as to make it be only logarithmically divergent, we still need a cutoff for the four-fermion sector.\footnote{By using the partially dressed tadpole of Fig. (\ref{semitadpole}) rather than the fully dressed tadpole of Fig. (\ref{lw3}), one is not able to take advantage of the fact that at $\alpha=\pi/3$, the four-fermion interaction would be power-counting renormalizable  since at that value $d_{\theta}(\alpha)=2$. Thus even at $\alpha=\pi/3$ one would still need a cutoff when a quenched ladder Abelian gluon model is coupled to a four-fermion interaction. However, when a critical scaling JBW electrodynamics with $d_{\theta}(\alpha)=2$ is coupled to a four-fermion interaction, no cutoff is needed.}  So this time we take the asymptotic solution to be of the form $B(p^2)=m(p^2/\Lambda^2)^{(\nu-1)/2}$, and obtain
\begin{eqnarray}
\nu=\pm\left(1-\frac{3\alpha}{\pi}\right)^{1/2},\qquad g\langle \Omega_m|\bar{\psi}\psi|\Omega_m\rangle+\frac{3\alpha m}{2\pi(\nu-1)}=0.
\label{M129}
\end{eqnarray}
With $\nu-1$ being negative, $g\langle \Omega_m|\bar{\psi}\psi|\Omega_m\rangle$ is given by its leading term according to 
\begin{eqnarray}
g\langle \Omega_m|\bar{\psi}\psi|\Omega_m\rangle=-\frac{mg\Lambda^2}{2\pi^2(1+\nu)},
\label{M130}
\end{eqnarray}
an expression that limits to the leading term in (\ref{M60}) as $\alpha \rightarrow 0$ if we take $\nu=+(1-3\alpha/\pi)^{1/2}$
As noted in \cite{Kondo1989,Bardeen1990} and references therein, broken symmetry solutions thus lie on the  critical surface\footnote{To be on this surface requires that  $\alpha < \pi/3$, since otherwise $g$ would have to be complex and the four-fermion interaction Lagrangian would not be Hermitian.}  
\begin{eqnarray}
-g\Lambda^2=\pi^2(1+(1-3\alpha/\pi)^{1/2})^2.
\label{M131}
\end{eqnarray}
While the quenched ladder approximation on its own has no dynamical symmetry breaking solutions if $\alpha \leq \pi/3$, now we see that we can get broken symmetry solutions in the $\alpha\leq \pi/3$ region provided the Abelian gluon model is accompanied by a four-fermion interaction with an appropriately chosen value for $-g\Lambda^2$. Thus as we make $\alpha$ smaller and smaller, we have to make $-g\Lambda^2$ be bigger and bigger, while at  $\alpha=\pi/3$ itself, we still need a minimum $-g\Lambda^2=\pi^2$. In contrast, in the JBW case where the coupling constant is not free to vary but must satisfy $\beta(\alpha)=0$ identically, not only is there not any quadratic $\Lambda^2$ term in $\langle \Omega_m|\bar{\psi}\psi|\Omega_m\rangle$ to begin with (cf. (\ref{M104})), as we have seen, symmetry breaking occurs no matter how small $g$ might be as long as it is negative (viz. attractive). Thus while a quenched ladder Abelian gluon model coupled to a four-fermion interaction and a critical scaling JBW electrodynamics coupled to a four-fermion interaction can both exhibit dynamical symmetry breaking at $d_{\theta}(\alpha)=2$, only the JBW case can do so for an arbitrarily weakly coupled four-fermion interaction.\footnote{Because of studies of models such as the quenched ladder approximation to the Abelian gluon theory in the $\alpha > \pi/3$ region, it is thought that dynamical symmetry breaking can only occur for strong coupling. And with the weak interaction symmetry breaking scale (viz. the value of $\langle S|\hat{\phi}|S\rangle$) being much bigger than the strong interaction chiral symmetry breaking scale (viz. $f_{\pi}$) that is to be produced by QCD, a non-Abelian technicolor gauge theory of strength greater than QCD has been invoked in order to break the weak interaction symmetry  dynamically. This breaking induces high mass (TeV or so region) technifermions and should lead to an equally high mass dynamical Higgs boson. While theoretical attempts to then bring the Higgs mass down to the 125 GeV value that it is now known to have are currently ongoing, no satisfactory solution to this problem has yet been found, with attempts to have the Higgs boson emerge as a hoped-for  relatively light (viz. pseudo) Goldstone boson that could be associated with  a spontaneous breakdown of scale symmetry (cf. a dilaton) have yet to succeed. However, as we have seen, none of this may be necessary, since our study here shows that dynamical symmetry breaking can occur even with weak coupling, to potentially make theories such as technicolor unnecessary.}

\subsection{Weak Coupling Versus Strong Coupling}

In the literature there are various arguments that indicate that in order to get dynamical symmetry breaking one needs strong coupling. The ones we have encountered here are based on the point-coupled NJL model (need $-g\Lambda^2>1$), the quenched, planar graph approximation to the Abelian gluon model (need $\alpha > \pi/3$), and a quenched, planar graph approximation Abelian gluon model coupled to a point-coupled NJL model (need $-g\Lambda^2=\pi^2(1+(1-3\alpha/\pi)^{1/2})^2$). However, the very first example of dynamical symmetry breaking that was presented in the literature was the BCS model, where symmetry breaking and Cooper pairing would occur no matter how weak the coupling constant is as long as it is attractive (cf. (\ref{M34})), with the Pauli blocking due to the filled Fermi sea eliminating the need for the additional binding that a two-body system on its own does not possess. To address this dichotomy between strong and weak coupling we have shown that there actually is no Goldstone boson in the Abelian gluon model when the non-planar graphs are included, no matter whether the coupling constant is weak or strong, with the symmetry breaking wisdom based on the quenched ladder approximation not being reliable. Then, when we couple a critical scaling Abelian model with $\beta(\alpha)=0$, $\gamma_{\theta}(\alpha)=-1$ to a four-fermion theory with coupling $g$, we do find a Goldstone boson no matter how small $g$ might be.  As with BCS theory, in a critical scaling Abelian gluon model coupled to a four-fermion interaction, the driver is not the strength of the coupling constant but the existence of an infinite number of degrees of freedom, with the vacuum energy density $\tilde{\epsilon}(m)$ given in (\ref{M106}) being more negative when $m$ is non-zero than when $m$ is zero simply because the negative energy states in the Dirac sea are all occupied, with a set of massive Dirac sea states having lower energy density than a set of massless ones. Now if chiral-symmetry breaking massive fermions are explicitly favored in the mean-field sector, then because of the underlying chiral symmetry of the four-fermion plus gauge theory model the residual interaction must generate a Goldstone boson. But since the lowering of the energy density in the mean-field sector occurs for any $g$ with the same sign as $\langle\Omega_m|\bar{\psi}\psi|\Omega_m\rangle=m/g$,  Goldstone boson generation in the residual interaction sector must occur for any $g$ no matter how weak it may be.\footnote{For the point-coupled NJL model we should thus choose the  $-g>4\pi^2/\Lambda^2\rightarrow 0$ realization of (\ref{M60}) discussed in Sec. (6.2) and not the $-g\Lambda^2/4\pi^2>1$ one.}  As we thus see, strong coupling is not in fact needed for dynamical symmetry breaking, and it can be obtained even when the coupling is weak.

\subsection{Anomalous Dimensions and the Renormalizability of the Four-Fermion Interaction}

\begin{figure}[htpb]
\begin{center}
\begin{minipage}[t]{9 cm}
\epsfig{file=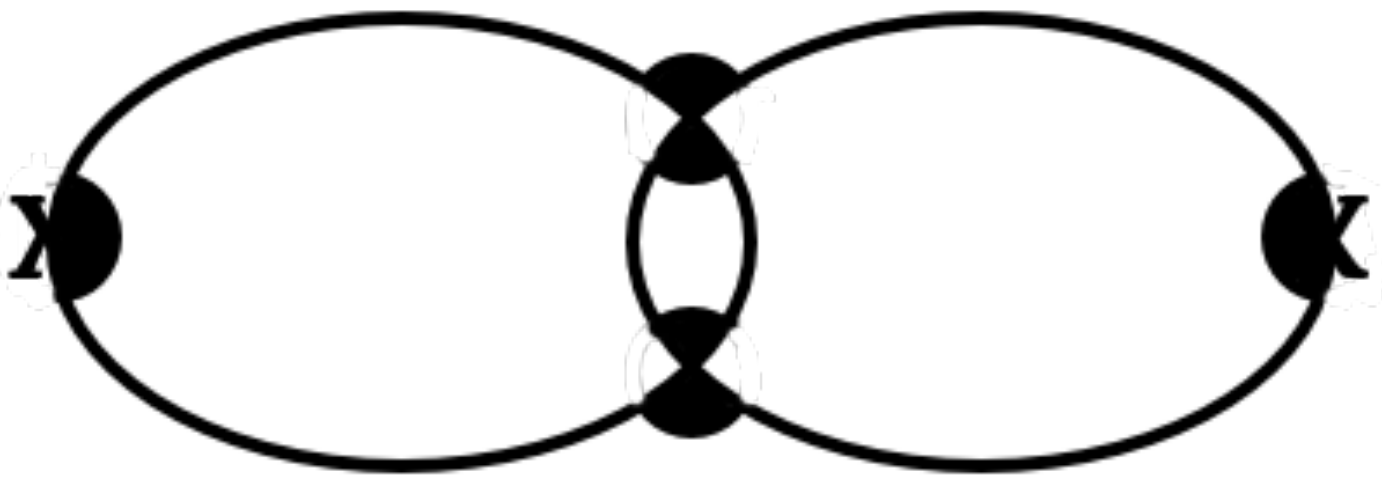,scale=0.5}
\end{minipage}
\begin{minipage}[t]{16.5 cm}
\caption{Order $g^2$ contribution to $\Pi_{\rm S}(q^2,M=0)$. The blobs denote $\tilde{\Gamma}_{\rm S}(p,p+q,q)$ with appropriate momenta.}
\label{design1}
\end{minipage}
\end{center}
\end{figure}

\begin{figure}[htpb]
\begin{center}
\begin{minipage}[t]{9 cm}
\epsfig{file=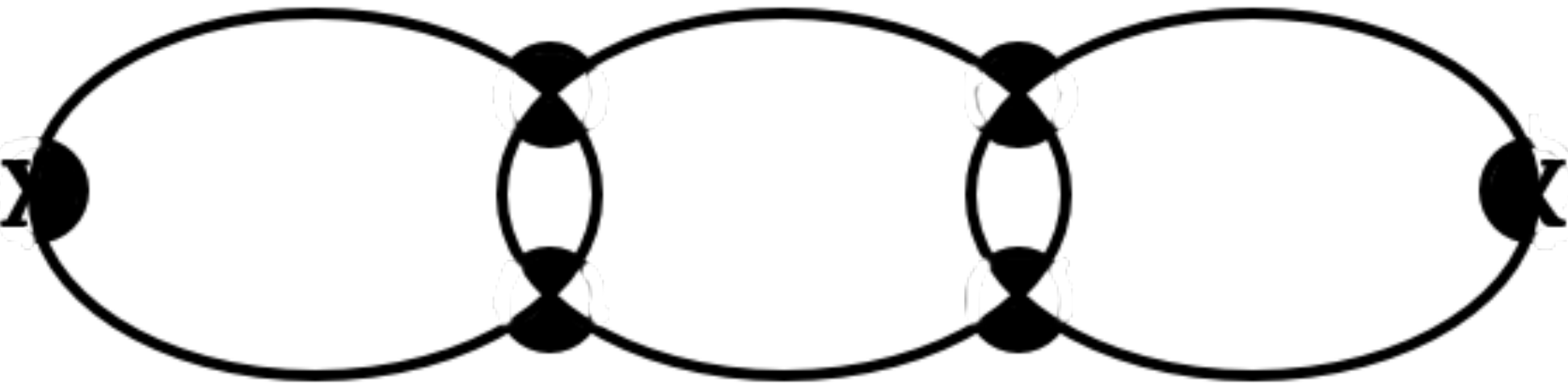,scale=0.5}
\end{minipage}
\begin{minipage}[t]{16.5 cm}
\caption{Order $g^4$ contribution to $\Pi_{\rm S}(q^2,M=0)$. The blobs denote $\tilde{\Gamma}_{\rm S}(p,p+q,q)$ with appropriate momenta.}
\label{design2}
\end{minipage}
\end{center}
\end{figure}

In our analysis of dynamical symmetry breaking we have seen that the condition $d_{\theta}(\alpha)=2$ (viz. $\gamma_{\theta}(\alpha)=-1$) has played a central role. Also we have indicated that this condition would lead to a power-counting renormalizable four-fermion interaction, since with this condition the $(\bar{\psi}\psi)^2$ interaction (and analogously $(\bar{\psi}i\gamma^5\psi)^2$) is effectively acting as a dimension four operator.  In the literature various authors have suggested that if $d_{\theta}$ is reduced to two (or less of course) then the four-fermion theory would be become renormalizable (see \cite{Mannheim1975}, \cite{Leung1986}, \cite {Miransky1989}, \cite{Kondo1993}, \cite{Harada1994}). Now in Sec. (8.4) we showed that the $T_{\rm S}(q^2,M)=1/(g^{-1}-\Pi_{\rm S}(q^2,M))$ and $T_{\rm P}(q^2,M)=1/(g^{-1}-\Pi_{\rm P}(q^2,M))$ scattering amplitudes were finite to lowest order in the four--fermion coupling constant $g$ that appears in the $I_{\rm QED-FF}$ action given in (\ref{M88}). However, to actually establish renormalizability one needs to extend this result to higher orders in $g$, so as to incorporate graphs such as those in Figs. (\ref{design1}) and (\ref{design2}) and their higher order generalizations. Thus as well as dress  $\Pi_{\rm S}(q^2,M)$ and  $\Pi_{\rm P}(q^2,M)$ with QED contributions, we also need to dress them with higher order four-fermion  contributions. 

To this end we note that since, according to (\ref{M98}), the leading asymptotic behavior of $\tilde{S}_{\mu}(p)$ is $1/\slashed{p}$ if $\gamma_{\theta}(\alpha)=-1$, the ultraviolet divergence structure of $\Pi_{\rm S}(q^2,M)$ is the same as that of $\Pi_{\rm S}(q^2,M=0)$, and thus, because of the underlying chiral symmetry, also the same as that of $\Pi_{\rm P}(q^2,M)$ and $\Pi_{\rm P}(q^2,M=0)$. We can thus determine the ultraviolet structure of the graphs in Figs. (\ref{design1}) and (\ref{design2}) on so on by evaluating them with massless fermion propagators. To determine exactly where to put the dressed $\tilde{\Gamma}_{\rm S}(p,p+q,q)$ vertex and its pseudoscalar analog $\tilde{\Gamma}_{\rm P}(p,p+q,q)$, we note that in the path integral ${\int}D[\bar{\psi}]D[\psi]D[A_{\mu}]\exp(iI_{\rm QED-FF})$ we can add in a dummy Gaussian integration $\int D[\sigma]\exp[i\int d^4x (\sigma-g\bar{\psi}\psi)^2/2g]$ (and analogously for $\bar{\psi}i\gamma^5\psi$). When combined with the four-fermion terms in $I_{\rm QED-FF}$ this leads to a net contribution of the form $\int D[\sigma]\exp(iI_{\sigma})$ where $I_{\sigma}=\int d^4x[-\sigma\bar{\psi}\psi+\sigma^2/2g]$, to thus effectively break up the point four-fermion interactions into $\sigma$ (and $\pi$) mediated Yukawa interactions with zero-momentum, point-like propagators $1/(-m_{\sigma}^2)=g$, $1/(-m_{\pi}^2)=g$. The scalar and pseudoscalar Yukawa vertices are then dressed with $\tilde{\Gamma}_{\rm S}(p,p+q,q)$ and $\tilde{\Gamma}_{\rm P}(p,p+q,q)$. This gives the pattern of  vertex dressings exhibited in Figs. (\ref{design1}) and (\ref{design2}).

Given Figs. (\ref{design1}) and (\ref{design2}) and their higher order analogs, with $\gamma_{\theta}(\alpha)=-1$ the Green's functions $\Pi_{\rm S}(q^2,M)$ and $\Pi_{\rm P}(q^2,M)$ are found \cite{Mannheim2016d} to only diverge as a single logarithm to all orders in $g$ (i.e. no higher powers of logarithms). Similarly, we find that dressing the tadpole graph of Fig. (\ref{lw3}) to all orders in $g$ causes $g^{-1}=\langle \Omega_M|\bar{\psi}\psi|\Omega_M\rangle/M$ to diverge as the self-same single logarithm with the self-same coefficient. In consequence, the scalar and pseudoscalar fermion-antifermion scattering amplitudes are not just renormalizable, they are completely finite to all orders in $g$.

The reason why we get this automatic cancellation of ultraviolet divergences is that in the expansions given in Figs.  (\ref{lw2}) and (\ref{lw6}) the only ultraviolet divergent graphs are those with two $\bar{\psi}\psi$ insertions, with $\epsilon(m)$, $\Pi_{\rm S}(q^2,M)$ and $\Pi_{\rm P}(q^2,M)$ thus all having the identical ultraviolet divergence structure (the massless theory $\Pi_{\rm S}(q^2,M=0)$ can be recognized as the first graph in the summation given in Fig. (\ref{lw6})). Moreover, this continues to be the case even after the Green's functions are dressed to all orders in $g$ as per Figs. (\ref{design1}) and (\ref{design2}) and so on. The divergent part of $\epsilon(m)$ is given by $(1/2)G^{(2)}_0(q_{\mu}=0,m=0)m^2$, with the divergent part of $\epsilon^{\prime}(m)$ thus being given by $G^{(2)}_0(q_{\mu}=0,m=0)m$. And with $G^{(2)}_0(q_{\mu}=0,m=0)$, $\Pi_{\rm S}(q_{\mu}=0,M=0)$, and $\Pi_{\rm P}(q_{\mu}=0,M=0)$ all being identically equal in the massless theory, on identifying $\epsilon^{\prime}(m)$ with $m/g$ at $m=M$, the cancellations automatically follow.

It is important to note the role played by dynamical symmetry breaking. As far as ultraviolet divergences are concerned, we find that with $\Pi_{\rm S}(q^2,M)$ and $\Pi_{\rm P}(q^2,M)$ only diverging as the same single logarithm to all orders in $g$, each of them only needs one  common subtraction. We could thus pick $g^{-1}$ so as to provide the needed subtraction, with $T_{\rm S}(q^2,M)=1/(g^{-1}-\Pi_{\rm S}(q^2,M))$ and $T_{\rm P}(q^2,M)=1/(g^{-1}-\Pi_{\rm P}(q^2,M))$  then both being rendered finite. However, with dynamical symmetry breaking (an infrared effect), the $g^{-1}$ that obeys the Hartree-Fock condition $M/g=\langle \Omega_M|\bar{\psi}\psi|\Omega_M\rangle$ is precisely found to be the one that is needed to render the scattering amplitudes finite. This shows the power  of dynamical symmetry breaking.

While the renormalizabilty we have found here bears some similarity to the renormalizabilty that occurs in a Yukawa coupled scalar field theory, there is a crucial difference. In the case of the action $I=\int d^4x (i\bar{\psi}\gamma^{\mu}\partial_{\mu}\psi+(1/2)\partial_{\mu}S\partial^{\mu}S-(1/2)m_S^2S^2-hS\bar{\psi}\psi)$ with a quantum scalar field $S$ and Yukawa coupling constant $h$, one can introduce a scalar field self-energy  $\Pi(q^2)$ and a scalar field propagator $D(q^2)=1/(q^2-h^2\Pi(q^2))$, with the fermion loop contributions to $\Pi(q^2)$ being identical with the $\Pi_{\rm S}(q^2,M=0)$ graphs associated with the four-fermion theory. Now in general, after removing a quadratic divergence, $\Pi(q^2)$ will contain not just a single logarithm but also higher powers of logarithms. Now while one can cancel a single logarithm by a subtraction, one cannot cancel higher powers of logarithms by subtraction [${\rm ln}^2(\Lambda^2/q^2)-{\rm ln}^2(\Lambda^2/m^2)={\rm ln}(m^2/q^2)({\rm ln}(\Lambda^2/q^2)+{\rm ln}(\Lambda^2/m^2))]$. Thus in addition one must introduce a wave function renormalization for the scalar field propagator, with it being $\tilde{D}(q^2)=Z^{-1}D(q^2)$ that is then finite. Thus in the Yukawa case one needs both subtractions and multiplicative renormalizations. However, in the four-fermion case there is no analog of wave function renormalization for the $\bar{\psi}\psi$ Green's functions, as they are not coupled to an elementary scalar quantum field. Thus one only has subtraction at one's disposal in order to renormalize the scalar and pseudoscalar channel fermion-antifermion scattering amplitudes in the four-fermion case. Thus in the four-fermion case we can only allow a single logarithmic divergence and not any higher powers of logarithms, and that is precisely what we find. Now in and of itself power counting renormalizability does not exclude higher powers of logarithms. (For QED for instance, in (\ref{M86}) we found that $m_0=m(\Lambda/m)^{\gamma_{\theta}(\alpha)}=m\exp[\gamma_{\theta}(\alpha){\rm ln}(\Lambda/m)]=m(1+\gamma_{\theta}(\alpha){\rm ln}(\Lambda/m)+\gamma^2_{\theta}(\alpha){\rm ln}^2(\Lambda/m)/2+...$.) However, for the four-fermion theory to be renormalizable, only a single logarithm is allowed, just as we have found.

Since our introduction of a dummy $\sigma$ field path integration led to the generation of a Yukawa $\sigma\bar{\psi}\psi$ term, the path integration only lacks a $\sigma$ field kinetic energy, since it otherwise could have ben associated with a field theory containing a generic quantum scalar field action $I_S=\int d^4x (i\bar{\psi}\gamma^{\mu}\partial_{\mu}\psi+(1/2)\partial_{\mu}S\partial^{\mu}S-hS\bar{\psi}\psi)$ that is renormalizable. However, a point-coupled four-fermion theory is not renormalizable, and thus it is puzzling that it could be related to a Yukawa theory that is. However, the two theories differ in the form of the scalar field propagator. The dummy $\sigma$ propagator is given by $1/(-m_{\sigma}^2)$, while the propagator of the quantum scalar field $S$ is given by $1(q^2-m_S^2)$. For large $q^2$ the scalar $S$ field propagator is suppressed at large $q^2$, while the dummy $\sigma$ field propagator is not, with the scalar $S$ field theory only being renormalizable because of the $1/q^2$ suppression, i.e. because of two extra powers of convergence. Without an analogous suppression the four-fermion theory is not renormalizable unless the $\bar{\psi}\psi$ vertices can be suppressed, and indeed it is the suppression supplied by dressing the vertices with $\gamma_{\theta}=-1$ that converts the quadratic divergence of a point-coupled $\Pi_{\rm S}(q^2,M=0)$ into a logarithmic divergence, with one obtaining two extra powers of convergence.

With the four-fermion $(\bar{\psi}\psi)^2$ and $(\bar{\psi}i\gamma^5\psi)^2$ interactions having been made renormalizable by the $\gamma_{\theta}(\alpha)$ dressings, we are now able to both couple the four-fermion $(\bar{\psi}\psi)^2+(\bar{\psi}i\gamma^5\psi)^2$ interaction to gravity and include it in a potential theory of everything without affecting renormalizability, and this we will do in the following in Secs. (10) and (11). (With conserved currents remaining canonical in a conformal invariant world, for  $\bar{\psi}\gamma_{\mu}\psi\bar{\psi}\gamma^{\mu}\psi$ type interactions we must still replace them by $\bar{\psi}\gamma_{\mu}\psi A^{\mu}$ type couplings to intermediate vector bosons in order to obtain renormalizability.) However, if a dynamical Higgs boson is to replace an elementary Higgs boson we will need to augment  the dummy $I_{\sigma}$ action with some kinetic energy, and will need to see what the difference is between a dummy $\sigma$ field and an elementary quantum scalar Higgs field. We now address these issues, and show that a $\sigma$ field kinetic energy term is not generated by adding it on to $I_{\sigma}$ but by doing a path integration over the fermions.

\section{Why Does an Elementary Higgs Model  Work so Well in Weak Interactions if the Higgs Boson is Dynamical?}

In the highly successful standard $SU(2)_{L}\times U(1)$ theory of electroweak interactions the Higgs boson is taken to be an elementary field that appears in the fundamental Lagrangian. Any approach which seeks to replace this elementary Higgs boson by a dynamical one has to explain why the standard model with its elementary Higgs boson works as well as it does, has to recover its tested predictions, and has to determine whether there might be any observable differences. To this end we note that the path integral associated with the massless fermion action $I_{\rm QED-FF}$ of interest to us in this article is of the form
\begin{eqnarray}
Z(\bar{\eta}, \eta,J_{\mu})&=&\int D[\bar{\psi}]D[\psi]D[A_{\mu}]\exp\bigg{[}i\int d^4x \bigg{(}-\frac{1}{4}F_{\mu\nu}F^{\mu\nu}+\bar{\psi}i\gamma^{\mu}\partial_{\mu}\psi
-e\bar{\psi}\gamma^{\mu}A_{\mu}\psi 
\nonumber\\
&-&\frac{g}{2}(\bar{\psi}\psi)^2-\frac{g}{2}(\bar{\psi}i\gamma^5\psi)^2+\bar{\eta}\psi+\bar{\psi}\eta +J_{\mu}A^{\mu}\bigg{)}\bigg{]}.
\label{M132}
\end{eqnarray}
As it is not central to the discussion here, no non-Abelian structure is considered.\footnote{Given the centrality of chiral symmetry to our analysis of dynamical symmetry breaking, in an actual application to weak interactions one would have to take the weak interaction to have a chiral $SU(2)_{L}\times SU(2)_{R}\times U(1)$ structure of the type described in e.g. \cite{Mannheim1980} and references therein. Incidentally, we note that in analog to Cooper pairing, in \cite{Mannheim1980} it was shown that right-handed neutrino pairing would break the $SU(2)_{R}$ sector.}  On introducing real dummy fields $\sigma$ and $\pi$ and doing a Gaussian path integration on them, we can rewrite the path integral as 
\begin{eqnarray}
&&Z(\bar{\eta}, \eta,J_{\mu})
=\int D[\bar{\psi}]D[\psi]D[A_{\mu}]D[\sigma]D[\pi]\exp\bigg{[}i\int d^4x \bigg{(}-\frac{1}{4}F_{\mu\nu}F^{\mu\nu}+\bar{\psi}i\gamma^{\mu} \partial_{\mu}\psi
-e\bar{\psi}\gamma^{\mu}A_{\mu}\psi
\nonumber\\
&&-\frac{g}{2}(\bar{\psi}\psi)^2+\frac{g}{2}\left(\frac{\sigma}{g}-\bar{\psi}\psi\right)^2
-\frac{g}{2}(\bar{\psi}i\gamma^5\psi)^2+\frac{g}{2}\left(\frac{\pi}{g}-\bar{\psi}i\gamma^5\psi\right)^2+\bar{\eta}\psi+\bar{\psi}\eta+J_{\mu}A^{\mu}\bigg{)}\bigg{]},
\label{M133}
\end{eqnarray}
and thus as
\begin{eqnarray}
Z(\bar{\eta}, \eta,J_{\mu})
&=&\int D[\bar{\psi}]D[\psi]D[A_{\mu}]D[\sigma]D[\pi]\exp\bigg{[}i\int d^4x \bigg{(}-\frac{1}{4}F_{\mu\nu}F^{\mu\nu}+
\bar{\psi}\gamma^{\mu}i\partial_{\mu}\psi
-e\bar{\psi}\gamma^{\mu}A_{\mu}\psi 
\nonumber\\
&-&\sigma\bar{\psi}\psi+\frac{\sigma^2}{2g}-\pi\bar{\psi}i\gamma^5\psi +\frac{\pi^2}{2g}+\bar{\eta}\psi+\bar{\psi}\eta+J_{\mu}A^{\mu}\bigg{)}\bigg{]}.
\label{M134}
\end{eqnarray}
We recognize the action that is obtained in (\ref{M134}) as being of precisely the same form as the mean-field action $I_{\rm QED-MF}$ given in (\ref{M88}) above as generalized to include a pseudoscalar sector, with $\sigma(x)$ replacing $m(x)$. Since the functional variation with respect to the sources of $Z(\bar{\eta}, \eta,J_{\mu})$ as given in (\ref{M132})  generates the Green's functions associated with $I_{\rm QED-FF}$, the functional variation with respect to the sources of $Z(\bar{\eta}, \eta,J_{\mu})$ as given in (\ref{M134}) does so too. We can thus generate Green's functions in a theory that contains no elementary scalar fields using a generating functional associated with one that does. While $Z(\bar{\eta}, \eta,J_{\mu})$ as given in (\ref{M134}) looks very much like the generating functional of an elementary Higgs theory, it differs from it in three ways: there is no kinetic energy term for the $\sigma(x)$ or $\pi(x)$ fields, no double-well potential energy term for them either, and most crucially as we shall see, no $J(x)\sigma(x)$ or $J_5(x)\pi(x)$ source terms for them. 

To generate kinetic energy and potential energy terms for $\sigma(x)$ and $\pi(x)$, we now require that there be critical scaling in the QED sector with the dynamical dimensions of $\bar{\psi}\psi$ and  $\bar{\psi}i\gamma^5\psi$ being reduced from three to two.\footnote{Because of the chiral symmetry, both $\bar{\psi}\psi$ and  $\bar{\psi}i\gamma^5\psi$ have the same short-distance behavior.} Thus path integration on $A_{\mu}$ serves to replace point couplings by dressed couplings, with figures such as Figs. (\ref{lw1}), (\ref{baretadpole}), and (\ref{lw5}) being replaced by Figs. (\ref{lw2}), (\ref{lw3}), and (\ref{lw6}). Path integration in the fermion sector is straightforward since all the terms in (\ref{M134}) are linear in $\bar{\psi}$ and $\psi$, with the path integration thus being equivalent to a one-loop Feynman diagram (as evaluated with dressed vertices). Following path integration in the fermion sector, on introducing $\tilde{\Gamma}_{\rm S}(x,m=0)$ as the Fourier transform of $\tilde{\Gamma}_{\rm S}(p,p,0,m=0)$,
we obtain an effective action in the $\sigma$ sector, which, in analog to (\ref{M12}), is of the form \cite{Mannheim1978}  
\begin{eqnarray}
Z(\bar{\eta},\eta,J_{\mu})&=&\exp\left[i{\rm Tr}{\rm ln}\left(\frac{i\slashed{\partial}_x-\int d^4x^{\prime}\sigma(x^{\prime})\tilde{\Gamma}_{\rm S}(x-x^{\prime},m=0)}{i\slashed{\partial}_x}\right)\right]
\nonumber\\
&=&\int D[\sigma]\exp[iI_{\rm EFF}(\sigma)]=\int D[\sigma]\exp\left[i\int d^4 x\left(-\tilde{\epsilon}(\sigma)+\frac{Z(\sigma)}{2}\partial_{\mu}\sigma\partial^{\mu}\sigma+...\right)\right], 
\label{M135}
\end{eqnarray}
where according to (\ref{M107}) 
\begin{eqnarray}
I_{\rm EFF}(\sigma)=\int d^4 x\left[-\frac{\sigma^2(x)\mu^2}{16\pi^2}\left[{\rm ln}\left(\frac{\sigma^2(x)}{M^2}\right)-1\right]
+\frac{3\mu}{256\pi \sigma(x)}\partial_{\mu}\sigma(x)\partial^{\mu}\sigma(x)+....\right].
\label{M136}
\end{eqnarray}
We recognize $I_{\rm EFF}(\sigma)$ as being in the form of a Higgs action with both a double-well potential and a kinetic energy term for $\sigma(x)$.\footnote{With $\int d^4p\exp(ip\cdot x)(-p^2)^{-\lambda}=i\pi^2 2^{4-2\lambda}\Gamma(2-\lambda)(-x^2)^{\lambda-2}/\Gamma(\lambda)$ and with $\sigma(x)$ being real, the quantity $f(x)=\int d^4x^{\prime}\sigma(x^{\prime})\tilde{\Gamma}_{\rm S}(x-x^{\prime},m=0)$ is pure imaginary when $\lambda=1/2$ (viz. $\gamma_{\theta}(\alpha)=-1)$. Also, if $\sigma(x)$ is an even function of $x$, the quantity $f(x)$ is an even function of $x$ also. Thus just as in our discussion of  $I_{\rm EFF}$ of the NJL model as given in (\ref{M62}), if $\sigma(x)$ is either constant or symmetric, in the JBW case $I_{\rm EFF}(\sigma)$ as given in (\ref{M135}) and (\ref{M136}) is real to all orders in derivatives of $\sigma(x)$.} Analogously, in the pseudoscalar case we note that the massless graphs of Fig. (\ref{lw2}) are unchanged if we replace zero-momentum $\bar{\psi}\psi$ insertions by zero-momentum $\bar{\psi}i\gamma^5\psi$ insertions. However, there are changes once the insertions carry momentum as in Fig. (\ref{lw6}), and with $\Pi^{\prime}_{\rm P}(q^2=0,m)$ evaluating to $\Pi^{\prime}_{\rm P}(q^2=0,m)=-7\mu/128\pi m$  \cite{Mannheim2015}, we thus obtain the effective action
\begin{eqnarray}
I_{\rm EFF}(\pi)=\int d^4 x\left[-\frac{\pi^2(x)\mu^2}{16\pi^2}\left[{\rm ln}\left(\frac{\pi^2(x)}{M^2}\right)-1\right]
+\frac{7\mu}{256\pi \pi(x)}\partial_{\mu}\pi(x)\partial^{\mu}\pi(x)+....\right].
\label{M137}
\end{eqnarray}
in the pseudoscalar sector.

Ignoring the higher-derivative terms in (\ref{M135}), the stationarity condition for $I_{\rm EFF}(\sigma)$ is of the form
\begin{eqnarray}
-\tilde{\epsilon}^{\prime}(\sigma)-Z(\sigma)\Box\sigma=0.
\label{M138}
\end{eqnarray}
Expanding around the stationary minimum at $\sigma=M$ then gives rise to fluctuations with  squared mass $\tilde{\epsilon}^{\prime\prime}(M)/Z(M)=32\mu M/3 \pi$. Since this value is real, it cannot and does not correspond to the position of the pole in $T_{\rm S}(q^2,M)$, which in (\ref{M125}) was shown to be complex.\footnote{As just noted, for a constant $\sigma(x)$ or for a static, spherically symmetric $\sigma(x)$, the entire, all-derivative $I_{\rm EFF}(\sigma)$ would equally be real, with fluctuations around its exact, all-order, static, spherically symmetric  minimum only yielding real mass squared fluctuations and not complex ones.} Whether the Higgs mass is real or complex is thus a key discriminant between elementary and dynamical Higgs bosons.

While the $Z(\bar{\eta},\eta,J_{\mu})$ path integral in (\ref{M134}) looks like that associated with elementary scalar and pseudoscalar fields, there is one key difference: there are no $J(x)\sigma(x)$ or $J_5(x)\pi(x)$ source terms for $\sigma(x)$ and $\pi(x)$. Specifically, if the scalar and pseudoscalar fields were to be elementary, the path integral would be associated with $Z(\bar{\eta}, \eta, J_{\mu}, J,J_5)$ instead, and then the Higgs squared mass would be given by the real $\tilde{\epsilon}^{\prime\prime}(\sigma)/Z(\sigma)$ as evaluated at the minimum of the potential (assuming we ignore higher-derivative terms in (\ref{M136})).\footnote{As constructed, it is tempting to think of $\sigma(x)$ and $\pi(x)$ as being related to Higgs and Goldstone fields. However, they cannot be since lacking sources they act as c-numbers not q-numbers, to thus be the analogs of mean-field order parameters rather than quantum fields. Moreover, we could evaluate the action of (\ref{M134}) in the quenched ladder approximation to the Abelian gluon model in the $\alpha \leq \pi/3$ region. Since, as described above, there are no Higgs or Goldstone bound states in this region, $\sigma(x)$ and $\pi(x)$ could thus not correspond to them. (Moreover, even in the event that there is dynamical symmetry breaking we still could not identify the squared mass of $\pi(x)$ as $\tilde{\epsilon}^{\prime\prime}(\pi)/Z(\pi)=32\mu M/7 \pi$ as evaluated at the minimum of the potential as the Goldstone boson squared mass since  $\tilde{\epsilon}^{\prime\prime}(\pi)/Z(\pi)$ is not zero.) Regardless of whether or not dynamical symmetry breaking takes place in a chiral four-fermion theory, we can always introduce dummy $\sigma(x)$ and $\pi(x)$  variables. From the perspective of (\ref{M134}), whether or not dynamical symmetry breaking does in fact take place depends on the non-perturbative structure of the theory as summed to all orders in off-shell $\sigma(x)$ and $\pi(x)$ exchange diagrams.} With one and the same Lagrangian, the off-shell scalar field (internal exchange and loop diagram) contributions to Green's functions with external fermion legs as generated by either $Z(\bar{\eta}, \eta,J_{\mu})$ or  $Z(\bar{\eta}, \eta, J_{\mu}, J,J_5)$ would be identical, with it being the all-order iteration of internal $\sigma$ and $\pi$ exchange diagrams that would generate the dynamical Goldstone and Higgs bosons that are not present in the $\sigma$ field and $\pi$ field sector actions themselves. However, $Z(\bar{\eta}, \eta, J_{\mu}, J,J_5)$ would also allow for Green's functions with external boson legs as well. Thus elementary and dynamical Higgs bosons only differ when the Higgs field goes on shell, while not differing off shell at all. Since all tests of the standard model prior to the recent actual discovery of the Higgs boson only involved off-shell Higgs physics, we thus recover all prior standard Higgs results.\footnote{In principle at least that is, since one of course still has to do explicit calculations.} However, in the on-shell Higgs region that has only recently begun to be  explored, there will be differences. With the Higgs width being an on-shell property of the Higgs field, again we see that it is in the width of the Higgs boson that one could potentially distinguish between a dynamical Higgs boson and an elementary one.

\section{Mass Generation and the Cosmological Constant Problem}

\subsection{From a Dynamical Higgs Boson to Conformal Gravity}

The discussion of dynamical Higgs boson generation that we have presented so far has been formulated in flat spacetime. In our study of massless QED coupled to a four-fermion interaction we have found that the dynamical symmetry breaking mechanism that gives fermions mass  produces a dynamical Higgs boson at the same time. Since the fermions would be massless in the absence of any symmetry breaking, there is an underlying conformal symmetry, to give the critical scaling that we realized with anomalous dimensions. With the dimension $d_{\theta}(\alpha)=1+\gamma_{\theta}(\alpha)$ of $\bar{\psi}\psi$ being reduced from three to two, we found that the four-fermion interaction was softened sufficiently to lead to a vacuum energy density  and scalar and pseudoscalar channel fermion-antifermion scattering amplitudes that were finite to all orders in the four-fermion coupling constant $g$. Now even though we have softened the vacuum energy density sufficiently to make it finite, as such it still makes a large contribution to the cosmological constant. Hence once we couple the theory to gravity we will have a cosmological constant problem that we will have to deal with. To address this issue we note that if there is an underlying conformal symmetry then it will be possessed by the energy-momentum tensor. Then since the energy-momentum tensor is the source of gravity, in order for both the gravity and matter sectors to share common symmetries (as one should require of two sectors that are coupled to each other), we should extend the conformal symmetry to the gravity sector as well. Then, with all of the $SU(3)\times SU(2)\times U(1)$ strong, electromagnetic and weak interactions being based on actions with dimensionless coupling constants, we propose that every fundamental interaction in nature is to have an underlying conformal symmetry.\footnote{The only dimensionful term that appears in the fundamental $SU(3)\times SU(2)\times U(1)$  action is the $-\mu^2\phi^2/2$ term in the potential of an elementary Higgs field. Such a term is absent if the Higgs boson is dynamical.} We are thus led to consider the gravity sector to be based on the conformal invariant conformal gravity theory that has been advanced by the present author as a candidate alternative to standard gravity. We shall thus explore conformal gravity in the following, and shall see that with it one can solve the cosmological constant problem that dynamical symmetry breaking  produces.  In this exploration we will find that when $\gamma_{\theta}(\alpha)=-1$ the four-fermion interaction plays a double role -- it is needed to generate a dynamical Higgs boson and is needed to cancel vacuum energy density infinities that couple to gravity. We begin with a general discussion of the vacuum energy density problem.

\subsection{The Vacuum Energy Density Problem}

In the above we have discussed the connection between dynamical mass generation and the vacuum energy density. Since the cosmological constant problem involves the vacuum energy density, the cosmological constant problem  is intimately tied in with mass generation, and so we turn now to a more detailed analysis of the structure of the vacuum energy density. There are two separate issues for the vacuum energy density. First, simply because a matter field energy-momentum tensor is composed  of products of quantum fields at the same spacetime point, there is a zero-point problem. This problem already occurs in a massless theory with a normal vacuum, and also occurs in a massive fermion theory with a normal vacuum (i.e. kinematic fermion mass). And second, when one generates mass via symmetry breaking, not only is there still a  zero-point vacuum energy density contribution, in addition a cosmological constant term is generated. 

To illustrate the issues that are involved, it is convenient to first look at the zero-point, vacuum expectation value of the energy-momentum tensor 
\begin{eqnarray}
T^{\mu\nu}_{\rm M}=i\hbar \bar{\psi}\gamma^{\mu}\partial^{\nu}\psi 
\label{M139}
\end{eqnarray}
of a free fermion matter field of mass $m=0$ in flat, four-dimensional spacetime, with the fermion obeying the massless Dirac equation.\footnote{The suffix ${\rm M}$ in  the various $T^{\mu\nu}_{\rm M}$ considered here and in the various $\rho_{\rm M}$ and $p_{\rm M}$ considered below denotes ``matter field", where the matter fields are those fields that appear in the $T^{\mu\nu}_{\rm M}$ source terms of gravitational equations of motion.} Since the fermion is massless the energy-momentum tensor is traceless, i.e. $\eta_{\mu\nu}T^{\mu\nu}_{\rm M}=0$. With $k^{\mu}=(k,\bar{k})$, $|\bar{k}|=k$, following a Feynman contour integration in the complex frequency plane the vacuum matrix element evaluates to 
\begin{eqnarray}
\langle \Omega_0 |T^{\mu\nu}_{\rm M}|\Omega_0\rangle= -\frac{2\hbar}{(2\pi)^3}\int_{-\infty}^{\infty}d^3k\frac{k^{\mu}k^{\nu}}{k}.
\label{M140}
\end{eqnarray}
With its $k^{\mu}k^{\nu}$ structure  $\langle \Omega_0 |T^{\mu\nu}_{\rm M}|\Omega_0\rangle$ has the generic form of a perfect fluid with a timelike fluid velocity vector $U^{\mu}=(1,0,0,0)$, viz.
\begin{eqnarray}
\langle \Omega_0 |T^{\mu\nu}_{\rm M}|\Omega_0\rangle= (\rho_{\rm M}+p_{\rm M})U^{\mu}U^{\nu}+p_{\rm M}\eta^{\mu\nu},
\label{M141}
\end{eqnarray}
where
\begin{eqnarray}
\rho_{\rm M}=\langle \Omega_0|T^{00}_{\rm M}|\Omega_0\rangle= -\frac{2\hbar}{(2\pi)^3}\int_{-\infty}^{\infty}d^3kk,
\label{M142}
\end{eqnarray}
\begin{eqnarray}
p_{\rm M}&=&\langle \Omega_0 |T^{11}_{\rm M}|\Omega_0\rangle=\langle \Omega_0 |T^{22}_{\rm M}|\Omega_0\rangle=\langle \Omega_0 |T^{33}_{\rm M}|\Omega_0\rangle
= -\frac{2\hbar}{3(2\pi)^3}\int_{-\infty}^{\infty}d^3k k.
\label{M143}
\end{eqnarray}
The zero-point energy density $\rho_{\rm M}$ and the zero-point pressure $p_{\rm M}$ are related by the tracelessness condition
\begin{eqnarray}
\eta_{\mu\nu}\langle \Omega_0 |T^{\mu\nu}_{\rm M}|\Omega_0\rangle= 3p_{\rm M}-\rho_{\rm M}=0
\label{M144}
\end{eqnarray}
since $\eta_{\mu\nu}k^{\mu}k^{\nu}=0$. (We use ${\rm diag}[\eta_{\mu\nu}]=(-1,1,1,1)$ here and in the discussion of gravity below.) Since  $p_{\rm M}$ is not equal to $-\rho_{\rm M}$, the zero-point  energy-momentum tensor does not have the form of a cosmological constant term, to underscore that fact that the zero-point problem is distinct from the cosmological constant problem. With both $\rho_{\rm M}$ and $p_{\rm M}$ being divergent, in terms of a 3-momentum cutoff $K$ the divergences can be parametrized as the quartic divergences
\begin{eqnarray}
\rho_{\rm M}=-\frac{\hbar K^4}{4\pi^2},\qquad p _{\rm M}=-\frac{\hbar K^4}{12\pi^2}.
\label{M145}
\end{eqnarray}
These divergences would have to be canceled  in some way. 

Moreover, additional infinities are encountered if the fermion has a mass. For a free massive fermion with a kinematic mass in flat spacetime, one is still in a normal vacuum $|\Omega_0\rangle$ in which $\langle \Omega_0|\bar{\psi}\psi|\Omega_0\rangle=0$. In this case the form of the energy-momentum tensor  remains as given in (\ref{M139}), but since the Dirac equation becomes that of a massive fermion, the energy-momentum tensor is no longer traceless, with it instead obeying  
$\eta_{\mu\nu}\langle \Omega_0 |T^{\mu\nu}_{\rm M}|\Omega_0\rangle=m\bar{\psi}\psi$. Thus with $k^{\mu}=((k^2+m^2/\hbar^2)^{1/2},\bar{k})$, one still has the generic structure
\begin{eqnarray}
\langle \Omega_0 |T^{\mu\nu}_{\rm M}|\Omega_0\rangle= -\frac{2\hbar}{(2\pi)^3}\int_{-\infty}^{\infty}d^3k\frac{k^{\mu}k^{\nu}}{(k^2+m^2/\hbar^2)^{1/2}}=(\rho_{\rm M}+p_{\rm M})U^{\mu}U^{\nu}+p_{\rm M}\eta^{\mu\nu},
\label{M146}
\end{eqnarray}
but where now $\rho_{\rm M}$ and $p_{\rm M}$ evaluate to
\begin{eqnarray}
\rho_{\rm M}&=&-\frac{\hbar K^4}{4\pi^2}- \frac{m^2K^2}{4\pi^2\hbar} +\frac{m^4}{16\pi^2\hbar^3}{\rm ln}\left(\frac{4\hbar^2K^2}{m^2}\right)
-\frac{m^4}{32\pi^2\hbar^3},
\nonumber \\
p _{\rm M}&=&-\frac{\hbar K^4}{12\pi^2}+ \frac{m^2K^2}{12\pi^2\hbar} -\frac{m^4}{16\pi^2\hbar^3}{\rm ln}\left(\frac{4\hbar^2K^2}{m^2}\right)
+\frac{7m^4}{96\pi^2\hbar^3}.
\label{M147}
\end{eqnarray}
And while $3p_{\rm M}-\rho_{\rm M}$ is no longer zero, $p_{\rm M}$ remains unequal to $-\rho_{\rm M}$. In addition to the previous quartic divergence, in (\ref{M147}) we also encounter quadratic and logarithmic divergences. These additional divergences would also need to be canceled in some way. Now while it is tempting to simply normal order these infinities away, and even though one can indeed do so in flat spacetime since there one can only measure energy density differences, once one couples to gravity one cannot throw energy density terms away since the hallmark of Einstein gravity is that gravity couple to all forms of energy density and not just to their finite parts. We shall discuss these points below when we do couple to gravity 

Now while one could normal order away vacuum infinities when one is in  flat spacetime, it is instructive to try to remove them by a dynamical scheme. We will discuss such a dynamical scheme below (actually critical  scaling with anomalous dimensions as  discussed above, but as then coupled to gravity), but first we will investigate what happens if one removes vacuum infinities by counterterms. To this end, we note that while physically unmotivated (in the sense that it corresponds to mathematical fields rather than real ones), a straightforward way to parametrize divergences is to use a Pauli-Villars regulator scheme. For (\ref{M147}) we follow \cite{Mannheim2011} and introduce a set of covariant Pauli-Villars regulator masses $M_i$, with each such regulator contributing an analog of (\ref{M147}) as multiplied by some overall factor $\eta_i$ (due to the fermionic or bosonic nature of the regulator or to a chosen Hilbert space metric signature\footnote{For negative norm (ghost state) regulators it is possible \cite{Bender2008a,Bender2008b,Mannheim2012} to find a quantization scheme in which the negative norms are only apparent, with the regulators  actually having positive norm, and we will discuss this point below in reference to the conformal gravity ghost problem.}) as per
\begin{eqnarray}
(\langle \Omega_0 |T^{\mu\nu}_{\rm M}|\Omega_0\rangle)_{\rm REG}= \langle \Omega_0 |T^{\mu\nu}_{\rm M}|\Omega_0\rangle+\sum \eta_i\langle \Omega_0|T^{\mu\nu}_{\rm M}(i)|\Omega_0\rangle.
\label{M148}
\end{eqnarray}
The choice
\begin{eqnarray}
1+\sum\eta_i=0, \qquad m^2+\sum \eta_iM_i^2=0,  \qquad m^4+\sum \eta_iM_i^4=0
\label{M149}
\end{eqnarray}
will not only then lead to finite regulated $\rho_{\rm REG}$ and $p_{\rm REG}$, it will give them the values
\begin{eqnarray}
\rho_{\rm REG}=-p_{\rm REG}=- \frac{\hbar}{16\pi^2}\left(m^4{\rm ln}m^2+\sum \eta_i M_i^4{\rm ln}M_i^2\right).
\label{M150}
\end{eqnarray}
As we see, the regulation procedure will not just make $\rho_{\rm REG}$ and $p_{\rm REG}$ be finite, it will make them be equal and opposite, with $p_{\rm REG}=-\rho_{\rm REG}$. We recognize such a form to be just that of a cosmological constant term, with a regulated $(\langle \Omega_0 |T^{\mu\nu }_{\rm M}|\Omega_0\rangle)_{\rm REG}$ then behaving as $-\rho_{\rm REG}\eta^{\mu\nu}$. The regulation procedure has thus converted a zero-point problem into a cosmological constant problem.\footnote{That this has to happen dates back to Sakharov, who pointed out that because of Lorentz invariance the only form that the vacuum expectation value of a well-defined energy-momentum tensor could have would be one in which it is equal to a spacetime constant times  the only available rank two tensor for the situation, namely the metric tensor $\eta_{\mu\nu}$.} Moreover, it has created a rather severe one since $\rho_{\rm REG}$ will be as large as the regulator masses that appear in (\ref{M150}). Thus in standard Einstein gravity where regulator masses are at the Planck scale, the contribution of the cosmological constant term to cosmic expansion would be 120 or so orders of magnitude larger than Hubble plot data or anisotropic temperature variation data in the cosmic microwave background could possibly tolerate. 

If the fermion gets its mass by the symmetry breaking associated with an elementary scalar field $S(x)$ with a fundamental double-well potential, there are additional vacuum problems. Consider a flat spacetime matter action 
\begin{eqnarray}
I_M=-\int d^4x\left[\frac{1}{2}\partial_{\mu}S\partial^{\mu}
S-\frac{1}{2}\mu^2S^2+\frac{1}{4}\lambda^2 S^4
+i\bar{\psi}\gamma^{\mu}\partial_\mu            
\psi -hS\bar{\psi}\psi\right],
\label{M151}
\end{eqnarray}                                 
where $h$ and $\lambda$ are real dimensionless coupling constants, and $\mu^2$ is the double-well potential mass parameter. Variation of this action with respect to  $\psi(x)$ and $S(x)$ yields the equations of motion
\begin{eqnarray}
i \gamma^{\mu}\partial_{\mu}                               
\psi - h S \psi = 0,
\label{M152}
\end{eqnarray}                                 
and 
\begin{eqnarray}
\partial_{\mu}\partial^{\mu}S+\mu^2S
-\lambda^2 S^3 +h\bar{\psi}\psi=0,
\label{M153}
\end{eqnarray}                                 
with the energy-momentum tensor being of the form 
\begin{eqnarray}
T^{\rm M}_{\mu \nu} = i \bar{\psi} \gamma_{\mu}\partial_{\nu}                                                              
\psi+\partial_{\mu}S\partial_{\nu}S
-\frac{1}{2}g_{\mu\nu}\partial_{\alpha}S\partial^{\alpha} S
-\frac{1}{2}g_{\mu\nu}\mu^2S^2
+\frac{1}{4}g_{\mu\nu}\lambda^2S^4.
\label{M154}
\end{eqnarray}                                 
In the presence of a spontaneously broken vacuum $|\Omega_M\rangle$, a non-zero constant expectation
value $S_0=\langle\Omega_M|S|\Omega_M\rangle$  for the scalar field obeys 
\begin{eqnarray}
\mu^2S_0-\lambda^2S_0^3+h\langle\Omega_M|\bar{\psi}\psi|\Omega_M\rangle=0,
\label{M155}
\end{eqnarray}        
the fermion obeys the massive fermion Dirac equation 
\begin{eqnarray}
i \gamma^{\mu}\partial_{\mu}\psi - hS_0 \psi = 0,
\label{M156}
\end{eqnarray}        
the vacuum matrix element of the energy-momentum tensor takes the form 
\begin{eqnarray}
\langle\Omega_M|T^{\rm M}_{\mu \nu}|\Omega_M\rangle = 
\langle\Omega_M|i \bar{\psi}\gamma_{\mu}\partial_{\nu}\psi|\Omega_M\rangle
-\frac{1}{2}g_{\mu\nu}\mu^2S_0^2
+\frac{1}{4}g_{\mu\nu}\lambda^2S_0^4,
\label{M157}
\end{eqnarray}                                 
with its trace being given by
\begin{eqnarray}
\eta_{\mu\nu}\langle\Omega_M|T^{\rm M}_{\mu \nu}|\Omega_M\rangle = 
hS_0\langle\Omega_M|\bar{\psi}\psi|\Omega_M\rangle
-2\mu^2S_0^2
+\lambda^2S_0^4.
\label{M158}
\end{eqnarray}                                 
Thus, as already noted in Sec. (4), the symmetry breaking induces a cosmological constant term in the energy-momentum tensor,\footnote{In Sec. (4) we only discussed the scalar field contribution. Now there is also an $S_0$-dependent fermionic contribution to  $T^{\rm M}_{\mu \nu}$.} and if the $S_0$ parameter is associated with the Higgs boson mass scale, the contribution  of this cosmological constant term to standard, Einstein-gravity-based  cosmology would be 60 or so orders of magnitude larger than cosmological data could possibly tolerate. And then, when radiative corrections are included, one would in addition generate zero-point contributions that are of order the Planck mass regulator scale.

Another example of the interplay between  dynamical mass generation and the cosmological constant may be found in the conformal invariant, curved space generalization of the above scalar field model that had been discussed in \cite{Mannheim2006}. Here the matter sector action is taken to be of the conformal form 
\begin{eqnarray}
I_M({\rm conf})=-\int d^4x(-g)^{1/2}\left[\frac{1}{2}\nabla_{\mu}S
\nabla^{\mu}S-\frac{1}{12}S^2R^\mu_{\phantom         
{\mu}\mu}
+\lambda S^4
+i\bar{\psi}\gamma^{c}V^{\mu}_c(x)[\partial_\mu+\Gamma_\mu(x)]             
\psi -hS\bar{\psi}\psi\right],
\label{M159}
\end{eqnarray}                                 
where $h$ and $\lambda$ are dimensionless coupling
constants.\footnote{Here the $\gamma^c$ form a set of fixed basis Dirac gamma matrices, the $V^{\mu}_c(x)$ are vierbeins, and the spin connection $\Gamma_{\mu}(x)$ is given by $\Gamma_{\mu}(x)=-(1/8)[\gamma_a,\gamma_b](V^b_{\nu}\partial_{\mu}V^{a\nu}+V^b_{\lambda}\Lambda^{\lambda}_{\phantom{\lambda}\nu\mu}V^{a\nu})$ where $\Lambda^{\lambda}_{\phantom{\lambda}\nu\mu}=(1/2)g^{\lambda \sigma}(\partial_{\nu}g_{\mu\sigma} +\partial_{\mu}g_{\nu\sigma}-\partial_{\sigma}g_{\nu\mu})$.}
As such, the $I_{\rm M}({\rm conf})$ action is the most
general curved space matter action for the $\psi(x)$ and $S(x)$ fields
that is invariant under both general coordinate transformations and the local conformal transformation
$S(x)\rightarrow e^{-\alpha(x)}S(x)$, $\psi(x)\rightarrow
e^{-3\alpha(x)/2}\psi(x)$,
$\bar{\psi}(x)\rightarrow e^{-3\alpha(x)/2}\bar{\psi}(x)$,
$V^a_{\mu}(x)\rightarrow e^{\alpha(x)}V^a_{\mu}(x)$,
$g_{\mu\nu}(x)\rightarrow e^{2\alpha(x)}g_{\mu\nu}(x)$. Variation of
this action with respect to 
$\psi(x)$ and
$S(x)$ yields the equations of motion
\begin{eqnarray}
i \gamma^{c}V^{\mu}_c(x)[\partial_{\mu} +\Gamma_\mu(x)]                              
\psi - h S \psi = 0,
\label{M160}
\end{eqnarray}                                 
and 
\begin{eqnarray}
\nabla_{\mu}\nabla^{\mu}S+\frac{1}{6}SR^\mu_{\phantom{\mu}\mu}
-4\lambda S^3 +h\bar{\psi}\psi=0,
\label{M161}
\end{eqnarray}                                 
while variation with respect to the metric yields an energy-momentum tensor 
\begin{eqnarray}
T^{\rm M}_{\mu \nu}({\rm conf})&=& i \bar{\psi} \gamma^{c}V_{\mu c}(x)[
\partial_{\nu}                    
+\Gamma_\nu(x)]                                                                 
\psi+\frac{2}{3}\nabla_{\mu} \nabla_{\nu} S
-\frac{1}{6}g_{\mu\nu}\nabla_{\alpha}S\nabla^{\alpha}S
-\frac{1}{3}S\nabla_{\mu}\nabla_{\nu}S
\nonumber \\             
&&+\frac{1}{12}g_{\mu\nu}S\nabla_{\alpha}\nabla^{\alpha}S                           
-\frac{1}{6}S^2\left(R_{\mu\nu}
-\frac{1}{4}g_{\mu\nu}R^\alpha_{\phantom{\alpha}\alpha}\right)         
-\frac{1}{4}g_{\mu\nu}h S\bar{\psi}\psi. 
\label{M162}
\end{eqnarray}                                 
Use of the matter field equations of motion
then confirms that this energy-momentum tensor obeys the tracelessness condition $g_{\mu\nu}T_{\rm M}^{\mu\nu}({\rm conf})=0$, just as it should do in  a conformal invariant theory.

In the presence of a spontaneously broken non-zero constant expectation
value $S_0$ for the scalar field, the energy-momentum tensor is then
found to simplify to  
\begin{eqnarray}
T^{\rm M}_{\mu \nu}({\rm conf})= i \bar{\psi}
\gamma^{c}V_{\mu c}(x)[\partial_{\nu}+\Gamma_\nu(x)]\psi
-\frac{1}{4}g_{\mu\nu}hS_0\bar{\psi}\psi
-\frac{1}{6} S_0^2\left(R_{\mu\nu}-\frac{1}{4}g_{\mu\nu}
R^\alpha_{\phantom{\alpha}\alpha}\right).
\label{M163}
\end{eqnarray}                                 
To appreciate the implications of this energy-momentum tensor, it suffices to take the flat space limit, viz. 
\begin{eqnarray}
T^{\rm M}_{\mu \nu}({\rm conf, flat}) = i \bar{\psi}
\gamma_{\mu}\partial_{\nu}\psi
-\frac{1}{4}\eta_{\mu\nu}hS_0\bar{\psi}\psi.
\label{M164}
\end{eqnarray}                                 
With the fermion now obeying 
\begin{eqnarray}
i \gamma^{\mu}\partial_{\mu}\psi - M \psi = 0,
\label{M165}
\end{eqnarray}                                 
where $M=hS_0$,  the tracelessness of the flat space energy-momentum tensor given in  (\ref{M164}) is manifest. In consequence, we see that it is possible for a fermion to acquire a mass without the trace of the energy-momentum tensor needing to be the non-zero $m\bar{\psi}\psi$ that it would be if the mass were kinematical. In addition we see that not only is a cosmological constant term induced in  $T^{\rm M}_{\mu \nu}({\rm conf})$ when there is mass generation, it is explicitly needed to maintain the tracelessness of $T^{\rm M}_{\mu \nu}({\rm conf})$ that is required by the underlying conformal invariance of the theory (the trace of the $R_{\mu\nu}-(1/4)g_{\mu\nu}R^\alpha_{\phantom{\alpha}\alpha}$ term in (\ref{M163}) is zero by itself). And as such, the  contribution of this cosmological constant term to $T^{\rm M}_{\mu \nu}({\rm conf})$ would have to be of the same order of magnitude as the contribution of the matter fields, and not orders of magnitude larger. As we thus see, mass generation and cosmological constant generation are intimately connected. Having seen how mass generation, a cosmological constant term, and vacuum energy density infinities can arise, we turn now to see how standard Einstein gravity handles them.

\subsection{Is the Standard Gravity Cosmological Constant Problem Properly Posed?}

Even though standard Einstein gravity is based on the familiar second-order equation of motion 
\begin{eqnarray}
\frac{1}{8\pi G}\left(R^{\mu\nu} -\frac{1}{2}g^{\mu\nu}R^{\alpha}_{\phantom{\alpha}\alpha}\right)=-T^{\mu\nu}_{\rm M},
\label{M166}
\end{eqnarray}
the actual status of this equation requires some clarification. Since the two sides of the equation are to be equal to each other, they must either both be quantum-mechanical or must both be classical.  However, since the gravity side is not well-defined quantum-mechanically, one takes the gravity side to be classical. Now at the time the Einstein equations were first introduced the energy-momentum tensor side was taken to be classical too. However, with Chandrasekhar showing that white dwarf stars  were stabilized by the Pauli degeneracy of the Fermi sea of the electrons within the star, it became clear not so much that the gravitational sources were quantum-mechanical, but that gravity was aware of this, and that the quantum-mechanical nature of its source was relevant to gravitational astrophysics. With the discovery of the cosmic microwave background and of its black-body nature, it became clear that gravitational cosmology was equally aware of the quantum nature of gravitational sources.  To get round the fact that the gravity side of the Einstein equations is classical (${\rm CL}$) while the matter side is quantum-mechanical,  one replaces (\ref{M166}) by a hybrid 
\begin{eqnarray}
\frac{1}{8\pi G}\left(R^{\mu\nu} -\frac{1}{2}g^{\mu\nu}R^{\alpha}_{\phantom{\alpha}\alpha}\right)_{\rm CL}=-\langle \psi|T^{\mu\nu}_{\rm M}|\psi \rangle,
\label{M167}
\end{eqnarray}
in appropriate states $\psi$. However, since the matter term in  (\ref{M167}) consists of products of quantum matter fields at the same spacetime point, the matter term has a zero-point problem.  But since the gravity side of (\ref{M167}) is finite, it cannot be equal to something that is infinite. Thus one must find a mechanism to cancel infinities on the matter side, and must find one that does so via the matter side alone. However instead, in the literature one commonly ignores the fact that the hallmark of Einstein gravity is that gravity is to couple to all forms of energy density rather than only to energy density differences, and subtracts off (i.e. normal orders away) the zero-point infinity by hand and replaces (\ref{M167}) by the finite 
\begin{eqnarray}
\frac{1}{8\pi G}\left(R^{\mu\nu} -\frac{1}{2}g^{\mu\nu}R^{\alpha}_{\phantom{\alpha}\alpha}\right)_{\rm CL}=-\left(\langle \psi|T^{\mu\nu}_{\rm M}|\psi \rangle\right)_{\rm FIN}.
\label{M168}
\end{eqnarray}
(${\rm FIN}$ denotes finite.) Thus in treating the contribution of the electron Fermi sea to white dwarf stars or the contribution of the cosmic microwave background to cosmic evolution, one uses an energy operator of the generic form $H=\sum(a^{\dagger}(\bar{k})a(\bar{k})+1/2)\hbar\omega_k$, and then by hand discards the $H=\sum \hbar \omega_k/2 $ term. And then, after all this is done, the finite part of $\langle \psi |T^{\mu\nu}_{\rm M}|\psi \rangle$ or of the vacuum $\langle \Omega|T^{\mu\nu}_{\rm M}|\Omega \rangle$ still has an uncanceled and as yet uncontrolled cosmological constant contribution that still needs to be dealt with.

The present author is not aware of any formal derivation of (\ref{M168}) starting from a consistent quantum gravity theory,\footnote{If one starts with a path integral over both metric paths and matter field paths, formally (\ref{M167}) would correspond to a path integration over the matter fields and a stationary variation on the metric.  However, when one performs the path integration over all the other metric paths one finds that the path integral does not actually exist, with there being infinities not just in the vacuum zero-point sector but in scattering amplitudes as well. (From the path integral perspective the shortcoming of the Einstein-Hilbert action is that in Euclidean time this action is negative on some metric paths but positive on others, with the Euclidean path integral then not being bounded.)} and notes that since it is (\ref{M168}) that is conventionally used in astrophysics and cosmology, it would not appear to yet be on a fully secure footing.  Thus in the gravity literature one starts with (\ref{M168}) as a given, and then tries to solve the cosmological constant problem associated with the fact that the right-hand side of (\ref{M168})  is 60 to 120 orders of magnitude larger than the cosmology associated with (\ref{M168}) could possibly tolerate. It appears to us that, as currently presented, the standard gravity cosmological constant problem is not properly posed, as it is based on a starting point for which there would not appear to be any justification.\footnote{We are not questioning the validity of the Einstein equations per se here but only their use as per (\ref{M168}).}  Absent any such justification, it is not clear whether there is actually any significance in trying to make sense of the implications of (\ref{M168}) in the first place. 

Despite this note of caution, since a justification for (\ref{M168}) might still emerge,\footnote{With (\ref{M168}) taken as a given, it does have a lot of success in fitting data, though to do so it requires a large amount of so-far undetected dark matter, and an even larger amount of so-far not understood dark energy.} it is nonetheless of interest to try to make sense of (\ref{M167}) and (\ref{M168}). Of particular interest is the approach based on supersymmetry since it does lead to testable 
predictions.\footnote{Reviews of supersymmetry may be found in e.g. \cite{Wess1992,Weinberg2000,Polchinski1998,Shifman2012}.} The basic idea behind supersymmetry is the existence of a symmetry between bosons and fermions that is exact at the level of the Lagrangian and only broken dynamically. For gravity the utility of such a symmetry is quite extensive, as it addresses the vacuum energy density and cosmological constant problems, provides potential dark matter candidates, admits of a local supergravity extension, and is central to the construction of the string-theory based approach to quantum gravity. Also, and of particular interest to us here, it addresses the elementary Higgs field self-energy problem (the hierarchy problem  -- so-called since a hierarchy of mass scales is involved), in which radiative corrections to the Higgs self-energy would, if not controlled in some way, lead to a Higgs boson mass at the Planck mass regulator mass scale rather than at the 125 GeV scale that it has now been found to have.\footnote{Since theories with an elementary Higgs field are renormalizable, the Higgs boson mass can certainly be made finite  without the need to introduce counterterms that are not in the form of the terms that are already present in the fundamental Lagrangian. However, without the protection of some symmetry (gauge invariance for instance protects otherwise quadratically divergent gauge boson masses), the mass that would be generated would be at a high scale. This in principle distinction between scalar fields and gauge fields could be viewed as supporting the position expressed in this article, namely that the Higgs boson is a dynamical field rather than the elementary one that a gauge field is, and as we have seen above, the mass of a dynamically generated  Higgs boson is nicely naturally of order the symmetry breaking mass scale rather than of any high mass regulator scale.}

With the key difference between boson and fermion loops being an overall minus sign due to their differing permutation symmetry statistics, such loops can potentially cancel each others' infinities. Thus for the fermion contribution to the vacuum zero-point energy density given in  (\ref{M146}),  a cancellation of the mass-independent quartic divergence is immediately provided by a boson loop, with the mass-dependent quadratic and logarithmic divergences being canceled as well if the fermions and bosons are degenerate in mass, i.e.  if the supersymmetry is exact. As regards the cosmological constant, it is actually zero if the supersymmetry is exact. Specifically, in a supersymmetric theory one has a generic anticommutator of the form $\{Q^{\alpha},Q_{\alpha}^{\dagger}\}=H$, where the $Q_{\alpha}$ are Grassmann supercharges and $H$ is the Hamiltonian. If the supercharges annihilate $|\Omega_0\rangle$ (viz. unbroken supersymmetry), then $\langle \Omega_0|H|\Omega_0\rangle$ is zero, the energy of the vacuum is zero, and the cosmological constant is thus zero too. Finally, in the event of an exact supersymmetry, the boson and fermion loop contributions to the mass-dependent quadratic divergence in the self-energy of an elementary Higgs field also cancel each other identically,\footnote{The fermionic contribution is given by the same $\Pi_{\rm S}(q^2,M)$ as used in (\ref{M67}) in the NJL model.} to thereby provide a candidate solution to the hierarchy problem.

Attractive as these cancellations are, they do not survive once the supersymmetry is broken and counterpart fermions and bosons stop being degenerate in mass. Moreover, actual experimental detection of any of the required superpartners  of the standard fermions and bosons has so far proven elusive. Now until quite recently one could account for such lack of detection by breaking the mass degeneracy between ordinary particles and their superpartners, and endowing the superpartners with ever higher masses or ever weaker couplings to ordinary matter. However, in so doing the degree of lack of cancellation of infinities would become greater and greater, as would the size of the cosmological constant term that would then be generated by the loss of unbroken supersymmetry, with such lack of detection to date of any superparticles leading to a cosmological constant term that would be at least 60 or so orders of magnitude larger than the standard gravity (\ref{M168}) could possibly tolerate. Moreover, in order to cancel the quadratic self-energy divergence that an elementary Higgs field would have, one would need a supersymmetric particle with a mass reasonably close to that of the Higgs boson itself. And with the Higgs boson mass now having been determined, one should thus anticipate finding a superparticle in the same 125 GeV mass region, with the cancellation not being able to succeed  if the requisite superparticle mass is made too large. However, no evidence for any such particle has emerged in an exploration of this mass region at the LHC, or in decays such as $B^0_{\rm s}\rightarrow \mu^{+}+\mu^{-}$  that were thought to be particularly favorable for supersymmetry \cite{Aaij2013,Khachatryan2015}.  Not only was no evidence for supersymmetry found in the original  LHC run 1 at a beam center of mass energy of 7 to 8 TeV, at the even higher energies in the subsequent current LHC run 2 at a beam center of mass energy of 13 TeV, and with a far more extensive search of possible relevant channels, no sign of any superparticles up to masses quite significantly above 125 GeV was found at all.\footnote{Run 2 data presented by the ATLAS, CMS, and LHCb collaborations at the 38th International Conference on High Energy Physics in Chicago in August 2016 may be found at www.ichep2016.org and in the conference proceedings.} The lack of detection of any superparticles in the currently available energy region thus poses a challenge to supersymmetry not just in general but to its proposed solution to the hierarchy problem in particular. And while the superparticle search at the LHC is still ongoing, and while the so far unsuccessful accompanying underground searches for supersymmetric dark matter candidates are continuing, nonetheless the situation is disquieting enough that one should at least contemplate whether it might be possible to dispense with supersymmetry altogether. However, since we have seen that supersymmetry does control infinities very well as long as it remains unbroken, in any alternate approach we should again look for some underlying symmetry to control infinities. Thus given the role that conformal invariance played in our treatment above of dynamical Higgs boson generation in a critical scaling QED coupled to a four-fermion interaction, we are thus motivated to consider conformal invariance as that requisite symmetry. On extending conformal symmetry to the gravity sector, we are thus led to consideration of conformal gravity, and as we shall see, it will not only enable us to control infinities when the symmetry is unbroken, it will continue to be able to control them when the symmetry is broken dynamically. Moreover, if the Higgs boson is generated dynamically, there would then be no hierarchy problem to begin with, as it is only an elementary Higgs field that would have a quadratically divergent self-energy problem in the first place.\footnote{For some other discussion of the zero-point energy density problem in cosmology see \cite{Volovik2005,Branchina2010}.}

\subsection{The Consistency of Quantum Conformal Gravity}

In 1918  while working on a possible metrication (geometrization) of electromagnetism  \cite{Weyl1918a} Weyl discovered a tensor, the Weyl or conformal tensor \cite{Weyl1918b} 
\begin{eqnarray}
C_{\lambda\mu\nu\kappa}= R_{\lambda\mu\nu\kappa}
-\frac{1}{2}\left(g_{\lambda\nu}R_{\mu\kappa}-
g_{\lambda\kappa}R_{\mu\nu}-
g_{\mu\nu}R_{\lambda\kappa}+
g_{\mu\kappa}R_{\lambda\nu}\right)
+\frac{1}{6}R^{\alpha}_{\phantom{\alpha}\alpha}\left(
g_{\lambda\nu}g_{\mu\kappa}-
g_{\lambda\kappa}g_{\mu\nu}\right),
\label{M169}
\end{eqnarray}
that has the remarkable property that under local conformal transformations of the form $g_{\mu\nu}(x)\rightarrow e^{2\alpha(x)}g_{\mu\nu}(x)$ with spacetime dependent $\alpha(x)$, the Weyl tensor transforms as 
\begin{eqnarray}
C^{\lambda}_{\phantom{\lambda}\mu\nu\kappa}(x)\rightarrow 
C^{\lambda}_{\phantom{\lambda}\mu\nu\kappa}(x),
\label{M170}
\end{eqnarray}
with all derivatives of $\alpha(x)$ being found to drop out identically.
As such, the Weyl tensor $C^{\lambda}_{\phantom{\lambda}\mu\nu\kappa}$
bears the same relation to local conformal transformations as the
Maxwell tensor $F_{\mu\kappa}=\nabla_{\mu}A_{\kappa}-\nabla_{\kappa}A_{\mu}$ does to local gauge transformations,
with the kinematic relation $g^{\mu\kappa}F_{\mu\kappa}=0$ having as a
counterpart the kinematic
$g^{\mu\kappa}C^{\lambda}_{\phantom{\lambda}\mu\nu\kappa}=0$, with the
Weyl tensor being the traceless piece of the Riemann tensor.

Conformal gravity is the unique pure metric theory of gravity in four spacetime dimensions that possesses this local conformal symmetry, with the pure gravitational sector of the theory being given by the action (see e.g. \cite{Mannheim2006,Mannheim2012})
\begin{eqnarray}
I_{\rm W}&=&-\alpha_g\int d^4x (-g)^{1/2}C_{\lambda\mu\nu\kappa} C^{\lambda\mu\nu\kappa}
=-\alpha_g\int d^4x (-g)^{1/2}\left[R_{\lambda\mu\nu\kappa}
R^{\lambda\mu\nu\kappa}-2R_{\mu\kappa}R^{\mu\kappa}+\frac{1}{3}
(R^{\alpha}_{\phantom{\alpha}\alpha})^2\right]
\nonumber \\
&=&-2\alpha_g\int d^4x
(-g)^{1/2}\left[R_{\mu\kappa}R^{\mu\kappa}-\frac{1}{3}
(R^{\alpha}_{\phantom{\alpha}\alpha})^2\right],
\label{M171}
\end{eqnarray}
where $\alpha_g$ is a dimensionless gravitational coupling  constant, and where the last equality follows since the quantity $(-g)^{1/2}\left[R_{\lambda\mu\nu\kappa}R^{\lambda\mu\nu\kappa}-4R_{\mu\kappa}R^{\mu\kappa}+
(R^{\alpha}_{\phantom{\alpha}\alpha})^2\right]$ is a total divergence (the Gauss-Bonnet theorem). 
We note that absent from the $I_{\rm W}$ action is any fundamental cosmological constant term since the action $-\int d^4x(-g)^{1/2}\Lambda$ is not conformal invariant.\footnote{Also excluded is the Einstein-Hilbert action $I_{\rm EH}=-(1/16 \pi G)\int d^4x (-g)^{1/2}R^{\alpha}_{\phantom{\alpha}\alpha}$, a point we return to below.} Conformal invariance thus provides a good starting point to address the cosmological constant problem. And as we shall see below, the conformal theory continues to be able to control the cosmological constant term even after the conformal symmetry is dynamically broken by giving the dimensionful 
$\bar{\psi}\psi$ a non-vanishing vacuum expectation value.

Functional variation of the $I_{\rm W}$  action with respect to the metric defines a gravitational rank two tensor (see e.g. \cite{Mannheim2006,Mannheim2012})
\begin{eqnarray}
W^{\mu \nu}&=&
\frac{1}{2}g^{\mu\nu}\nabla_{\beta}\nabla^{\beta}R^{\alpha}_{\phantom{\alpha}\alpha}+
\nabla_{\beta}\nabla^{\beta}R^{\mu\nu}                    
-\nabla_{\beta}\nabla^{\nu}R^{\mu\beta}                        
-\nabla_{\beta}\nabla^{\mu}R^{\nu \beta}    
-2R^{\mu\beta}R^{\nu}_{\phantom{\nu}\beta}                                 
+\frac{1}{2}g^{\mu\nu}R_{\alpha\beta}R^{\alpha\beta}
\nonumber\\                      
&-&\frac{2}{3}g^{\mu\nu}\nabla_{\beta}\nabla^{\beta}R^{\alpha}_{\phantom{\alpha}\alpha} 
+\frac{2}{3}\nabla^{\nu}\nabla^{\mu}R^{\alpha}_{\phantom{\alpha}\alpha}                           
+\frac{2}{3} R^{\alpha}_{\phantom{\alpha}\alpha}R^{\mu\nu}                              
-\frac{1}{6}g^{\mu\nu}(R^{\alpha}_{\phantom{\alpha}\alpha})^2
\label{M172}
\end{eqnarray}        
that is covariantly conserved ($\nabla_{\mu}W^{\mu\nu}=0$) and covariantly traceless ($g_{\mu\nu}W^{\mu\nu}=0$). On introducing a conformal invariant matter action $I_{\rm M}$ (such as the one given in (\ref{M159}) or the massless QED action of interest to us in this article), variation of $I_{\rm W}+I_{\rm M}$ with respect to the metric yields a fourth-order derivative equation of motion for the metric of the form
\begin{eqnarray}
-4\alpha_g W^{\mu\nu}=-T^{\mu\nu}_{\rm M}.
\label{M173}
\end{eqnarray}
If we define $-4\alpha_g W^{\mu\nu}$ to be the energy-momentum tensor $T^{\mu\nu}_{\rm GRAV}$ of gravity (i.e. the variation with respect to the metric of the pure gravitational sector of the action), and introduce an energy-momentum tensor for the universe as a whole, we can rewrite (\ref{M173}) as
\begin{eqnarray}
T^{\mu\nu}_{\rm UNIV}=T^{\mu\nu}_{\rm GRAV}+T^{\mu\nu}_{\rm M}=0,
\label{M174}
\end{eqnarray}
to thus put the gravity and matter sectors on an equal footing, while showing that the  total energy-momentum tensor of the universe is zero.

Unlike standard Einstein gravity with its dimensionful Newtonian coupling constant $G$, because of its dimensionless coupling constant $\alpha_g$, as a quantum theory of gravity conformal gravity is power-counting renormalizable.\footnote{With the metric $g_{\mu\nu}$ being dimensionless, in an expansion around flat spacetime of the dimension four quantity $C_{\lambda\mu\nu\kappa} C^{\lambda\mu\nu\kappa}$ as a power series in a gravitational fluctuation $h_{\mu\nu}=g_{\mu\nu}-\eta_{\mu\nu}$, each term will contain $h_{\mu\nu}$  a specific number of times together with exactly four derivatives since it is the derivatives that carry the dimension of the $C_{\lambda\mu\nu\kappa} C^{\lambda\mu\nu\kappa}$ term. The term that is quadratic in $h_{\mu\nu}$ will thus give a $1/k^4$ propagator, and each time we work to one more order in $h_{\mu\nu}$ we add an extra $1/k^4$ propagator and a compensating factor of $k^{\mu}k^{\nu}k^{\sigma}k^{\tau}$ in the vertex. With equal numbers of powers of $k^{\mu}$ being added in numerator and denominator, the ultraviolet behavior is not modified, and renormalizability is thereby maintained.} However, with $W^{\mu\nu}$ being a fourth-order derivative function of the metric, conformal gravity had been thought to possess negative norm ghost states and not be unitary. To see the nature of the difficulty, we expand to lowest order around a flat background, and ignoring $\mu$ or $\nu$ indices, write (\ref{M173}) in the generic $\Box^2\phi=j$ form  \cite{Mannheim2006}, with the associated propagator being writable as  
\begin{eqnarray}
D(k^2)=\frac{1}{k^4}=\lim_{M\rightarrow 0}\frac{1}{M^2}\left(\frac{1}{k^2}-\frac{1}{k^2+M^2}\right).
\label{M175}
\end{eqnarray}
With propagators ordinarily being identified as Green's functions of the form $D(k^2)=\int d^4x\exp[-ik\cdot x]\langle \Omega|T[\phi(x)\phi(0)]|\Omega\rangle$, the relative minus sign in (\ref{M175}) would suggest that the two sets of propagators on the right-hand side of (\ref{M175}) would be quantized so that one would have normal positive metric signature and the other negative ghost signature (i.e. respectively positive and negative residues at the poles in the $k_0$ plane), with the states obeying $\sum |n\rangle\langle n|-\sum |m\rangle\langle m|=I$. However, drawing such a conclusion is too hasty \cite{Bender2008a,Bender2008b}, since from the structure of a c-number $D(k^2)$ one cannot simply read off the structure of the underlying q-number Hilbert space. Rather, one has to construct the quantum Hilbert space first and from it then determine the c-number propagator, and not the other way round. And when  Bender and Mannheim constructed the requisite quantum Hilbert space in this particular case, from an analysis of asymptotic boundary conditions (which only gave convergent wave functions when the fields were continued into the complex plane), they found that the quantum Hamiltonian of the theory (essentially the Hamiltonian of the Pais-Uhlenbeck fourth-order oscillator theory \cite{Pais1950}) was not Hermitian. However, Bender and Mannheim also found that the Hamiltonian had an antilinear $PT$ symmetry that required all energy eigenvalues to nonetheless be real (as must be the case here since all poles in (\ref{M175}) are on the real $k_0$ axis).\footnote{With Hermiticity only being a sufficient condition for the reality of eigenvalues, as noted in \cite{Mannheim2015b} and references therein the necessary condition is that the Hamiltonian possess an antilinear symmetry. With complex Lorentz invariance requiring that the antilinear symmetry be $CPT$,  the $CPT$ theorem can be established without the need to assume Hermiticity of the Hamiltonian \cite{Mannheim2015b,Mannheim2016b}.  However, since the electrically neutral metric tensor is $C$ even, for conformal gravity (or for the Pais-Uhlenbeck theory) $CPT$ symmetry reduces to $PT$ symmetry.}  When a Hamiltonian is Hermitian one can introduce right-eigenvectors that obey $H|R\rangle=E|R\rangle$, with their conjugates being left-eigenvectors that obey $\langle R|H=\langle R|E$ if $E$ is real. However, when a Hamiltonian is not Hermitian but its eigenvalues are nonetheless real, the conjugates of its right-eigenvectors obey  $\langle R|H^{\dagger}=\langle R|E$, and are thus not left-eigenvectors, as the left-eigenvectors must obey $\langle L|H=\langle L|E$ if $E$ is real. In the non-Hermitian case it is the $\langle L(t)|R(t)\rangle=\langle L(t=0)|\exp(iHt)\exp(-iHt)|R(t=0)\rangle=\langle L(t=0)|R(t=0)\rangle$ norm that is time independent and not the $\langle R|R\rangle$ norm, with Bender and Mannheim showing that no $\langle L|R\rangle$ norm had negative signature in the fourth-order case. Analogously, in the non-Hermitian case one must distinguish between left-vacua and right-vacua, and identify $D(k^2)$ not as $\int d^4x\exp[-ik\cdot x]\langle \Omega_R|T[\phi(x)\phi(0)]|\Omega_R\rangle$, but as $\int d^4x\exp[-ik\cdot x]\langle \Omega_L|T[\phi(x)\phi(0)]|\Omega_R\rangle$ instead. With the set of all $\langle R|$ being complete we can set $\langle L|=\langle R|V$ where $V$ is some general operator, and set $D(k^2)=\int d^4x\exp[-ik\cdot x]\langle \Omega_R|VT[\phi(x)\phi(0)]|\Omega_R\rangle$.\footnote{In \cite{Mannheim2015b} and references therein it was shown that the operator $V$ is the operator that effects $VHV^{-1}=H^{\dagger}$.}  It is then through the presence of the operator $V$ that the negative minus sign in (\ref{M175}) is generated, and not through any  negative ghost signature associated with the states. In this way that the fourth-order conformal gravity theory turns out to be unitary after all.\footnote{The analysis given in \cite{Bender2008a,Bender2008b} shows that a Pauli-Villars propagator, viz. one precisely of the form given in (\ref{M175}), is unitary too, as it too is associated with the Pais-Uhlenbeck theory.}

\subsection{The Conformal Gravity Cancellation of Infinities}

With conformal gravity being both renormalizable and ghost free, we can now treat it as a bona fide quantum theory of gravity. Thus unlike in the Einstein case, we can now treat (\ref{M173}) as an equation in which both sides can consistently be taken to be quantum-mechanical. Then, since a conformal invariant gravity sector and a conformal invariant and a thus dimensionless-coupling-constant-based matter sector are both renormalizable, the vanishing of $T^{\mu\nu}_{\rm UNIV}$ holds for both bare quantum fields and dressed ones alike, to thus persist following radiative corrections.\footnote{Both $T^{\mu\nu}_{\rm GRAV}$ and $T^{\mu\nu}_{\rm M}$ separately contain trace anomalies, since their tracelessness is due to conformal invariance Ward identities that are violated by radiative corrections. However, since the stationary vanishing of their $T^{\mu\nu}_{\rm UNIV}$ sum is neither due to a Ward identity or violated by radiative corrections, the gravity sector and matter sector trace anomalies thus have to cancel each other identically order by order in perturbation theory, with $T^{\mu\nu}_{\rm UNIV}$ being anomaly free  \cite{Mannheim2012,Mannheim2012b,Mannheim2015}.} Thus we can now quantize the gravitational field consistently, and will then obtain a zero-point contribution to $T^{\mu\nu}_{\rm GRAV}$. With the graviton being bosonic, the gravity zero-point contribution to $T^{\mu\nu}_{\rm GRAV}$ and the fermion zero-point contribution to $T^{\mu\nu}_{\rm M}$ will have opposite signs,  with the vanishing of $T^{\mu\nu}_{\rm UNIV}$ in (\ref{M174}) causing the zero-point contributions to cancel each other identically. As we thus see, in essence in conformal gravity the graviton itself performs the task that a bosonic superpartner of the fermion does in the supersymmetry case. However, because $T^{\mu\nu}_{\rm UNIV}$ will continue to be zero even after the conformal symmetry is broken,\footnote{Since symmetry breaking is an infrared, long range order, effect,  the mutual cancellation of $T^{\mu\nu}_{\rm GRAV}$ and $T^{\mu\nu}_{\rm M}$ radiative corrections due to the vanishing of $T^{\mu\nu}_{\rm UNIV}$ is not affected.} unlike in the broken supersymmetry case, in the broken conformal symmetry case the zero-point cancellations will persist, with the induced cosmological constant term that accompanies mass generation being constrained by the continuing vanishing of $T^{\mu\nu}_{\rm UNIV}$. It is in this way then that the cosmological constant term is brought under control. 

To see how these cancellations work in practice, we note first  \cite{Mannheim2011,Mannheim2012} that we cannot quantize conformal gravity as a stand-alone theory since its coupling to a quantized matter source according to $T^{\mu\nu}_{\rm GRAV}+T^{\mu\nu}_{\rm M}=0$ does not permit the wave function renormalization constant for the gravitational field (cf. $Z(k)$ below) to be independently specified. Rather, it is fixed by the coupling to the matter fields. As noted in \cite{Mannheim2011,Mannheim2012}, while we can quantize matter fields via the familiar stand-alone canonical procedure, gravity is quantized entirely by virtue of it being coupled to a matter source that is quantized.\footnote{In a canonical quantization of fields one quantizes solutions to the equations  of motion of the field without reference to the zero-point contribution to the energy-momentum tensor of the fields. However, for gravity the equation of motion is expressly a condition on the energy-momentum tensor itself (cf. \ref{M174}), and one cannot consistently set $T^{\mu\nu}_{\rm GRAV}=0$ since it has a non-vanishing zero-point contribution. Hence one must instead set $T^{\mu\nu}_{\rm GRAV}+T^{\mu\nu}_{\rm M}=0$.} In \cite{Mannheim2011,Mannheim2012} it was proposed that, unlike in Einstein gravity, there be no intrinsic classical gravity at all, with gravity being produced entirely by quantum effects. In such a situation one should expand the theory as a power series in Planck's constant rather than as a power series in the gravitational coupling constant, with there thus being no term of order $\hbar^0$ in the expansion at all. To obtain the first non-trivial term in the expansion, viz. the term of order $\hbar$, we expand around flat spacetime. Then, on taking  the vacuum expectation value of $T^{\mu\nu}_{\rm GRAV}$ in the unbroken conformal symmetry vacuum $|\Omega_0 \rangle$ (viz. the same vacuum as used in (\ref{M140})), we obtain a quartically divergent zero-point energy density in the gravity sector of the form \cite{Mannheim2012}
\begin{eqnarray}
\langle \Omega_0|T^{\mu\nu}_{\rm GRAV}|\Omega_0 \rangle=\frac{2\hbar}{(2\pi)^3} \int_{-\infty}^{\infty}d^3k\frac{Z(k)k^{\mu}k^{\nu}}{k},\qquad \langle \Omega_0|T^{00}_{\rm GRAV}|\Omega_0 \rangle=\frac{\hbar}{\pi^2} \int_{0}^{K}dk k^3Z(k)
\label{M176}
\end{eqnarray}
where $Z(k=|\bar{k}|)$ is the as yet to be determined gravitational field wave function renormalization constant, as defined \cite{Mannheim2012} as the coefficient of the delta function in canonical commutation relations for the momentum modes of the gravitational field.\footnote{With $d^3k/k$ being Lorentz invariant and with $k^{\mu}$ being the only available vector, the $d^3kk^{\mu}k^{\nu}/k$ form for  the rank two tensor $\langle \Omega_0|T^{\mu\nu}_{\rm GRAV}|\Omega_0 \rangle$ then follows in any covariant theory that possesses no fundamental mass scale.  Compared to a second-order theory evaluation of $\langle \Omega_0|T^{\mu\nu}_{\rm GRAV}|\Omega_0 \rangle$, a fourth-order theory evaluation would put an extra factor of $k^2$ in both denominator (as per (\ref{M175})) and numerator (as per power counting), to still lead to (\ref{M176}).} Inserting (\ref{M176}) and (\ref{M140}) into (\ref{M174}) then yields 
\begin{eqnarray}
Z(k)=1.
\label{M177}
\end{eqnarray}
Thus, we simultaneously fix the gravity sector renormalization constant and effect a complete cancellation of the quartically divergent zero-point terms. To underscore that $Z(k)$ cannot be assigned independently but is determined by the structure of the matter source to which gravity is coupled,  we note that if the gravitational source consists of $M$ massless gauge bosons and $N$ massless two-component fermions, the vanishing of $\langle \Omega_0|T^{\mu\nu}_{\rm UNIV}|\Omega_0 \rangle$ then entails that $2Z(k)+M-N=0$ \cite{Mannheim2012}, with gravity adjusting to whatever its source is.\footnote{If there is no matter field source at all, i.e. stand-alone gravity, then $T^{\mu\nu}_{\rm GRAV}$ would be zero, and one would have $Z(k)=0$ and no quantization of the gravitational field at all.}

When the conformal symmetry is broken and a cosmological constant term is generated, $Z(k)$ will again adjust to its source. This will then allow gravity itself to cancel the cosmological constant term that is induced. However, to consistently implement such a cancellation we need the matter source to also have a conformal structure, and this is precisely what was found above when we studied dynamical symmetry breaking in a critical scaling theory with $\gamma_{\theta}(\alpha)=-1$. In (\ref{M94}) we had evaluated the associated $\epsilon(m)$. However, this $\epsilon(m)$ itself was the energy density difference $i/(2\pi)^4\int d^4p\left[{\rm Tr~ln}(\tilde{S}^{-1}_{\mu}(p))-{\rm Tr~ln}(\slashed{ p}+i\epsilon)\right]$. In taking the matrix element of $T^{\mu\nu}_{\rm M}$  in the self-consistent, Hartree-Fock vacuum $|\Omega_M \rangle$, in $\langle \Omega_M|T^{00}_{\rm M}|\Omega_M\rangle$ only the quantity $i/(2\pi)^4\int d^4p{\rm Tr~ln}(\tilde{S}^{-1}_{\mu}(p))$ appears. And this time it is the energy density itself and not an energy density difference that appears, just as has to be the case if we couple to gravity. Thus while the $\gamma_{\theta}(\alpha)=-1$ condition had converted the quadratic divergence in the point-coupled $\epsilon(m)$ of (\ref{M61}) into the  logarithmic divergence given in (\ref{M103}), now the previously canceled quartic divergence has returned. With the mean field also generating the $-m^2/2g$ term, we can now identify the mean-field vacuum energy density of the massless fermion $I_{\rm QED-FF}$ action of (\ref{M87}) to be $i/(2\pi)^4\int d^4p{\rm Tr~ln}(\tilde{S}^{-1}_{\mu}(p))-m^2/2g$, with the $m^2/2g$ term then canceling the logarithmic divergence just as in (\ref{M106}), doing so while not affecting the quartic divergence at all. Consequently, $\langle \Omega_M|T^{00}_{\rm M}|\Omega_M\rangle$ now has a quartic divergence and a finite piece. And since we can also set  $\langle \Omega_M|T^{00}_{\rm M}|\Omega_M\rangle=\epsilon(m)-m^2/2g+i/(2\pi)^4\int d^4p{\rm Tr~ln}(\slashed{ p}+i\epsilon)=\tilde{\epsilon}(m)+i/(2\pi)^4\int d^4p{\rm Tr~ln}(\slashed{ p}+i\epsilon)$, the vanishing of $T^{\mu\nu}_{\rm UNIV}$ then entails that 
\begin{eqnarray}
\langle \Omega_M|T^{00}_{\rm GRAV}|\Omega_M \rangle &=&\frac{\hbar}{\pi^2}\int_0^K dk k^3Z(k)=-\langle \Omega_M|T^{00}_{\rm M}|\Omega_M \rangle
\nonumber\\
&=&\frac{\hbar}{\pi^2}\int_0^K dk k^3+\frac{\mu^2M^2}{16\pi^2\hbar^3}=\frac{\hbar K^4}{4\pi^2}+\frac{\mu^2M^2}{16\pi^2\hbar^3},
\label{M178}
\end{eqnarray}
as evaluated at the $m=M$ minimum of $\tilde{\epsilon}(m)=\epsilon(m)-m^2/2g$ as per (\ref{M106}).
\begin{figure}[htpb]
\begin{center}
\begin{minipage}[t]{10 cm}
\epsfig{file=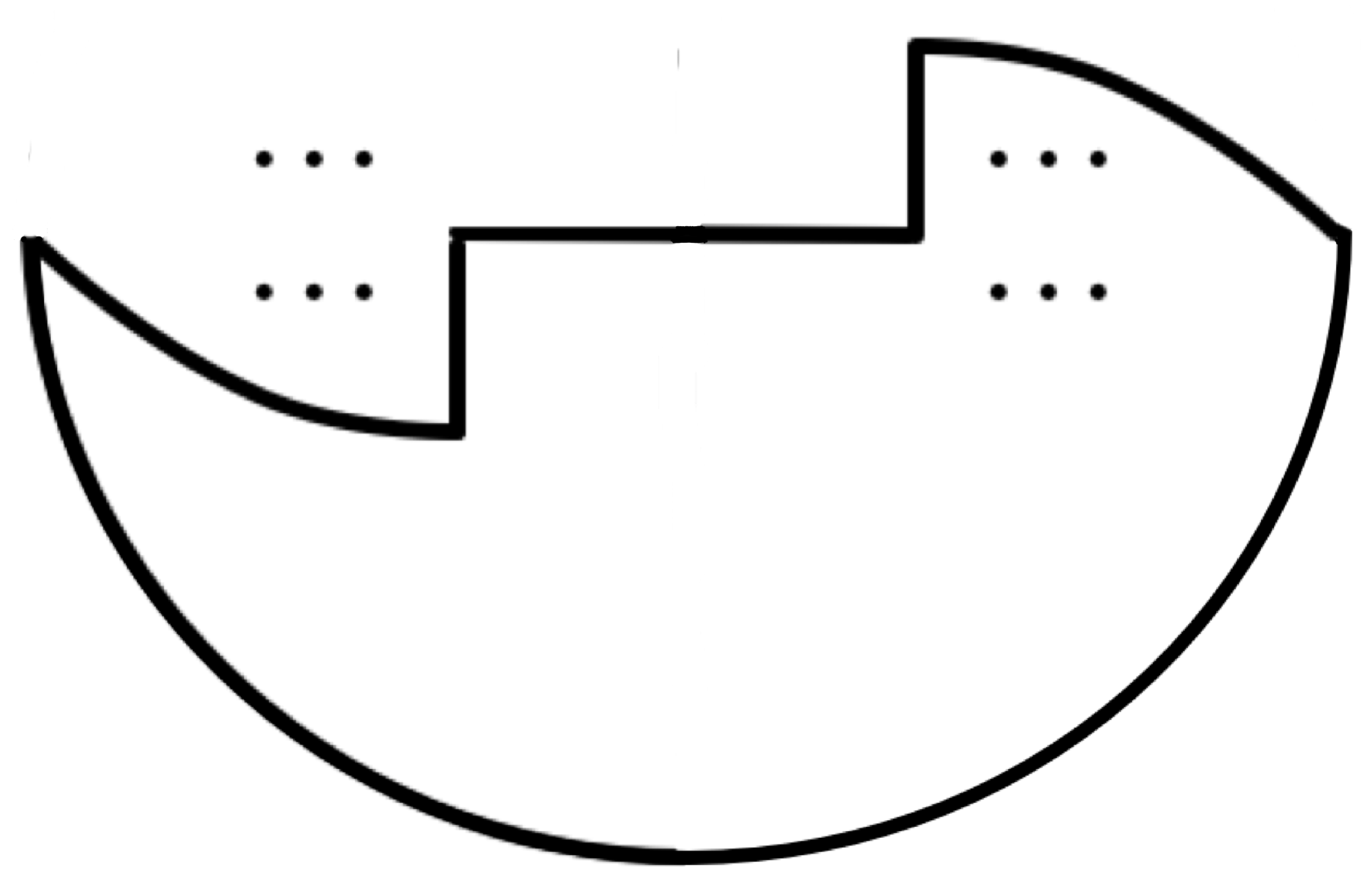,width=4.0in,height=2.0in}
\end{minipage}
\begin{minipage}[t]{17.5 cm}
\caption{The deformed, complex $p_0$ plane contour needed for a Feynman contour integral involving $\tilde{S}_{\mu}(p)$. Poles are shown as dots.}
\label{lw10}
\end{minipage}
\end{center}
\end{figure}

To extract out an explicit value for $Z(k)$ we need to write the right-hand side of (\ref{M178}) as a three-dimensional integral.  With $\langle \Omega_m|\bar{\psi}\psi|\Omega_m\rangle =\epsilon^{\prime}(m)=m/g$, through the use of (\ref{M95}) we can set 
\begin{eqnarray}
\langle \Omega_M|T^{00}_{\rm M}|\Omega_M\rangle=
i\int \frac{d^4p}{(2\pi)^4}\left({\rm Tr~ln}(\tilde{S}^{-1}_{\mu}(p))+
\frac{m}{2}{\rm Tr}[\tilde{\Gamma}_{\rm S}(p,p,0,m=0)\tilde{S}_{\mu}(p)]\right).
\label{M179}
\end{eqnarray}
On performing the $p_0$ integration, we will obtain the three-dimensional integral we seek. However, the poles of $\tilde{S}_{\mu}(p)$ are not on the real $p_0$ axis. We must thus deform the Feynman contour. As discussed in \cite{Mannheim2015}, if we first continue $\tilde{\Gamma}_{\rm S}(p,p,0,m=0)$ as 
\begin{eqnarray}
\tilde{\Gamma}_{\rm S}(p,p,0,m=0)=\left(\frac{-p^2-i\epsilon}{\mu^2}\right)^{\gamma_{\theta}(\alpha)/2}\rightarrow \left(\frac{p^2+i\epsilon}{\nu^2}\right)^{\gamma_{\theta}(\alpha)/2},
\label{M180}
\end{eqnarray}
all the $p_0$ plane poles in $\langle \Omega_M|T^{00}_{\rm M}|\Omega_M\rangle$ with $Re[p_0]>0$
would then be below the real $p_0$ axis and within the Feynman contour, while all the poles with $Re[p_0]<0$ would be above the real $p_0$ axis and thus be outside the Feynman contour. To continue back to $\mu$ we deform the Feynman contour as shown in Fig. (\ref{lw10}),  so that we still include all poles with $Re[p_0]>0$ while still excluding all poles with $Re[p_0]<0$. At $\gamma_{\theta}(\alpha)=-1$, the $\tilde{S}_{\mu}(p)$ propagator has two sets of poles, which are at $p_0^2=p^2+im\mu$, $p_0^2=p^2-im\mu$.\footnote{As had been noted above, as introduced in (\ref{M98}),  $\tilde{S}_{\mu}(p)$ is only the propagator in the $I_{\rm QED-MF}$ mean-field approximation to the $I_{\rm QED-FF}$ action given in (\ref{M87}). Its pole structure will be modified by the residual $I_{\rm QED-RI}$ interaction.} The $p_0$ plane contour integral is then readily done, and with it having already been done with the $\nu$ contour in \cite{Mannheim2012},  we find that with the $\mu$ contour and with $m=M$,    (\ref{M178}) takes the form given by \cite{Mannheim2015}
\begin{eqnarray}
\frac{\hbar}{\pi^2}\int_0^K dk k^3Z(k)&=&\frac{\hbar}{\pi^2}\int_0^K dk k^2\bigg{[}(k^2+iM\mu/\hbar^2)^{1/2}-\frac{iM\mu}{4\hbar^2(k^2+iM\mu/\hbar^2)^{1/2}} 
\nonumber\\
&+&(k^2-iM\mu/\hbar^2)^{1/2}+\frac{iM\mu}{4\hbar^2(k^2-iM\mu/\hbar^2)^{1/2}}\bigg{]}.
\label{M181}
\end{eqnarray}
Consequently,  $Z(k)$ is given by \cite{Mannheim2015}
\begin{eqnarray}
kZ(k)&=&(k^2+iM^2/\hbar^2)^{1/2}-\frac{iM^2}{4\hbar^2(k^2+iM^2/\hbar^2)^{1/2}} 
\nonumber\\
&+&(k^2-iM^2/\hbar^2)^{1/2}+\frac{iM^2}{4\hbar^2(k^2-iM^2/\hbar^2)^{1/2}}.
\label{M182}
\end{eqnarray}
For convenience we have set $\mu=M$ here. As we see, $Z(k)$ is again determined by the dynamics, and even though the gravitational modes themselves remain massless, $Z(k)$ adjusts to the fact that the fermion has mass. With this readjustment, the zero-point vacuum energy density in the gravity sector identically cancels both the quartic divergence and the induced finite cosmological constant term in the matter sector (cf. the $\hbar K^4/4\pi^2$ and $\mu^2M^2/16\pi^2\hbar^3$ terms in (\ref{M178})). 

To establish that there is a  conformal structure associated with (\ref{M178}), we compare it with (\ref{M164}) and (\ref{M165}). We immediately recognize (\ref{M181}) as being a sum over two separate mass sectors  of the time-time components of matrix elements of energy-momentum tensors each one of which behaves as $\langle \Omega_M|T^{\mu\nu}_{\rm M}({\rm conf,~flat})|\Omega_M\rangle$. Since we are only working to lowest order in $\hbar$ in our application here  of (\ref{M174}), it suffices to restrict $T^{\mu\nu}_{\rm M}$ to flat to this order, with curvature corrections to $T^{\mu\nu}_{\rm M}$ only appearing in higher order in $\hbar$.  Then, since $T^{\mu\nu}_{\rm M}({\rm conf,~flat})=i\bar{\psi}\gamma^{\mu}\partial^{\nu}\psi-(1/4)\eta^{\mu\nu}M\bar{\psi}\psi$ is traceless, the conformal structure of  (\ref{M178}) is established.\footnote{While a point-coupled four-fermion interaction would not in and of itself be conformal invariant, the effect of critical scaling is to the spread out the four-fermion vertex. According to (\ref{M90}) for a given $d_{\theta}$ the vertex would be spread out to  give a non-local (NL)  action of the form $I({\rm NL})=\int d^4xd^4x^{\prime}\bar{\psi}(x)\psi(x)[(x-x^{\prime})^2]^{-d_{\theta}}\bar{\psi}(x^{\prime})\psi(x^{\prime})$. With $\bar{\psi}\psi$ having dynamical dimension $d_{\theta}$, under a conformal transformation $I({\rm NL})$ would transform into 
$I({\rm NL})=\int d^4xd^4x^{\prime}e^{2\alpha d_{\theta}}\bar{\psi}(e^{\alpha}x)\psi(e^{\alpha}x)[(x-x^{\prime})^2]^{-d_{\theta}}\bar{\psi}(e^{\alpha}x^{\prime})\psi(e^{\alpha}x^{\prime})$, and thus under a change of variable into 
 $I({\rm NL})=\int d^4yd^4y^{\prime}e^{-8\alpha}e^{2\alpha d_{\theta}}e^{2\alpha d_{\theta}}\bar{\psi}(y)\psi(y)[(y-y^{\prime})^2]^{-d_{\theta}}\bar{\psi}(y^{\prime})\psi(y^{\prime})$. Consequently, $I({\rm NL})$ is conformal invariant when $d_{\theta}=2$, just as desired. Thus the non-local action $\int d^4x\left[-(1/4)F_{\mu\nu}F^{\mu\nu}+\bar{\psi}\gamma^{\mu}(i\partial_{\mu}-eA_{\mu})\psi\right] -(g/2)I({\rm NL})$ is conformal invariant, as is its curved space generalization. We can thus discuss the problem from the perspective of the standard local mean-field approach used in (\ref{M87}) and (\ref{M88}) or from the perspective of the non-local $I({\rm NL})$. When $d_{\theta}=2$ the two approaches are equivalent, with the decomposition into two separate mass sectors as exhibited in (\ref{M181}) allowing us to treat the vacuum energy density of one non-local theory as that of two local ones.} Given this result, we see, and we emphasize that we see,  the explicit need for a cosmological constant term in order to maintain tracelessness, something that the mean-field theory provides in the form of the $-m^2/2g$ term. Thus not only do we need to couple a critical scaling QED to a four-fermion interaction in order to cancel the logarithmic divergence in the vacuum energy density, this same coupling generates a cosmological constant term that serves to maintain conformal invariance. Moreover, not only does conformal invariance forbid the presence of any cosmological constant term in the fundamental action, this same conformal invariance also controls the cosmological constant term that is induced by the symmetry breaking. Since the residual interaction associated with the four-fermion interaction then generates dynamical Goldstone and Higgs bosons, we see that critical scaling, conformal symmetry, four-fermion interactions, dynamical Goldstone and Higgs boson generation, and control of the cosmological constant are all intimately tied together.

Now all of the above cancellations occur no matter how big the $\mu^2M^2\equiv M^4$ term in (\ref{M178}) might be (even if $M$ is of Higgs mass scale, grandunified scale, or Planck scale), and none of it is observable since the cancellations all occur in the vacuum, i.e. they are due entirely to the occupied negative energy states in the Dirac sea. However, what one measures in actual astrophysical phenomena is not properties of the vacuum but the behavior of the positive energy modes that can  be excited out of it. To be more precise, we note that since all of the infinities in $T^{\mu\nu}_{\rm GRAV}$ and $T^{\mu\nu}_{\rm M}$  are due to the infinite number of modes in the vacuum sector, if we decompose $T^{\mu\nu}_{\rm GRAV}$ and $T^{\mu\nu}_{\rm M}$ into finite particle (PART)  and divergent vacuum (VAC) parts according to $T^{\mu\nu}_{\rm GRAV}=(T^{\mu\nu}_{\rm GRAV})_{\rm PART}+(T^{\mu\nu}_{\rm GRAV})_{\rm VAC}$, $T^{\mu\nu}_{\rm M}=(T^{\mu\nu}_{\rm M})_{\rm PART}+(T^{\mu\nu}_{\rm M})_{\rm VAC}$, (\ref{M174}) will decompose into 
\begin{eqnarray}
&&(T^{\mu\nu}_{\rm GRAV})_{\rm VAC}+(T^{\mu\nu}_{\rm M})_{\rm VAC}=0,
\label{M183}
\end{eqnarray}
and
\begin{eqnarray}
&&(T^{\mu\nu}_{\rm GRAV})_{\rm PART}+(T^{\mu\nu}_{\rm M})_{\rm PART}=0.
\label{M184}
\end{eqnarray}
All of the vacuum energy density infinities and even the finite part in (\ref{M178})  are taken care of by (\ref{M183}), and for astrophysics and cosmology we can then use the completely infinity-free (\ref{M184}). In this way for studying white dwarfs or the cosmic microwave background  we can now use $H=\sum a^{\dagger}(\bar{k})a(\bar{k})\hbar \omega_k$ alone after all, as the zero-point contribution has already been taken care of by gravity itself and does not appear in (\ref{M184}) at all. Moreover, when we do excite positive energy modes out of the vacuum we will generate a new cosmological constant contribution, and it is this term that is measured in cosmology. Cosmology thus only sees the change in the vacuum energy density due to adding in positive energy modes and does not see the full negative energy mode vacuum energy density itself, i.e. in (\ref{M184}) one is sensitive not to  $\langle \Omega_M|T^{\mu\nu}_{\rm M}|\Omega_M\rangle$, and not even to  $\langle \Omega_M|bT^{\mu\nu}_{\rm M}b^{\dagger}|\Omega_M\rangle$, but only to their difference  $\langle \Omega_M|bT^{\mu\nu}_{\rm M}b^{\dagger}|\Omega_M\rangle -\langle \Omega_M|T^{\mu\nu}_{\rm M}|\Omega_M\rangle$.
Also gravity sees this effect mode by mode, i.e. gravity mode by fermion mode. Thus, for instance,  in the application of $T^{\rm M}_{\mu\nu}({\rm conf})$ of (\ref{M163}) to cosmology that was described in \cite{Mannheim2006} and is briefly discussed below, the quantity $S_0$ is not $\langle \Omega_M|S|\Omega_M\rangle$ but the much smaller $\langle \Omega_M|bSb^{\dagger}|\Omega_M\rangle- \langle \Omega_M|S|\Omega_M\rangle$. In contrast, if one uses the Einstein gravity (\ref{M168}) and the matter field source given in (\ref{M157}), then gravity sees an entire sum over fermion modes and sees the full and large $\langle \Omega_M|S|\Omega_M\rangle$. To summarize, if one wants to take care of the cosmological constant problem, one has to take care of the zero-point problem, and when one has a renormalizable theory of gravity, via an interplay with gravity itself one is then able to do so.

\section{Some Other Aspects of Conformal Symmetry and Conformal Gravity}

\subsection{Conformal gravity and the Dark Matter Problem}

With conformal symmetry as realized via critical scaling and anomalous dimensions being able to address some key issues in contemporary physics such as the generation of dynamical Goldstone and Higgs bosons, control of the cosmological constant, and the construction of a consistent quantum theory of gravity, viz. conformal gravity,  it is of interest to see how conformal gravity fares in addressing some other issues of concern to contemporary physics. In two recent papers \cite{Mannheim2006,Mannheim2012} the case was presented for considering conformal gravity as a possible alternative to standard Einstein gravity. In this section we provide a brief update. 

We had noted earlier that conformal invariance excludes the Einstein-Hilbert action. However, it does not exclude the Schwarzschild solution to Einstein gravity, since $R_{\mu\nu}=0$ is an exact exterior solution to (\ref{M173}) in any source-free region where $T^{\mu\nu}_{\rm M}=0$.\footnote{The Schwarzschild solution is not Riemann flat but only Ricci flat.  However, even though the Weyl tensor does depend on the Riemann tensor, because of the Gauss-Bonnet theorem the conformal gravity action $I_{\rm W}$ only depends on the Ricci tensor and Ricci scalar as per (\ref{M171}). The variation $W^{\mu\nu}$ of $I_{\rm W}$ as given in (\ref{M172}) thus only depends on the Ricci tensor and Ricci scalar and their derivatives, and all the terms in $W^{\mu\nu}$  vanish identically if $R_{\mu\nu}$ vanishes identically.} Moreover, to recover the successful results of a given theory one does not need to recover its equations, one only needs to recover its solutions, and one actually only needs to recover them in the kinematic region where they have been tested. However, while the Schwarzschild solution is a solution to conformal gravity, there are other solutions since the vanishing of $W^{\mu\nu}$ can be achieved without the vanishing of the Ricci tensor itself.

To determine what these other solutions might look like, Mannheim and Kazanas studied the geometry associated with a static, spherically symmetric source in the conformal gravity theory, to find \cite{Mannheim1989,Mannheim1994} that the coefficient $B(r)=-g_{00}(r)$ obeyed the exact, all-order classical fourth-order Poisson equation
\begin{eqnarray} 
\nabla^4B(r)=\frac{3}{4\alpha_gB(r)}(T^0_{\phantom{0}0}-T^r_{\phantom{r}r})=f(r).
\label{M185}
\end{eqnarray}
The general solution to this equation is given by 
\begin{eqnarray}
B(r)&=&-\frac{r}{2}\int_0^r
dr^{\prime}r^{\prime 2}f(r^{\prime})
-\frac{1}{6r}\int_0^r
dr^{\prime}r^{\prime 4}f(r^{\prime})
\nonumber \\
&&
-\frac{1}{2}\int_r^{\infty}
dr^{\prime}r^{\prime 3}f(r^{\prime})
-\frac{r^2}{6}\int_r^{\infty}
dr^{\prime}r^{\prime }f(r^{\prime})+B_0(r),
\label{M186}
\end{eqnarray}                                 
where $B_0(r)$ obeys $\nabla^4B_0(r)=0$. Since the integration in (\ref{M186}) extends all the way to $r=\infty$, the $B(r)$ potential receives contributions from material both inside and outside any system of interest. According to (\ref{M186}),  matter confined to the interior of a star of radius $r_0$ produces a potential exterior to the star of the form $V^*(r>r_0)=-\beta^*c^2/r+\gamma^*c^2 r/2$ per unit solar mass of star. We thus recover the Newtonian potential, while showing that in principle it does not need to be associated with the standard Einstein gravitational theory.\footnote{While the solution given by (\ref{M186}) limits to the exterior Schwarzschild solution for small $r$, there is no limit in which $W^{\mu\nu}$ becomes $R^{\mu\nu}-(1/2)g^{\mu\nu}R^{\alpha}_{\phantom{\alpha}\alpha}$. We thus recover the standard model solution but do not recover its equations, with Einstein gravity only being sufficient to give the Schwarzschild solution and Newton's Law of Gravity but not necessary.} However, we find that the potential gets modified at large distances where $r =O[(\beta^*/\gamma^*)^{1/2}]$, so that the solution recovers the Newton potential at short distances only, with the non-Ricci-flat solutions to conformal gravity only modifying the Ricci-flat one at large distances and not at small ones. We thus recover the solution to Einstein gravity in the kinematic solar system region where it has been tested with the use of known luminous sources alone, while for an appropriate galactic-determined value for $\gamma^*$, the solution will depart from Newton on precisely those galactic distances where in standard gravity one first has to resort to dark matter. 

Specifically, for galaxies, integrating the  $V^*(r)$ potential over a thin disk of stars with a surface brightness $\Sigma(R)=\Sigma_0\exp(-R/R_0)$ with scale length $R_0$ (the typical configuration for the stars in a spiral galaxy) yields the net local potential produced by the stars in the galaxy itself, and leads to a locally generated contribution to galactic circular velocities of the form \cite{Mannheim2006}
\begin{eqnarray}
v_{{\rm LOC}}^2&=&
\frac{N^*\beta^*c^2 R^2}{2R_0^3}\bigg{[}I_0\left(\frac{R}{2R_0}
\right)K_0\left(\frac{R}{2R_0}\right)
-I_1\left(\frac{R}{2R_0}\right)
K_1\left(\frac{R}{2R_0}\right)\bigg{]}
\nonumber\\
&+&\frac{N^*\gamma^* c^2R^2}{2R_0}I_1\left(\frac{R}{2R_0}\right)
K_1\left(\frac{R}{2R_0}\right),
\label{M187}
\end{eqnarray} 
where $N^*$ is the number of stars in the galaxy. 

There are two contributions due to material outside the galaxy, i.e. due to the rest of the universe. The first contribution is a linear potential term with coefficient $\gamma_0/2=(-k)^{1/2}$ coming from cosmology \cite{Mannheim2006}, a term that is associated with the $B_0(r)$ term, and due to writing a comoving Robertson-Walker geometry with negative curvature in the rest frame coordinate system of the galaxy (with $k>0$, $\gamma_0$ would be complex).  The second contribution arises from the integral from $r$ to $\infty$ term in (\ref{M186}) due to cosmological inhomogeneities such as clusters of galaxies, and is of a quadratic potential form with coefficient $\kappa$  \cite{Mannheim2011b}. Since both of these external contributions come from the universe as a whole, they are  both independent of any particular galaxy of interest, and thus act universally on every galaxy, with every galaxy seeing the same $\gamma_0$-dependent and $\kappa$-dependent potentials. When all these internal and external contributions are combined, the total rotational velocities in galaxies are given by
\begin{eqnarray} 
v_{{\rm TOT}}^2=v_{{\rm LOC}}^2+\frac{\gamma_0c^2 R}{2}-\kappa c^2 R^2.
\label{M188}
\end{eqnarray}
\begin{figure}[htpb]
\begin{center}
\epsfig{file=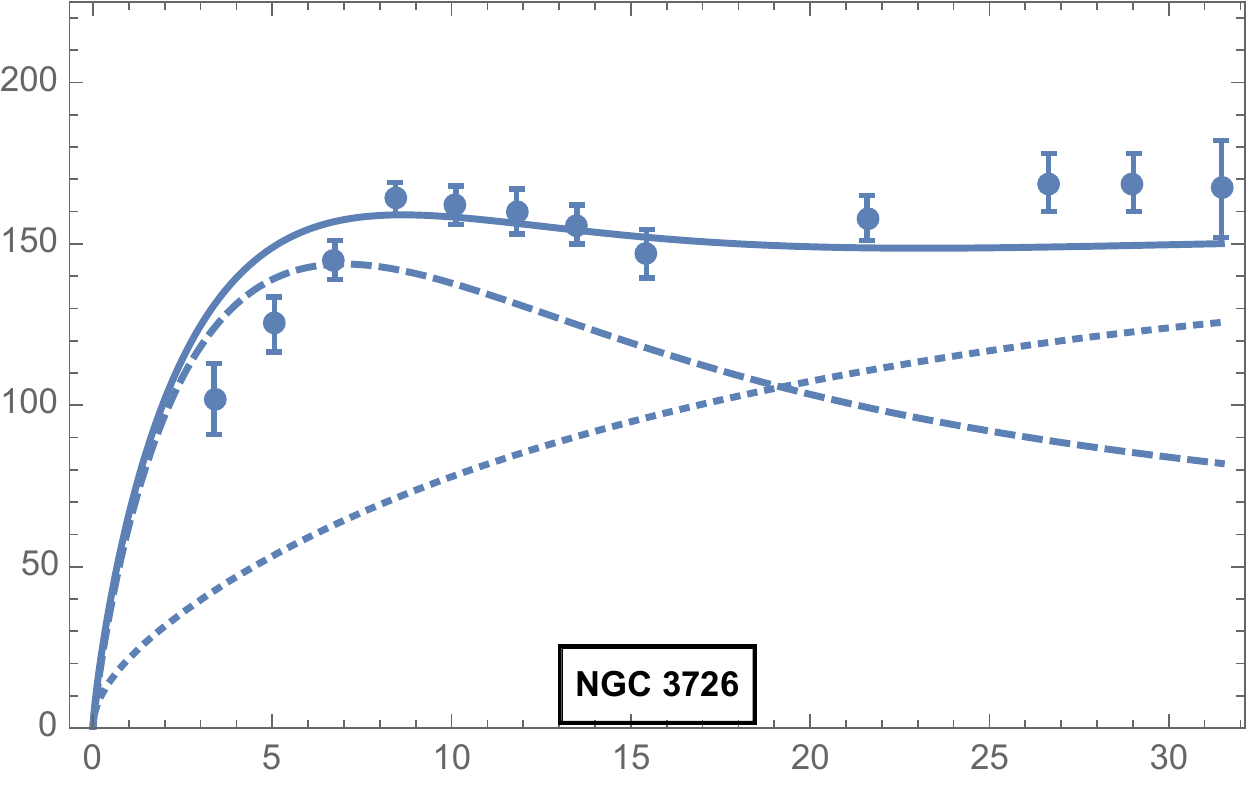,scale=0.4}\qquad
\epsfig{file=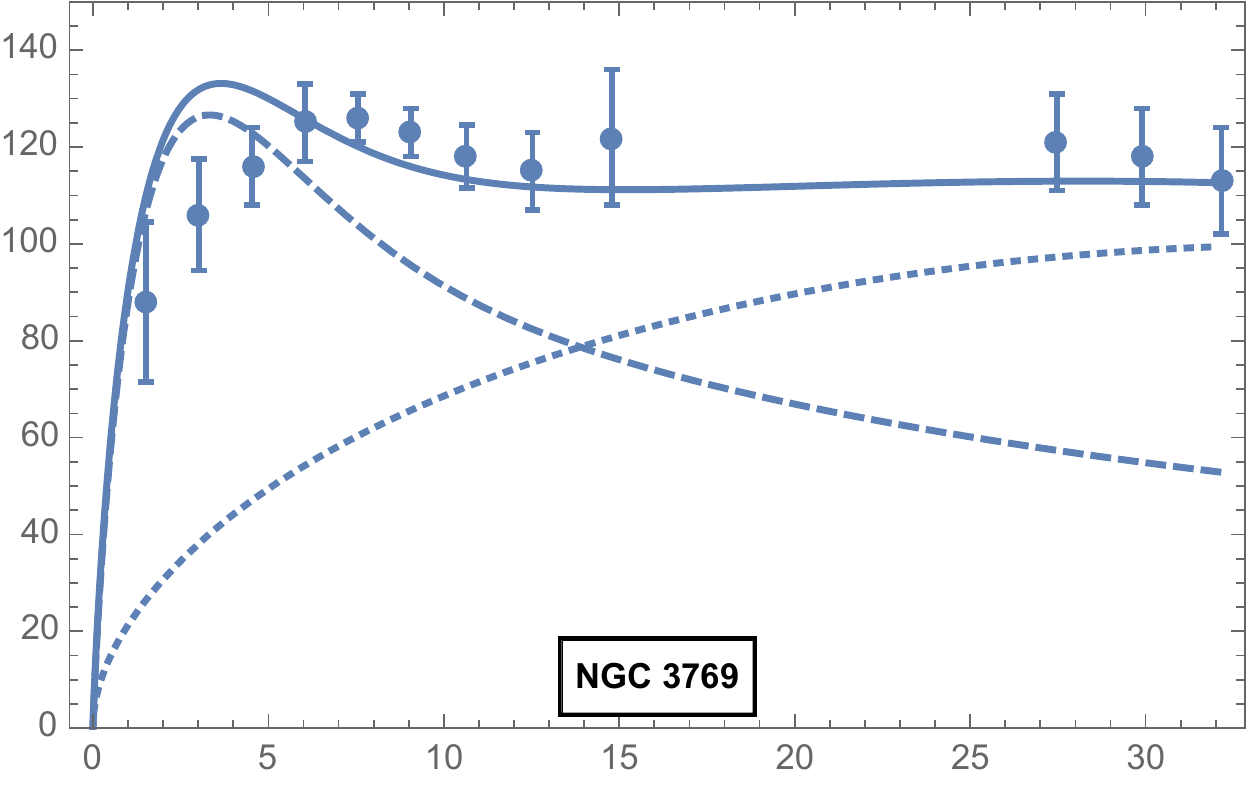,scale=0.4}\qquad
\epsfig{file=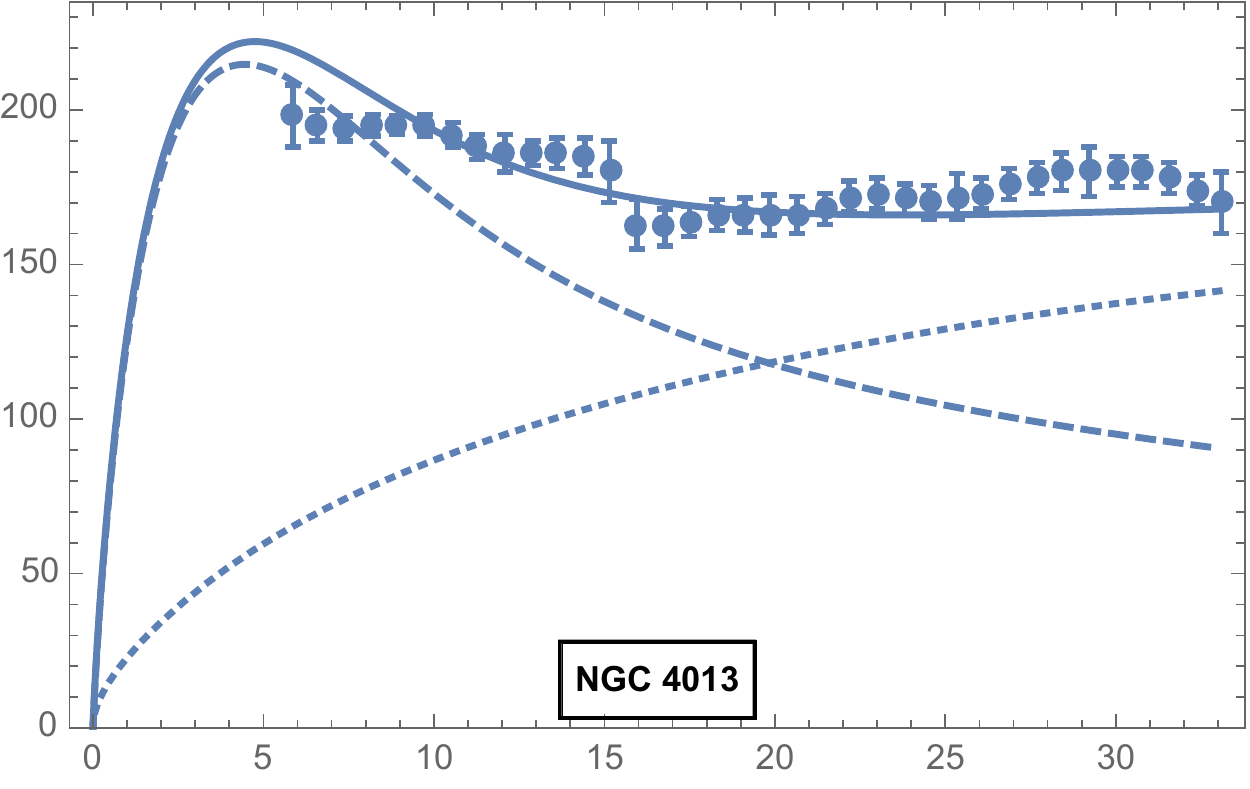,scale=0.4}\\
\epsfig{file=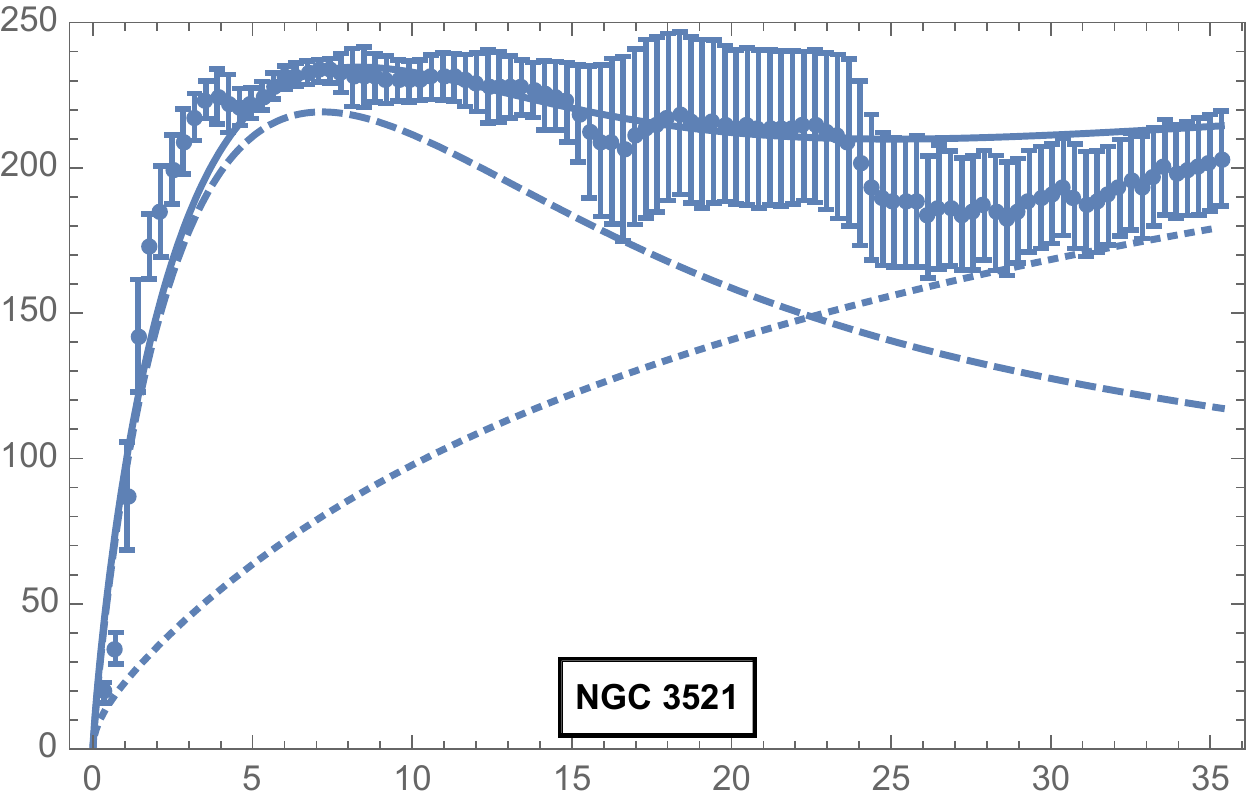,scale=0.4}\qquad
\epsfig{file=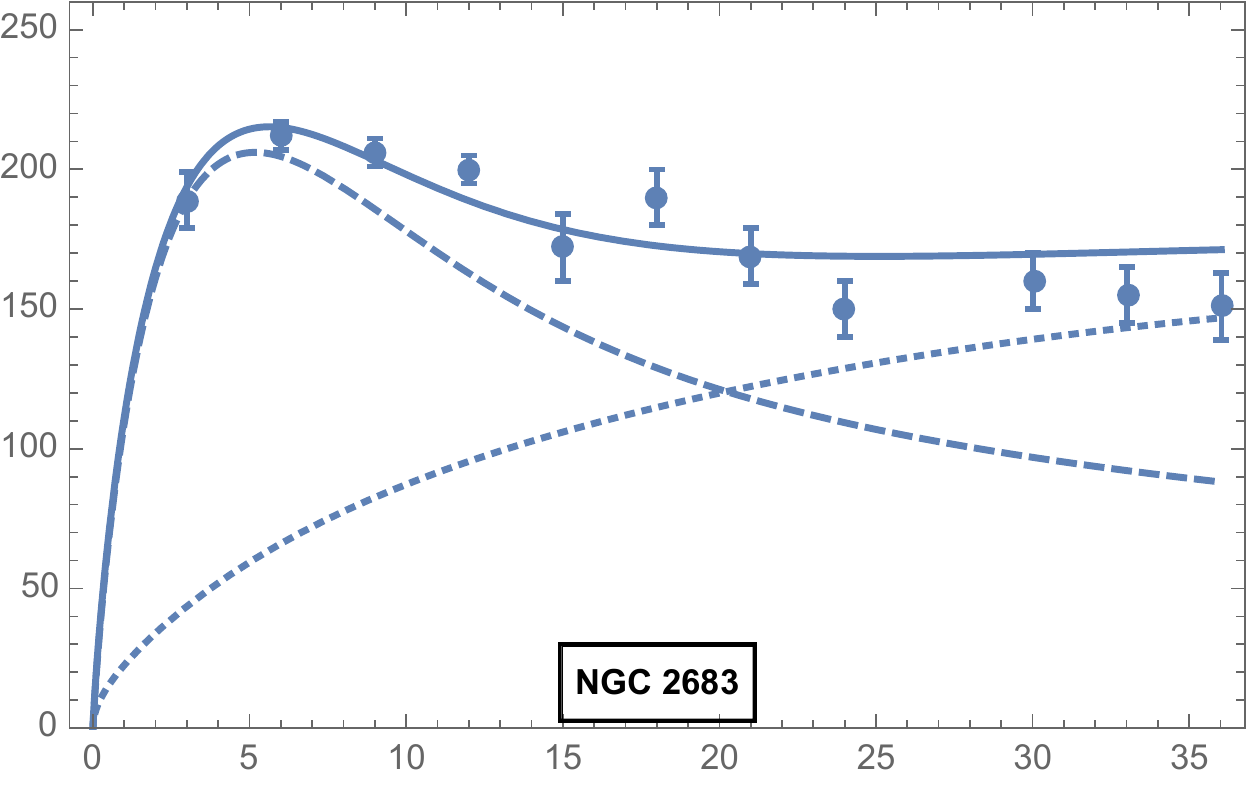,scale=0.4}\qquad
\epsfig{file=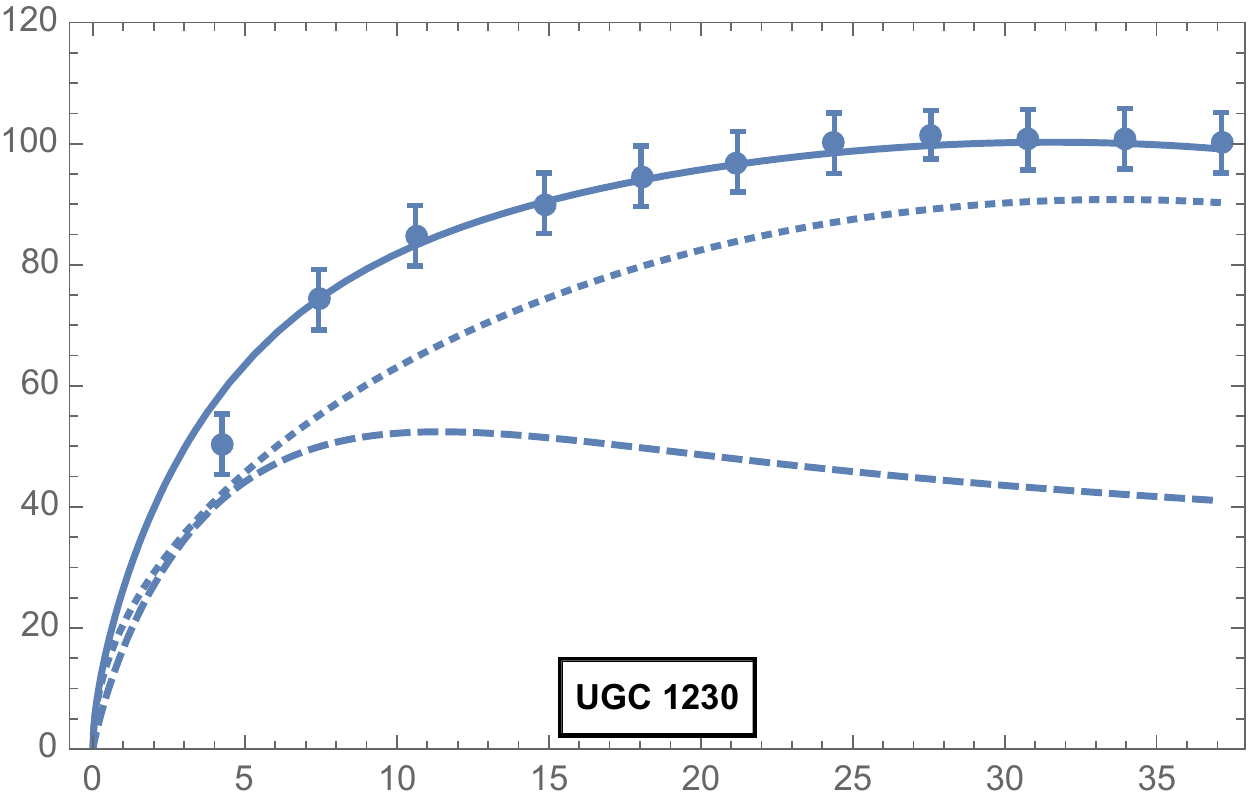,scale=0.4}\\
\epsfig{file=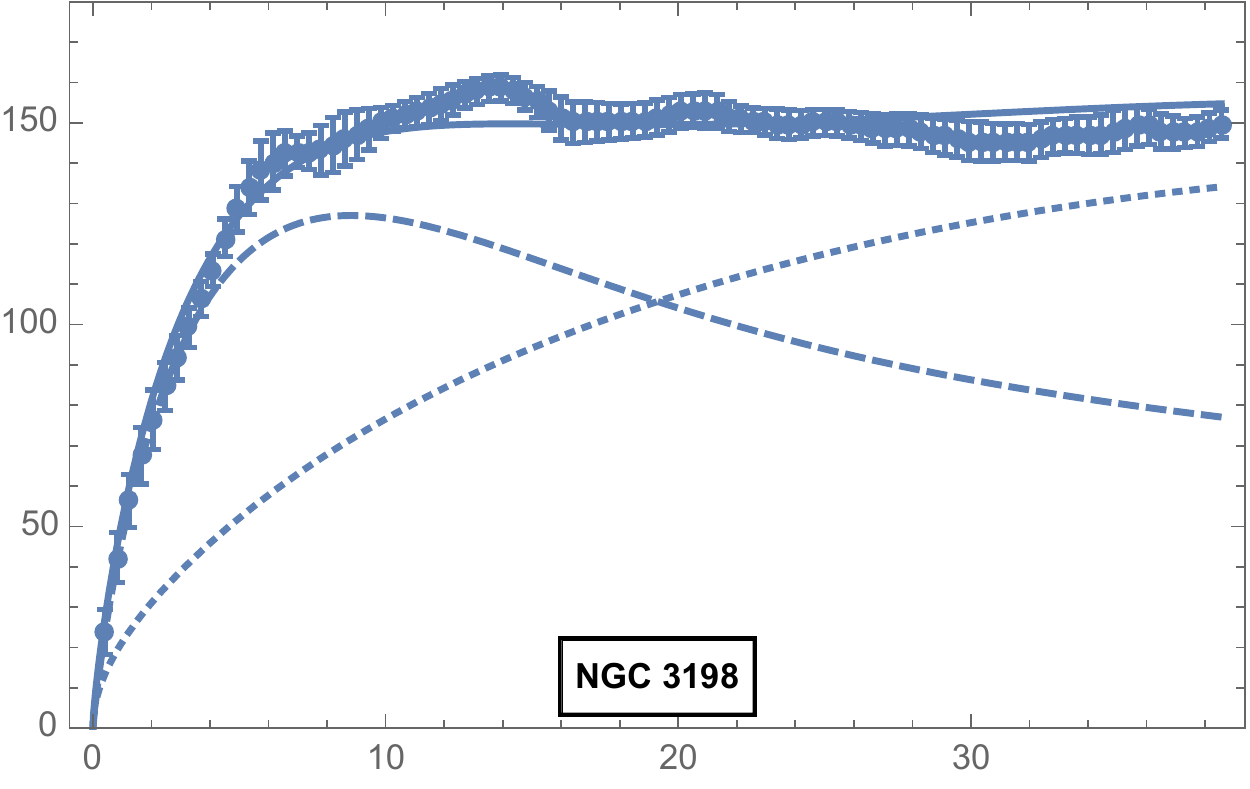,scale=0.4}\qquad
\epsfig{file=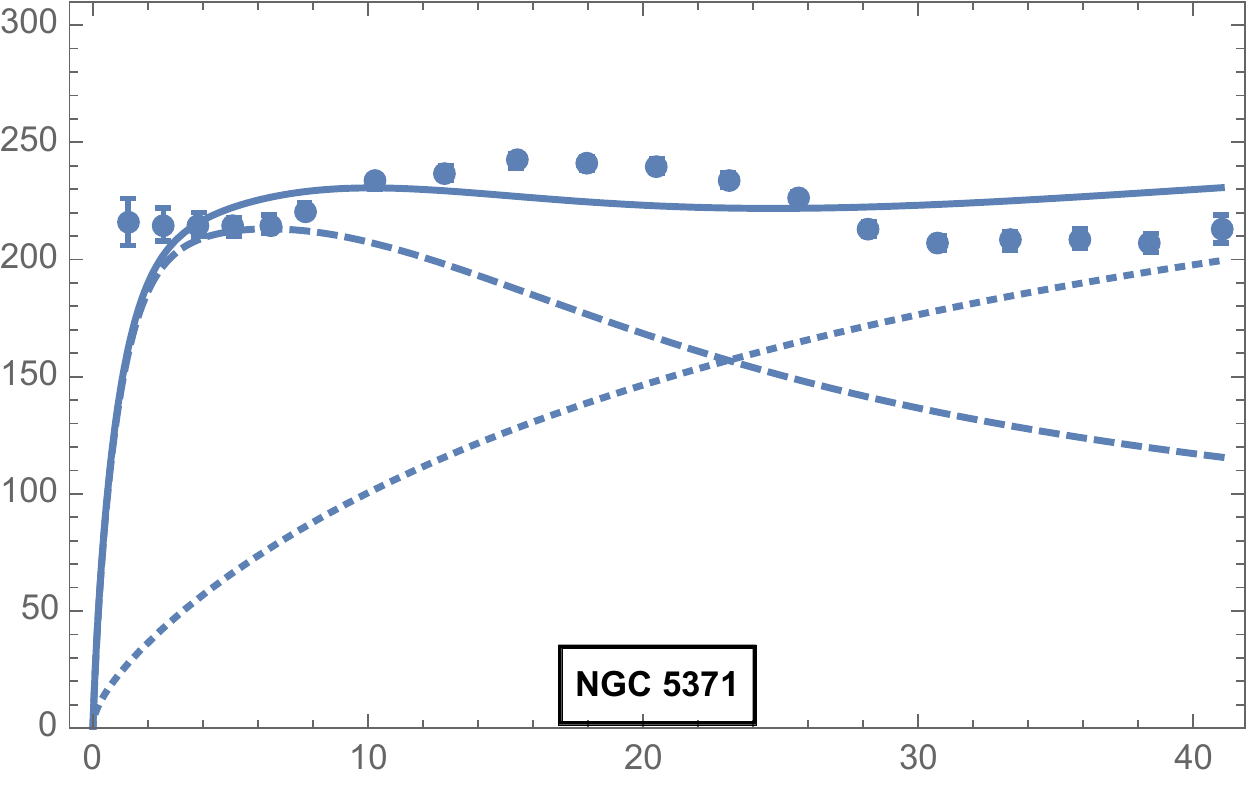,scale=0.4}\qquad
\epsfig{file=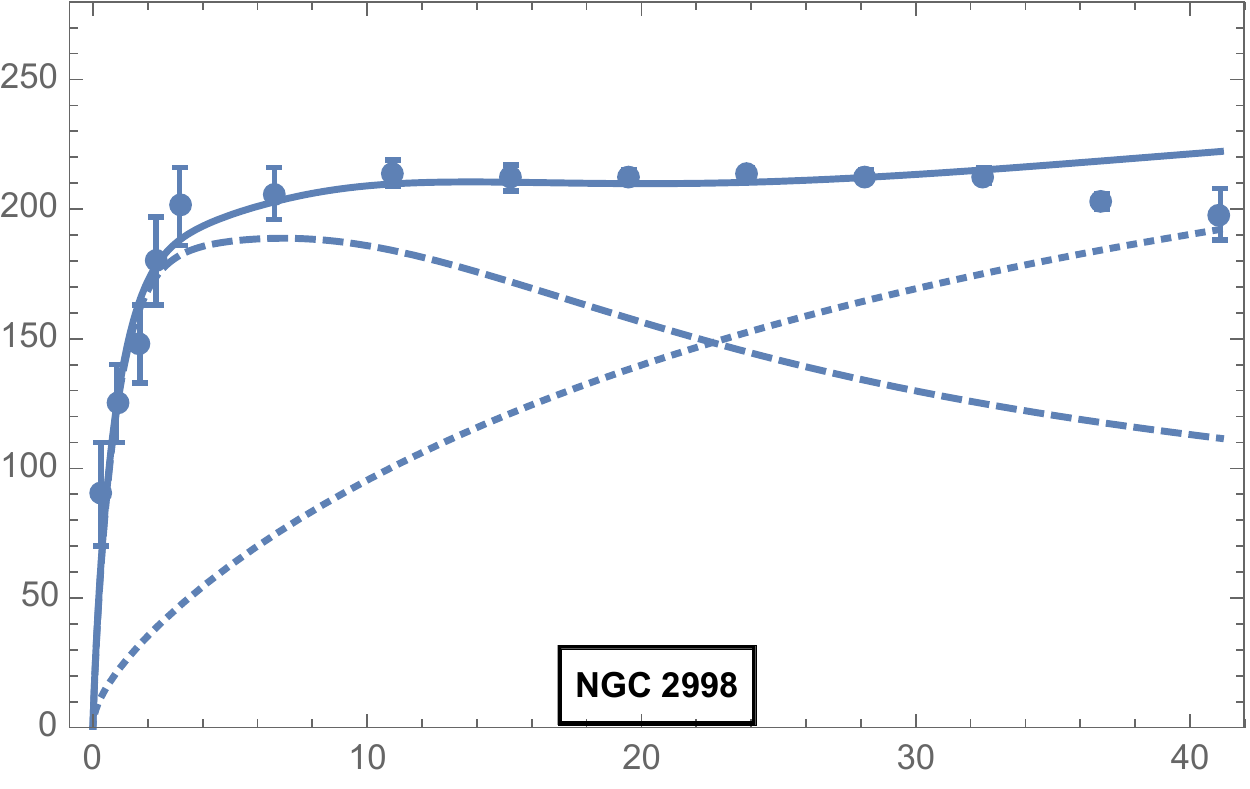,scale=0.4}\\
\epsfig{file=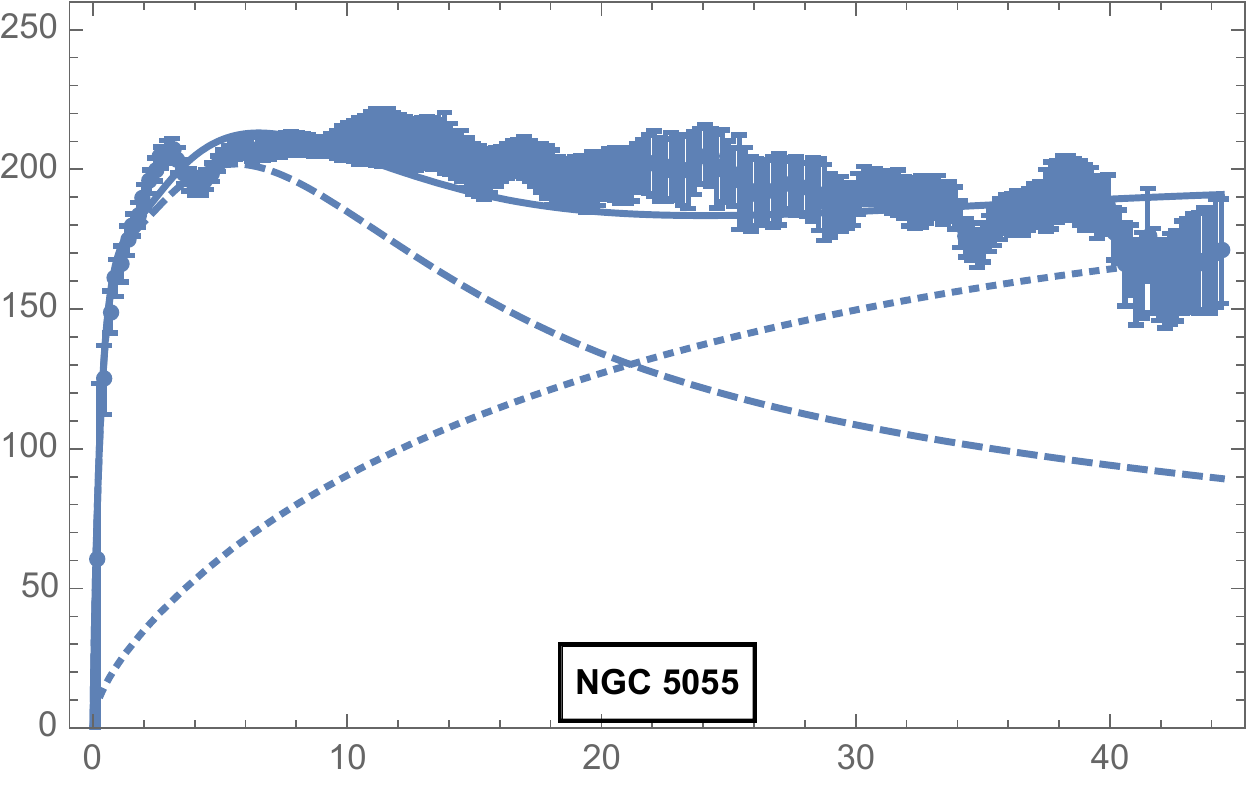,scale=0.4}\qquad
\epsfig{file=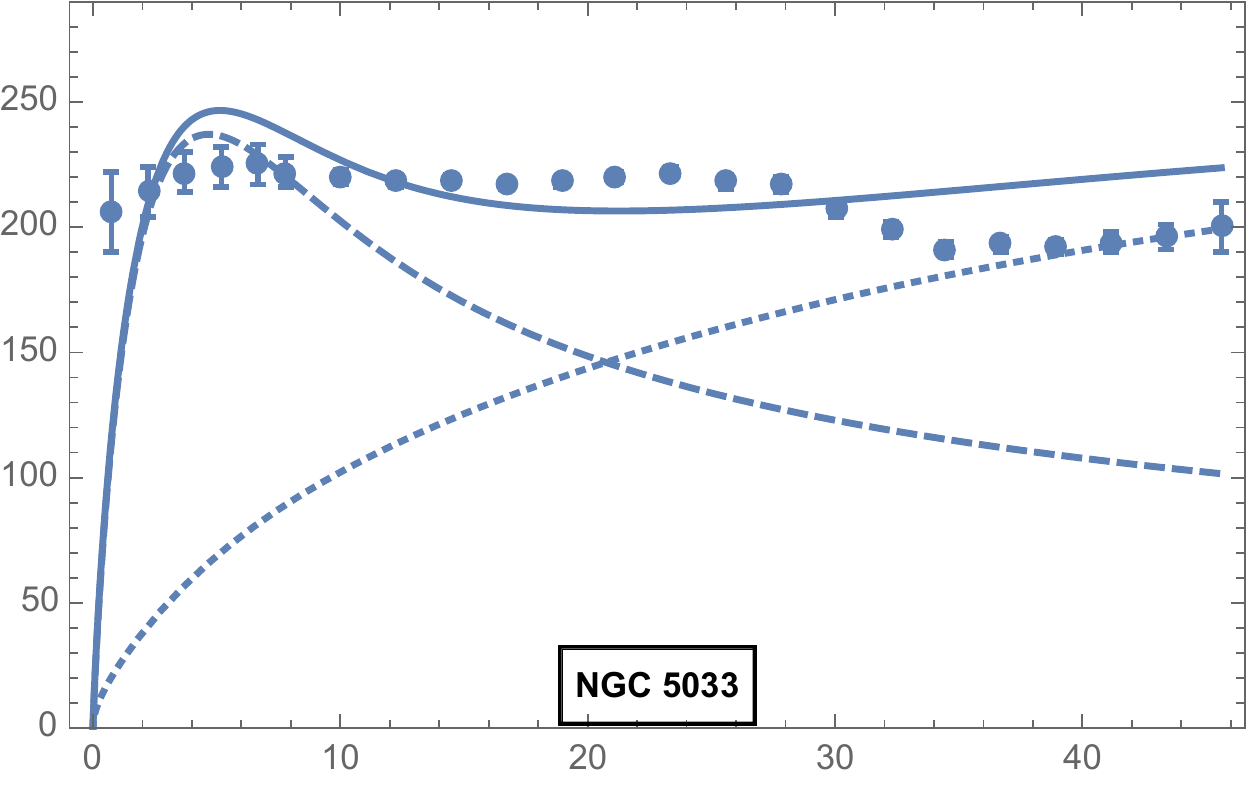,scale=0.4}\qquad
\epsfig{file=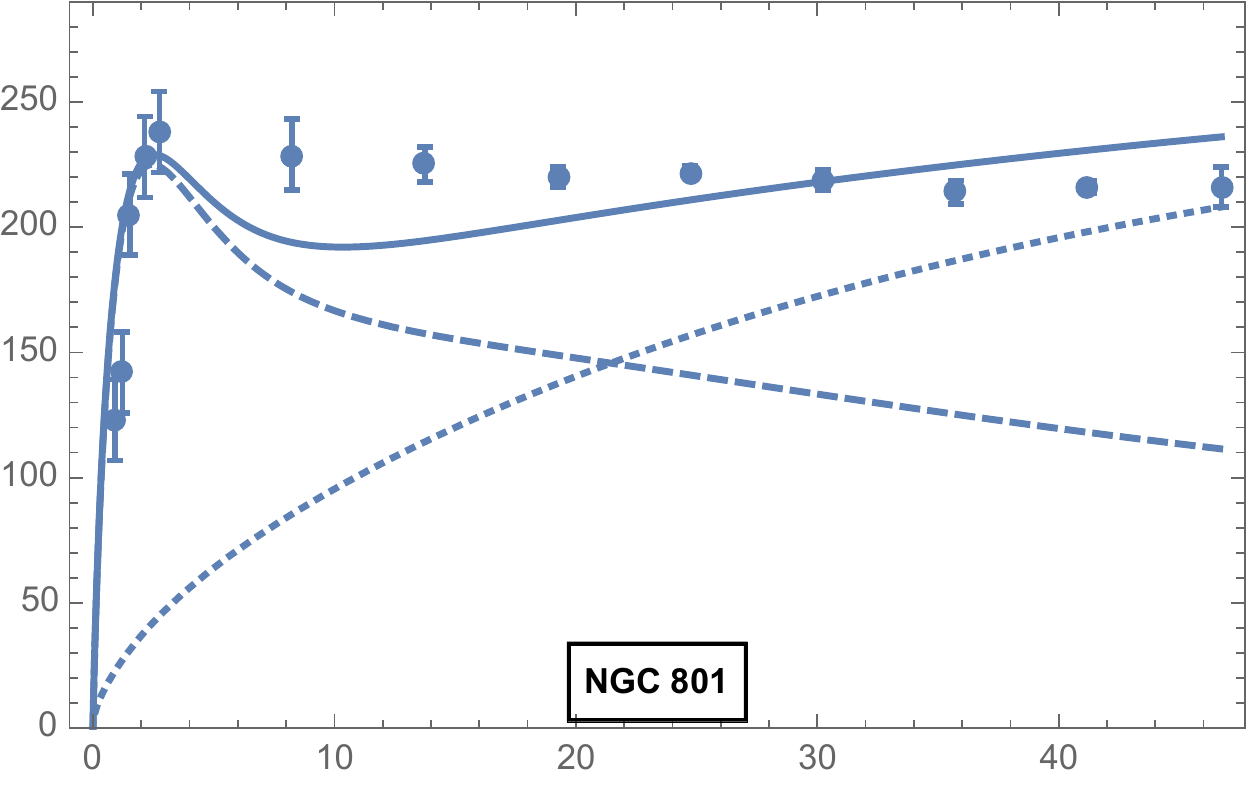,scale=0.4}\\
\epsfig{file=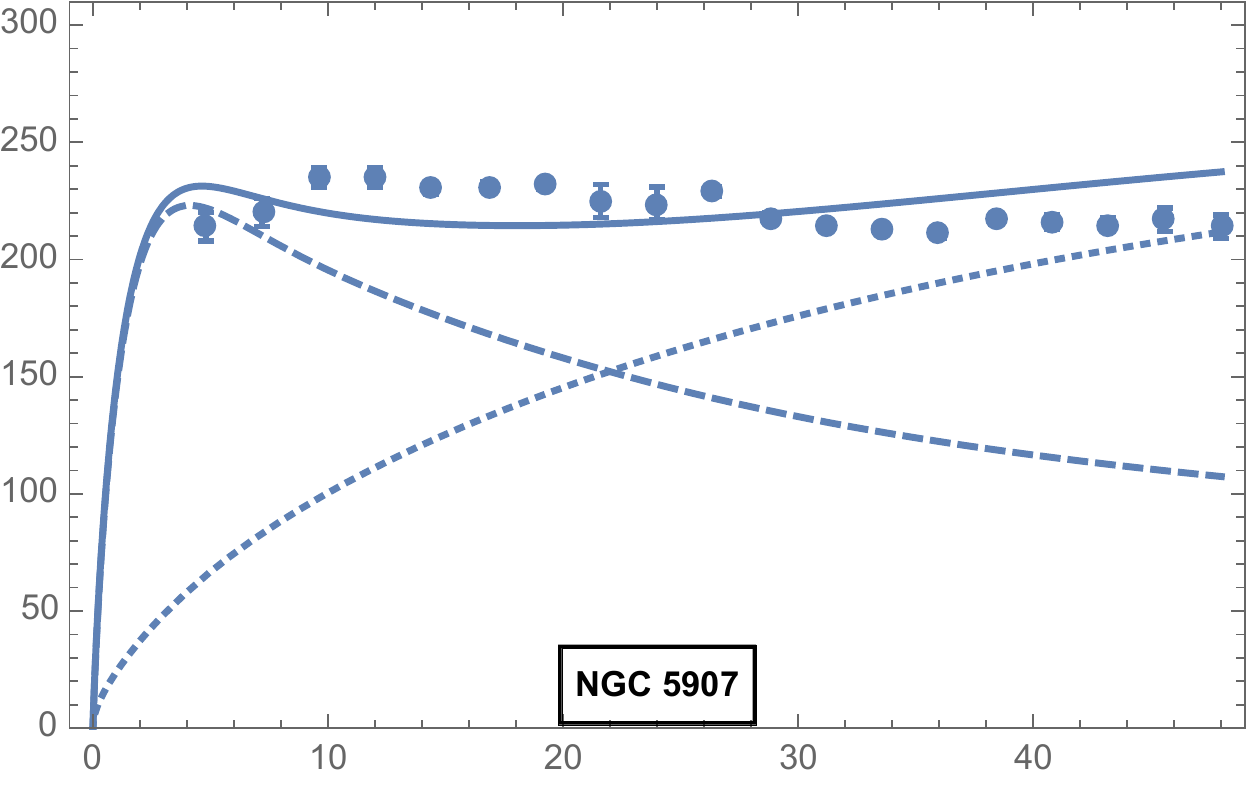,scale=0.4}\qquad
\epsfig{file=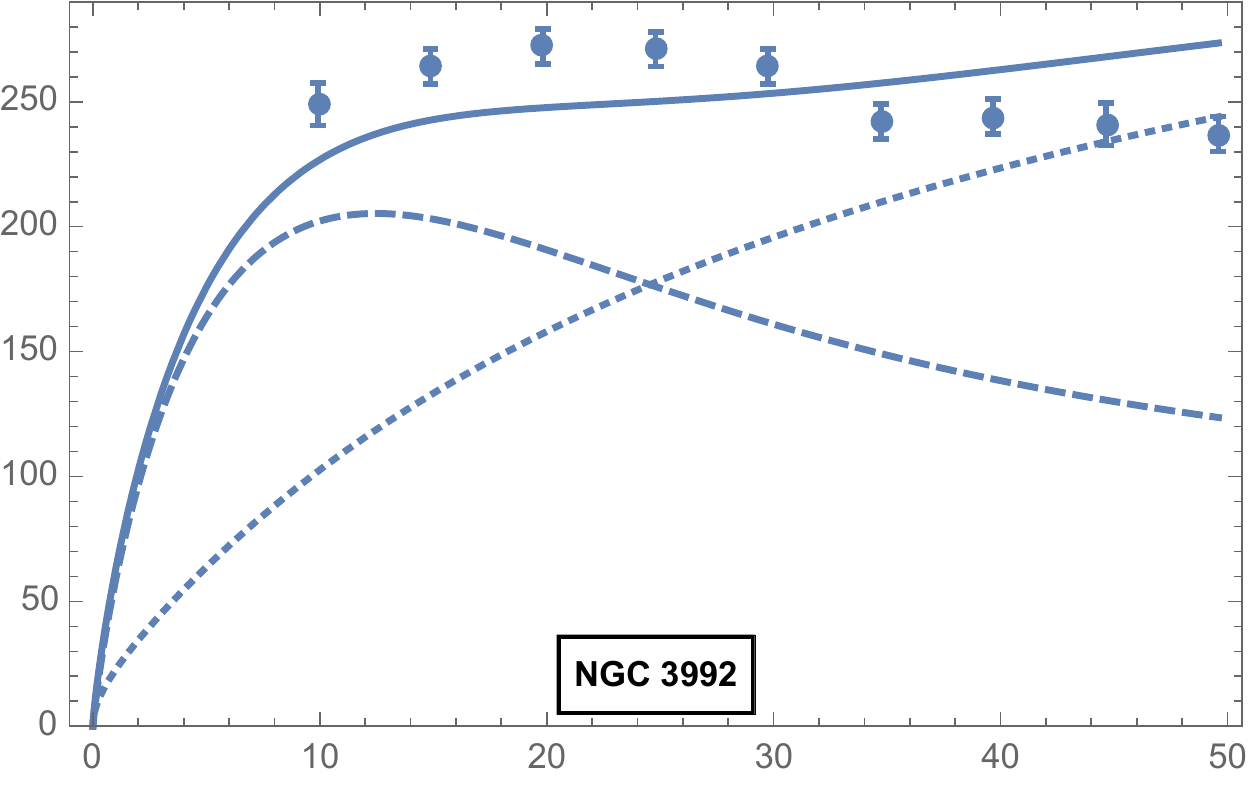,scale=0.4}\qquad
\epsfig{file=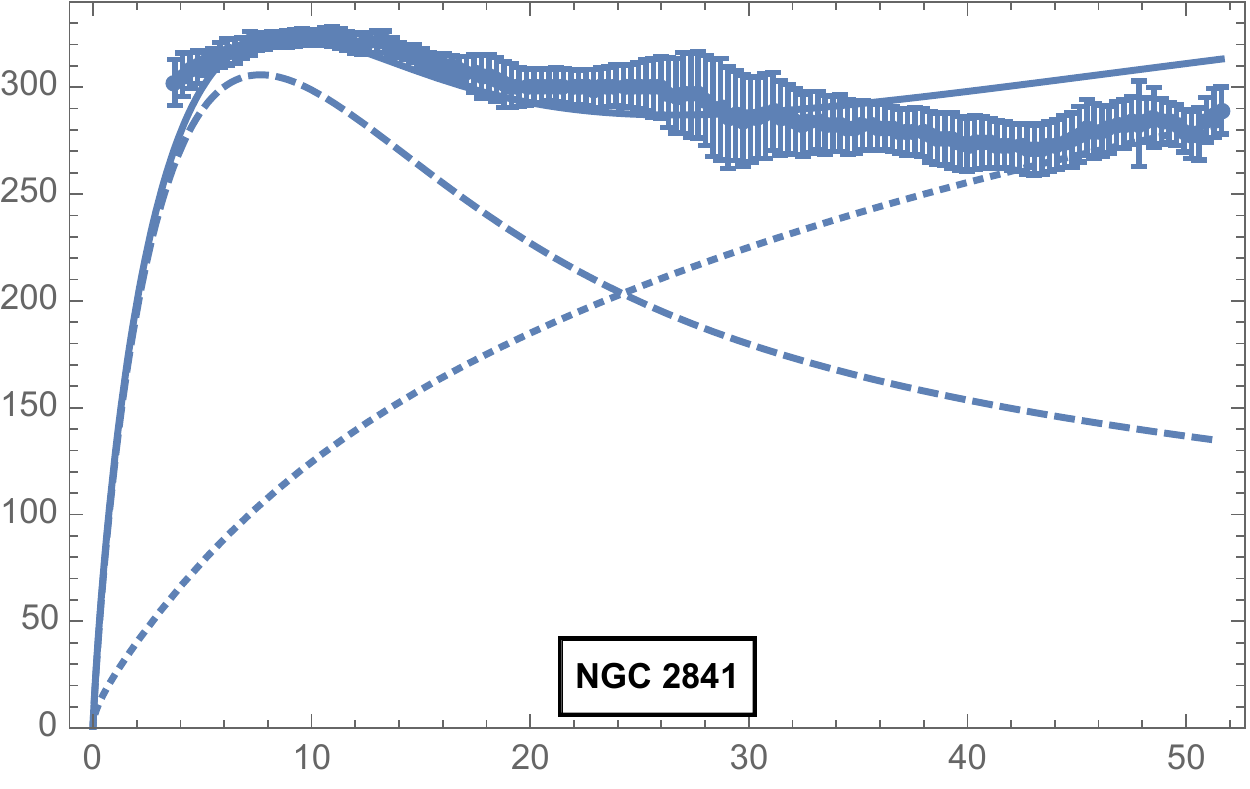,scale=0.4}\\
\epsfig{file=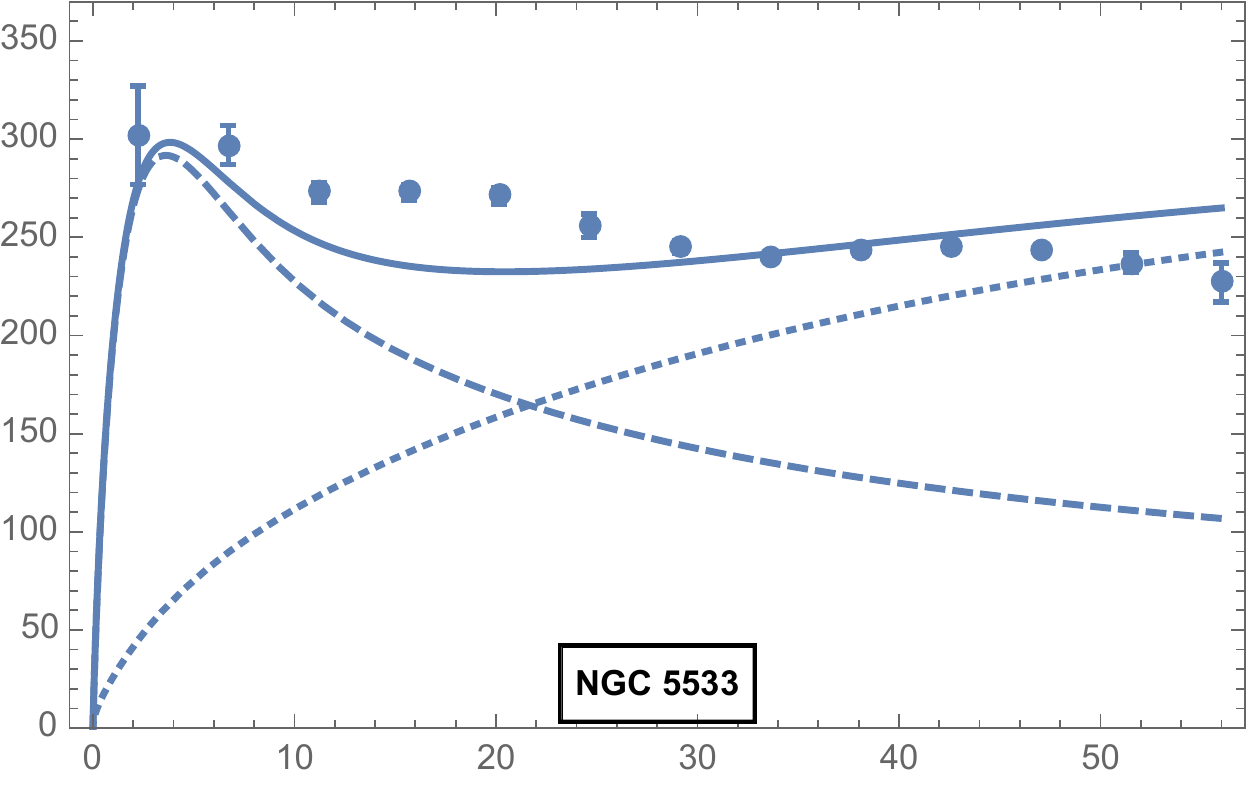,scale=0.4}\qquad
\epsfig{file=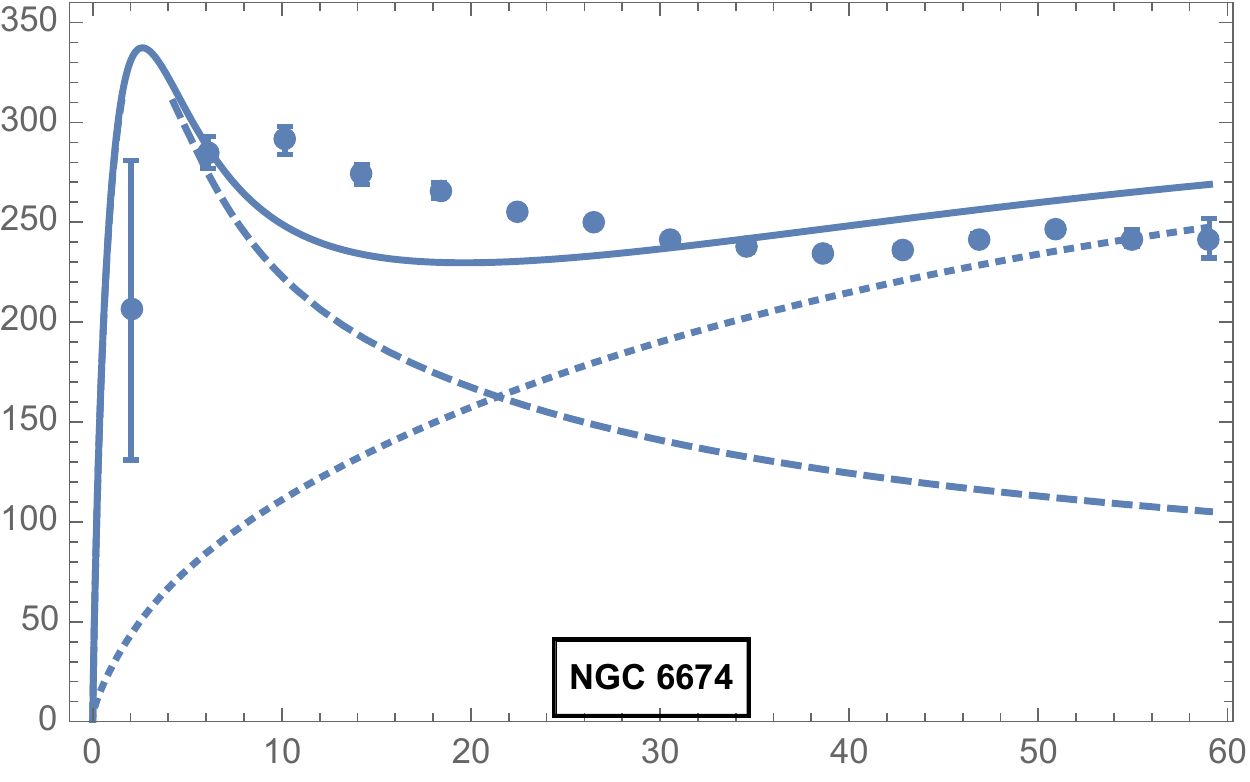,scale=0.4}\qquad
\epsfig{file=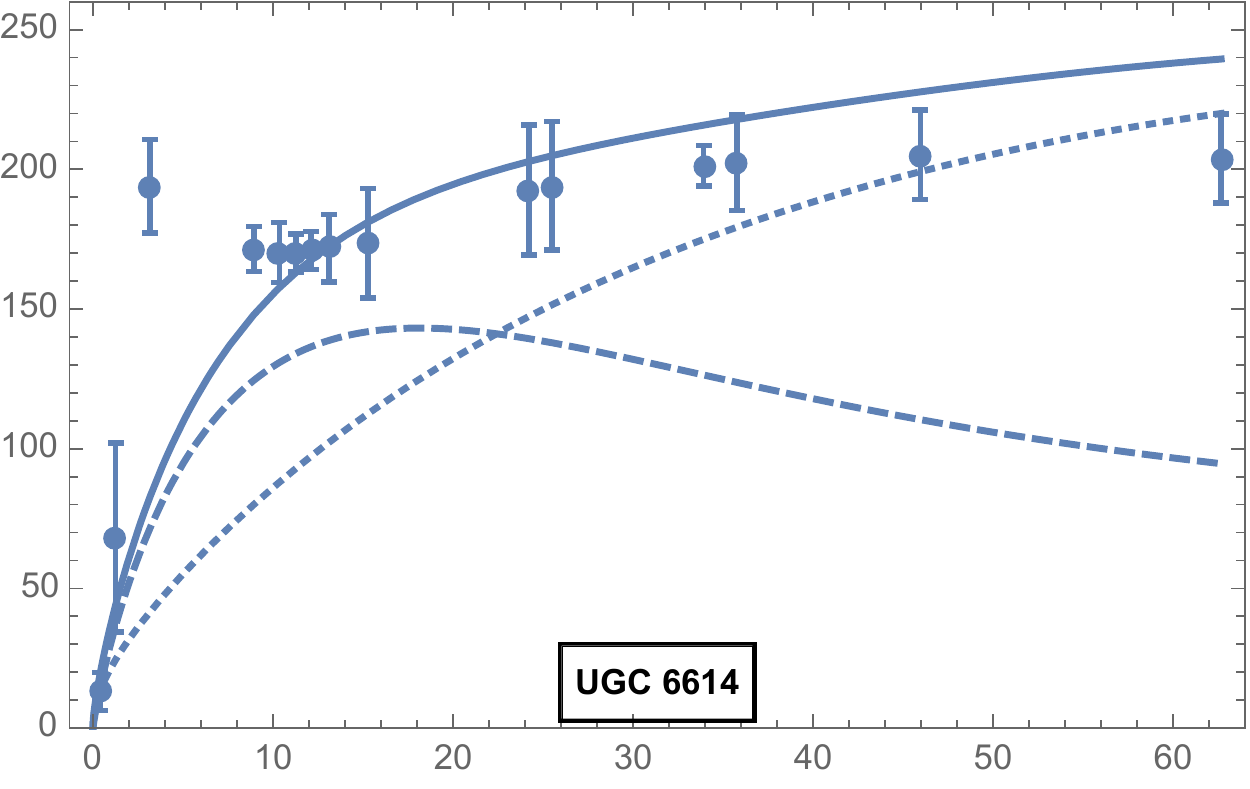,scale=0.4}\\
\epsfig{file=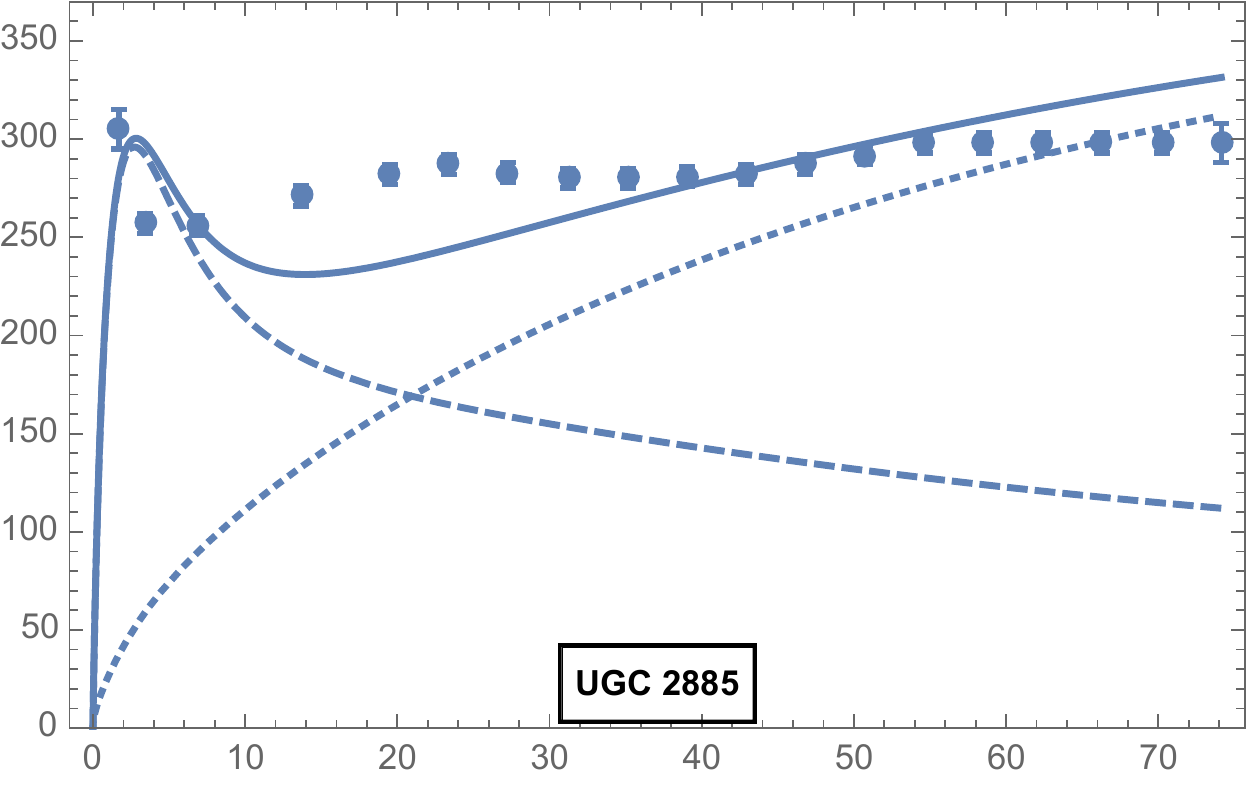,scale=0.4}\qquad
\epsfig{file=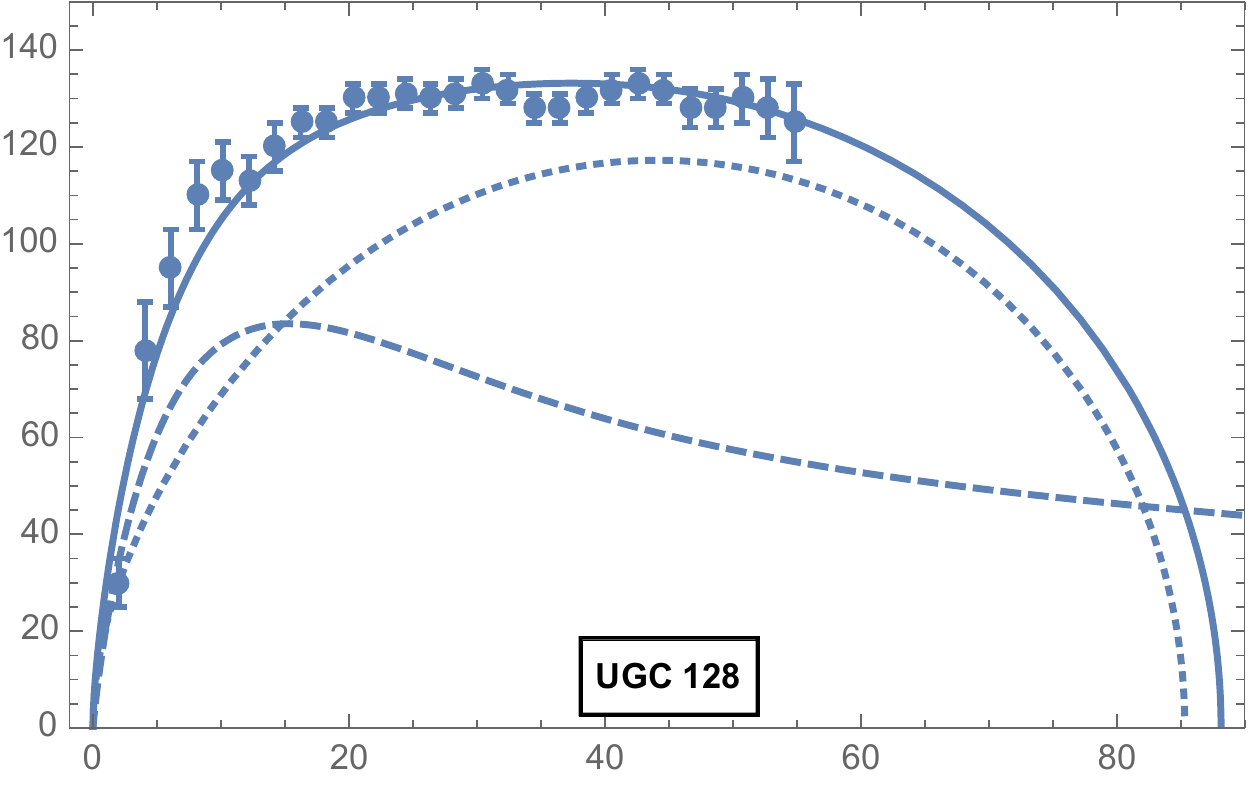,scale=0.4}\qquad
\epsfig{file=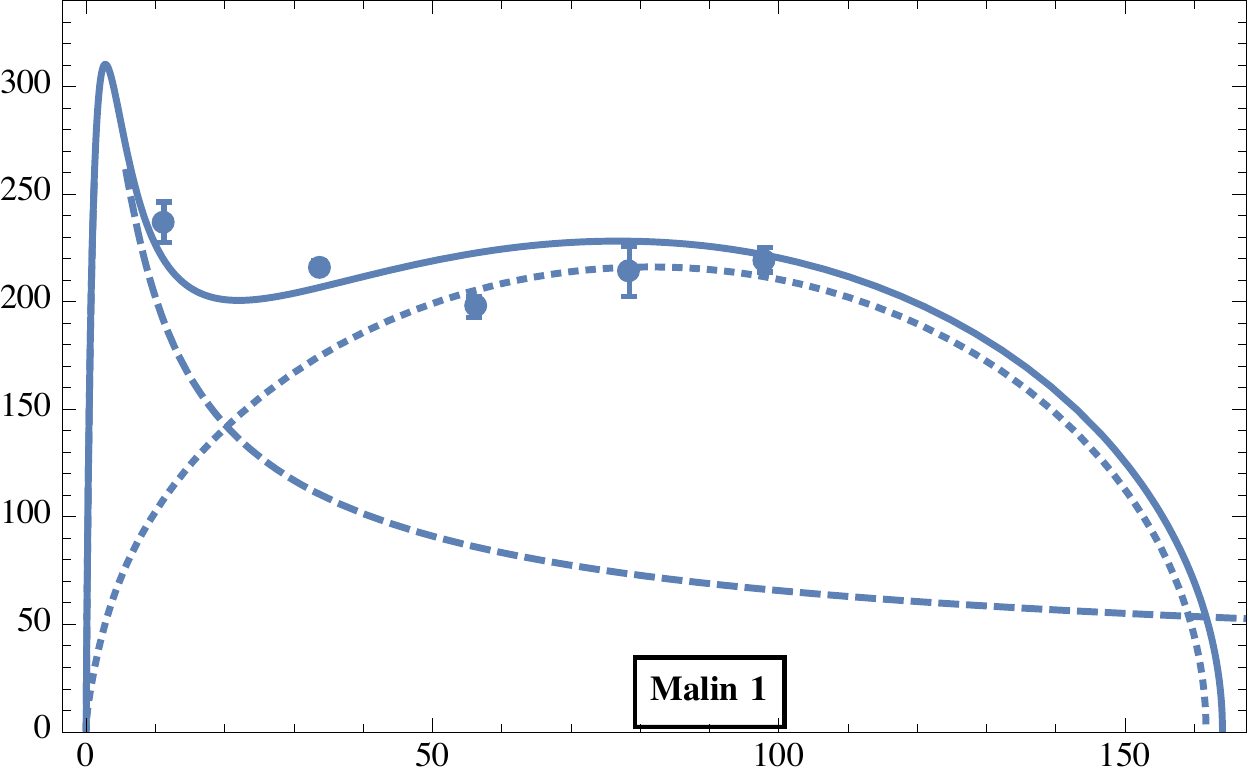,scale=0.4}
\caption{Fitting to the rotational velocities (in ${\rm km}~{\rm sec}^{-1}$ on the vertical axis) of  the 21 large galaxy sample as plotted as a function of radial distance (in ${\rm kpc}$ on the horizontal axis).}
\label{largegalaxies2}
\end{center}
\end{figure}

When the present author first studied rotation velocity curves in the conformal gravity theory,  the data that were available did not go out far enough in distance for the $-\kappa c^2 R^2$ term to be significant, with the $v_{{\rm TOT}}^2=v_{{\rm LOC}}^2+\gamma_0c^2 R/2$ formula providing a very good fit (see e.g. \cite{Mannheim2006}) to a then available 11 galaxy sample without the need for any dark matter. However, with 141 spiral galaxies now being available and with 21 of them going out far enough in distance to be sensitive to the $-\kappa c^2 R^2$ term,
Mannheim and O'Brien \cite{Mannheim2011b,Mannheim2012c,OBrien2012,Mannheim2013} applied the full $v_{{\rm TOT}}^2=v_{{\rm LOC}}^2+\gamma_0c^2 R/2-\kappa c^2 R^2$ formula to the set of 141 rotation curves, and found very good fitting with parameters
\begin{eqnarray}
\beta^*&=&1.48\times 10^5 {\rm cm},\qquad \gamma^*=5.42\times 10^{-41} {\rm cm}^{-1},\qquad
\nonumber\\
\gamma_0&=&3.06\times
10^{-30} {\rm cm}^{-1},\qquad \kappa = 9.54\times 10^{-54} {\rm cm}^{-2},
\label{M189}
\end{eqnarray} 
with no dark matter being needed. With $(\beta^*/\gamma^*)^{1/2}$ being found to be of order  $10^{23}$ cm, we see that the effect of the linear $\gamma^*$ potential does indeed only become significant on galactic distance scales, as do the $\gamma_0$ and $\kappa$ terms also. Consequently, without our requiring it a priori, solar system phenomenology is left intact.

Even though there is only one free parameter per galaxy, viz. $N^*$, a parameter that is common to all galactic rotation curve fits, and even though there is basically no flexibility, (\ref{M188}) fully captures the essence of the data. Without the quadratic $-\kappa c^2 R^2$ term, the linear potential would eventually start to dominate over the Newtonian potential and the typically flat galactic rotation curves would be expected to begin to rise. However, no rise was found in the 21 galaxies in the 141 galaxy sample that went out the furthest in distance, with each one possessing 10 or so points where  a rise should have been seen. That these 200 or so points were then controlled by one single quadratic term with a universal coefficient that acted on every single one of the 21 galaxies in exactly the same way is quite remarkable since one single parameter accounted for no less than 200 or so data points.\footnote{Fits to the 21 largest galaxies are shown in Fig. (\ref{largegalaxies2}). For each galaxy
we exhibit the contributions due to the luminous Newtonian term alone (dashed curve), the two linear terms and the quadratic terms combined (dotted curve), with the full curve showing the total contribution. No dark matter is assumed.}  For comparison, dark matter fits to galactic rotation curves typically require two free parameters per galactic dark matter halo (in addition to an $N^*$ for each galaxy), to thus need 282 more free parameters than the conformal theory in order to fit the 141 galaxy sample. If dark matter is to be correct, then dark matter theory should be able to derive (\ref{M188}) as it does describe the galactic data. For the moment though there is no indication that it can do so, or that it can predict the needed 282 halo parameters from first principles and show that they have any kind of  universal relationship such as the one exhibited in (\ref{M188}). Moreover, there is also no indication that supersymmetry, the most favored dark matter option, can do so either.\footnote{The success of any alternate theory of gravity in fitting astrophysical data cannot in and of itself exclude the possible existence of dark matter per se, with supersymmetry (whose existence is not in and of itself excluded by conformal gravity) for instance providing some potential dark matter candidates. However, the conformal gravity fits to galactic rotation curves completely account for all the rotation curve data, to thus leave little room for any dark matter contribution that might appear alongside the conformal gravity contribution.}

As a final comment on the conformal gravity fitting to galactic rotation curves, we note that as we go to even larger galactic distances than available in the current 141 galaxy sample, the quadratic $-\kappa c^2 R^2$ term will eventually start to dominate and galactic rotation curves must then start to fall. Moreover, since $v_{{\rm TOT}}^2$ can never be negative, there would have to be a largest radial distance beyond which there would be no bound orbits.\footnote{In Fig. (\ref{largegalaxies2}) we exhibit this effect for the two largest galaxies in the sample, UGC 128 and Malin 1.} There would thus have to be a universal largest size that galaxies could be, of order $\gamma_0/\kappa\sim 10^{23}$ cm, with the global physics that generates the $\gamma_0$ and $\kappa$ potential terms thus imposing a natural limit on the size of galaxies. Beyond the testable prediction that galactic rotation curves must eventually start to fall, it is of interest to note that, in contrast to the $1/r$ Newtonian potential where there is no upper limit to the possible size of bound orbits, in conformal gravity galactic orbits could not be of indefinitely large radius. 

Now while much more still needs to be done in conformal gravity (a full list of the challenges that conformal gravity currently faces may be found in \cite{Mannheim2012b}), two of the most urgent issues are gravitational lensing and fluctuations in the cosmic microwave background. For lensing the impact of the recently identified $\kappa$ term needs to be worked out since one is dealing with  light coming in from a non-asymptotically flat background. For fluctuations in the cosmic microwave background a first step has recently been taken with a conformal cosmology fluctuation theory having been presented in \cite{Mannheim2012b}. It would be of interest to ascertain whether conformal gravity could account for lensing or the temperature variations detected in the cosmic microwave background  without any need for dark matter.

\begin{figure}[htpb]
\begin{center}
\begin{minipage}[t]{8 cm}
\epsfig{file=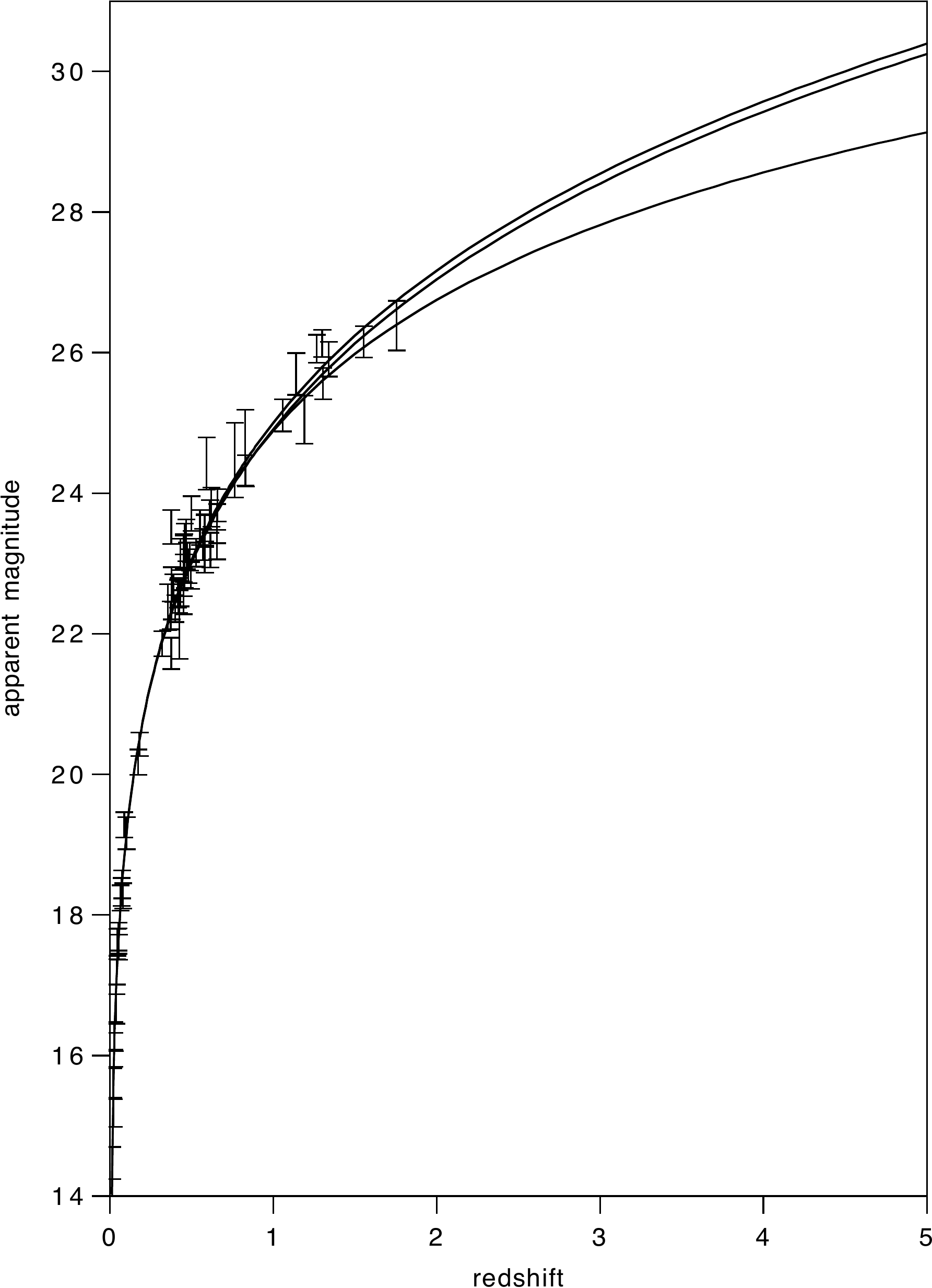,scale=0.6}
\end{minipage}
\begin{minipage}[t]{16.5 cm}
\caption{Hubble plot expectations for $q_0=-0.37$ (highest curve) and
$q_0=0$ (middle curve) conformal gravity and for
$\Omega_{M}(t_0)=0.3$,
$\Omega_{\Lambda}(t_0)=0.7$ standard gravity (lowest curve).}
\label{accelerating}
\end{minipage}
\end{center}
\end{figure}

Also of cosmological interest is to note that through use of the conformal invariant matter action $I_{\rm M}({\rm conf})$ given in (\ref{M159}),\footnote{As per the discussion of (\ref{M184}), the scalar field $S$ in $I_{\rm M}(\rm conf)$ is to be understood as an expectation value over positive frequency modes alone, viz. $S=\langle \Omega_M|b\bar{\psi}\psi b^{\dagger}|\Omega_M\rangle -\langle \Omega_M|\bar{\psi}\psi |\Omega_M \rangle$, an altogether smaller quantity than $\langle \Omega_M|\bar{\psi}\psi |\Omega_M\rangle$ itself.}  one can show that the Hubble plot associated with a conformal background Robertson-Walker cosmology is described by a luminosity function $d_L(z)$ of the form \cite{Mannheim2006}
\begin{eqnarray} 
d_L(z)=-\frac{c}{H_0}\frac{(1+z)^2}{q_0}\left(1-\left[1+q_0-\frac{q_0}{(1+z)^2}\right]^{1/2}\right),
\label{M190}
\end{eqnarray}
where $z$ is the redshift, and where $H_0$ and $q_0$ are the current era values of the Hubble and deceleration parameters. On taking the parameter $\lambda$ in $I_{\rm M}(\rm conf)$ to be negative (as required if $I_{\rm M}(\rm conf)$ is to be associated with a cosmological phase transition in which the free energy is lowered), and on taking the Robertson-Walker spatial  three curvature $k$ to be negative (as per the discussion above), in the conformal theory the structure of the  cosmological evolution equations is such that $q_0$ is constrained to lie in the interval $-1\leq q_0\leq 0$ no matter what the magnitudes of the parameters in $I_{\rm M}({\rm conf})$. Fits given in \cite{Mannheim2006} (and reproduced here as Fig. (\ref{accelerating})) to the accelerating universe Hubble plot data using (\ref{M190}) are found to lead to the value $q_0=-0.37$, a value that is nicely in the $-1\leq q_0\leq 0$ range. (The neither accelerating nor decelerating $q_0=0$ plot is included for comparison purposes.) The fits are every bit as good as the standard $\Omega_M=0.3$, $\Omega_{\Lambda}=0.7$ dark matter plus dark energy paradigm, with the conformal gravity theory being able to fit the data with no need for dark matter. And, with $q_0$ automatically being both small and negative, none of the 60 to 120 order of magnitude fine tuning associated with the standard paradigm is needed. As constructed, the conformal theory is naturally accelerating at all epochs (which is why it required no fine tuning in the first place) while in contrast, the standard paradigm is fine tuned to only be accelerating at late epochs. Consequently, at redshifts above one or so the conformal theory and standard theory Hubble plot expectations start to differ, and as shown in Fig. (\ref{accelerating}), begin to do so quite markedly by a redshift of two or so. Study of the Hubble plot at a redshift of two or so could thus be very instructive.

\subsection{Conformal Invariance and the Metrication of Electromagnetism}

Another recent development in conformal gravity has been a study \cite{Mannheim2014,Mannheim2016c} of a metrication of the fundamental forces, the very same objective that originally led Weyl to conformal symmetry in the first place. Just two years after Einstein metricated gravity with his development of general relativity, Weyl \cite{Weyl1918a} attempted to metricate electromagnetism by proposing that the standard Riemannian geometry Levi-Civita connection  
\begin{eqnarray}
\Lambda^{\lambda}_{\mu\nu}=\frac{1}{2}g^{\lambda\alpha}(\partial_{\mu}g_{\nu\alpha} +\partial_{\nu}g_{\mu\alpha}-\partial_{\alpha}g_{\nu\mu} )
\label{M191}
\end{eqnarray}
be augmented with a second connection, the Weyl connection 
\begin{eqnarray}
W^{\lambda}_{\mu\nu}=-g^{\lambda\alpha}(g_{\nu\alpha}A_{\mu} +g_{\mu\alpha}A_{\nu}-g_{\nu\mu}A_{\alpha})
\label{M192}
\end{eqnarray}
that expressly depended on the electromagnetic potential $A_{\mu}$, to thus put $A_{\mu}$ into the geometry. With this augmentation, the generalized connection 
\begin{eqnarray}
\tilde{\Gamma}^{\lambda}_{\mu\nu}=\Lambda^{\lambda}_{\mu\nu}+W^{\lambda}_{\mu\nu}
\label{M193}
\end{eqnarray}
has the remarkable property of being left invariant  under a joint local transformation on the metric and the electromagnetic potential of the form
\begin{eqnarray}
g_{\mu\nu}(x)\rightarrow e^{2\beta(x)}g_{\mu\nu}(x),\qquad A_{\mu}(x)\rightarrow A_{\mu}(x)+\partial_{\mu} \beta(x), 
\label{M194}
\end{eqnarray}
for any  spacetime-dependent $\beta(x)$. Consequently, a generalized Riemann  tensor or a generalized Weyl tensor built out of this generalized connection would be invariant under (\ref{M194}) also. However, with the covariant derivative of the metric associated with this generalized connection being the non-zero  $\tilde{\nabla}_{\sigma}g^{\mu\nu}=-2g^{\mu\nu}A_{\sigma}$, parallel transport is then path (and thus history) dependent, with the theory thus being untenable.

Much later it was realized that the theory had another drawback. When $W^{\lambda}_{\mu\nu}$ is inserted into the generalized Dirac action 
\begin{eqnarray}
I_{\rm D}=\frac{1}{2}\int d^4x(-g)^{1/2}i\bar{\psi}\gamma^{c}V^{\mu}_c(x)(\partial_{\mu}+\tilde{\Gamma}_{\mu})\psi+{\rm H.~c.}, 
\label{M195}
\end{eqnarray}
where $\tilde{\Gamma}_{\mu}(x)=-(1/8)[\gamma_a,\gamma_b](V^b_{\nu}\partial_{\mu}V^{a\nu}+V^b_{\lambda}\tilde{\Gamma}^{\lambda}_{\phantom{\lambda}\nu\mu}V^{a\nu})$ is constructed with the generalized $\tilde{\Gamma}^{\lambda}_{\mu\nu} $ connection given in (\ref{M193}), the Weyl connection is found to drop out identically  \cite{Hayashi1977,Mannheim2014}. Thus the Weyl connection never could have provided a geometric description of electromagnetism in the first place.

To rectify this, we note that the Weyl prescription for constructing $\tilde{\Gamma}^{\lambda}_{\mu\nu}$ is to replace the $\partial_{\mu}$ factor in $\Lambda^{\lambda}_{\mu\nu}$ by $\partial_{\mu}-2A_{\mu}$. However, we recall that to  obtain the standard minimal electromagnetic gauge coupling we must use not $\partial_{\mu}-A_{\mu}$ but $\partial_{\mu}-iA_{\mu}$, with the factor of $i$ actually  being required because of Hermiticity or $CPT$ invariance. Thus it is suggested  \cite{Mannheim2014} to proceed analogously with  the Weyl connection by replacing $W^{\lambda}_{\mu\nu}$ by 
\begin{eqnarray} 
V^{\lambda}_{\mu\nu}=-\frac{2i}{3}g^{\lambda\alpha}(g_{\nu\alpha}A_{\mu} +g_{\mu\alpha}A_{\nu}-g_{\nu\mu}A_{\alpha}).
\label{M196}
\end{eqnarray}
Then, on using the generalized connection 
\begin{eqnarray}
\tilde{\Gamma}^{\lambda}_{\mu\nu}=\Lambda^{\lambda}_{\mu\nu}+V^{\lambda}_{\mu\nu}
\label{M197}
\end{eqnarray}
in a generalized spin connection $\tilde{\Gamma}_{\mu}$ that is determined from it, following some algebra we find that the generalized Dirac action takes none other than the form \cite{Mannheim2014} 
\begin{eqnarray}
I_{\rm D}=\int d^4x(-g)^{1/2}i\bar{\psi}\gamma^{c}V^{\mu}_c(x)(\partial_{\mu}+\Gamma_{\mu}-iA_{\mu})\psi
\label{M198}
\end{eqnarray}
With the action given in (\ref{M198}) being the standard one that is obtained by standard local electromagnetic gauge invariance (viz. invariance under $\psi(x)\rightarrow e^{i\alpha(x)}\psi(x)$, $A_{\mu}(x)\rightarrow A_{\mu}(x)+\partial_{\mu}\alpha(x)$), the metrication of electromagnetism is thereby achieved. The $I_{\rm D}$  action thus has a dual characterization -- it can be generated via local gauge invariance or via a generalized geometric connection. The two viewpoints are equivalent.

As constructed, the action $I_{\rm D}$ is not just locally gauge invariant, it is locally conformal invariant too, as it is left invariant under 
\begin{eqnarray}
\psi(x)\rightarrow e^{-3\beta(x)/2}\psi(x),~~ g_{\mu\nu}(x)\rightarrow e^{2\beta(x)}g_{\mu\nu}(x),~~
V^a_{\mu}(x)\rightarrow e^{\beta(x)}V^a_{\mu}(x),~~ A_{\mu}(x)\rightarrow A_{\mu}(x). 
\label{M199}
\end{eqnarray}
However, comparing with (\ref{M194}), we see that in (\ref{M199}), this time $A_{\mu}(x)$ is not to transform at all. In consequence,  neither a generalized Riemann  tensor or a generalized Weyl tensor built out of the connection given in (\ref{M197}) would be invariant under (\ref{M199}). Moreover, neither would the standard Riemann  tensor as constructed via the standard Levi-Civita connection alone be invariant either. However, the conformal Weyl  tensor as constructed in (\ref{M169}) via the standard Levi-Civita connection alone would be locally conformal invariant under (\ref{M199}).  Consequently, given the local conformal invariance that the Dirac action possesses, the only action that we could write down in the pure gravity sector that would have this invariance too would be the $I_{\rm W}$ action given in (\ref{M171}). We are thus naturally led to conformal gravity. And since conformal gravity is based on a strictly Riemannian geometry with the standard Levi-Civita connection, there is no parallel transport problem, any more than there would have been if we had introduced $A_{\mu}$ into the Dirac action and obtained (\ref{M198}) by local electromagnetic gauge invariance in a theory with a Levi-Civita based $\Gamma_{\mu}$. If (\ref{M198}) is associated with a strictly Riemannian geometry when $A_{\mu}$ is introduced via local electromagnetic gauge invariance, it must equally be associated with a strictly Riemannian geometry when $A_{\mu}$ is introduced via the generalized connection given in (\ref{M197}). Invariance under (\ref{M199}) thus forces the gravity sector to be strictly Riemannian, and there is no parallel transport problem. The local conformal structure of $I_{\rm D}$ thus prevents $V^{\lambda}_{\mu\nu}$ with its otherwise seemingly problematic factor of $i$ from coupling in the gravity sector, to thereby render the gravity sector real. And, with there being no parallel transport problem, even with its non-zero $V^{\lambda}_{\mu\nu}$, the theory is viable.

\subsection{Conformal Gravity and the Metrication of the Fundamental Forces}

To generalize metrication to incorporate axial symmetry, we allow for torsion and introduce the antisymmetric Cartan torsion tensor  $Q^{\lambda}_{\phantom{\alpha}\mu\nu}$ and the associated contorsion $K^{\lambda}_{\phantom{\alpha}\mu\nu}$ according to 
\begin{eqnarray}
Q^{\lambda}_{\phantom{\alpha}\mu\nu}&=&\Gamma^{\lambda}_{\phantom{\alpha}\mu\nu}-\Gamma^{\lambda}_{\phantom{\alpha}\nu\mu}=-Q^{\lambda}_{\phantom{\alpha}\nu\mu},\qquad K^{\lambda}_{\phantom{\alpha}\mu\nu}=\frac{1}{2}g^{\lambda\alpha}(Q_{\mu\nu\alpha}+Q_{\nu\mu\alpha}-Q_{\alpha\nu\mu}).
\label{M200}
\end{eqnarray}
Then with 
\begin{eqnarray}
S^{\mu}=\frac{1}{8}(-g)^{-1/2}\epsilon^{\mu\alpha\beta\gamma}Q_{\alpha\beta\gamma},
\label{M201}
\end{eqnarray}
insertion of 
\begin{eqnarray}
\tilde{\Gamma}^{\lambda}_{\mu\nu}=\Lambda^{\lambda}_{\mu\nu}+V^{\lambda}_{\mu\nu}+K^{\lambda}_{\mu\nu}
\label{M202}
\end{eqnarray}
into $\Gamma^{\mu}$ is found \cite{Shapiro2002} to change $I_{\rm D}$ of (\ref{M198}) into  
\begin{eqnarray}
I_{\rm D}=\int d^4x(-g)^{1/2}i\bar{\psi}\gamma^{c}V^{\mu}_c(x)(\partial_{\mu}+\Gamma_\mu-iA_{\mu} -i\gamma^5S_{\mu})\psi.
\label{M203}
\end{eqnarray}
This action is invariant  under $\psi(x)\rightarrow \exp[i\gamma^5 \delta(x)]\psi(x)$, $S_{\mu}(x)\rightarrow S_{\mu}(x)+\partial_{\mu}\delta(x)$, with local axial symmetry thus being metricated too. The action is also invariant under local conformal transformations as well, if, just like $A_{\mu}(x)$, $S_{\mu}(x)$ does not transform under a local conformal transformation at all (viz. $S_{\mu}(x)\rightarrow S_{\mu}(x)$). It is our view that rather than being something arcane, torsion manifests itself as an axial gauge boson, one that would then have escaped detection if it acquires a large enough Higgs mechanism mass. Thus if we seek a metrication of the fundamental forces through the Weyl and torsion connections, we are led to a quite far reaching conclusion: not only must the fundamental forces be described by local gauge theories, they must be described by spontaneously broken ones.

The extension to the non-Abelian case is also direct. If for instance we put the fermions into the fundamental representation of  $SU(N)\times SU(N)$ with $SU(N)$ generators $T^i$ that obey  $[T^i,T^j]=if^{ijk}T^k$, replace $A_{\mu}$ by $g_VT^iA^{i}_{\mu}$, replace $Q_{\alpha\beta\gamma}$ by $g_AT^iQ^i_{\alpha\beta\gamma}$, and thus replace $S_{\mu}$ by $g_AT^iS^{i}_{\mu}$ in the connections, we obtain a locally $SU(N)\times SU(N)$ invariant Dirac action of the form \cite{Mannheim2014,Mannheim2015}
\begin{eqnarray}
J_{\rm D}=\int d^4x(-g)^{1/2}i\bar{\psi}\gamma^{c}V^{\mu}_c(\partial_{\mu}+\Gamma_{\mu}
-ig_VT^iA^i_{\mu} -ig_A\gamma^5T^iS^i_{\mu})\psi. 
\label{M204}
\end{eqnarray}
We can thus metricate chirally-symmetric non-Abelian theories (on setting $g_A=g_V$). As constructed, the action $J_{\rm D}$ is quite remarkable, as it not only possesses some of the key local invariances in physics, viz. general coordinate invariance, local Lorentz invariance, local vector non-Abelian gauge symmetry,  local axial-vector non-Abelian gauge symmetry, and local conformal symmetry, every one of these invariances can  have a geometric origin.

\subsection{The Inevitability of Conformal Gravity}

When the fermion spin connection $\Gamma_{\mu}(x)$ was first introduced into physics, its purpose was to implement local Lorentz invariance, since while $i\bar{\psi}\gamma^{\mu}\partial_{\mu}\psi$ was invariant under global Lorentz transformations  on $\psi$ of the form $\psi\rightarrow \exp (iw^{\mu\nu}M_{\mu\nu})\psi$ with a spacetime-independent $w^{\mu\nu}$, on its own $i\bar{\psi}\gamma^{\mu}\partial_{\mu}\psi$ was not invariant under local ones in which $w^{\mu\nu}$ depended on the spacetime coordinates. However, $i\bar{\psi}\gamma^{c}V^{\mu}_c(x)(\partial_{\mu}+\Gamma_{\mu})\psi$ is. Now while it was not intended, the standard curved space massless Dirac action $\int d^4x(-g)^{1/2}i\bar{\psi}\gamma^{c}V^{\mu}_c(x)(\partial_{\mu}+\Gamma_{\mu})\psi$ with a Levi-Civita based spin connection just happens to be invariant under the local conformal transformations given in (\ref{M199}).\footnote{In flat spacetime the action $\int d^4x i\bar{\psi}\gamma^{\mu}\partial_{\mu}\psi$ is invariant under the global $\psi(x_{\mu})\rightarrow e^{-3\beta/2}\psi(e^{\beta}x_{\mu})$, as the change of variable $x_{\mu}^{\prime}=e^{\beta}x_{\mu}$ reveals. The spin connection is then needed when $\beta$ is taken to depend on the spacetime coordinates.} Thus, as we see, we are essentially  getting local conformal invariance for free, without needing to ask for it a priori. Moreover, we can interpret $\Gamma_{\mu}(x)$ as the gauge field of local conformal invariance in exactly the same way as $A_{\mu}(x)$ acts as the gauge field of local gauge invariance, with both the real and the imaginary parts of the $-3\beta/2+i\alpha$ phase of the fermion being gauged. Now while the conformal invariance of the massless Dirac action would be lost if the fermion were to be massive, in order to maintain conformal invariance at the level of the Lagrangian we would need mass to be induced dynamically, precisely just as we have been considering in this article. With the same analysis holding for all of the fundamental  strong, electromagnetic, and weak interactions since they are all based on local gauge theories with dimensionless coupling constants, they would all be fully conformal invariant at the level of the Lagrangian if all masses are induced dynamically. We thus see that if there are no fundamental mass terms at all and if all mass is to come from vacuum breaking, conformal invariance is a quite natural invariance for physics.\footnote{In the standard $SU(2)\times U(1)$ theory of electroweak interactions, there is just one  fundamental mass scale in the Lagrangian, namely that associated with the $\mu^2\phi^2/2$ term in the Lagrangian of an elementary Higgs field. In a critical scaling theory with a dynamical Higgs field, such non-conformal-invariant terms  do not appear in the fundamental Lagrangian.} In this way we see an intimate connection between conformal symmetry and dynamical mass generation.

Further support for the relevance of conformal gravity may be found by doing a fermionic path integration of the Dirac action given in (\ref{M204}). If we set $\int D[\psi]D[\bar{\psi}]\exp(iJ_{\rm D})=\exp(iI_{\rm EFF})$, then with $J_{\rm D}$ being linear in both $\psi$ and $\bar{\psi}$ the fermion path integral can be done analytically. On doing the fermion path integration we generate an effective action $I_{\rm EFF}$ whose leading term is of the form   
\begin{eqnarray}
I_{\rm EFF}&=&\int d^4x(-g)^{1/2}C\bigg{[}\frac{1}{20}\left[R_{\mu\nu}R^{\mu\nu}-\frac{1}{3}(R^{\alpha}_{\phantom{\alpha}\alpha})^2\right]
+\frac{1}{3}G_{\mu\nu}^iG^{\mu\nu}_i+\frac{1}{3}S_{\mu\nu}^iS^{\mu\nu}_i\bigg{]},
\label{M205}
\end{eqnarray}
where $C$ is a log divergent constant, and $G^{\mu\nu}_i$ and $S^{\mu\nu}_i$ are the standard non-Abelian antisymmetric rank two  tensors associated with $A_{\mu}^i$ and $S_{\mu}^i$.\footnote{The $g_{\mu\nu}$-dependent and $A_{\mu}^i$-dependent components of $I_{\rm EFF}$ and the value of the constant $C$ may be found in \cite{tHooft2010a,tHooft2010b,tHooft2011} and the $S_{\mu}^i$-dependent  component may be found in \cite{Shapiro2002}.} We recognize the terms in (\ref{M205}) as respectively being of the form of none other than the bosonic $g_{\mu\nu}$, $A_{\mu}^i$, and $S_{\mu}^i$ sectors of the conformal gravity, vector Yang-Mills, and axial-vector Yang-Mills theories. That $I_{\rm EFF}$ must have the form that it has is because it must possess all the local symmetries that $J_{\rm D}$ itself possesses. As we see, the $\Lambda^{\lambda}_{\mu\nu}$, $V^{\lambda}_{\mu\nu}$, and $K^{\lambda}_{\mu\nu}$ contributions to the generalized $\tilde{\Gamma}_{\mu}$ lead to three distinct and independent sectors in (\ref{M205}), and despite the presence of non-Abelian terms in $\tilde{\Gamma}_{\mu}$, the resulting conformal gravitational sector expressly has no internal symmetry dependence, just as one would want. Thus save only for the gravity sector, we recognize (\ref{M204}) and (\ref{M205}) as the standard action used in fundamental physics, only now all generated geometrically. Geometry and conformal invariance thus lead us to (\ref{M204}) and (\ref{M205}), in exactly the same way as conformal invariance and gauge invariance would also do. We note that we do not induce either the Einstein-Hilbert $I_{\rm  EH}=-(1/16\pi G)\int d^4x (-g)^{1/2}R^{\alpha}_{\phantom{\alpha}\alpha}$ or the cosmological constant $I_{\Lambda}=-\int d^4x (-g)^{1/2}\Lambda$ actions as neither action is conformal invariant ($G$ and $\Lambda$ both carry dimension). Rather, we expressly induce $I_{\rm W}$, and in fact cannot avoid doing so. In \cite{Mannheim2012} we have made the case for local conformal gravity, while in \cite{tHooft2015a} 't Hooft has made the case for local conformal symmetry, with 't Hooft even noting \cite{tHooft2015b} that a conformal structure for gravity seems to be inevitable, just as we find here.

To explicitly see how the conformal structure is maintained in the presence of mass, we note that if we do introduce a mass term into the action and augment  $J_{\rm D}$ with $\delta J_{\rm D}=-\int d^4x(-g)^{1/2}M(x)\bar{\psi}(x)\psi(x)$, then $I_{\rm EFF}$ is modified to $I_{\rm EFF}+\delta I_{\rm EFF}$, with the fermion path integration yielding (ignoring internal symmetry indices) \cite{tHooft2010a,Mannheim2012}
\begin{eqnarray}
 \delta I_{\rm EFF}&=&\int d^4x (-g)^{1/2}C\bigg{[}-M^4(x)+\frac{1}{6}M^2(x) R^{\alpha}_{\phantom{\alpha}\alpha}
 \nonumber\\
&-&(\partial_{\mu}+iA_{\mu}+i\gamma^5S_{\mu})M(x)(\partial^{\mu}-iA^{\mu}-i\gamma^5S_{\mu})M(x)\bigg{]}.
\label{M206}
\end{eqnarray}
Comparing with (\ref{M159}), we see that we have naturally generated an $M(x)$ dependence that through the $M^2(x) R^{\alpha}_{\phantom{\alpha}\alpha}/6$ term is conformally coupled to the geometry, just as must be the case since in the fermion path integration the $\delta J_{\rm D}$ mass term acts in the same locally conformal invariant manner as a Yukawa-coupled scalar field.   Now while we have generated $I_{\rm EFF}$ and  $\delta I_{\rm EFF}$ by doing a path integration over the fermions, this is equivalent to doing a one fermion loop Feynman diagram calculation. On now comparing with the one-loop mean-field $I_{\rm EFF}$ given in (\ref{M62}) and (\ref{M63}), we recognize (\ref{M206}) as being equivalent to the mean-field action when we evaluate (\ref{M62}) at the energy density minimum where $m(x)=M$, as then generalized to curved space. On dressing the point fermion loop diagrams with photon exchange dressings (i.e. on additionally doing a path integration of $\exp(iJ_{\rm D})$ over $A_{\mu}$), we see that if we have critical scaling and $\gamma_{\theta}(\alpha)=-1$,  (\ref{M206}) will be dressed into (\ref{M107}), as appropriately generalized. Thus if mass is generated by a mean-field procedure in which there is a four-fermion interaction, the mean field will precisely have the form given in (\ref{M206}), as then dressed by a critical scaling that makes the four-fermion interaction be both power-counting renormalizable and conformal invariant. Below we discuss the four-fermion interaction terms in more detail. 

The very fact that fermion path integration generated the conformal gravity action in (\ref{M205}) reinforces the result of  \cite{Bender2008a,Bender2008b} that conformal gravity is ghost free and unitary. Specifically, since the relevant fermion path integration is equivalent to a one loop Feynman diagram, and since one cannot change the structure of a Hilbert space in perturbation theory, it must be the case that either $J_{\rm D}$ and $I_{\rm EFF}$ both have states  of negative norm or neither does. But the $J_{\rm D}$ action is the standard fermion sector action that is used in particle physics all the time, and it is completely free of states with negative norm. Thus conformal gravity must be free of negative norm states also,\footnote{As noted in \cite{Mannheim2016c}, to establish unitarity in a conformal gravity path integral formulation requires continuing the path integral measure into the complex plane, just as one always has to do \cite{Mannheim2015b} in non-Hermitian theories such as conformal gravity because of the need to obtain well-behaved asymptotic boundary conditions (cf. Sec. (10.4)), with such a continuation being needed  in order to enable the path integral to exist.}  and must thus be as consistent at the quantum level as the $J_{\rm D}$ action from which it can be derived.

\subsection{A Theory of Everything}

Since there is a log divergence in (\ref{M205}), and since (\ref{M205}) is obtained via a fermion loop  radiative correction, we can cancel the divergence by a counterterm. To do so we augment the Dirac action $J_{\rm D}$ with a fundamental  Yang-Mills gauge field ($I_{\rm YM}$) action and a conformal ($I_{\rm W}$) metric sector action of the form
\begin{eqnarray}
&&I_{\rm W}+I_{\rm YM}=\int d^4x(-g)^{1/2}\bigg{[}-2\alpha_g\bigg{(}R_{\mu\nu}R^{\mu\nu}
-\frac{1}{3}(R^{\alpha}_{\phantom{\alpha}\alpha})^2\bigg{)}-\frac{1}{4}G_{\mu\nu}^iG^{\mu\nu}_i-\frac{1}{4}S_{\mu\nu}^iS^{\mu\nu}_i\bigg{]}.
\label{M207}
\end{eqnarray}
Even while this action will now take care of the infinities in (\ref{M205}) and its higher order radiative corrections, there will still be uncanceled vacuum energy density contributions. To cancel them we introduce an $SU(N)\times SU(N)$ invariant four-fermion action with coupling $g_{\rm FF}$, viz. 
\begin{eqnarray}
I_{\rm FF}&=&-\int d^4x(-g)^{1/2}\frac{g_{\rm FF}}{2}\bigg{[}\bar{\psi}T^i\psi\bar{\psi}T^i\psi
+\bar{\psi}i\gamma^5T^i\psi\bar{\psi}i\gamma^5T^i\psi\bigg{]},
\label{M208}
\end{eqnarray}
and can thus we write down a fundamental action for a conformal invariant universe, viz.
\begin{eqnarray}
I_{\rm UNIV}=J_{\rm D}+I_{\rm W}+I_{\rm YM}+I_{\rm FF},
\label{M209}
\end{eqnarray}
with there being no $\bar{\psi}V_{\mu}^i\psi\bar{\psi}V^{\mu}_i\psi$ type four-fermion interactions as they have been replaced via the introduction of intermediate vector bosons, vector bosons which themselves couple conformally.
If the dynamics associated with (\ref{M209}) leads to an $I_{\rm FF}$ with dynamical dimension equal to four, the $I_{\rm UNIV}$ action will then provide a fully conformal invariant, renormalizable, and consistent action for the universe in which all mass is generated in the vacuum by dynamical symmetry breaking, with all bound states being generated by the residual interaction associated with  $I_{\rm FF}$.\footnote{While we have formally introduced $I_{\rm UNIV}$ here, doing actual mass generation calculations that would yield quark and lepton mass spectra is for the future. However, as our derivation of  $T_{\rm P}(q^2,M)$ and $T_{\rm S}(q^2,M)$ in (\ref{M117}) and (\ref{M126}) shows, in the dynamical case the pseudoscalar and scalar bound state Yukawa couplings  to fermion-antifermion pairs are determined by the dynamics and do not need to be taken as input parameters as in the elementary Higgs case. Thus in the dynamical case the free parameters associated with the quark and lepton mass matrices are potentially calculable.} As regards the four-fermion $I_{\rm FF}$ term, we note that  it makes the vacuum energy density finite when $d_{\theta}=2$, while also making the four-fermion interaction contribution to the scalar and pseudoscalar channel $T_{\rm S}(q^2,M)$ and $T_{\rm P}(q^2,M)$ scattering amplitudes be finite too, just as discussed in Secs. (8.4) and (8.8). However, there are also gauge boson and graviton contributions to scattering amplitudes, and the contributions that they make via $I_{\rm W}+I_{\rm YM}$ are also all renormalizable. Moreover, if the beta functions associated with all three of the conformal gravity, vector Yang-Mills, and axial-vector Yang-Mills actions all vanish, and if all the associated non-Abelian $d_{\theta}$ are all equal to two, not only would the full $I_{\rm UNIV}$ theory be renormalizable, the theory would even be finite and exist without renormalization \cite{Mannheim2015} [just as the Schwinger-Dyson equation and the fermion-antifermion scattering amplitudes (cf. (\ref{M73}), (\ref{M117}), (\ref{M126})) exist without renormalization in the Abelian case]. Thus for weak  interactions we make the four-fermion interaction model of weak interactions renormalizable by replacing it by a gauge theory whose gauge bosons become massive through the Higgs mechanism, and in the presence of gravity we make the vacuum energy density finite by introducing an $I_{\rm FF}$ four-fermion interaction that is made renormalizable by anomalous dimensions. There are thus two ways to control the divergences of four-fermion interactions, either replace them by a gauge theory or soften them via anomalous dimensions, and both play a role in physics. Given its content, we can identify (\ref{M209}) as a candidate theory of everything, one that is potentially finite, and one that is formulated in the four spacetime dimensions for which there is evidence.\footnote{Since string theory requires both supersymmetry and extra dimensions, if neither actually exists, then rather than being a possible theory of everything, string theory would instead be a theory of more than everything.}

\section{Comparing Conformal Symmetry and Supersymmetry}
 
It is of interest to compare conformal symmetry and supersymmetry, as they both seek to address many of the same issues. In order to compare them, we note first that unlike supersymmetry, conformal symmetry does not need to be postulated. Specifically, if we start with fermions alone and couple them to curved space via the requirement of local Lorentz invariance, we will directly be led to the $\int d^4x(-g)^{1/2}i\bar{\psi}\gamma^{c}V^{\mu}_c(x)(\partial_{\mu}+\Gamma_{\mu})\psi$ action. As noted above, this action is not just locally Lorentz invariant but is locally conformal invariant too. However, it  is not in any way supersymmetric. Thus for the $\int d^4x(-g)^{1/2}i\bar{\psi}\gamma^{c}V^{\mu}_c(x)(\partial_{\mu}+\Gamma_{\mu})\psi$ action local conformal invariance is output not input  (as is then the conformal gravity action generated in (\ref{M205}) by fermion path integration), and thus never needs to be postulated. However one cannot get from $i\bar{\psi}\gamma^{c}V^{\mu}_c(x)(\partial_{\mu}+\Gamma_{\mu})\psi$ to a supersymmetric action without postulating the existence of bosonic superpartners of the fermions. The reason why it is fermions and conformal symmetry rather than fermions, bosons and supersymmetry that is  singled out is that the 15-dimensional $O(4,2)$ conformal group is the full symmetry of the light cone on which all particles move in the absence of mass. And with the 15-dimensional $SU(2,2)$ being the covering group of the conformal group,  the fundamental representation of the conformal group is  the four-component spinor, all on its own with no bosonic superpartner being needed. Moreover, if all mass scales are to be generated dynamically, there must be an underlying conformal symmetry in the fundamental action, with it only being broken in the vacuum.

In terms of what conformal symmetry and supersymmetry aspire to do, we note that both theories seek to address the vacuum energy density problem, the cosmological constant problem, the Higgs self-energy problem, the dark matter problem, and the quantum gravity problem.  While supersymmetry can address the first three of these problems when it is unbroken, as we had noted above, it runs into difficulty once the supersymmetry is broken. While supersymmetry can provide possible dark matter candidates, it is not yet able to derive a formula such as the observationally-tested  conformal gravity formula given in  (\ref{M188}) for galactic rotation curves, or provide a formula that requires no fine tuning for the Hubble plot such as the conformal gravity one given in  (\ref{M190}). And as to quantum gravity, to construct a consistent string theory of quantum gravity, one needs not just supersymmetry, one in addition needs six extra spacetime dimensions. And then beyond all this, superparticles of course need to found, and yet to date there is no sign of any of them, either at the LHC or in underground dark matter searches. 

As we have seen in this article, for every one of these five major problems conformal symmetry appears to be doing better than supersymmetry, primarily because it is able to retain control of these problems even after the conformal symmetry is spontaneously broken. Through conformal symmetry one is able to resolve these problems without the need for any dark matter, any dark energy, or any fine tuning of parameters. In addition, through conformal symmetry one can construct a consistent, unitary, and renormalizable theory of quantum gravity without any new particles or any extra spacetime dimensions. And one can do so while making some readily testable predictions such as the extension of (\ref{M188}) and (\ref{M190}) beyond the domains in which they have currently been tested.  Determining whether the Higgs boson is elementary or dynamical could go a long way to establishing whether it is supersymmetry (as needed for the self-energy hierarchy problem of an elementary Higgs boson) or conformal symmetry  (critical scaling, anomalous dimensions, a dynamically generated Higgs boson with no hierarchy problem, and a renormalizable four-fermion interaction) that might be favored.\footnote{While critical scaling and anomalous dimensions were key components of the conformal bootstrap program that was developed in the 1970s by workers such as Migdal and Polyakov, it is of interest to note that with the current focus on conformal field theories in the literature, the conformal bootstrap program has recently come into vogue again (for a review see e.g. \cite{Simmons-Duffin2016}). Author's note: it is also of interest to note that Robert Brout and Francois Englert were interested \cite{Brout1973,Englert1974} in the conformal bootstrap program when I was a post-doc with them in Brussels in the early seventies just at the time that condensed matter renormalization group ideas were coming into particle physics, and my own interest in JBW electrodynamics and also \cite{Mannheim1975b} in the conformal bootstrap  was stimulated by my interaction with them at that time.}

\section{Summary}

In this article we have addressed the question of whether the Higgs boson is elementary or composite. Even though the standard elementary Higgs boson electroweak model works extremely well, there are nonetheless some basic concerns with the model: (i) radiative corrections to the Higgs self-energy have a high scale hierarchy problem, (ii) there is an enormous number of free parameters in the coupling of the Higgs boson to the quarks and leptons, (iii) in its spontaneously broken phase the Higgs potential provides a large contribution to the cosmological constant, (iv) as with any quantum field theory there is a zero-point problem, though this only becomes of concern when the theory is coupled to gravity, as it ultimately must be, and (v) the whole notion of an elementary Higgs boson stands in sharp contrast to the BCS theory of superconductivity or the generation of a Goldstone boson pion in strong interactions since neither of these two well-established cases involves any elementary Higgs boson at all, with the breaking being done dynamically by fermion bilinear condensates. As we have shown, all of these concerns can be addressed if the symmetry breaking is dynamical: (i) a bound state Higgs boson has no hierarchy problem, (ii) when the Higgs boson is a bound state, the same dynamics that fixes its mass as the position of the pole in a fermion-antifermion scattering amplitude also fixes its coupling to the fermions as the residue at the pole, (iii,iv) when the Higgs generation mechanism involves the four-fermion interaction then on coupling to conformal gravity one can resolve both the cosmological constant problem and the zero-point problem, and (v) with a dynamical Higgs boson electroweak and strong interaction symmetry breaking are treated equivalently. 

\section{The Moral of the Story}

With the vacuum of quantum field theory being a dynamical one,  in a sense Einstein's ether has reemerged. Only it has reemerged not as the mechanical ether of classical physics that was excluded by the Michelson-Morley experiment, but as a dynamical, quantum-field-theoretic one full of Dirac's negative energy particles, an infinite number of such particles whose dynamics can spontaneously break symmetries. The type of physics that would be taking place in this vacuum depends on how symmetries are broken, i.e. on whether the breaking is by  elementary Higgs fields or by dynamical composites. If the symmetry is broken by an elementary Higgs field, then the Higgs boson gives mass to fundamental gauge bosons and fermions alike. However, if the breaking is done dynamically, then it is the structure of an ordered vacuum itself that generates masses, with the mass generation mechanism in turn then producing the Higgs boson. In the dynamical case then mass produces Higgs rather than Higgs produces mass. In the dynamical case we should not be thinking of the Higgs boson as being the ``god particle". Rather, if anything, we should be thinking of the  vacuum as being the ``god vacuum".

\section{Acknowledgments}
The author wishes to thank Michael Mannheim, Brendan Pratt, and Candost Akkaya for help in preparing the figures, and Drs. Joshua Erlich, Valery P. Gusynin, and Juha Javanainen for comments.

\end{document}